\documentclass[12pt,3p]{elsarticle}
\biboptions{sort&compress}
\usepackage[utf8]{inputenc}
\usepackage[usenames]{color}
\usepackage{xcolor}

\usepackage[colorlinks=true,citecolor=blue,linkcolor=blue,urlcolor=blue]{hyperref}
\usepackage{graphicx}
\usepackage{braket}
\usepackage{amsfonts}
\usepackage{amsmath}
\usepackage{import}
\usepackage{subfiles}

\newcommand{\eref}[1]{Eq.~(\ref{#1})} 
\newcommand{\fref}[1]{Fig.~\ref{#1}}

\usepackage[backgroundcolor=white,textsize=tiny]{todonotes}

\begin{document}

\title{Persistent currents in ultracold gases}
\author[1]{J. Polo\corref{cor1}}
\ead{juan.polo@tii.ae }

\affiliation[1]{organization={Quantum Research Center, Technology Innovation Institute}, addressline={Abu Dhabi},
postcode={P.O. Box 9639}, country={UAE}}
\author[1]{W.J. Chetcuti}

\author[1]{T. Haug}

\author[4]{A. Minguzzi\corref{cor1}}
\ead{anna.minguzzi@lpmmc.cnrs.fr }
\affiliation[4]{organization={Universite Grenoble-Alpes}, addressline={CNRS, LPMMC},city={Grenoble},
postcode={38000}, country={France}}
\author[5]{K. Wright}
\affiliation[5]{organization={Department of Physics and Astronomy, Dartmouth College}, addressline={6127 Wilder Laboratory},city={Hannover},
postcode={NH 03766}, country={USA}}
\author[1,6,7]{L. Amico}
\affiliation[6]{organization={Dipartimento di Fisica e Astronomia `Ettore Majorana', Universita di Catania},  country={Italy}}
\affiliation[7]{organization={INFN-Sezione di Catania}, addressline={Via S. Sofia 64},city={Catania},postcode={95127}, country={Italy}}

\begin{abstract}
Persistent currents flowing in spatially closed tracks define one of the most iconic concepts in mesoscopic physics. They have been studied in solid-state platforms such as superfluids, superconductors and metals. Cold atoms trapped in magneto-optical toroidal circuits and driven by suitable artificial gauge fields allow us to study persistent currents with unprecedented control and flexibility of the system's physical conditions. Here, we review persistent currents of ultracold matter. Capitalizing on the remarkable progress in driving different atomic species to quantum degeneracy, persistent currents of single or multicomponent bosons/fermions, and their mixtures can be addressed within the present experimental know-how.  This way, fundamental concepts of quantum science and many-body physics, like macroscopic quantum coherence, solitons, vortex dynamics, fermionic pairing and BEC-BCS crossover can be studied from a novel perspective. Finally, we discuss how persistent currents can form the basis of new technological applications like matter-wave gyroscopes and interferometers.        
\end{abstract}


\cortext[cor1]{Corresponding authors}

\maketitle
\tableofcontents

\section{Introduction}\label{sec:Intro}

Persistent currents are permanent flows of quantum matter occurring in annular geometries pierced by an effective magnetic field. The existence of persistent currents was discovered in superconducting circuits at the turn of the 20$^{\mathrm{th}}$ century~\cite{onsager1961magnetic,byers1961theoretical,bloch1965off,deaver1961experimental,doll1961experimental}. Such a notion represented a milestone for the development of the concept of  superfluidity~\cite{reppy1964persistent,gammel1984persistent,pekola1985persistent,leggett1999superfluidity}. Rather than zero viscosity or zero resistance, the persistent current arises from the  system being coherent beyond of a microscopic length scale ~\cite{yang1962concept,anderson1966considerations}. Indeed,  although smaller than the superconducting/superfluid counterparts,  dissipationless electric currents has been  observed also in normal metallic settings  as  the electrons' coherence length gets comparable with the characteristic size of the ring~\cite{buttiker1983josephson,buttiker1984quantum,webb1985observation,saminadayar2004equilibrium,zvyagin1995persistent,bouchiat1991persistent,bouchiat1991persistentb,szopa1991persistent,riedel1991persistent,imry2002intro,ambegaokar1990coherence}.   As a function of the threaded magnetic flux,  specific Aharonov-Bohm oscillations can emerge~\cite{aharonov1959significance,gefen1984quantum,bouchiat1991persistent,buttiker1983josephson,webb1985observation,buttiker1984quantum}.  Persistent currents can only be observed in the quantum regime at very low temperatures where decoherence effects coming for example from disorder and thermal fluctuations are  negligible - see for instance~\cite{mohanty1999persistent,chakraborty1994electron,entin2006persistent,matveev2002persistent,riedel1993mesoscopic,cheung1989persistent,kulik2010persistent,yerin2021genesis}. As a result, detecting persistent currents in normal metals has turned out to be quite challenging~\cite{levy1990magnet,mohanty1999persistent,bluhm2009persistent,bleszynski2009persistent}.

Soon after the first experiments on bosonic ultracold atoms~\cite{anderson1995observation,davis1995bose-einstein,cornell2002nobel,ketterle2002nobel}, attention was devoted to their superfluid features~\cite{pethick2008bose,pitaevskii2016bose,ginzburg2004nobel,leggett2004nobel,abrikosov2004nobel, leggett2008quantum}. Despite being charge-neutral in nature, ultracold atoms can be set in motion through the application of artificial gauge fields that can be realized with different approaches~\cite{dalibard2011artificial,goldman2014light}, ranging from  stirring  that exploits the equivalence between the Lorentz and Coriolis forces~\cite{fetter2009rotating,dalibard2016introduction,wright2013driving,cai2022persistent} to suitable Raman transitions~\cite{andersen2006quantized,ramanathan2011superflow,moulder2012quantized,ryu2007observation} and phase imprinting techniques employing nowadays commercially available devices for light sculpting such as Spatial Light Modulators (SLM) or Digital Micromirror Devices (DMD)  ~\cite{kumar2018producing,del_pace2022imprinting,franke2008advances,gauthier2021dynamic,gauthier2019quantitative,rubinsztein2016roadmap,henderson2009experimental}. 
Ultracold atoms provide a platform to realize persistent currents with  very specific capabilities that are hardly, if not impossible, to be  accessed by standard schemes, as  solid-state physics implementations. To start with, ultracold atoms have paradigmatically robust coherence properties and control of the physical conditions~\cite{bloch2008quantum,gardiner2017quantum,lewenstein2012ultracold}. Sophisticated high-resolution image systems allow to monitor the system with a remarkably high precision - see for example~\cite{bakr2009quantum,sherson2010single,Endres2016atom,bergschneider2018spin,qian2021super,kwon2020strongly,del_pace2022imprinting}. Dynamics of ultracold matter can be studied to an unprecedented  degree since the present experimental know-how allows to change the system's physical conditions during the course of the experiment `on the fly'~\cite{gauthier2021dynamic,barredo2018synthetic,gaunt2012robust}. Furthermore, the wide range of atomic species that can be used in cold atom experiments allows to study persistent current with a large variety of different quantum fluids, as $N$-component bosons/fermions or  mixtures thereof~\cite{scazza2014observation,taie2022observation,takahashi2022quantum,myatt1997production,hall1998measurements,schreck2001quasipure,roati2002fermi,lahaye2009physics}. 
Indeed,  atomic persistent currents together with their quantization properties have been addressed directly in a number of experiments.
Specifically, the earliest experimental studies of matter-wave currents in ultracold systems involved weakly interacting bosonic atoms~\cite{ryu2007observation,ramanathan2011superflow,moulder2012quantized,wright2013driving,neely2013characteristics,pandey2019hyper,wolf2021stationary}, but work in this area recently expanded to include fermionic-pair superfluids with tunable interactions~\cite{cai2022persistent,del_pace2022imprinting,pecci2021probing,pecci2022coherence}.  Persistent currents of $N$-component gases~\cite{white20161emergence,chetcuti2021persistent,chetcuti2023probe,chetcuti2022interference,osterloh2023exact,consiglio2022variational,chetcuti2023persistent,pecci2023persistent,ferraretto2023enhancement,richaud2021interaction,richaud2022mimicking,wu2013mean,smyrnakis2014persistent,polo2024static}, i.e.\ atoms with $N$ internal degrees of freedom, and of Bose-Bose~\cite{matsushita2018mixture,spehner2021persistent,brauneis2023emergence} or Bose-Fermi mixtures~\cite{suga2014persistent} have been investigated theoretically. Close networks of Rydberg atoms provide a conceptual extension of the field and have attracted recent attention~\cite{Endres2016atom,schymik2020enhanced,barredo2015coherent,barredo2018synthetic}. In such proposals, rather than of matter, a controlled current can occur in terms of atomic excitations~\cite{perciavalle2022controlled,perciavalle2024quantum,perciavalle2023coherent,lienhard2020realization,wu2022manipulating,li2022coherent,bornet2024enhancing,han2024tuning}.

These facts, together with the aforementioned enhanced features of system's control and flexibility, open the way to current-based simulators in which persistent currents of ultracold atoms can be harnessed as a general tool for studying many-body systems~\cite{polo2018damping,polo2019oscillations,cai2022persistent,del_pace2022imprinting,pecci2021probing,chetcuti2021persistent,chetcuti2023probe,richaud2022mimicking,kohn2020superfluid,domanti2023coherence,tengstrand2021persistent,perezobiol2022coherent,haug2019aharonov,haug2019andreev,haug2018mesoscopic,cherny2009decay}. Indeed, a persistent  current produced in response to an applied flux is sensitive to the presence of extended and/or 
bound states ~\cite{byers1961theoretical,naldesi2019rise,naldesi2022enhancing,pecci2021probing,chetcuti2023probe}, and highlight the role of impurities \cite{cominotti2014optimal,aghamalyan2015coherent,haug2018readout,polo2022quantum}.   In the context of quantum simulation, we note that the periodic boundary conditions, smoothly implementing the translational symmetry,  can substantially help  the convergence of the results in a finite-size physical system
to the thermodynamical limit one~\cite{fisher1972scaling}.

Finally,  persistent currents of neutral matter-waves convey a technological significance. In particular they provide a key ingredient  of Atomtronics, the  quantum technology of ultracold atoms propagating  in magneto-optical guides~\cite{amico2017focus,amico2021roadmap,amico2022atomtronic,pepino2021entropy,polo2024perspective}. 
Indeed, ring-shaped degenerate gases interrupted by few localized weak links define the  atomic analogs of  Superconducting Quantum Intereference Devices (SQUIDs) of quantum electronics~\cite{tinkham2004introduction}. The ultracold atoms counterpart of SQUIDs  have been experimentally realized~\cite{ryu2013experimental,aghamalyan2015coherent,haug2018readout,ryu2020quantum,eckel2014interferometric}. In virtue of the notoriously mild decoherence properties of such  platforms,  central questions for quantum science such as the  macroscopic quantum tunneling/coherence~\cite{leggett1987macroscopic}, have been studied intensively~\cite{hallwood2006macroscopic,hallwood2010robust,amico2014superfluid,solenov2010macroscopic,nunnenkamp2011superposition,schenke2011nonadiabatic,aghamalyan2015coherent,aghamalyan2016atomtronic,haug2018readout,perezobiol2022coherent}. 
At the same time, atomic SQUIDs are very promising in quantum sensing~\cite{degen2017quantum}  as compact   Sagnac interferometers~\cite{barrett2014sagnac,olson2007cold-atom, helm2015sagnac,kim2022one} or  gyroscopes~\cite{kasevich1997precision, krzyzanowska2022matter,burke2009scalable,moan2020quantum,qi2017magnetically,wu2007demonstration,kim2022one,beydler2024guided}.

The review is organized as follows. In Sec.~\ref{sec:Methods}, we  discuss how persistent currents can arise in mesoscopic cold atom rings and address few relevant features on the concept of superfluidity and phase coherence.  Sec.~\ref{sec:framework} introduces the theoretical framework  used to understand persistent currents in bosonic and fermionic systems. The ensuing sections (Secs.~\ref{sec:PerCurrBose} and~\ref{sec:PerCurrFerm}) address persistent currents in bosonic and fermionic systems, respectively. In Sec.~\ref{sec:atomtronics}, we discuss how persistent currents can set the basis for relevant atomtronics applications in  quantum technologies such as rotation sensing, persistent current-based qubits and controlled manipulations of Rydberg atoms. Finally, we conclude our review in Sec.~\ref{sec:Conc} by presenting few remarks  and expounding on the future prospects for the field.

\section{Concept of persistent currents in cold atoms}\label{sec:Methods}

Persistent currents are a manifestation of quantum coherence. On the operational side, we note that such phenomenon is achieved by promoting the ground state of the system at rest to  an excited  state in which particles circulate in the toroidal track. As such, persistent currents should be understood as metastable states with a very long life time~\cite{leggett2008quantum,leggett1999superfluidity}. This important concept relies  on having a finite critical velocity in the system. In particular, for the ideal Bose gas that has a zero critical velocity, the dissipation-free current decays once the synthetic gauge field is removed. Indeed, the persistent current can be expressed as a thermodynamic equilibrium function only in a frame in which the synthetic field is at rest - see Sec.~\ref{sec:PerCurrDev}. For  persistent currents achieved by stirring, for example, the latter  is provided  by the frame rotating with the stirring laser potential - see Sec.~\ref{sec:ColdAtomsMotion}. Cold atoms settings have allowed to  highlight the stability features of persistent currents  with  remarkable insights- see  Sec.~\ref{VortexDynamicsBose}. 

In this section, we provide the main ingredients of the persistent current phenomena in cold atoms. Relations with superfluidity and phase coherence give important insights on the concept of persistent currents. Since the current is comprised of electrically neutral particles, specific schemes for applying effective magnetic fields have been experimentally realized. Persistent currents are nowadays accessible in several experimental laboratories through suitable interference protocols.

\subsection{Persistent currents in ultracold atoms}\label{sec:PerCurrDev}

The purpose of this section is to demonstrate that the phenomenon of persistent currents arises as a manifestation of the phase of the wavefunction. Here, we illustrate how subjecting cold atoms confined in mesoscopic rings to a synthetic gauge field triggers quantized matter-wave currents. Our approach follows the logic carried out in~\cite{viefers2004quantum} adapting it to neutral atoms. 

\begin{figure}[h!]
\includegraphics[width=0.49\textwidth]{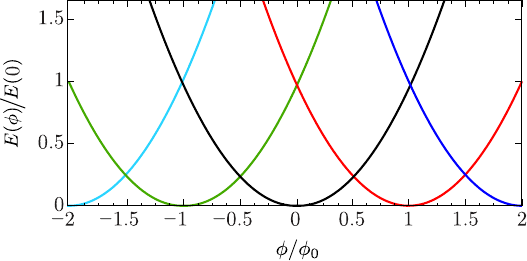}
\includegraphics[width=0.49\textwidth]{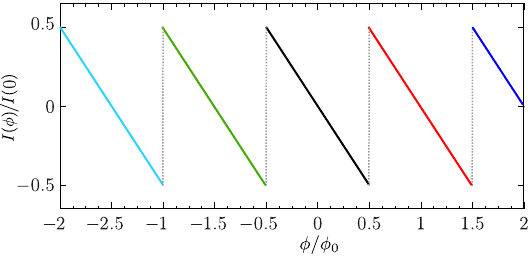}
\centering
\caption{Single particle energy spectrum $\epsilon_{n} (\phi)$ and the corresponding persistent current $I_{n}(\phi)$ as a function of the flux $\phi/\phi_{0}$. Figure (\textbf{a}) depicts the first $n=5$ single particle eigenenergies for a ring of size $L=20$. The ground-state persistent current in the system corresponding to the first four energy level crossings is depicted in Figure (\textbf{b}). The discontinuous jumps in the current (dotted lines) manifest due to the energy levels crossings exhibited by the system at $\phi = \frac{j}{2}\phi_{0}$ for integer $j$ represented by crosses. It can be clearly seen that both the energy spectrum and the current are periodic in $\phi_{0}$ as discussed in the text.}
\label{fig:Single_particle_spectrum_current}
\end{figure}

In the presence of a magnetic field $\mathbf{B}$, the canonical  momentum of the electrons  is modified due to its coupling with the magnetic vector potential $\mathbf{A}$ through their charge $q$, with $\mathbf{B} = \nabla\times\mathbf{A}$. Employing the Lagrangian formalism the canonical momentum can be demonstrated to be   $\textbf{P} = \textbf{p} +q\mathbf{A}$, where $\textbf{p}=m \textbf v$ is the kinetic momentum. The corresponding Hamiltonian expressed in terms of canonical momenta and coordinates reads 
\begin{equation}\label{eq:ClassicHamField}
    \mathcal{H}_{\mathbf{B}\mathrm{-field}} = \frac{1}{2m}[ \textbf{P} - q\mathbf{A}(\mathbf{r})]^{2}  +q\Phi(\mathbf{r}), 
\end{equation}
where $\mathbf{P}=-\imath\hbar\nabla$ and  $\Phi (\mathbf{r})$ is the scalar potential that can be taken to be identically zero in the radiative gauge \cite{landau2013classical}. Naturally, the kinetic momentum of ultracold neutral atoms is not affected by external magnetic fields. However, it is possible to engineer synthetic gauge fields through various means \cite{lin2009synthetic,dalibard2011artificial,goldman2014light} such that the behavior of charged particles subjected to Lorentz forces in presence of a magnetic field is emulated in a cold atom system, leading to a Hamiltonian bearing a similar form to \eref{eq:ClassicHamField}
\begin{equation}\label{eq:ClassicHamFieldartifical}
    \mathcal{H}_{\mathrm{artificial}} = \frac{1}{2m}[ \textbf{P} - \chi\mathbf{A}(\mathbf{r})]^{2}.
\end{equation}
The specific form of $\chi$ will depend on the scheme utilized to generate the synthetic gauge field, some of which will be discussed in Sec.~\ref{sec:ColdAtomsMotion}. To understand the implications of the artificial gauge field, we first revisit the time-independent Schrödinger equation for a single quantum particle having a mass $m$ residing in a one-dimensional ring of radius $R$. In this geometry, $\nabla = \frac{1}{R}\frac{\partial}{\partial\varphi}$
where $\varphi$ corresponds to the azimuthal polar coordinate.
\begin{figure}[h!]
\includegraphics[width=0.3\textwidth]{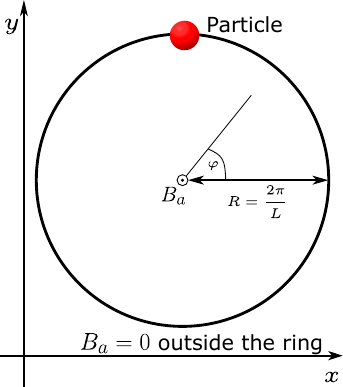}
\centering
\caption{Schematic of a single particle of mass $m$ confined in a ring of radius $R$ enclosing an artificial magnetic field $\mathbf{B}_a$ coming out of the page. The corresponding magnetic vector potential $\mathbf{A}$ is taken to lie in the $x$-$y$ plane such that it is along the ring parallel to the azimuthal angle $\varphi$. In this manner, the field is zero at the ring's radius setting the stage to define an instance of the Aharonov-Bohm effect.}
\label{fig:schem}
\end{figure}
Introducing an artificial magnetic field along the $z$ direction $\mathbf{B}_a=B_a\mathbf{z}$, the Schrödinger equation now reads
\begin{equation}\label{eq:TDS_flux}
    \frac{1}{2m}\bigg(\frac{-\imath\hbar}{R}\frac{\partial }{\partial\varphi} - \chi A_{\varphi} \bigg)^{2}\psi_{n}(\varphi) = \epsilon_{n}\psi_{n}(\varphi),
\end{equation}
where we expressed the vector potential as  $\mathbf{A} = \frac{1}{2}\mathbf{B}_{a}\times\mathbf{r}=A_{\varphi}\mathbf{\hat{\varphi}}$, where  $A_{\varphi}$ is determined  by imposing the circular integral $R\oint A_{\varphi}\mathrm{d}\varphi= \phi$, with $\phi$ being the magnetic flux threading the ring:  $\phi = \pi R^{2} B_{a}$. The field is chosen such that it is null at the ring's radius, as indicated in Fig.~\ref{fig:schem}. We remark that even if the field extends beyond its parameter, the effects will only depend on the flux piercing the ring. The eigenfunctions $\psi_{n} = \frac{1}{\sqrt{2\pi R}} e^{\imath n \varphi}$ satisfying \eref{eq:TDS_flux} are plane waves where $n$ is an integer fixed by imposing periodic boundary conditions on the wavefunction, $\psi(\varphi) = \psi(\varphi + 2\pi)$, and associated with an angular momentum $n\hbar$. The corresponding energies are:
\begin{equation}\label{eq:energy_flux}
    \epsilon_{n}(\phi) = \frac{\hbar^{2}}{2mR^{2}}\bigg( n-\frac{\phi}{\phi_{0}}\bigg)^{2},
\end{equation}
where we defined the elementary flux quantum as $\phi_{0}=\frac{h}{\chi}$. Plotting the spectrum against the flux (see Fig.~\ref{fig:Single_particle_spectrum_current}(\textbf{a})), one finds that 
$\epsilon_{n}$ is is piece-wise parabolic, with a symmetry $\phi\leftrightarrow -\phi$. Therefore, $\epsilon_{n}$ turns out to be a periodic function of $\phi$ with period $\phi_{0}$. Specifically, there are crossings between levels of different $n$ at $\phi=\frac{j}{2}\phi_{0}$ or $\phi=j\phi_{0}$ for integer $j$ depending on the particle number parity: as the flux piercing the ring increases, there is a shift in the angular momentum of the circulating particles counteracting it, causing these intersections.

In what follows, we make the connection with the Aharonov-Bohm effect, which  generally results from a specific quantum interference experienced by particles propagating in a solenoidal magnetic field~\cite{aharonov1959significance,ehrenberg1949refractive} (for the experimental demonstration of the effect see~\cite{chambers1960shift,tonomura1986evidence}). To this end, we perform a gauge transformation that, as we shall see, removes the vector potential from the Hamiltonian and incorporates a phase factor in the wavefunction. This is implemented by rotating the Hamiltonian in Eq.~\eqref{eq:TDS_flux} using $\mathcal{U}K\mathcal{U}^{-1}\mathcal{U}K\mathcal{U}^{-1}\mathcal{U}\psi = -\frac{2m}{\hbar^2}\epsilon_n \; \mathcal{U}\psi$ where 
$K= \left(\frac{1}{R}\frac{\partial }{\partial\varphi} - \frac{\imath\chi}{\hbar } A_{\varphi} \right)$
and  $\mathcal{U}=e^{-\frac{\imath}{\hbar} \chi R\int A \mathrm{d}\varphi}$: 
\begin{equation}\label{eq:nullify}
    e^{-\frac{\imath}{\hbar} \chi R\int A d\varphi}\left(\frac{1}{R}\frac{\partial }{\partial\varphi} - \frac{\imath\chi}{\hbar } A_{\varphi} \right)e^{\frac{\imath}{\hbar} \chi R\int A d\varphi} = \frac{1}{R}\frac{\partial}{\partial\varphi}.
\end{equation}
Correspondingly, the wavefunction acquires a phase factor $
\psi \rightarrow \psi'=e^{-\frac{i}{\hbar} \chi R\int A \mathrm{d}\varphi}\psi
$, which coincides with the Aharonov-Bohm phase. Additionally, such a phase also modifies the boundary conditions of the system \begin{equation}
\label{BlochWavefunction}
\psi'(\varphi+2\pi)=e^{-\frac{i}{\hbar}\chi R\oint A \mathrm{d}\varphi}\psi'(\varphi) \,. \end{equation} The response of the system to the change in the boundary conditions was studied by Kohn~\cite{kohn1964theory} and further discussed by Thouless and co-workers~\cite{thouless1977maximum,edwards1972numerical}. We note that making a correspondence between the crystal  momenta and the effective magnetic flux $\phi$, the expression of the wavefunction Eq.~\eqref{BlochWavefunction} makes the basis for the formal analogy between particles in ring pierced by a magnetic field and the Bloch theory describing particles in periodic potentials~\cite{ashcroft1976solid}. In this sense, as originally noted  by Büttiker, Imry and Landauer~\cite{buttiker1983josephson,imry2002intro}, and discussed by Leggett~\cite{leggett1991theorem}, $\epsilon(\phi)$ provides the `band structure' for the problem -- See also Sec.~\ref{sec:bloch} 

To obtain the current $I(\phi) = -\frac{1}{L}\int^{L}_{0}j_{n}(x,\phi)\mathrm{d}x$ we refer to the probability current density:
\begin{equation}\label{eq:ProbCurr}
    j_{n}(x,\phi) = \frac{\imath\hbar}{2m}\bigg[[\psi_{n}^{*}(\varphi)K\psi_{n}(\varphi) - \psi_{n}(\varphi)K^{*}\psi^{*}_{n}(\varphi)\bigg] = -\frac{h}{mL^2} \bigg(n-\frac{\phi}{\phi_{0}}\bigg),
\end{equation}
which results in:
\begin{equation}
\label{eq:current7}
    I(\phi) =  \frac{ h}{mL^{2}}\bigg(n-\frac{\phi}{\phi_{0}}\bigg).
\end{equation}
We note that Eq.~\eqref{eq:current7} can also be obtained through the derivative of the energy Eq.~\eqref{eq:energy_flux}:
\begin{equation}\label{eq:CurrentFromEnergy}
    I(\phi) = -\frac{1}{h}\frac{\partial \epsilon_{n} (\phi)}{\partial\tilde{\phi}}, 
\end{equation}
where $\tilde{\phi}=\phi/\phi_{0}$. A relation similar to Eq.~\eqref{eq:CurrentFromEnergy} holds for finite temperatures. Specifically, at finite temperatures, the current emerges from the  occupation of the eigenstates given by their thermal distribution and thus can be obtained from the the Helmholtz free energy $F$ as
$I_{th}(\phi) = -\frac{1}{h}\frac{\partial F (\phi)}{\partial\tilde{\phi}}$. Clearly, at zero temperature, only the ground-state level of the system is occupied such that the expression for the current reads 
\begin{equation}\label{eq:ZeroTempCurrDef}
    I(\phi) = -\frac{1}{h}\frac{\partial E_{0} (\phi)}{\partial\tilde{\phi}}, 
\end{equation}
where $E_{0}$ denotes the ground-state energy of the system and $I$ is the corresponding circulating current.

In summary, {\it persistent currents arise from the coupling between the effective magnetic field and the phase of the wavefunction}. This coupling leads to a sizable dependence of the system's energy on $\phi$, for example quantified by the $\phi$-bandwidth. Moreover, {\it in the frame in which the synthetic field is at rest}, the persistent current is an equilibrium quantity, thus, it never decays as long as $\phi$ is present. In real systems, it should be clear that a persistent current can emerge only if the phase can be preserved during the particles' motion on a characteristic length scale~\cite{thouless1977maximum,imry1986physics,leggett1991dephasing}. Such a property, defining the notion of phase coherence,  will be further elaborated in the next section. In particular, here we note that for ultracold atoms the persistent current can be started also by a suitable engineering of the phase of the `macroscopic' wavefunction (see also Sec.~\ref{sec:ColdAtomsMotion}).

Although the theory  discussed above refers to the single-particle case, the  significance and most of general aspects of the persistent current we worked out can be applied to interacting many-body systems. For a system of $N_{p}$ particles $I(\phi) = \sum_{\{n\}} \frac{h}{mL^{2}} \big(n - \frac{\phi}{\phi_{0}}\big)$,
which is obtained by adding the different contributions of the single-particle currents over the particle distribution $\{n\}$. The corresponding angular momentum per particle is then defined as $\ell = \sum_{\{n\}}n/N_{p}$, which is also commonly referred to in experiments as the \textit{winding number}. Additionally,  we note that the flux dependence of the system is encoded solely in the kinetic term. This reflects the notion that through a linear combination of Slater determinants, one can obtain any circulating current state~\cite{viefers2004quantum}. 

As in the  single-particle case, the ground-state persistent current as a function of the flux threading the system gives a characteristic sawtooth shape with a period fixed by the elementary flux quantum --Figure~\ref{fig:Single_particle_spectrum_current}(\textbf{b}). Discontinuous jumps in the current indicate the intersection of two energy parabolas with different angular momenta $\ell$, corresponding to a superposition of these states. This fact should be kept in mind when utilizing the analytical expression, as $\{n\}$ should be changed with increasing flux to get the configuration with the lowest energy, giving the sawtooth current landscape. In the presence of many-body interactions, the persistent current profile, i.e.\ its magnitude, parity, and periodicity, displays specific dependencies reflecting important aspects of the system. As we will see in Secs.~\ref{sec:PerCurrBose} and~\ref{sec:PerCurrFerm}, such a feature makes persistent currents an invaluable diagnostic tool to investigate many-body systems.

The occurrence of persistent currents is not limited to rings in the continuous limit, but can also be extended to lattice systems as well. Performing lattice regularization on the general many-body Hamiltonian (see Eq.~\eqref{eq:GenHam} in Sec.~\ref{sec:ModelsBose}) for a system with $N_{s}$ lattice sites, by expanding the field operators in terms of the Wannier functions $\Psi_{\alpha}(\mathbf{r}) = \sum_{j}w(\mathbf{r}-\mathbf{r}_{j})a_{j,\alpha}$, we get the tight-binding Hamiltonian\footnote{We restricted to the one-band model assuming that the interactions between the bands are weak and to the tight-binding approximation suppressing hoppings beyond nearest-neighbours.} $\mathcal{H}_{TB} = -\tilde{J}\sum_{\alpha}\sum^{N_{s}}_{j}(a^{\dagger}_{j,\alpha}a_{j+1,\alpha}+ \mathrm{h.c.})$, where the hopping amplitude denoted by $\tilde{J}$ encompasses the flux dependence of the system such that $\tilde{J} = \int w^{*}(\mathbf{r}-\mathbf{r}_{i})\big[\frac{1}{2m}(-\imath\hbar\nabla (\mathbf{r}) + \chi\mathbf{A}(\mathbf{r}))^{2} + V_{\mathrm{ext}} \big]w(\mathbf{r}-\mathbf{r}_{i+1})\mathrm{d}\mathbf{r}$. Going a step further, the gauge field can be incorporated into the Wannier functions allowing for the vector potential to be sufficiently smooth on the atomic scale by introducing the transformation $\tilde{w}(\mathbf{r}-\mathbf{r}_{j}) = e^{-\imath\Phi(\mathbf{r})}w(\mathbf{r}-\mathbf{r}_{j})\approx e^{-\imath\Phi(\mathbf{r}_{j})}w(\mathbf{r}-\mathbf{r}_{j})$ where $\Phi(\mathbf{r}) = \int_{\mathbf{r}_{j}}^{\mathbf{r}} \mathbf{A}(\mathbf{r})\mathrm{d}\mathbf{r}$ with $\mathbf{r}_{j}$ corresponding to the coordinate of lattice site $j$. Consequently, applying this to our model with nearest-neighbour hoppings, the tunnelling parameter acquires a complex phase $\tilde{J}=Je^{\imath\Phi}$ recasting the Hamiltonian into the following form 
\begin{equation}\label{eq:fluxtightbindingHam}
    \mathcal{H}_{TB} = -J\sum\limits^{N_{s}}_{j=1}(a^{\dagger}_{j}a_{j+1}e^{\frac{2\imath\pi}{N_{s}}\frac{\phi}{\phi_{0}}}+ \mathrm{h.c.}).
\end{equation}
The phase factor is called the Peierls phase factor, with the technique employed to get it being known as the Peierls substitution~\cite{peierls1933zur,essler2005one}. For lattice models, the persistent current can be evaluated through the current operator
\begin{equation}\label{eq:CurrOperator}
    \mathcal{I}(\phi) = \frac{2\imath\pi J}{h N_{s}}\sum\limits_{j=1}^{N_{s}}( a_{j}^{\dagger}a_{j+1}e^{\frac{2\imath\pi}{N_{s}}\frac{\phi}{\phi_{0}}} - \mathrm{h.c.}),
\end{equation}
which is readily obtained through the Hellmann-Feynmann theorem.\footnote{$\mathrm{d}E/\mathrm{d}\phi = \langle \Psi (\phi)|\mathrm{d}\mathcal{H}/\mathrm{d}\phi|\Psi (\phi)\rangle$} As we mentioned previously, persistent currents define an instance of the Aharonov-Bohm effect in a closed loop. Such a statement implies that matter-wave currents originate on account of the large coherence length of the circulating particles with respect to the system's size, i.e.\ a pure mesoscopic feature~\cite{imry2002intro} (accounted for through the $\frac{1}{N_{s}}$ correction).  Furthermore, we point out that the mechanism behind persistent currents in lattice systems, i.e.\ coherent tunneling, is distinct to that of the continuous regime (hydrodynamics mechanism). Nonetheless, it is known \textit{a posteriori} that these two mechanisms coincide since in the dilute regime of vanishing lattice spacing, the lattice model tends to its continuous counterpart (see Sec.~\ref{sec:FermHubbHam}).

\subsection{Phase coherence, superfluidity and persistent currents}\label{sec:TheoryPhase}

In this section, we introduce fundamental notions in  degenerate gases, such as as phase coherence, condensed and superfluid fractions. Although distinct, they can all lead to quantized dissipationless matter flow. 

Phase coherence is the ability of the quantum wavefunction to retain its phase information.   While it is an obvious  property to address for single or non-interacting particles, the concept is more subtle for systems of interacting particles~\cite{leggett1991concept,leggett2008quantum}. 

The concept of phase coherence can be discussed phenomenologically through a `macroscopic' wavefunction describing all particles condensed in a single quantum state $\psi_c(x)=|\psi(x)| e^{i\theta(x)}$. This picture is relevant  either to  bosonic condensates or to  superconductors at zero temperature.  In this case, the expression of the current density in Eq.~\eqref{eq:ProbCurr} provides the superfluid velocity $v=j(x,\phi)/|\psi(x)|^2$: 
\begin{equation}
v_s=\frac{\hbar}{m}\left(\nabla \theta-\frac{\chi}{\hbar} A\right)
\end{equation}
and, in turn, a persistent current. We  note that the persistent current in systems with a macroscopic phase coherence can also arise because of a suitable  phase gradient  $\nabla \theta(x)$ of the condensate wavefunction. Such a feature is exploited in ultracold atoms set up through the protocol of phase imprinting (see Sec.~\ref{sec:ColdAtomsMotion}).

In many-body settings, phase coherence can be quantified  in terms of one- and two-body density matrices \cite{yang1962concept}: 
\begin{equation}
\rho_1(x,y)=\langle  \Psi^\dagger(x)  \Psi(y)\rangle \; , 
\end{equation}
\begin{equation}
\rho_2(x,x',y,y')=\langle  \Psi^\dagger(x)  \Psi^\dagger(x')  \hat \Psi(y) \hat \Psi(y')\rangle ,
\end{equation}
where $\Psi^\dagger (x)$ ($\Psi(x)$) are field operators creating (annihilating) a boson or a fermion at position $x$ -  while we omit vector notations, the arguments in this section apply to any spatial dimensions.\footnote{All the above relations are readily generalized to lattice systems, by considering the discretized versions of $\rho_2$
according to $\rho_2(j,l,m,n)=\langle \hat c_{j,\uparrow}^\dagger  \hat c_{l,\downarrow}^\dagger \hat c_{m,\downarrow} \hat c_{n,\uparrow}\rangle$.} The eigenvectors of $\rho_1(x,y)$ define the {\it natural orbitals} and the eigenvalues fix their population. Off-Diagonal Long-Range Order (ODLRO) in bosonic systems occurs if one of the eigenvalues is $\lambda_1=O(N)$.  In this case, the largest eigenvalue of $\rho_2(x,x',y,y')$ is $O(N^2)$. Instead for fermionic system, ODLRO can occur without a macroscopic occupation of the one-particle density matrix orbital and it is reflected in an $O(N)$  eigenvalue of  $\rho_2(x,x',y,y')$. Such a difference reflects the fundamental feature that bosonic systems can indeed  Bose-Einstein condense and be described by the  condensate  wave-function  $\psi_c(x)$ above. As we discuss further below, corresponds to the mostly represented  natural orbital of $\rho_1$; the  phase $\phi$ coincides with the single-particle phase. In the case of fermionic systems, the Pauli principle prevents the occupancy of a single quantum level by particles with the same spin, and therefore the many-body coherence needs to be achieved from the single particles' coherence through higher order correlations \cite{leggett2008quantum}. In finite systems  or in the presence of interactions,  quantum fluctuations can reduce the scaling from $\lambda_1(N)$ and $\lambda_2(N)$ to $O(N^\alpha)$ with $0<\alpha<1$. 

Recently, interesting features on the interplay between single particle and interacting many-body coherences in a Fermi gas were demonstrated to occur in the persistent current~\cite{pecci2022single}. With reference to the read-out protocols discussed in Sec.~\ref{sec:Interference}, the intermediate-time interference can reflect single particle phase coherence, while many-body quantum coherence and long-range order, emerge at long interference times.  Remarkably, the off diagonal long-range order displays  a stepwise dependence  on the external effective magnetic field.  

The condensate fraction is defined as $f_c=N_c/N$, where $N_c$ is the effective  population of the mostly represented natural orbital of $\rho_1$.  Notably, the   superfluid fraction with $f_s$ defines the portion of the system with mass $M_s=m N f_s$ moving with a superfluid velocity $v(x)=(\hbar/m) \nabla \phi(x)$. We remark that, even though the superfluid velocity is expressed through the  phase of the condensate $\phi$, $f_s$ and $f_c$ can be very distinct. A particularly striking example in which $f_s$ and $f_c$ are indeed very different is provided by $^4$He at zero temperature in which $f_s=N/V$ with a condensate fraction $f_c$ that is about the $10\%$ of $N$ in a volume $V$~\cite{leggett1999superfluidity,yang1962concept}.  

In a ring configuration,  a superfluid current can be achieved through a linearly increasing phase along the ring: $\phi(x)=x/L$ with $L$ corresponding to the system length. Such a phase implies that the many-body wave function experiences twisted boundary conditions:
\begin{equation}
\psi(x_1,\dots, x_k+L,\dots,x_N)=e^{i\phi}\psi(x_1,\dots,x_N).
\label{twisted}
\end{equation}
For moderate $\phi$ and assuming that the superfluid motion only implies an increase of the total kinetic energy of the system $E_\phi-E_{\phi=0}>0$, $f_s$ can be  demonstrated to be related to the phase-stiffness~\cite{leggett1973topics,fisher1973helicity}:
\begin{equation}
f_s={{2m L^2}\over{\hbar^2N}} {{E_\phi-E_{\phi=0}}\over{\phi^2}}.
\label{superfluid}
\end{equation}
Note how such an expression does not involve the notion of the condensate fraction $f_c$. We also point out that, the superfluid behavior expressed by Eq.~\eqref{superfluid} cannot describe any dynamical instability coming from the  critical velocity~\cite{roth2003superfluidity,lieb2005superfluidity}.   

Even though persistent currents and the superfluid fraction are related quantities they define different concepts. To make the above statement  explicit, we refer to a system described by the general Hamiltonian
\begin{equation}
H=-{{\hbar^2}\over{2m}}  {{ \partial^2 }\over{\partial x^2}}  + H_{GI},
\end{equation}
in which $H_{GI}$ is assumed to be invariant by a local gauge transformation (conserving the total particle number); it may contain interactions and local potentials. We assume that Eq.~\ref{twisted} holds. By relabelling $\psi \rightarrow \psi e^{-i\phi}$,  the phase twist can be moved  to the canonical momentum:$-i \hbar{{ \partial }\over{\partial x}}\rightarrow -i \hbar{{ \partial }\over{\partial x}} -\phi$. In this case, it can be proved that  
\begin{equation}
\label{eq:EnergyMagnetic}
E_\phi -E_{\phi=0}= {{1}\over{2m}}+\sum_{\nu\neq 0} {{|\langle\psi_0|\mathcal{I}|\psi_{\nu}\rangle|^2}\over{E_\nu-E_0}},
\end{equation}
where $I$ is the current operator defined in Sec.~\ref{sec:PerCurrDev}. Therefore
\begin{equation}
f_s={{2m L^2}\over{\phi^2 \hbar^2N}} \left({{1}\over{2m}}+\sum_{\nu\neq 0} {{|\langle\psi_0|\mathcal{I}|\psi_{\nu}\rangle|^2}\over{E_\nu-E_0}} \right ).
\label{eq:SuperfluidPersistent}
\end{equation}
Eq.~\ref{eq:SuperfluidPersistent} shows how the superfluid fraction is indeed related to matrix elements of the current operator and as such {\it is not} an equilibrium  thermodynamic quantity. 

To close the section, we note that the superfluid fraction is related to the Drude weight that at zero temperature reads  $D_\omega \doteq L \pi \left ({{\partial^2 E(\phi)}/{\partial \phi^2}}\right )_{\phi=0}$ which is the real part of the linear response to an effective magnetic field~\cite{kohn1964theory}. Noting that $\partial^2E(\phi)/\partial \phi^2$ can be discretized as $(E_\phi -E_{\phi=0})/\phi^2$, by Eqs.~\eqref{eq:EnergyMagnetic} and~\eqref{eq:SuperfluidPersistent}, the Drude weight is related to the superfluid fraction: $D_\omega={ {\pi \hbar^2 N}\over{2 mL}}  f_s $ -- see also~\cite{hetenyi2012current,hetenyi2014drude}. Noting that the first order derivative of the energy is the current $\mathcal{I}$, $ D_\omega$  can be calculated as the response of the persistent current to an applied effective magnetic field: 
\begin{equation}
D_\omega =L \pi \left ({{\partial \mathcal{I}(\phi)}\over{\partial \phi}}\right )_{\phi=0}.
\end{equation}
In a mesoscopic system, a finite Drude weight reflects a dissipationless flow; normal systems, instead, are characterized by a vanishing Drude weight~\cite{shastry1990twisted,castella1995integrability,fye1991drude}. Response in systems with spin degrees of freedom (as fermionic ones) can be studied by applying a non-scalar effective magnetic field~\cite{amico2005quantum,shastry1990twisted}. Recently, Drude weight in a cigar-shaped cold atom system was measured experimentally~\cite{schuttelkopf2024characterising}. Non-equilibrium transport can be studied by non-linear Drude weights~\cite{watanabe2020general,watanabe2020generalized,urichuk2022nonlinear}.  We finally comment that the Drude weight has been related to the so-called quantum metric of the ground state~\cite{salerno2023drude}.

\subsection{Setting cold atoms in motion }\label{sec:ColdAtomsMotion}

 Given their charge-neutral nature, cold atoms do not experience a Lorentz force when placed in a magnetic field. However, they are affected by electric and magnetic dipole forces in the presence of field gradients, and this makes it possible to engineer an interaction with external fields such that the center-of-mass motion of an atom mimics that of a charged particle in a magnetic field. While there are still some important practical limits on what can be accomplished with these techniques, there is a large and still-expanding variety of methods for subjecting cold atom systems to artificial gauge potentials~\cite{dalibard2011artificial, goldman2014light}.  

The specific means able to set the condensate in motion heavily depend on the fundamental properties of the order parameter representing the condensate wave function. For the cases in which the order parameter is a simple complex scalar field (as in the bosonic  cold atoms), the winding number is conserved as long as the condensate fraction does not vanish. In the cases in which the condensate fraction is characterized by more complicated symmetries (as in fermionic gases), winding numbers can change even without the order parameter presenting any node~\cite{leggett1991dephasing}. As we discussed in Sect.\ref{sec:TheoryPhase}, the velocity of the matter-wave is not just the gradient of the phase of the condensate wave function but it contains also the gauge field term: $v_s=(\hbar/m)(\nabla \phi-\chi A)$ in which $A$ is the artificial vector potential.  If the latter is applied sufficiently gradually with $\phi$ staying constant,  a flow of matter-wave can be started as a response to the Aharonov-Bohm flux (see Sec.\ref{sec:PerCurrDev}). Alternatively, the flow can be started as a specific manipulation of the phase of the condensate wavefunction even without acting on the artificial gauge field. The latter cases correspond to phase imprinting protocols. 

Finally, we note that an important difference arises in the nature of the persistent current,  depending on the way the effective magnetic field is realized. Specifically, if the motion is started by a stirring laser beam, then the magnetic field is produced as a Coriolis force in the rotating frame of the stirred potential. In this case, the persistent current depends on the stirring rotation displaying a characteristic sawtooth shape in the co-rotating frame-see Sec.~\ref{sec:PerCurrDev}. In contrast, Raman or phase imprinting protocols are able to generate an artificial field in the lab frame. Nevertheless, the labframe persistent current versus magnetic field is a staircase in both instances -see Sec.~\ref{sec:ColdAtomsMotion}.

\subsubsection{Rotation}
One method of inducing rotation in the system is to stir the quantum fluid, typically carried out by a moving barrier~\cite{fetter2009rotating, wright2013driving, neely2013characteristics, cai2022persistent}. It should be emphasized however that such an operation leads to rotations following a quite different mechanism to stirring a spoon in a classical liquid (that in fact would not set a superfluid in motion). Indeed, the quantum gas is stirred by producing a dark soliton-type, namely a spatial region in the annulus in which the condensed fraction gets suppressed to a point where phase slips can occur with negligible energy cost. By moving the dark soliton-type,  to compensate for the phase kinking or slippage in the depletion region,  the condensate wavefunction develops a nonzero phase gradient ultimately putting the fluid in motion. Such a protocol can be realized, for example, by driving the condensate with a focused blue-detuned laser beam (see Fig.~\ref{fig:wrightstirring} for an example). In what follows, we demonstrate how to instigate a persistent current flow in a cold atoms platform through stirring. 

Let us take $N_{p}$ particles of mass $m$, that can be either bosonic or fermionic in nature, trapped in a ring-shaped potential of radius $R$, interacting via a two-body contact interaction of strength $g$. Such a system can be modeled by the first quantized version of the Hamiltonian in Eq.~\eqref{eq:GenHam} that takes the form
\begin{equation}\label{eq:FirstQuantHam}
    \mathcal{H}_{0} = -\sum\limits_{j=1}^{N_{p}}\frac{\hbar^{2}}{2m}\frac{\partial^{2}}{\partial x_{j}^{2}} +V_{\mathrm{ext}}(x) + g\sum\limits_{i<j}^{N_{p}}\delta (x_{i}-x_{j}).
\end{equation}
In order to rotate the condensate, we place a time-dependent potential barrier $V(x-\Omega Rt)$ that moves at an angular velocity $\Omega$, such that the total system is described by this Hamiltonian $\mathcal{H}(\Omega,t) = \mathcal{H}_{0} + V(x-\Omega Rt)$. The barrier is assumed to be switched on adiabatically, so no high-energy excitations are generated in the condensate. By performing a suitable unitary transformation $\mathcal{H}_{rot} = \mathcal{U}^{\dagger}(t)\mathcal{H}(\Omega,t)\mathcal{U}(t) +\imath\partial_{t}\mathcal{U}(t)$ with $\mathcal{U}=\mathrm{exp}(\imath L_{z}\Omega t/\hbar)$, we can switch over to the rotating reference frame with a frequency $\Omega$ as that of the potential barrier, thereby eliminating the time dependence of the Hamiltonian leaving us with 
$\mathcal{H}_{rot} = \mathcal{H}_{0} + V(x) - \Omega\hat{L}_{z}$, where $L_{z}$ is the $z$-component of the angular momentum perpendicular to the $x$-$y$ plane in which the ring is taken to lie in~\cite{fetter2009rotating,polo2020exact}. Translating to polar coordinates as in Sec.~\ref{sec:PerCurrDev} and expressing the angular momentum as $-\Omega L_{z} = \imath\hbar \Omega\frac{\partial}{\partial\varphi}$, the Hamiltonian can be written as
\begin{equation}\label{eq:FieldColdHam}
    \mathcal{H}_{rot} = \frac{1}{2mR^{2}}\sum\limits_{j=1}^{N_{p}}\bigg(-\imath\hbar\frac{\partial}{\partial\varphi_{j}}- m\Omega R^{2} \bigg)^{2} +\frac{g}{R}\sum\limits_{i<j}^{N_{p}}\delta (\varphi_{i}-\varphi_{j}) - \frac{1}{2}m\Omega^{2}R^{2}.
\end{equation}
The presence of the rotating barrier introduces a shift in the momentum operator and, in turn, in the system's total energy. Suppose we ignore the interaction and look at the expression for the single-particle case, it is clear that the Hamiltonians of Eqs.~\eqref{eq:ClassicHamField} and~\eqref{eq:FieldColdHam} are of a similar qualitative structure by choosing a gauge field\footnote{Note that we take the gauge in~\eqref{eq:ClassicHamField} to be zero and ignore the last term in~\eqref{eq:FieldColdHam} as this corresponds to a uniform shift in the energy spectrum.}  $\chi\mathbf{A} = m\Omega R^{2}$.Therefore, the behaviour of a charged particle in a magnetic field can be emulated in a cold atoms setting by incorporating a synthetic gauge field $m\Omega R$. The analogy can be extended even further to the exerted forces, with the Lorentz force being mimicked by the Coriolis force, which originates as a response to the induced rotation. 
\begin{figure}[h!]
\includegraphics[width=1\textwidth]{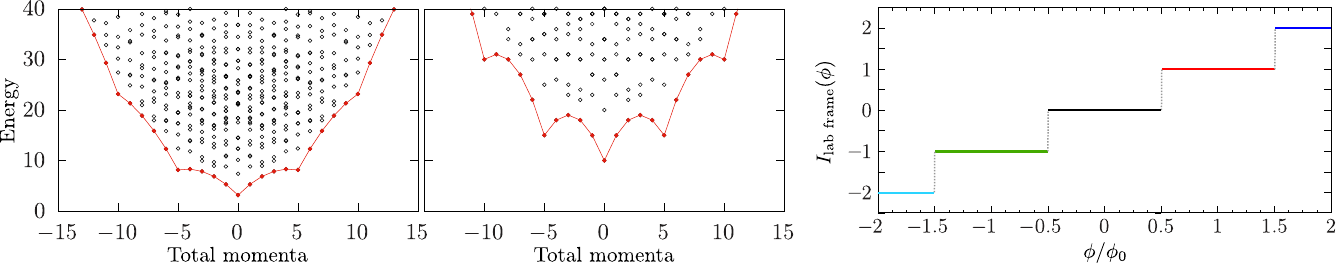}
\centering
    \put(-345,25){(\textbf{a})}
    \put(-25,25){(\textbf{b})}
\caption{Energy and corresponding persistent current in the lab frame. Panels \textbf{(a)} and \textbf{(b)} depict the energy of the system against the total angular momentum for $N_{p}=5$ weakly and strongly repulsive bosons respectively (the latter corresponding to spinless fermions). The energy is calculated through $E = \hbar^{2}/(2mR^{2})\sum_{\{n\}}n^{2}$. These figures show that energy states corresponding to larger angular momentum are metastable as the minima found at momenta $\ell n_p$ have a lower effective energy barrier allowing states to {\it slip} to lower momentum states through thermal or quantum fluctuations. Panel (\textbf{c}) displays the quantized nature of these metastable currents states through characteristic step-like behaviour.}
\label{fig:Lab_current}
\end{figure}

 From here, it becomes straightforward to show how persistent currents in ultracold atoms arise in the presence of a synthetic gauge field. Following the steps outlined in Sec.~\ref{sec:PerCurrDev}, by replacing $\chi A=m\Omega R^{2}$ one obtains the same expression for the current as in Eq.~\eqref{eq:ZeroTempCurrDef} and Aharonov-Bohm phase for the circulating particle akin to those of a normal metallic ring. Naturally, the elementary flux quantum for the two platforms is different, with the one for cold atoms being defined as $\Omega_{0}=\hbar/(mR^{2})$. \noindent Going forward, we will denote the artificial gauge field acting on the system by $\phi$ and the corresponding bare flux quantum by $\phi_{0}$ as defined in Sec.~\ref{sec:PerCurrDev}, the latter of which will be taken to have a unit value unless explicitly stated. 

So far, our calculations of the persistent current have been carried out in the rotating frame. Naturally, in an experimental setting, we need to look at the current in the non-rotating frame, which can be ontained by performing a Galilean boost. Put simply, this amounts to incorporating the velocity of the ``moving frame'' $v$ to that of every particle. Resorting to the standard definition of the current, i.e.\ the total number of particles flowing in the system at a given time $t$ such that $I = \frac{N_{p}}{\Delta t}$ with $\Delta t = \frac{L}{v}$, the current in the lab frame can be defined as 
\begin{equation}\label{eq:LabCurrentConti}
    I_{lab}(\phi) = I_{0}(\phi) +\frac{N_{p}v}{L} \rightarrow I_{lab}(\phi) = I_{0}(\phi) + \frac{2\pi\hbar N_{p}}{mL^{2}}\frac{\phi}{\phi_{0}},
\end{equation}
wherein we expressed the velocity as $v = 2\pi R\phi$. Alternatively, one can define the current through the continuity equation $\partial_{t} \hat{n} + \nabla \cdot \mathcal{I}_{lab}=0$ with $\hat{n}$ and $\mathcal{I}_{lab}$ corresponding to the density and current operators respectively. In the case of a one-dimensional lattice system modeled by Eq.~\eqref{eq:fluxtightbindingHam}, the current operator in the lab frame becomes
\begin{equation}\label{eq:LabCurrentLat}
    \mathcal{I}_{lab}(\phi) = -\imath\sum\limits_{j=1}^{N_{p}}(a_{j}^{\dagger}a_{j+1} -a_{j+1}^{\dagger}a_{j}),
\end{equation}
which at a specific site $j$ is the difference between the incoming and outgoing particles. In contrast to the rotating frame where the current is a sawtooth, the one in the lab frame is given by discrete steps -- see Fig.~\ref{fig:Lab_current}(\textbf{b}). The ideal Bose gas is a special case. Due to the spectrum of the energy against the total momenta, the ground-state of the system being at $\ell=0$, the current obtained is not metastable as it decays once the source imparting rotation is removed. For this reason, we showcase the spectrum for weakly repulsive interactions ($c=10$) by obtaining the momenta through Bethe ansatz, to be discussed in more detail in Sec.~\ref{sec:BATonks}. The plateaus reflect the quantized nature of the angular momentum associated with the circulating particles, which increases on each subsequent jump.\footnote{In the case of fermions, the parity of the current is captured through these jumps, occurring at integer of half-odd integer values of the flux.} It must be stressed that in the non-rotating frame, we get a better picture of the trade-off between the stability of the system and the magnitude of the current (circulation values) through  the energy of the system (see Fig.~\ref{fig:Lab_current}(\textbf{a})). To offset the instability of the system at higher rotation rates, one expects the formation of excitations, such as vortices, that reduce the total energy~\cite{wright2013driving,donnelly1966stability}. 

\subsubsection{Phase imprinting}
An alternative method to induce persistent currents is to directly imprint the phase into the cold atom condensate by using lasers~\cite{dalfovo1999theory,bolda1998detection,dobrek1999optical}. One can understand phase imprinting instructively by regarding the example of a single particle in a ring~\cite{andrelczyk2001optical}, parameterizing the polar angle $x\in[0,2\pi]$. Now, we apply a potential that increases linearly with $x$ as $V(x)=x$ where we choose unitless numbers. We assume the potential is large compared to the kinetic energy of the atom, which can be achieved by applying a strong potential for a very short time. In this limit, the phase of the wavefunction evolves with the potential $V(x)$ according to the Schr\"odinger equation, which gives us $\ket{\psi(x,t)}\sim \exp(-i V(x) t)\ket{\psi(x,0)}$. Subsequently, we select a final imprinting time $T=\ell$ in unitless numbers, where $\ell$ is the integer phase winding we want to achieve. The wavefunction in polar coordinates is given by $\ket{\psi(x,T)}\sim \exp(-ix\ell)\ket{\psi(x,0)}$, having now acquired a phase winding of $\ell$. By tuning the potential strength or evolution time, one can use this approach to create arbitrary phase windings $\ell$. This concept has first been applied to generate dark solitons~\cite{burger1999dark,denschlag2000generating} and vortices~\cite{gajda1999optical} into cold atoms. This approach using tailored time-dependent laser potentials has been applied in many cold atom experiments to create persistent currents in rings~\cite{moulder2012quantized,kumar2018producing,del_pace2022imprinting}.  Through these methods, matter-wave currents in ultracold atomic gases have been experimentally realized in both bosonic~\cite{ryu2007observation,ramanathan2011superflow,wolf2021stationary} and very recently in fermionic systems~\cite{cai2022persistent,del_pace2022imprinting}.
 
\subsubsection{Engineering persistent current by Floquet driving}
We now present another strategy to induce persistent currents into cold atom systems which relies on periodically varying the Hamiltonian $H(t)$ in time $t$. In particular, one chooses $H(t+T)=H(t)$ such that it returns to its original form after period $T$. This periodic Floquet driving of the quantum system can reveal a much richer physics than the undriven Hamiltonian~\cite{eckardt2017colloquium,weitenberg2021tailoring}. 

For example, periodic driving allows one to control the nature of quantum tunneling in cold atom lattices~\cite{aidelsburger2013realization}. In undriven systems, the tunneling rate is real-valued, given due to the fact that it is a time-reversal symmetric process. However, by driving the system periodically in time, the effective dynamics over one period of driving can fundamentally change the nature of tunneling. In particular, one can achieve complex-valued tunneling, breaking time-reversal symmetry effectively~\cite{aidelsburger2013realization}. The driving requires precise control over the dynamics, which nowadays is routinely achievable in  cold atom experiments~\cite{eckardt2017colloquium,weitenberg2021tailoring}. As such, Floquet driving can generate  effective dynamics that is not possible with the bare time-independent Hamiltonian by itself.

Such an idea has been adopted to create persistent currents in a ring lattice~\cite{roushan2017chiral}. Specifically, we start from a physical system with {\it no artificial gauge field}, e.g. no rotation, phase imprinting, or any other method of current generation. The goal is to generate persistent current states by suitable driving of the system.  At the formal level, the goal is to achieve  an effective Hamiltonian with complex tunneling terms $H_\text{eff}=\sum_j Je^{-i\theta_j}a_{j}^{\dagger}a_{j+1} +\text{h.c.}$ with controllable phases $\theta_j$ in which the undriven Hamiltonian is given by $H=\sum_j Ja_{j}^{\dagger}a_{j+1} +\text{h.c.}$. This can be achieved by modulating the on-site potential $H_\text{on-site}(t)=\sum_j\mu_j(t)a_{j}^{\dagger}a_{j}$ periodically in time. Here, the time-dependent potential $\mu_j(t)$ must break time-reversal symmetry to achieve complex phases in the effective tunneling~\cite{eckardt2010frustrated,struck2012tunable}.  In particular, the driving must break reflection symmetry for suitable time $\tau$, i.e. $\mu_j(t-\tau)\neq \mu_j(-t-\tau)$, as well as shift (anti)symmetry, i.e. $\mu_j(t-T/2)\neq-\mu_j(t)$. Integrating over one period of a sufficiently fast dynamics, the effective Hamiltonian $H_\text{eff}$ can acquire a non-trivial gauge field $\theta_j$. 

In two-dimensional cold atom lattice systems, the periodic modulation has been achieved by creating tilted lattices. Here, the tunneling is initially suppressed, then the lattice is accelerated by periodic modulation of the laser field in such a way that the tunneling is restored due to resonant Raman processes~\cite{maciej2012tunable,atala2014observation,kennedy2015observation}. This way, effective Hamiltonians with non-trivial gauge fields have been created. Furthermore, one can even prepare the ground-states of the effective Hamiltonian~\cite{maciej2012tunable,atala2014observation,kennedy2015observation}.

This driving of the local potential even extends to systems with non-regular lattices. In particular, Ref.~\cite{struck2012tunable} proposes to drive the on-site potential $\mu_j(t)$ with two frequencies in order to break time-reversal symmetry. By appropriately tuning the frequencies, this can create arbitrary phases $\theta_j$ in the tunneling term between each lattice site individually. To achieve a well-defined effective Hamiltonian, the time-dependent control over the potential of system must be faster than the bare tunneling rate. For light-shaped cold atom traps, digital micromirror devices (DMDs) offer to control potentials with a speed which is orders of magnitude faster than the dynamics of the atoms~\cite{rubinsztein2016roadmap}. Thus, this approach is well suited to create fast periodic potential modulations that can engineer effective ring Hamiltonians with flux.

\subsubsection{Machine learning} 
A key challenge in experiments is to prepare specific persistent current states of high quality. In particular, there are multiple criteria that one wants to achieve at the same time. It is desirable to generate the persistent current in as short time as possible to avoid the impact due to noise and atom losses, and also minimize excitations in the condensate induced by the preparation protocol. Moreover, one wants to be able to prepare specific currents of different phase windings in a deterministic manner. 

Naturally, these criteria can be improved by optimization of the experimental protocol that generates the persistent current. To give an example, the protocol for generating currents via a rotating barrier involves optimizing parameters such as width, amplitude, and velocity of the barrier, as well as the ramping protocol in time~\cite{simjanovski2023optimizing}. 

Experimentalists commonly optimized such parameters by carefully tuning the experimental apparatus by hand. This can be a cumbersome and time-consuming process~\cite{mathey2016realizing}. Moreover, this process often does not find the best solutions as humans tend to have natural biases towards certain patterns in the parameters, which may be sub-optimal. 

Thus, it is natural to ask whether we can harness the recent advances in artificial intelligence to improve optimization tasks~\cite{goodfellow2016deep}. Here, a classical computer learns from past experiences or even directly controls the parameters of the protocol in order to learn the best way to generate currents.
There is promise in machine learning to find non-conventional solutions that can outperform those found by human experts, which prompts a  wide applications in quantum physics~\cite{carleo2019machine,krenn2020computer,bharti2020machine}.

In cold atoms, machine learning found widespread application to many tasks such as enhancing rotation sensing in lattice-based cold atom interferometers~\cite{chih2022train}, read-out of quantum vortices in cold atom condensates~\cite{metz2021deep} or optimizing experimental preparation sequences~\cite{wigley2016fast}.

\begin{figure}
\centering\includegraphics[width=\linewidth]{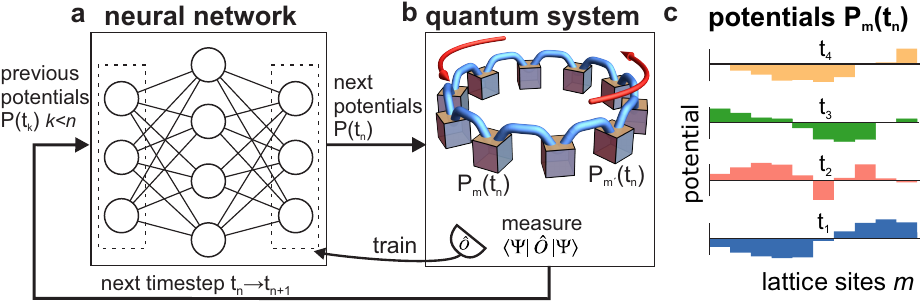}
    \caption{ An example for  deep reinforcement learning to optimize the preparation protocol for generating atomic current states in ring lattices. \textbf{(a)} Deep neural network which suggests lattice potentials for each timestep of the protocol. Neural network is trained using observables from the quantum system. \textbf{(b)} Atomic ring lattice where suggested potential is applied to create current state. The final state is measured and observables such as the momentum distribution are used to train neural network. \textbf{ (c)} Optimised protocol for potential to generate current, showing lattice potential for different lattice sites $m$ and discrete timesteps $t_n$. Figure from Ref.~\cite{haug2021machine}.}
    \label{fig:machinelearning}
\end{figure}

For atomic currents, the first theoretical and numerical study has been done in Ref.~\cite{haug2021machine}. The authors propose to improve the generation of persistent currents using deep reinforcement learning. An example of this learning setup is shown in Fig.~\ref{fig:machinelearning}.
A goal is to find a fast protocol that creates a specific atomic current state in an atomic ring lattice. In such a ring lattice, one can control the potential of each lattice site in time. Experimentally, this can be achieved by time-dependent control of  the potential of the ring lattice by light-shaping techniques~\cite{gauthier2016direct}. 
By changing the potential in a controlled way, specific current states can be generated. The authors find specific protocols that generate such currents in very short times using reinforcement learning~\cite{goodfellow2016deep}. Here, a machine learning algorithm directly controls the protocol to generate currents and learns from experience to achieve the target current. The particular method used is called  proximal policy learning~\cite{schulman2017proximal} where two deep neural networks, called actor and critic, work together to find good protocols. The actor network decides on the potential to choose at each timestep, while the critic evaluates the former's decisions by assigning a score. After the protocol chosen by the actor has been run, the current of the system is measured  and a reward is given to the neural network depending on how close the actual value is to the desired one. The weights of the network are adjusted depending on the reward. This learning process is repeated until it converges to a good solution.

Key advantages of reinforcement learning are that it is agnostic of the underlying physical process, robust to noise, and can be directly implemented in experiment as the neural network can learn from readily accessible observables such as the momentum distribution acquired from time-of-flight measurements.
This way, Ref.~\cite{haug2021machine} finds protocols which are an order of magnitude faster than the conventional protocol of driving a barrier. Additionally, novel current states composed of entangled superpositions of three momentum states could be generated, which could be of interest for sensing purposes.

An experimental demonstration of machine learning to optimise atomic currents has been done in Ref.~\cite{simjanovski2023optimizing}. Here, machine learning is applied to optimize the experimental protocol for inducing currents in a ring cold atom system. The authors use the M-LOOP package~\cite{wigley2016fast}, which utilizes a Gaussian learner.  This reinforcement learning model fits the observed experimental data to a Gaussian model. The next parameter to be explored is picked where the uncertainty of the model is large, thus  using past information efficiently and training the model in a cost-effective manner. Ref.~\cite{simjanovski2023optimizing} goal optimizes the stirring protocol to create specific phase windings. It optimizes four parameters, namely the stirring time of the barrier, the timeframe over which the barrier is removed, the angular velocity and a non-linear coefficient for the acceleration of the barrier. The cost function either targets a particular phase winding, or maximizes its absolute value. A penalty term is added to discourage long preparation times as well as stray vortex excitations. The authors find that the optimized protocols show a wide variance in the final parameters, indicating that the ansatz is underconstrained and robust to perturbations.

\subsubsection{Experimental methods}\label{sec:ExpMethods}

All experimental methods for controlling the current state of an ultracold quantum gas ultimately depend on some means of exchanging energy and momentum between the atoms and a suitably engineered electromagnetic field. This can be achieved by coupling to optical, microwave, RF, or DC magnetic fields in a wide variety of configurations and combinations that we will now briefly review. Engineering a suitable interaction for a real experimental setting with a particular atomic species requires detailed consideration of the atomic level structure and other energy scales involved, which we will not attempt to do here. In the discussion that follows we will focus on mechanisms that allow deterministic control of the current around a ring, but it is important to reiterate that persistent currents form spontaneously under a wide range of experimental conditions, and the ability to initialize a system in a known current state is as important as being able to control it subsequently.

The earliest experiments with ring-shaped BECs used diabatic off-resonant two-photon interactions to suddenly change the center-of-mass motion of the atoms~\cite{ryu2007observation, ramanathan2011superflow, moulder2012quantized}. The basis for this class of techniques is that when an atom absorbs a photon from one field and is stimulated to emit a photon into a field of different momentum, the atom's motion must change to satisfy conservation of momentum. If the angular momentum state of the two fields is different, then the angular momentum of the atoms is altered during the interaction. In early experiments this was always done on a time scale much shorter than the density response time of the BEC, or the trap oscillation period. This diabatic approach should be functionally distinguished from the adiabatic Raman dressing techniques that can be used to generate artificial gauge potentials~\cite{lin2011synthetic}, though the two methods are in some ways closely related. 

\begin{figure}[h!]
    \centering
    \includegraphics[width = 0.7\textwidth]{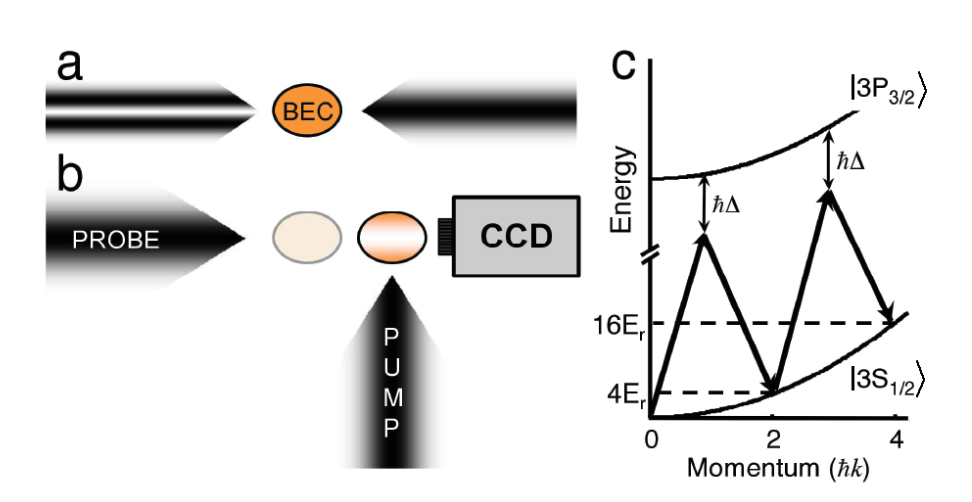}
    \caption{Transfer of orbital angular momentum to atoms: \textbf{(a) }Counterpropagating  Laguerre-Gaussian and Gaussian laser beams are applied to a BEC. \textbf{(b)} The atoms that have undergone the Raman transition (right cloud) have separated from those that did not (left cloud). \textbf{(c)} Diagram illustrating energy and momentum conservation of the 2-photon Raman process for one and two consecutive pulses.
    Figure from Ref.~\cite{andersen2006quantized}.}
    \label{fig:Andersen2006Fig1}
\end{figure}

When counter-propagating laser fields are used to couple between distinguishable states of linear momentum without changing the internal state, the technique is referred to as Bragg scattering~\cite{kozuma1999coherent, stamperkurn1999phonons}. If two beams of differing orbital angular momentum are used, the recoil affects both the linear and orbital angular momentum of the atoms. It is necessary to use a large linear momentum transfer in order for the initial and final states to be clearly separated in energy. If the experimental objective is to change only the orbital angular momentum state, it is therefore necessary to use a sequence of at least two Bragg pulse pairs with different momentum kicks, where the change in linear momentum cancels out, but the change in orbital angular momentum does not. This process was the basis for the first experiments creating persistent currents at NIST,  as shown in Fig.~\ref{fig:Andersen2006Fig1}, taken from Ref.~\cite{andersen2006quantized}. 

Alternatively, it is possible to use two optical fields to couple between two different internal states in what is referred to as a Raman process. In this case, it is possible for the levels to be energetically resolved without a large momentum transfer, making it possible to use collinearly propagating beams. As long as the energy difference between the two selected levels is not too large (as is generally true for coupling between different magnetic sublevels in the electronic ground state) the linear momentum imparted to the atoms by a single collinear pulse pair is small enough to be neglected. This Raman-beam  approach to orbital angular momentum control of a BEC was first demonstrated by Wright et al.~\cite{WrightSculptingPRL2009} with a simply-connected BEC, and was later used for some of the persistent current experiments with ring-BECs at NIST~\cite{ramanathan2011superflow} and at Cambridge~\cite{moulder2012quantized}.  

There are some important factors to be considered in the use of these kinds of two-photon techniques. With suitable laser intensities and moderate detunings from the $D$ lines of alkali metal atoms, the $\pi$-pulse times can be made as short as a few microseconds, but attention must be paid to keeping the single-photon scattering rate low enough to prevent unacceptable levels of heating and loss. A second challenge in using two-photon techniques is that it not generally safe to ignore the detuning and state-dependent AC Stark shifts that occur when applying such fields, especially for single-photon detunings on the order of the hyperfine splittings. Finally, any realistic experimental design must account for the possibility of spatially-varying Rabi frequencies and AC Stark shifts, because the phase singularities present in laser fields with nonzero orbital angular momentum necessarily make them spatially inhomogeneous~\cite{WrightSculptingPRL2009}. These drawbacks were less important in the earliest experiments~\cite{ramanathan2011superflow, moulder2012quantized}, when techniques for detecting the circulation state of the ring were less well developed, and it was advantageous to use a technique where the atomic orbital angular momentum was guaranteed to change by a quantized amount, as long as it reached the target state. 

\begin{figure}[h!]
    \centering
    \includegraphics[width = 0.7\textwidth]{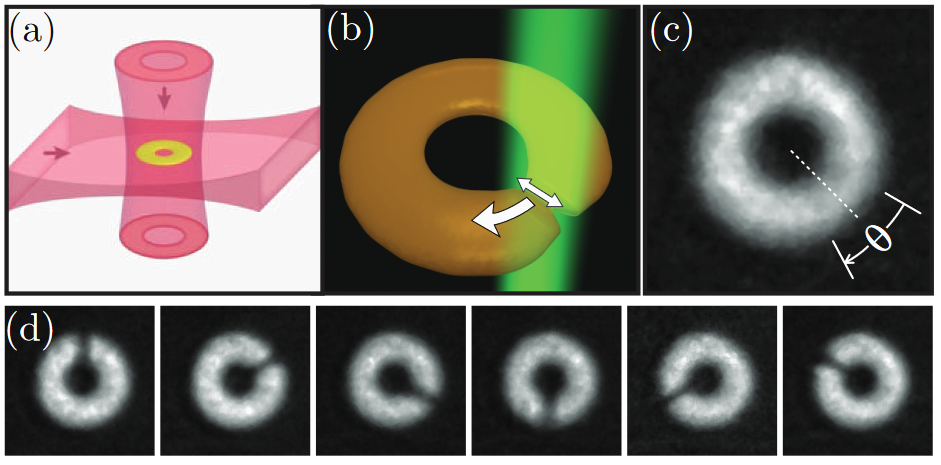}
    \caption{ Experimental stirring of ring condensate via weak link. \textbf{(a)} Schematic showing the attractive optical dipole trap. \textbf{(b)} Geometry of the barrier beam. \textbf{(c) }In situ absorption image of the ring condensate, viewed from above. \textbf{(d) }In situ images showing the effect of barrier at different times. Figure from Ref.~\cite{wright2013driving}.}
    \label{fig:wrightstirring}
\end{figure}

Techniques for creating persistent currents with a single far-detuned single-frequency laser field are comparatively simpler in many ways, and they have become more commonly used in recent experiments. Here, we will separately consider the two very different approaches of stirring and phase imprinting. The idea of stirring a superfluid to induce circulation is conceptually simple: create a localized disturbance in the potential and create a flow by moving it faster than the superfluid critical velocity. We note that quantum phase slips dynamics across the potential disturbance determines the final winding number of the superfluid flow - see Sec.~\ref{sec:PerCurrBose}.  In practice, there are many details about the creation and evolution of various types of excitations (phonons, solitons, vortices) in the superfluid that can affect the success and reliability of such an approach.  One has the choice of whether to apply a repulsive or attractive potential defect, and its size and shape compared to the channel width. There are also questions of how to turn it on, how it should move, and how to turn it off. Which technique is optimal for a given situation is highly dependent on system characteristics and experimental goals.

The first experiment creating a persistent current in a ring BEC via a stirring technique used a repulsive (blue detuned optical) barrier that was thin in the azimuthal direction and large enough radially to span the width of the ring, as shown in Fig.~\ref{fig:wrightstirring} taken from Ref.~\cite{wright2013driving}. The position of the spot was controlled using a 2D acousto-optic deflector, which allowed for the use of shaped, time-averaged potentials by rapidly scanning a tightly focused spot over a defined trajectory. In the case of some of the early ring BEC work done at Los Alamos National laboratory (LANL), the ring trap was a time-averaged ``painted'' potential generated by a scanned red-detuned beam, and movable potential barriers were formed by shutting off the scanned beam at one or more points, instead of by adding a repulsive beam~\cite{ryu2013experimental}.

Turning now to the subject of phase imprinting using optical fields, we note that the basic idea of using phase imprinting to create vortices in a simply-connected BEC was introduced in the work of Dobrek \emph{et al.}~\cite{DobrekOpticalPRA1999}. They proposed applying a far-detuned optical field with an approximately helicoidal intensity gradient (and Stark shift), for a time shorter than the characteristic density response time $\tau_\phi=\hbar/\mu$. This effectively imprints an azimuthally varying phase $\phi(\mathbf{r}) = \tau_\phi U_0(\mathbf{r}) /\hbar$ on the atomic cloud, where $U_0(\mathbf{r})$ is close to linear as a function of angle, except in a small ``kink'' region that must be present, and which cannot be made narrower than the resolution of the optical projection system.

It is natural to consider adaptation of this technique to generate persistent currents in a toroidal BEC. The first effort at modeling the evolution of a quasi-1D BEC after such a phase imprinting event seemed to suggest that counterflow occurring in the kink region would create large density waves and generally prevent the system from evolving into a well-defined non-zero current state after the impulse~\cite{zheng2003classical}. Further numerical simulations looked more specifically at the dynamics of density waves and dark solitons generated in the vicinity of the kink, and suggested that controlled angular momentum transfer was possible if the kink in the imprinting potential was localized to a region of lower density~\cite{swingle2005generation}. 
A wider range of conditions for phase imprinting was considered by Kumar et al.~\cite{kumar2018producing}, who used 2D GPE simulations to investigate the effects of finite optical resolution on the phase imprinting process, looking for an optimal approach to reaching a target current state. They found that temporarily reducing the density of the BEC in the vicinity of the dislocation in the imprinted phase pattern resulted in less excitation and heating of the BEC, and significantly increased the reliability of the technique for inducing a well-defined change in quantized flow, as long as the damping of density waves is sufficiently rapid .

\begin{figure}[h!]
    \centering
    \includegraphics[width = 0.7\textwidth]{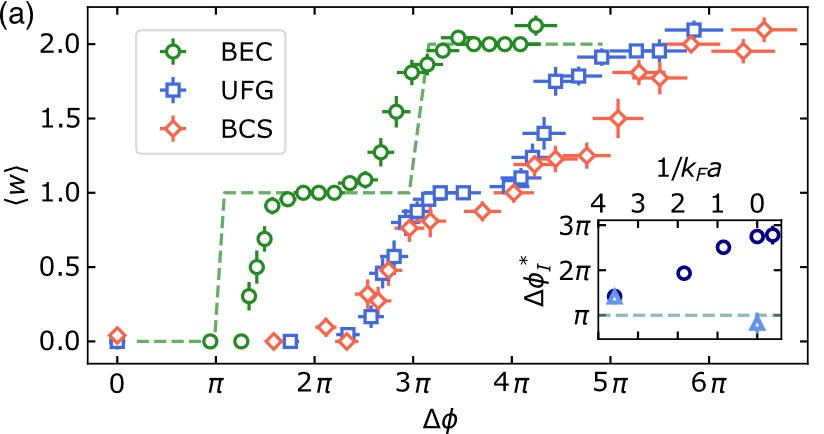}
    \caption{Winding number around a ring of fermionic superfluid 60 ms after imprinting with a phase jump $\Delta\phi$. The data labeled BEC were taken with the interaction parameter at $1/k_Fa=3.6$, and BCS at $1/k_Fa=-0.4$. The inset shows a comparison of the  phase imprinting on a fermionic superfluid, for varying interaction strength.
     Figure from Ref.~\cite{del_pace2022imprinting}.}
    \label{fig:del_pace_imprinting}
\end{figure}

The first experimental implementation of this technique was in the work done at LENS by del Pace et al~\cite{del_pace2022imprinting}. Because that work was done using a superfluid of fermionic $^6$Li atoms they were able to investigate how the response to phase imprinting changed with interaction strength, with some interesting differences observed between the BEC, unitary, and BCS regimes. Most importantly, they found that it was possible to reliably imprint a phase winding on the superfluid ring, even without modifying the density profile with a repulsive barrier as had been suggested in previous theoretical work. The dynamics at the phase dislocation were found to be important, leading to differences in the (higher chemical potential) unitary and BCS regimes, where it was necessary to apply a larger phase gradient to achieve a given winding number in the superfluid (see Fig.~\ref{fig:del_pace_imprinting}). Inserting a repulsive optical barrier at the location of the gradient discontinuity reduced the unwanted density excitations as expected, but was found to degrade the reliability of initializing the system in the $w=1$ state. With up to 5 repeated applications of the phase imprinting pulse, they were also able to realize higher winding number states of up to $w=9$. Further details of this work pertaining to lifetime and decay will be described in section~\ref{sec:ExpResFerm}.

Before concluding this section, we point out that essentially any method that has been used to generate vortices in a simply-connected system can potentially be adapted to generate a persistent current around an annulus. The technique of magnetic phase imprinting was proposed by Isoshima et al.~\cite{isoshima2000creation} and was first implemented by the Ketterle group~\cite{leanhardt2002imprinting} to generate vortices. It involves adiabatically inverting the direction of the axial part of a magnetic (or hybrid magnetic-optical) type trap, which creates a cross disgyration in the BEC as the axial field goes through zero. If the atoms are polarized in a single internal spin state with projection $m$ on the local field, and they adiabatically follow the field reversal during the transformation, the winding number of the current must necessarily change by $2m$. While this technique has not yet been used with a toroidally trapped condensate, it remains a possibility for the future.

\section{Theoretical framework}\label{sec:framework}

This section covers the main theoretical tools needed to understand and characterize persistent currents in both bosonic and fermionic gases. We start from the mean-field model for bosons,  
which can describe weakly interacting condensates confined in annular traps with several radial and axial modes occupied. We then focus on the one-dimensional limit corresponding to very tight transverse trapping confinement both in radial and axial direction. 
While early experimental results can be described by mean-field, quantum fluctuations are expected to become increasingly relevant in future experiments. 
In this regime, we introduce the theoretical frameworks allowing us to account for these quantum fluctuations and beyond mean-field effects, both in an exact way as well as through effective quantum field theory. 
We then turn to fermionic systems and present the main formalism needed to describe the BCS-BEC crossover in Fermi gases. Also for fermionic systems, we then focus on the one-dimensional limit, in particular on the SU(2) and SU(N) Fermi Hubbard models, where striking beyond-mean field effects emerge in the persistent currents. We finally present and dwell on the theory for the readout of the persistent currents in cold atomic systems through interference patterns. Examples of both 1D and 3D dynamics will be provided in the forthcoming Secs.~\ref{sec:PerCurrBose} and~\ref{sec:PerCurrFerm}.


\subsection{Bosons}

Bosonic particles are characterized by having a wavefunction completely symmetric under particle exchange. This important property ensures the possibility of occupying the same single-particle state in the case of non-interacting bosons, leading to Bose-Einstein condensation, which corresponds to the creation of a giant matter-wave. Condensates display  off-diagonal long-range order (ODLRO), that corresponds to large-distance phase coherence. In the case of interacting bosonic fluids, quantum depletion of the condensate state usually occurs. Still, they display the phenomenon of superfluidity, which may occur even in the absence of Bose-Einstein condensation -see Sec.~\ref{sec:TheoryPhase}. In low dimensions, phase fluctuations and correlations are enhanced, and coherence is degraded especially at large interactions. In this case, only quasi-ODLRO is possible. In this section, we present the main theoretical tools that we have at hand to describe  interacting Bose fluids in a ring geometry, ranging from mean-field approaches to effective field theories and lattice models.

\subsubsection{Mean-field theories}\label{sec:ModelsBose}

The general many-body Hamiltonian describing bosonic fluids pierced by a synthetic gauge field in second quantization reads
\begin{eqnarray}\label{eq:GenHam}
    \mathcal{H} = \int \Psi^{\dagger}(\mathbf{r})\bigg[\frac{1}{2m}(-\imath\hbar\nabla + \mathbf{A}(\mathbf{r}))^{2} +V_{\mathrm{toroidal}}(\mathbf{r})+V_{local}(\mathbf{r})\bigg]\Psi(\mathbf{r})\mathrm{d}\mathbf{r}  \nonumber \\ + \frac{1}{2}\int \Psi^{\dagger}(\mathbf{r})\Psi^{\dagger}(\mathbf{r'})\mathcal{V}(\mathbf{r}-\mathbf{r'})\Psi(\mathbf{r'})\Psi(\mathbf{r})\mathrm{d}\mathbf{r}\mathrm{d}\mathbf{r'},
\end{eqnarray}
where $\Psi^{\dagger}(\mathbf{r})$ ($\Psi (\mathbf{r})$) are the field operators that create (annihilate) a bosonic particle at position $\mathbf{r}$ with $\mathcal{V}(\mathbf{r}-\mathbf{r'})$ being the interatomic potential of two-body collisions. The potential $V_{local}(\mathbf{r})$ takes into account a local offset that may be due to impurities and $\mathbf{A}(\mathbf{r})$ is the vector potential.

Starting from Eq.~\eqref{eq:GenHam} one can derive a simpler description of the bosonic particles in the limit of large particle number $N_p$, weak interactions and zero temperatures,  where,  to a very good approximation, all bosons macroscopically occupy the same quantum state. At these low temperatures and for a large enough sample, the system becomes very diluted and one can simplify the generic interactions to be only two-body $s$-wave collisions. The interacting potential can then be approximated by a pseudo-potential description $\mathcal{V}(\mathbf{r}-\mathbf{r'}) = g_{3D} \delta(\mathbf{r}-\mathbf{r'})$ with  strength $g_{3D} = 4\pi\hbar^2 a_s / m$ characterized by the $s$-wave scattering length, $a_s$,  which can be positive ($a_s > 0$) or negative ($a_s < 0$) indicating that the effective interactions are repulsive or  attractive, respectively. These scattering lengths can, depending on the atomic species, be tuned in magnitude and sign using magnetic~\cite{inouye1998observation,roberts1998resonant} or optical~\cite{theis2004tuning,fedichev1996influence} Feshbach resonances.

Introducing the pseudo-potential into Eq.~\eqref{eq:GenHam} and using the mean-field description of the bosonic field operators $\Psi(\mathbf{r},t)=\braket{\Psi(\mathbf{r},t)}+\delta \Psi(\mathbf{r},t)$ where both thermal and quantum fluctuations are neglected, $\delta \Psi(\mathbf{r},t)\rightarrow 0$, one arrives to the celebrated Gross-Pitaevskii equation (GPE) proposed by Gross \cite{gross1957unified} and Pitaevskii \cite{ginzburg1958sov}, which reads:
\begin{equation}
\label{eq:GPE}
    \imath\hbar\frac{\partial \mathbf{\psi}(\mathbf{r},t)}{\partial t} = \bigg[-\frac{\hbar^{2}}{2m}\nabla^{2} + V_{\mathrm{ext}}(\mathbf{r})+g_{3D}|\mathbf{\psi}(\mathbf{r},t)|^{2}\bigg]\mathbf{\psi} (\mathbf{r},t),
\end{equation}
where $\mathbf{\psi}(\mathbf{r},t)\equiv\braket{\psi(\mathbf{r},t)}$ is the wavefunction of the condensate and $t$ denotes the time.

The GPE equation can describe bosonic particles in 3D; however, one can also be interested in investigating low-dimensional systems. One can achieve such a reduction of dimensionality by confining the particles to zero-point oscillations in one or two directions. This is typically realized by introducing a tightening harmonic potential with frequency $\omega_\perp$ such that $\mu \ll\hbar\omega_\perp$ \cite{petrov2003bose}, with $\mu$ being the chemical potential. We recall that for Bose-Einstein condensation to occur, it is required to have long-range order; however, despite not having full long-range order at lower dimensions, the system can still present quasi-ODLRO (i.e.\ algebraic decay of correlations; in such cases, we speak of `quasi-condensates', i.e.\ condensates with a fluctuating phase ~\cite{petrov2000regimes}).

\subsubsection{The Lieb-Liniger one dimensional model for interacting bosons}

There exist few theoretical models that allow us to exactly describe the many-body physics of interacting bosons beyond the mean-field description; however, such models can be found in lower dimensionalities. One-dimensional geometries are particularly appealing as they provide us with a wide variety of phenomena whilst being more manageable from a theoretical standpoint. Lieb \& Liniger (LL) used the microscopic description of a one-dimensional Bose gas interacting via a delta-like potential and found its exact solution in a seminal work \cite{lieb1963exact}. Here, we consider interacting bosons residing on a ring of size $L$, interacting with a contact potential of strength $g$, and threaded by an effective magnetic flux $\phi$:
\begin{equation}\label{eq:LLHam}
    \mathcal{H}_{LL} = \sum\limits_{j=1}^{N_{p}} \frac{1}{2m}\bigg(-\imath\hbar\frac{\partial}{\partial x_{j}} - \frac{2\pi}{L}\frac{\phi}{\phi_{0}}\bigg)^{2} + g \sum\limits_{i<j}^{N_{p}} \delta(x_{i} - x_{j}),
\end{equation}
with $x_{j}$ being the position of the $j$-th particle on the ring and $\phi_{0}$ corresponding to the bare flux quantum. The solution of this model provides us with a full many-body tractable model of interacting bosons in a finite-size ring that can sustain persistent currents. The Lieb-Liniger model is integrable through Bethe-ansatz~\cite{lieb1963exact} and will be discussed in detail in the following section. Its integrability and exact solution allow us to use this model as a test bench and reference model to understand the microscopic origin of the phenomena observed in more complex systems. For this reason, a plethora of works have and still do investigate bosonic systems using the Lieb-Liniger model \cite{piroli2016local,guillaume2017ground,piroli2016multiparticle,naldesi2019rise}. Despite its integrability through Bethe Ansatz, calculating high-order correlations using the exact solution becomes difficult. New techniques have been proposed, for example the use of quantum computers, in which one can prepare the quantum state to later perform measures on it \cite{dyke2021preparing,sopena2022algebraic,ruiz2024bethe}. 

The two models presented, GPE and LL, describe bosonic interacting particles in the continuum. However we remark again that the first is based on mean-field theory and therefore restricted to the limit of weak interactions and a large number of particles, while the latter describes the full-many body wavefunction exactly. Both theories can attain attractive, $a_s<0$, and repulsive $a_s>0$ interactions. For repulsive interactions, both models give a ground state with a homogeneous density along the ring because of the translational invariance present in the system. Intuitively, repulsion tends to spread particles apart, and therefore, a homogenous density keeps them as separated as possible without creating regions of higher density that would lead to increased local interactions. On the other hand, for attractive interactions, one could naively expect the localization of particles in space due to the negative two-body contact interactions. Such localization would, in principle, break translational invariance, and it is found in the mean-field description for sufficiently strong attractive interactions~\cite{kanamoto2003quantum}. This mean-field states were proposed as candidates to present time-crystal-like behavior and later disproved \cite{wilczek2012quantum,patrick2016comment}.
However, we want to clarify that the apparent breaking of translational invariance is only observed in single-shot density measurements where the particle's center-of-mass is localized at a given position. One can also access such localization by calculating expectation values of higher order correlators. We remark that, as pointed out in \cite{piroli2016local}, the alleged breaking of translational invariance is because one would need to average over the center-of-mass position of the mean-field solution in order to recover the correct translational invariant solution. This connection between
mean-field and many-body theories were also investigated in Ref.~\cite{calogero1975comparison}. Thus, here we speak of current states in the attractive regime, we consider the ground state to be a delocalized state of bound-state particles.

\subsubsection{The Bose-Hubbard model}\label{sec:BoseHubb}
Within 1D systems, another important set of models are the ones used to describe lattice systems. Bosonics systems trapped in optical lattices became relevant due to their analogy with condensed matter systems where atoms are fixed in a crystalline lattice structure. In bosonic systems, this is done through an external potential that effectively restricts the position where particles can reside. For sufficiently large external lattice potential $V=V_0\cos^2(x/a)$, with $V_0$ and $a$ being the lattice strength and lattice spacing respectively, a particular band structure appears. Finding such a strong periodicity in the system leads us to use the Bloch theorem to derive the model. Starting from the microscopic description of $N_p$ interacting bosons trapped in a lattice of $N_s$ sites pierced by an effective magnetic flux $\phi$ as in Eq.~\eqref{eq:GenHam}, one finds the so-called Bose-Hubbard model (BHM):

\begin{equation}\label{eq:BoseHubbard}
    \mathcal{H}_{BH} = \sum\limits^{N_{s}}_{j=1}\bigg[-J(b^{\dagger}_{j}b_{j+1}e^{\frac{2\imath\pi}{N_{s}}\frac{\phi}{\phi_{0}}} + \mathrm{h.c.}) + \frac{U}{2}n_{j}(n_{j}-1)\bigg]\,,
\end{equation}
where $b^{\dagger}_{j}$ ($b_{j}$) creates (annihilates) a boson, and $n_{j} = b_{j}^{\dagger}b_{j}$ is the local particle number operator for site $j$. The parameters $J$ and $U$ account for the hopping amplitude and on-site interaction when restricting the particles to occupy the lowest Bloch band in each lattice site respectively \cite{walters2013ab}. The elementary flux quantum is denoted by $\phi_{0}$ and is taken to be 1 unless explicitly stated. The BHM has been experimentally realized in experiments \cite{sherson2010single,bergschneider2018spin} with remarkable control on the parameters of the systems and read-out of the particles trapped in the lattice. With such a model, they probe the superfluid to Mott insulator transition in a highly controllable environment. This model also provides a full many-body description of the interacting bosons. However, due to its imposed restriction provided by the lattice, the Hilbert space dimensions become approachable from a numerical perspective, with techniques such as Exact Diagonalization or density matrix renormalization group (DMRG) \cite{white1992density,vidal2003efficient}.

Within the context of quantum technologies, lattice systems become an ideal playground environment to investigate atomic circuits where matter-waves are transported trough the lattice. 
Many proposals exist in this context, from fundamental studies of persistent currents in rings \cite{aghamalyan2016atomtronic,polo2022quantum} to devices that exploit these phenomena to change the behavior of matter-wave transport \cite{haug2019topological} or with applications to sensing \cite{naldesi2022enhancing}. These systems also provide us with a reference model where the many-body character of the system can be investigated and then compared to the continuum limit \cite{amico2004universality,naldesi2019rise,polo2020exact}.

\subsubsection{Quantum Phase Model}

The BHM can be effectively described with the quantum phase model (QPM) (also known as quantum rotor model)~\cite{fazio2001quantum,garcia2004variational} for large average occupation per site. This model was originally developed for understanding the behavior of arrays of Josephson junction where the phase dynamics of the superconducting order parameter is the most relevant parameter, while the fluctuations of the density are negligible for sufficiently low temperatures.  
An heuristic argument can be followed to map the BHM to the QPM. In the limit $\overline{n} = N_p/N_s \gg 1$ the number fluctuations on each site become small compared to those of the phase such that the annihilation operator can be written as $a_j\sim\sqrt{\overline{n}}e^{i\theta_j}$. The QPM can then be written as:
\begin{equation}
    H_{QPM} = \sum_{j=1}^{N_s} \left[ -2J_{QPM}\cos\left(\theta_j - \theta_{j+1} - \frac{2\pi\phi}{N_s\phi_0} \right) + \frac{U}{2} \delta n_j^2 \right].
\end{equation}
Here, $J_{QPM}=\overline{n} J$ and $\delta{n}_j = \overline{n} - n_j$ are the effective tunneling amplitude and the density fluctuations respectively, with $[\delta n_j,\theta_k] = i\hbar\delta_{j,k}$.
It's noteworthy that a careful examination of this mapping reveals significant inconsistencies~\cite{dubin1995mathematical} from the lower limit of the bosonic number operators. However, these can be taken care of by a proper definition of the basis \cite{garcia2004variational} or by a path integral treatment \cite{fisher1988quantum}.

The use of the QPM to investigate ultracold bosonic systems in optical lattices plays an important role specially in the experimental cases where the number of particles per sites are large \cite{sturm2017quantum,pezze2023stabilizing}. That is usually the case when the lattice is created by imaging setups such as DMD or SLMs that produce lattice sites of the order of diffraction limit, instead of the ones obtained by the interference of two counter-propagating laser fields that have a separation of half the wavelength.

\subsubsection{Exact solutions for interacting bosons in 1D: Bethe Ansatz and Tonks-Girardeau}\label{sec:BATonks}
An exact solution of the uniform Bose gas with delta interactions on a ring, i.e.\ the Lieb Liniger model  (\ref{eq:LLHam}) with periodic boundary conditions,  is provided by Bethe Ansatz \cite{lieb1963exact,lieb1963exactII}.  In such a case, it is possible to write the exact many-body wavefunction for the system as a linear combination of piece-wise plane waves
\begin{equation}
    \Psi(x_1,...x_N)=\sum_P a_P e^{i \sum_j k_j x_{P(j)}},
\label{eq:Psi-BA-bosons}
\end{equation}
where the wavevectors $k_j$ are denoted rapidities and $P$ corresponds to the $N!$ permutations of the ordering of the $N$ bosonic coordinates $x_1...x_N$. For each permutation $P$, we have a unique  {\it coordinate sector} corresponding to  $x_{P(1)}<x_{P(2)}<...<x_{P(N)}$. 
Within a given coordinate sector, the bosons are described by a free Schr\"odinger equation, whose solution is a many-body plane wave. Change from one coordinate sector to an adjacent one requires two particles to meet, and hence interact. The 
contact interactions can be replaced by the {\it cusp condition} 
\begin{equation}
\left(\partial_{x_{j+1}}-\partial_{x_{j}}\right)\Psi(x_{j+1}=x_j^+)= c \Psi(x_{j+1}=x_j),
\end{equation}
where, following the usual notation we have set $c=mg/\hbar^2$. The amplitudes $a_P$ are determined by imposing the cusp condition. For example, the relation between a coordinate sector P where particles, say 1 and 2 have plane wave momenta $p$ and $q$ in $P$ and another one denoted by Q, obtained by a permutation of the particles 1 and 2, is given by 
\begin{equation}
    a_Q=- a_P \frac{c-i(q-p)}{c+ i (q-p)} .
    \label{eq:BAamplitudes}
\end{equation}
Repeated use of Eq.~\eqref{eq:BAamplitudes} together with the imposition of the periodic boundary conditions, ie imposing that $\Psi(...,x_j+L,...)=\Psi(...x_j...)$ for any  $j$ yields a set of Bethe equations allowing us to determine the values of the rapidities in terms of the interaction strength and the ring circumference of length $L$. In the presence of an applied external artificial gauge field $\phi$, the many-body wavefunction has to satisfy twisted boundary conditions $\Psi(...,x_j+L,...)=e^{i 2\pi \phi/\phi_0}\Psi(...,x_j,....)$. This turns onto a set of modified Bethe equations for the rapidities which read 
\begin{equation}\label{eq:BArapidities}
k_j= \frac{2 \pi}{L} \left(I_j-\frac \phi \phi_0\right)+ 2 \sum_\ell \arctan\left(\frac{k_\ell- k_j}{c}\right).
\end{equation}
The quantum numbers $I_j$ are a set of (different) integer (semi-integers), which identify the state of the system for odd (even) number of bosons. For example, the ground state for the case of odd $N$ is determined by the `Fermi sphere' $I_j=\left\{-\frac{N-1}{2},...,0,...,\frac{N-1}{2} \right\}$. The $N$ Bethe equations need to be solved self-consistently to obtain the exact solution of the many-body wavefunction. We note that this solution holds for any non-zero interaction strength $g$, while the case of $g=0$ corresponds to a Bose-Einstein condensate where all the bosons occupy the same state at $k=0$ and cannot be described by the Bethe Ansatz. 

In the limit $g\rightarrow \infty$, corresponding to the {\it Tonks-Girardeau regime} of impenetrable bosons, the equations for the rapidities simplify since the second term on the R.H.S. of Eq.~\eqref{eq:BArapidities} vanishes. In such a case, the rapidities coincide with the wavevectors of a non-interacting Fermi gas with periodic (for odd $N$) or anti-periodic (for even $N$) boundary conditions. We notice that, while the rapidities are the same as the ones of a free Fermi gas, the amplitudes $a_P$ still ensure that the many-body wavefunction is symmetric under particle exchange, i.e.\ $\Psi(...x_i,x_j...)=\Psi(...x_j,x_i...)$  for any pair $\{i,j\}$ while for a Fermi gas one has $\Psi(...x_i,x_j...)=-\Psi(...x_j,x_i...)$. Due to the structure of its rapidities, a Tonks-Girardeau gas shares several properties, and in particular the local correlation functions, with its underlying Fermi gas. However, non-local correlations sharply differ. The Tonks-Girardeau wavefunction can be more generally written also for inhomogeneous systems as
\begin{equation}
    \Psi_{TG}(x_1,....x_N)=\prod_{1\le j<\ell\le N} {\rm sign} (x_j-x_\ell) \Psi_F(x_1,...,x_N), 
\end{equation}
with $\Psi_F$ being the wavefunction of the mapped Fermi gas in the same external potential (for details see e.g. the review \cite{minguzzi2022strongly}).

\subsubsection{Bosonizing the bosons: Luttinger liquid theory for 1D bosonic fluids}\label{sec:LuttingerBose}
The Luttinger liquid low-energy theory is a quantum hydrodynamic theory that describes systems with a linear gapless excitation spectrum. Although this requirement might seem stringent, many one-dimensional models present such properties and are now encompassed on what is known as Luttinger-liquids \cite{haldane1981luttinger,haldane1981effective}. This model was initially developed for fermionic systems under the so-called `bosonization' technique, however, it is also naturally extendable to 1D interacting bosons. Its main limitation is found in the fact that one can only describe the low-energy excitation spectrum. However, it can deal with weak to strong interacting systems and can be found to map to exactly solvable models such as the Sine-Gordon. Therefore, this theory becomes complementary to other approaches such as the GPE, Tonks-Girardeau limit, Bethe Ansatz, or DMRG, allowing us to capture the behavior of the system across different regimes of interactions. In what follows, we sketch out the steps leading to the Luttinger description of interacting bosons following the approach in~\cite{cazalilla2004bosonizing}.

From a historical perspective, this Hamiltonian was investigated already by Tomonaga~\cite{tomonaga1950remarks} in the context of 1D electron gases. Later investigated by Lieb and Mattis \cite{mattis1965exact}, Luther \cite{luther1974single}, Efetov and Larkin \cite{efetov1976correlation}, and reformulated by Haldane \cite{haldane1981effective,haldane1981luttinger}:
\begin{equation}\label{eq:BoseLuttingerHam}
    \mathcal{H}_{TLL} = \frac{\hbar vK}{2\pi}\int\limits_{0}^{L}\bigg[ (\partial_{x}\varphi (x))^{2} + \frac{1}{K^{2}}(\partial_{x}\theta (x))^{2} \bigg] \mathrm{d}x,
\end{equation}
where $\partial_{x}\theta (x)/\pi$ is the density-fluctuation field operator, conjugate to the phase operator $\varphi (x)$. The Luttinger liquid Hamiltonian is expressed in terms of two parameters: the velocity $v$ of the low-energy excitations and the dimensionless Luttinger parameter $K$ related to the compressibility of the cloud.

The Luttinger description for interacting bosons is given in terms of two bosonic fields that fulfill the following commutation relations ${[\partial_x\theta(x)/\pi,\phi(x')] \equiv -i\delta(x-x')}$. In homogeneous systems, this description leads to the following density operator \cite{cazalilla2004bosonizing}:
\begin{equation}
\rho(x) = [\rho_0+\partial_x\theta(x)/\pi] \sum_{m=-\infty}^{\infty} e^{2mi\theta(x)},
\end{equation}
where we explicitly see the form of the density fluctuations with respect to the bulk density $\rho_0$ and the exponential term that accounts for the fluctuations with momenta $q = \pm2\pi m\rho_0$. The phase field operator $\phi(x)$ appears in the bosonic field operator as $\Psi(x)=\sqrt{\rho(x)}e^{i\phi(x)}$.

In the context of ring geometries, one needs to consider periodic boundary conditions for the phase field $\phi(x+L)=\phi(x)+2\pi J$ and auxiliary field $\Theta(x+L) = \Theta(x) + \pi N$ where the latter is related to the density fluctuation field as $\Theta(x) = \theta(x)+N_0\pi x/L$ and where $N$ and $J$ are the particle number and angular momentum operators respectively and $N_0$ is the total average number of particles.

One can construct the bosonic and fermionic field operators by considering $\Psi(x)=\sqrt{\rho(x)}e^{i\phi(x)}$ and the aforementioned density operator \cite{cazalilla2004bosonizing}:
\begin{align}\label{eq:bosonized_fields}
    \Psi_B(x) \sim [\rho_0 +\partial_x\theta(x)/\pi]^{1/2} \sum_{m=-\infty}^{+\infty}{e^{-2mi\Theta(x)}e^{i\phi(x)}} ,\nonumber\\
    \Psi_F(x) \sim [\rho_0 +\partial_x\theta(x)/\pi]^{1/2} \sum_{m=-\infty}^{+\infty}{e^{-i(2m+1)\Theta(x)}e^{i\phi(x)}},
\end{align}
defining the auxiliary field $\Theta(x)\equiv \theta(x)+\pi\rho_0 x$, with subindices $B$ and $F$ indicating the bosonic and fermionic nature of the fields. We note that the fermionic operator has an extra term $e^{i\Theta(x)}$ that accounts for the anti-commutation (more details on the fermionic case will be discussed in Sec.~\ref{sec:fermilutt}). 

As an example, for bosonic particles trapped in a ring geometry, one can use the following mode expansion for the phase and density fluctuation fields:
\begin{align}
\theta(x)&=\theta_0+\frac{\pi}{\sqrt{L}}\sum_{\mu\ge1}\Phi_\mu(x)Q_\mu,\\
\partial_x\phi(x)&=\frac{2\pi}{L}(J-\Omega)+\frac{\sqrt{L}}{\hbar}\sum_{\mu\ge1}\Phi_\mu(x)P_\mu,
\end{align}
where $[Q_\mu,P_\mu]=i\hbar$ and $[\theta_0,J]=i/2$, with $\theta_0$ being the zero mode density fluctuation operator, and $J$ the angular momentum operator. The corresponding commutation relation among density and phase field operators can be rewritten as $[\frac{\theta(x)}{\pi},\partial_{x'}\phi(x')]=i\delta(x-x')$; thus, we can regard $\partial_{x}\phi(x)$ as the momentum conjugate to $\theta(x)$. Here it is assumed that the mode expansion functions $\Phi_\mu(x)$ form a complete orthonormal basis, i.e.\ $\sum_{\mu=0}\Phi^*_\mu(x)\Phi_\mu(x')=\delta(x-x')$ and $\int_0^L\mathrm{d}x\,\Phi_\mu(x)^* \Phi_{\mu'}(x)=\delta_{\mu,\mu'}$. The diagonalization of the $\mathcal{H}_{LL}$ in Eq.~\eqref{eq:LLHam} is achieved by considering $\partial_x^2\Phi_\mu(x)=-k_\mu^2\Phi_\mu(x)$.  and the effect of the boundary conditions on $\partial_x\Phi_\mu(x)|_{x=0}=\partial_x\Phi_\mu(x)|_{x=L}$ and $\Phi_\mu(0)=\Phi_\mu(L)$ leading to $\Phi_\mu(x)=\sqrt{\frac{1}{L}}\exp(i k_\mu x)$, with $k_\mu=2\pi\mu/L$ and where $\mu$ takes integer values This mode expansion to investigate ring geometries allows us to effectively calculate the dynamics of a few observables in the ring to considering the effects of weak impurities or perturbations in the system~\cite{polo2018damping}, where the phase coordinate is still a well-defined (low uncertainty) quantum variable. 

\subsection{Fermions}
Fermionic particles are anti-symmetric under particle exchange. Due to the Pauli principle, two fermions cannot occupy the same single-particle level. This major difference with respect to bosonic particles implies that their coherence properties are very different from the bosonic counterpart.  Still, in the presence of attractive interactions, two-component fermions can form bosonic pairs which can condense. Both in the repulsive and attractive cases, fermions can be strongly affected by interactions and give rise to exotic states of matter. 
Here, we review the main theoretical frameworks that have been used to compute the persistent currents in fermionic fluids.

\subsubsection{Mean-field and many-body theory for fermions in the BCS-BEC crossover}
We consider a two-component Fermi gas. In the case of attractive contact interactions with opposite spins, $s$-wave bound states are formed and the Fermi gas enters  a superfluid phase.\footnote{We will not cover here $p$-wave pairing which can occur among spin-polarized fermions with non-local interaction potentials, nor higher order type of pairing.}

Depending on the interaction strength, the nature of the bound  pairs goes from Cooper pairs to bound molecules which form a Bose-Einstein condensate. The simplest model covering the BCS to  crossover is the mean-field BCS theory. We start from the many-body grand canonical Hamiltonian ${\mathcal K}={\cal H}-\mu N $
\begin{equation}\label{eq:FermiMBHam}
    {\mathcal K} = 
    \sum_{k\sigma} \xi_k c_{k,\sigma}^\dagger c_{k\sigma} + v_0 
    \sum_{k,k',q,\sigma} c^\dagger_{k+q \sigma}  c^\dagger_{k-q\bar \sigma}   c_{k'\bar \sigma}   c_{k,\sigma},
\end{equation}
where $\xi_k=\epsilon_k-\mu$, with $\epsilon_k$ the free-particle dispersion,  $c_k$ and $c^\dagger_k$ are fermonic field operators satisfying anticommutation relations $\{c_k,c^\dagger_{k'} \}=\delta_{k,k'}$, $\sigma=\uparrow,\downarrow$ is the spin component, $\bar \sigma$ is the opposite component with respect to $\sigma$,  and $v_0$ is the contact interaction potential. The gap function is defined as 
\begin{equation}
\label{eq:gaphom}
\Delta= -v_0 \sum_{k} \langle c_{-k,\uparrow} c_{k,\downarrow} \rangle ,
\end{equation}
By taking the mean-field approximation of Eq.~(\ref{eq:FermiMBHam}) 
one gets the BCS Hamiltonian 
\begin{equation}\label{eq:FermiBCSHam}
    {\mathcal K}_{BCS} = \sum_{k\sigma} \xi_k c^\dagger_{k\sigma}  c_{k \sigma} -  \sum_{k} \Delta^* c_{-k\downarrow}   c_{k\uparrow}+ \Delta  c^\dagger_{k\uparrow}  c^\dagger_{- k \downarrow} - \Delta^2/v_0 ,
\end{equation}
Equation (\ref{eq:FermiBCSHam}) is diagonalized by a Bogoliubov transformation
\begin{eqnarray}
c_{k\uparrow}=u^{*}_k \alpha_k + v_k\beta^\dagger_{-k}, \nonumber \\
c_{k\downarrow}=-v_k\alpha^\dagger_{-k} + u_k^*\beta_{k}, 
\end{eqnarray}
where $|u_k|^2 + |v_k|^2=1$ in order to ensure the anti-commutation relations and we assumed for simplicity of notations that $u_k=u_{-k}$ and $v_k=v_{_k}$.
The amplitudes  $u_k$ and $v_k$ are determined by the Bogoliubov equations
\begin{equation}
\left(
\begin{array}{c c}
\xi_k & \Delta \\
\Delta & -\xi_k
\end{array}
\right)
\left(
\begin{array}{c}
u_k \\
v_k
\end{array}
\right) =E_k
\left(
\begin{array}{c}
u_k \\
v_k
\end{array}
\right)
\end{equation}
In the homogneous system, the resulting dispersion relation and amplitudes are
$E_k=\sqrt{\xi_k^2+ |\Delta^2|}$ and $u_k^2=1-v_k^2= (\xi_k/E_k+1)/2$. 
The gap equation (\ref{eq:gaphom}) at finite temperature reads
\begin{equation}
\label{eq:gap-bcs-hom}
\Delta= - v_0 \sum_k u_k v_k (1- 2 f(E_k)),
\end{equation}
where $f(E_k)=1/(e^{\beta E_k}+1)$ with $\beta=1/k_BT$ being the Fermi factor. The equation should be solved self-consistently to obtain the gap. Recalling that $u_kv_k=\Delta/2E_k$ one sees that Eq.(\ref{eq:gap-bcs-hom}) has an ultraviolet divergence, originating from the fact that short distances are not described well enough by the delta interaction potential. This divergence is readily cured by replacing the bare potential $v_0$  by the coupling strength $g=4 \pi \hbar^2 a_F/m$  involving the $s-$wave scattering length $a_F$ according to \cite{abrikosov1964methods}
\begin{equation}
\frac{1}{g}=\frac{1}{v_0}+\sum_k \frac{1}{2 \epsilon_k}.
\end{equation}

The above formalism can be extended to treat spatially inhomogeneous gases, subjected to an external confining potential. In such a case the grand-canonical Hamiltonian reads 
\begin{equation}\label{eq:FermiMBHamihom}
    {\mathcal K} = \sum_\sigma\int  \Psi_{\sigma}^\dagger(x) \left( -\frac{\hbar^2}{2m} \nabla^2 +V(x)-\mu\right)  \Psi_{\sigma} (x) + v_0 \int 
   \Psi^\dagger_{\uparrow}(x)  \Psi^\dagger_{\downarrow}(x)  \Psi_{\downarrow}(x)   \Psi_{\uparrow} (x) 
\end{equation}
The gap function is spatially dependent, 
\begin{equation}
\label{eq:gapinh}
\Delta(x)= -v_0 \langle \Psi_{\uparrow}(x) \Psi_{\downarrow}(x) \rangle .
\end{equation}
and is determined at the mean-field level by Bogoliubov amplitudes
\begin{equation}
\label{eq:gapinhbcs}
\Delta(x)= -v_0 \sum_\nu u_\nu(x)v_\nu(x) (1- 2 f(E_\nu)),
\end{equation}
obtained from the solution of the Bogoliubov-de Gennes equations
\begin{equation}
\left(
\begin{array}{c c}
{\cal H}(x) & \Delta(x) \\
\Delta(x) & - {\cal H}(x)
\end{array}
\right)
\left(
\begin{array}{c}
u_\nu(x) \\
v_\nu(x)
\end{array}
\right) =E_\nu
\left(
\begin{array}{c}
u_\nu(x) \\
v_\nu(x)
\end{array}
\right),
\end{equation}
The Bogoliubov amplitudes satisfy the orthonormality conditions
\begin{equation}
    \int dx \, u^*_\lambda(x) u_\nu(x)+ v^* _\lambda(x)v_\nu(x)=\delta_{\lambda,\nu},
\end{equation}
and
\begin{equation}
    \sum_\lambda u^* _\lambda(x)u_\lambda(x')+ v^* _\lambda(x)v_\lambda(x')=\delta(x-x').
\end{equation}
Also in the inhomogeneous system, the gap equation needs to be regularized. This can be achieved by using the contact pseudopotential~\cite{bruun1999BCS} or by suitable cut-off schemes~\cite{grasso2003hartree,simonucci2013temperature}.

The Bogoliubov-de Gennes equations reduce to the Ginzburg-Landau equation close to the critical temperature $T_c$ \cite{gorkov1959microscopic}, as well as to the Gross-Pitaevskii equation in the strong-coupling limit at zero temperature \cite{pieri2003derivation}. 
In the latter case, the  wavefunction $\psi(x)$ describes a condensate of tightly bound pairs of mass $2m$ and it is proportional to the gap function.

In addition to the gap function, the mean-field theory allows us also to estimate other experimentally accessible local observables such as the particle density and current in the superfluid Fermi gases 
\cite{spuntarelli2010solution}. 
For the purpose of this review, we stress that the inhomogeneous  Bogoliubov-de Gennes equations allow also for the description of the dynamics of Fermi superfluids in constrained geometries. Recent applications are the Josephson junctions descriptions in Refs.~\cite{spuntarelli2007josephson,piselli2020josephson}.

The BCS equations presented above give a qualitatively correct picture all through the BCS to BEC  crossover \cite{leggett1980diatomic,nozieres1985bose,sademelo1993bose}. However, they are not quantitative at, or close to the unitary regime of infinitely strong attractive interactions. In such a limit, quantum fluctuations play an important role and need to be accounted for by more sophisticated theories, such as theories including pairing fluctuations on top of coarse-grained Bogoliubov-de Gennes equations~\cite{piselli2023josephson,pisani2023inclusion,pisani2024critical} or density functional theories \cite{boulet2022local},  or numerical simulations, such as Quantum Monte Carlo methods \cite{astrakharchik2004equation,gandolfi2014quantum}. For more details on the BCS-theory applied to ultracold atomic gases see e.g. the review \cite{strinati2018BCS-BEC}.

\subsubsection{Fermi-Hubbard model}\label{sec:FermHubbHam}

The one-dimensional Fermi-Hubbard model represents the simplest Hamiltonian that captures the essence of strongly correlated fermions confined within a lattice structure. By providing an effective description for interacting electrons within solid-state materials, the Fermi-Hubbard model is a paradigmatic example that addresses the physical properties emerging from strong correlations ranging from quantum magnetism to superconductivity~\cite{rasetti1991hubbard,montorsi1992hubbard}. With the advent of cold atom technology, the Fermi-Hubbard Hamiltonian is one of the most emulated models in the field of cold atoms quantum simulators~\cite{tarruell2918quantum,esslinger2010fermi,lewenstein2007ultracold}. Such a model can be straightforwardly derived from the microscopic Hamiltonian by confining the fermions in an optical lattice through a suitable lattice potential, applying the tight-binding approximation and restricting to the lowest Bloch band, which holds when all energy scales are smaller than the spectral gap~\cite{essler2005one} as is the case at the typical operating conditions of cold atoms~\cite{jaksch2005cold}. For a system of $N_{p}$ SU(2) symmetric fermions residing in a ring-shaped lattice with $N_{s}$ sites and pierced with an effective magnetic flux $\phi$ reads
\begin{equation}\label{eq:FermiHubbard}
    \mathcal{H}_{FH} = -J\sum\limits_{\alpha=\uparrow,\downarrow}\sum\limits^{N_{s}}_{j=1}(c^{\dagger}_{j,\alpha}c_{j+1,\alpha}e^{\frac{2\imath\pi}{N_{s}}\frac{\phi}{\phi_{0}}} + \mathrm{h.c.}) + U\sum\limits_{j=1}^{N_{s}}n_{j,\uparrow}n_{j,\downarrow},
\end{equation}
where $c_{j,\alpha}^{\dagger}(c_{j,\alpha})$ creates (destroys) a fermion with colour $\alpha$ on-site $j$, $n_{j,\alpha} = c_{j,\alpha}^{\dagger}c_{j,\alpha}$ is the local particle number operator. The parameters $t$ and $U$ are real numbers that account for the hopping amplitude and on-site interaction respectively. As the name implies, the Fermi-Hubbard model is the fermionic counterpart to the Hamiltonian introduced in Sec.~\ref{sec:BoseHubb} having similar physics. For repulsive interactions $U>0$, a metallic-like behaviour is observed with a characteristic oscillation of the charge and spin-spin correlation functions due to the particles tending to avoid each other. In the case of attractive interactions $U<0$, the particles localize in space as they come together to form bound states with their size being restricted to two particles due to the Pauli exclusion principle. The latter behaviour is one instance of the differences arising between the fermionic and bosonic models due to the nature of their statistics. The restrictions set by the Pauli exclusion principle relax by considering SU($N$) fermions~\cite{gorshkov2010two,cazalilla2014ultracold,capponi2016phases}, which are multi-component fermions having $N$ different internal states, enhancing both the number and type of interactions: as $N\rightarrow\infty$, SU($N$) fermions mimic bosons in terms of occupation~\cite{frahm1995on}. Recent technological advancements in the field of ultracold atoms have led to the experimental realization of SU($N$)-symmetric fermionic systems  and the SU($N$) Hubbard model describing them by utilizing alkaline earth-like atoms~\cite{taie2010realization,pagano2014one,taie2012mott,scazza2014observation,sonderhouse2020thermodynamics,hofrichter2016direct}. The difference between the SU($N$) Hubbard Hamiltonian and its two-component counterpart originates from the effective interaction between two fermionic atoms. Instead of characterizing the effective interaction through the Lee-Huang-Yang potential, which is only applicable to bosons and two-component fermions, one utilizes a generalized version~\cite{yip19999zero,stamper2013spinor}. Since in the usual operating conditions of ultracold atoms the $s$-wave collisions dictate the main interactions, the interatomic potential is solely dependent on the internal angular momentum of the colliding atom pairs acquiring a higher dimensional SU(2) symmetry~\cite{cazalilla2014ultracold}. There exists a specific scenario, where all the scattering lengths of the system become equal and independent of the nuclear spin due to the absence of hyperfine coupling of the latter with the electron angular momentum, thereby enlarging the symmetry to the SU($N$) group~\cite{gorshkov2010two,capponi2016phases}. In the case of fermionic alkaline earth-like atoms, this SU($N$) symmetry emerges due to their inherent trait of having zero electron angular momentum in the ground-state. Consequently, these systems enabled the experimental realization of the SU($N$) Hubbard Hamiltonian having the following structure
\begin{equation}\label{eq:FermiHubbardSUN}
    \mathcal{H}_{SU(N)} = -J\sum\limits_{\alpha=1}^{N}\sum\limits^{N_{s}}_{j=1}(c^{\dagger}_{j,\alpha}c_{j+1,\alpha}e^{\frac{2\imath\pi}{N_{s}}\frac{\phi}{\phi_{0}}} + \mathrm{h.c.}) + U\sum\limits_{\alpha\neq\beta}^{N}\sum\limits_{j=1}^{N_{s}}n_{j,\alpha}n_{j,\beta},
\end{equation}
with $N$ being the number of colours/components. Although the SU($N$) Hubbard Hamiltonian has the same structure as the two-component model, the enhanced SU($N$) symmetry of the former, which can be readily observed through its commutation with the SU($N$) spin-permutation operators, presents richer and more exotic physics stemming from the interplay between the number of components and interactions. 

 In the continuous limit of vanishing lattice spacing, the SU($N$) Hubbard model tends to the Gaudin-Yang-Sutherland model, an integrable Hamiltonian describing the behavior of one-dimensional fermions in the continuum limit interacting via a delta-like potential thereby providing a complementary approach to the Lieb-Liniger model~\cite{sutherland1968further}. To perform the mapping, we start by defining the density of fermions in a ring of $N_{s}$ sites with lattice spacing $\Delta$ as $D=N_{p}/(N_{s}\Delta)$. Taking the filling fraction $\nu = N_{p}/N_{s}$, we have that it is related to the lattice spacing through $\nu = D\Delta$. Consequently, in the continuous regime of vanishing lattice spacing, i.e.\ $\Delta\rightarrow 0$, and finite particle density, the corresponding filling fraction must also be small. To preserve the anti-commutation relations in the continuous limits, the fermionic operators need to be rescaled in the following manner, $c_{j,\alpha}^{\dagger}=\sqrt{\Delta}\Psi_{\alpha}^{\dagger}(x_{j})$ where $x_{j}=j\Delta$ and $\Psi^{\dagger}_{\alpha}$ is the fermionic field operator. Substituting the expression for the fermionic operators into Eq.~\eqref{eq:FermiHubbardSUN}, the SU($N$) Hubbard Hamiltonian is mapped onto the Fermi gas quantum field theory $\mathcal{H}_{SU(N)} = t\Delta^{2}\mathcal{H}_{FG}-2N_{p}$ where $\mathcal{H}_{FG}$ is the Fermi gas quantum field Hamiltonian that reads~\cite{essler2005one}
\begin{equation}\label{eq:FermGasQuant}
    \mathcal{H}_{FG} = \int\limits_{0}^{L}\bigg[-(\partial_{x}\Psi^{\dagger}_{\alpha}(x))(\partial_{x}\Psi_{\alpha}(x)) + 2c\sum\limits_{\alpha < \beta}^{N}\Psi_{\alpha}^{\dagger}(x)\Psi_{\beta}^{\dagger}(x)\Psi_{\beta}(x)\Psi_{\alpha}(x)\bigg]\mathrm{d}x,
\end{equation}
with $\alpha$ and $\beta$ corresponding to the different colours interacting with a strength of $c=\frac{U}{4t\Delta}$. Through the eigenstates of Eq.~\eqref{eq:FermGasQuant} that are of the following form
\begin{equation}
    |\psi(\lambda)\rangle = \sum\limits^{N}_{\alpha_{1}\hdots \alpha_{N_{p}}}\int \chi(\mathbf{x}|\lambda)\Psi_{\alpha_{1}}^{\dagger}(x_{1})\hdots\Psi^{\dagger}_{\alpha_{N_{p}}}(x_{N_{p}})|0\rangle\mathrm{d}\mathbf{x},
\end{equation}
we observe that $\chi(\mathbf{x}|\lambda)$ are eigenfunctions of the Gaudin-Yang-Sutherland model that reads 
\begin{equation}\label{eq:GYSHam}
    \mathcal{H}_{GYS} = -\frac{\hbar^{2}}{2m}\sum\limits_{\alpha=1}^{N}\sum\limits_{i=1}^{N_{\alpha}}\frac{\partial^{2}}{\partial x_{i,\alpha}^{2}} + 2c\sum\limits_{i<j}\sum\limits_{\alpha,\beta}\delta(x_{i,\alpha}-x_{j,\beta}),
\end{equation}
where $N_{\alpha}$ denotes the number of particles with colour\footnote{The appropriate units can be found by expanding $t\Delta^{2}\mathcal{H}_{FG}$, such that the first term acquires a factor of $t\Delta^{2}$ with $U\Delta$ multiplying the second. Through the effective mass $m$, we have that $t\Delta^{2} = -\frac{\hbar^{2}}{2m}$ and that $c=U\Delta/2$ thereby giving us the correspondence between the lattice and continuous parameters~\cite{naldesi2023massive}.} $\alpha$.  An appealing feature of the Gaudin-Yang-Sutherland Hamiltonian is its integrability by Bethe ansatz for all $N$, at variance with its lattice counterpart as will be discussed in Sec.~\ref{sec:BA}. Consequently, through this mapping one can compare numerical results obtained for the SU($N$) Hubbard model with small filling fractions with the exact solution of the model. This allows a better understanding of the underlying system, which will be covered in more detail in Sec.~\ref{sec:PerCurrFerm}. Note that in the presence of an artificial field, one obtains a similar Hamiltonian to Eq.~\eqref{eq:LLHam}, in that the momenta are shifted by $\frac{2\pi\phi}{L\phi_{0}}$.

\subsubsection{Bethe Ansatz solution for the Fermi-Hubbard model}\label{sec:BA} 

Akin to its bosonic counterpart, the rich and exotic physics exhibited by the Fermi-Hubbard model stems from the interplay between its kinetic and potential terms. Due to this competition, the model is not diagonal in the Wannier or Bloch basis, barring certain limits~\cite{essler2005one,capponi2016phases}. Generally, this would entail resorting to numerical or perturbative approaches to grasp and navigate through the underlying many-body physics governing the system. However, unlike the Bose-Hubbard model, the one-dimensional fermionic Hamiltonian falls under the particular class of quantum systems exactly solvable by the Bethe ansatz equations~\cite{sutherland2004beautiful}.

 To properly understand the reasoning behind this, we need to revisit the logic behind the Bethe ansatz solution. As with all integrable models, a well-posed definition is rooted in the scattering of the system, constraining them to occur without diffraction~\cite{sutherland2004beautiful}. Specifically, Bethe ansatz solutions impose an additional requirement by necessitating that many-particle scatterings are factorizable in two-body ones in what is known as the Yang-Baxter relation~\cite{korepin1993quantum,faddeev1996algebraic}. Because of the Pauli exclusion principle, which prohibits more than two particles from occupying a given lattice site, the Fermi-Hubbard model guarantees this factorization and inherently meets the sufficient condition for Bethe ansatz integrability~\cite{deguchi2000thermodynamics}. In their seminal paper~\cite{lieb1968absence}, Lieb and Wu demonstrated the exact solvability of the model, where they built on the existing foundations of the integrability of the continuous Hamiltonian formulated by Gaudin~\cite{gaudin1967un} and Yang~\cite{yang1967some}. Naturally, due to bosonic statistics, the Bose-Hubbard model does not fulfil this stringent constraint~\cite{haldane1980solidification,choy1980some,choy1982failure}. Likewise, increasing the number of fermionic components to $N>2$, as is the case for the SU($N$) Fermi-Hubbard model, also spoils the integrability: the Pauli exclusion relaxes, allowing $N$ particles to inhabit and interact within a given site~\cite{frahm1995on,schlottmann1991spin}. Indeed, for $N\rightarrow\infty$, the relaxation is to the extent that the system emulates a bosonic one, which as discussed in Sec.~\ref{sec:BATonks}, is not integrable. Nevertheless, specific system parameters and filling fractions exist, preserving the SU($N$) Hubbard model's integrability~\cite{capponi2016phases}.
\begin{figure}[h!]
    \centering
    \includegraphics[width=0.8\textwidth]{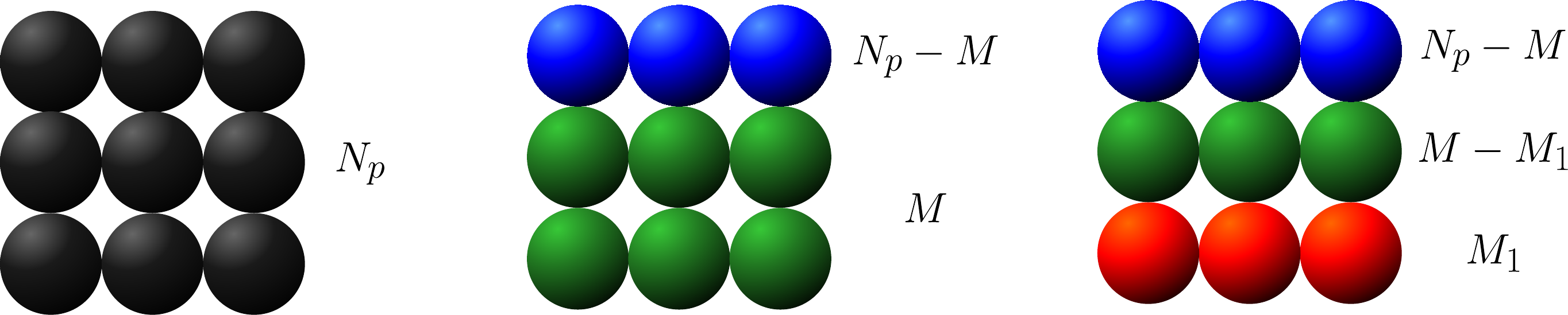}
    \caption{Diagram depicting the rationale of the nested Bethe ansatz approach for SU(3) symmetric fermions. In the first instance of the Bethe-Yang hypothesis, out of the initial $N_{p}=9$ particles (black), $N_{p}-M$ are considered to be in one colour (blue) whilst taking the rest $M=3$ to be in another (green). Another successive application, selects $M_{1}=3$ of these $M$ particles to be in a third distinct colour (red). Thus, the particle distribution in the three colours corresponds to $N_{p}-M$, $M-M_{1}$ and $M_{1}$.  }
    \label{fig:NestedBethe}
\end{figure}

Specifically, two integrable regimes fulfil the Yang-Baxter equation. The \textit{first regime} is found on taking the continuous limit with vanishing lattice spacing where model~\eqref{eq:FermiHubbard} tends to the Gaudin-Yang-Sutherland Hamiltonian in Eq.~\eqref{eq:GYSHam}~\cite{sutherland1968further,gu1989a}. A key characteristic in continuous systems is the separation of the center-of-mass dynamics and relative coordinates. On performing lattice regularization, this feature disappears accounting for the loss of integrability for the corresponding lattice theories~\cite{naldesi2022enhancing}. Such is the case of the Bose-Hubbard model as discussed in Sec.~\ref{sec:BATonks}. A more intuitive understanding is that the probability of having more than two particles meeting in the dilute limit vanishes~\cite{frahm1995on}. The \textit{second regime} is achieved for integer filling fractions of one particle per site subjected to strong repulsive interactions, described by the Lai-Sutherland Hamiltonian~\cite{lai1974lattice,sutherland1975model}. On account of the energy penalty they would incur, only one particle can inhabit a given site, and hopping is essentially virtual. Indeed, the system can be recast into a Lai-Sutherland anti-ferromagnet whose low-energy physics can be effectively described through the SU($N$) Heisenberg model obtained via second-order perturbation theory~\cite{capponi2016phases,cazalilla2014ultracold}. 

Sutherland discovered the exact solution of these models by generalizing the method formulated by Gaudin~\cite{gaudin1967un} and Yang~\cite{yang1967some} to $N$-component fermions. While for bosons, the essence of the problem lies in the exchange of momenta between two interacting particles, the situation becomes increasingly more complex when dealing with multicomponent systems having different internal states, such as SU($N$) fermions. In this case, the particles' interaction is described through a scattering matrix that accounts for transmission and reflection, which simply put corresponds to whether an exchange of colours between the particles has occurred. To tackle the two-component case with $M$ flipped spins, Gaudin and Yang employed a clever strategy of separating the problem into the spatial and spin parts. The spatial part involves taking the $N_{p}$ particles to be indistinguishable and employing the cusp condition like the bosonic case discussed in Sec.~\ref{sec:BATonks}. For the spin part, they realized that because of the fermions' anti-symmetric nature, it suffices to calculate the amplitude wavefunctions for a specific set of positions of the $M$ down spins. Consequently, the problem becomes more manageable by considering a fictitious spin model describing $M$ down spins residing on a chain of length $N_{p}$~\cite{pecci2022coherence}. Through this observation, Yang proposed an ansatz for the spin wavefunction $\varphi(y_{1},y_{2},\hdots,y_{M})$ corresponding to the amplitudes of the inhomogenous XXX Heisenberg model, with a newly introduced spin rapidity $\Lambda_{\alpha}$  analogous to the quasimomenta $k_{j}$ introduced in the spatial wavefunction (see~\cite{takahashi1999thermodynamics,essler2005one, deguchi2000thermodynamics} for a more detailed approach).  This so-called Bethe-Yang hypothesis can also be extended to multicomponent fermions, as demonstrated by Sutherland~\cite{sutherland1968further,sutherland1975model}.
Upon separating the $N_{p}$ particles into the $N$ different colours, the SU(3) case being depicted in Fig.~\ref{fig:NestedBethe}, one applies the Bethe-Yang hypothesis
although in a slightly different form (see~\cite{sutherland1968further}). Repeating the nested ansatz $N-1$ times, to account for the extra colours in the system, each successive application of the Bethe-Yang hypothesis sees the introduction of a new colour $M_{r}$ having an associated spin rapidity $\Lambda^{(r)}$.

The last step in constructing the Bethe equations is to impose periodic boundary conditions, i.e.\ $\psi(\hdots, x_{j},\hdots) = \psi(\hdots, x_{j}+L,\hdots) $. To account for the presence of the flux, one needs to employ the twisted boundary conditions outlined in Sec.~\ref{sec:BATonks}~\cite{shastry1990twisted}. For SU($N$) symmetric fermions modeled by the Gaudin-Yang-Sutherland Hamiltonian the Bethe equations read~\cite{frahm1995on,chetcuti2021persistent}
\begin{equation}\label{eq:BAcharge}
    e^{\imath(k_{j}L-\Phi)} = \prod\limits_{\alpha = 1}^{M_{1}}\frac{k_{j}-\Lambda_{\alpha}^{(1)}+\imath c}{k_{j}-\Lambda_{\alpha}^{(1)}-\imath c}, \hspace{2mm} j=1,\hdots , N_{p} 
\end{equation}
\begin{equation}\label{eq:BAspin}
    \prod\limits_{\substack{\beta = 1\\\beta\neq\alpha}}^{M_{r}}\frac{\Lambda_{\alpha}^{(r)}-\Lambda_{\beta}^{(r)}+2\imath c}{\Lambda_{\alpha}^{(r)}-\Lambda_{\beta}^{(r)}-2\imath c} = \prod\limits_{\beta = 1}^{M_{r}-1}\frac{\Lambda_{\alpha}^{(r)}-\Lambda_{\beta}^{(r+1)}+\imath c}{\Lambda_{\alpha}^{(r)}-\Lambda_{\beta}^{(r+1)}-\imath c}\cdot \prod\limits_{\beta = 1}^{M_{r}+1}\frac{\Lambda_{\alpha}^{(r)}-\Lambda_{\beta}^{(r+1)}+\imath c}{\Lambda_{\alpha}^{(r)}-\Lambda_{\beta}^{(r+1)}-\imath c}, \hspace{2mm} \alpha = 1,\hdots , M_{r},
\end{equation}
for $r=1,\hdots, N-1$, with $M_{0}=N_{p}$ and $\Lambda_{\beta}^{(0)}=k_{\beta}$ with\footnote{In the absence of a magnetic flux, one recovers the Bethe ansatz equations corresponding to a field-free Hamiltonian. Note that the flux dependence is incorporated solely into the first equation as it is colour independent. If this were not the case, then the solvability of the model could not be guaranteed~\cite{amico1998one}.} $\Phi=2\pi\phi$. The number of particles in any given colour is given by $M_{r}$ having the associated charge and spin rapidities denoted by $k_{j}$ and $\Lambda_{\alpha}^{(r)}$. The power of the Bethe-Yang hypothesis is the reduction of the problem's dimensionality, becoming smaller on each successive application, earning it the moniker of \textit{nested Bethe ansatz}. On top of the $N_{p}$ equations as in the bosonic/one-component case, there are additional $\sum_{r}M_{r}$ equations for $N$-component systems to account for the various internal states. For the lattice model, the Bethe equations are of the same form albeit slightly different since $k_{j}\rightarrow\sin k_{j}$, which originates during the lattice regularization of the continuous system due to the different dispersion relations of the two theories~\cite{amico2004universality,essler2005one}. Considering the logarithmic form of the Bethe equations and defining $\theta_{z}(x-y) = -2\arctan(\frac{x-y}{z})$, we find
\begin{equation}{\label{eq:LogBAcharge}}
k_{j}L -\sum\limits_{\alpha = 1}^{M_{1}}\theta_{c}(k_{j}-\Lambda_{\alpha})  =2\pi I_{j}+\Phi,  
\end{equation}
\begin{equation}\label{eq:LogBAspin1}
-\sum\limits_{j=1}^{N_{p}}\theta_{c}(\Lambda_{\alpha}-k_{j}) - \sum\limits_{a=1}^{M_{2}}\theta_{c}(\Lambda_{\alpha}-\lambda_{a}) + \sum\limits_{\beta=1}^{M_{1}}\theta_{2c}(\Lambda_{\alpha}-\Lambda_{\beta}) = 2\pi J_{\alpha},
\end{equation}
\begin{equation}\label{eq:LogBAspin2}
-\sum\limits_{\alpha=1}^{M_{1}}\theta_{c}(\lambda_{a}-\Lambda_{\alpha}) + \sum\limits_{b=1}^{M_{2}}\theta_{2c}(\lambda_{a}-\lambda_{b}) = 2\pi L_{a},
\end{equation}
which sees the introduction of the quantum numbers pertaining to the charge $I_{j}$ and spin rapidities $\{J_{\alpha}, L_{a}\}$ parameterising them.\footnote{For the sake of convenience, we opted to denote $\Lambda_{\alpha} = \Lambda_{\alpha}^{(1)}$ and $\lambda_{a} = \Lambda_{\alpha}^{(2)}$.} Manipulation of these integer or half-odd integer numbers, mutually distinct by construction, and centered around zero for the ground-state, generates the whole spectrum ($E=-2\sum_{j}\cos(k_{j})$ or $E = \sum_{j}k_{j}^{2}$ for lattice and continuous systems respectively), which plays a pivotal role in a deeper comprehension of the system's behaviour~\cite{takahashi1999thermodynamics,deguchi2000thermodynamics}.

When subjected to attractive interactions, the system sees the formation of bound states in its ground and low-lying excited states. As discussed in Sec.~\ref{sec:BATonks}, the spectral parameters of the Bethe ansatz equations admit complex solutions due to the presence of these bound states, which are accounted for through the \textit{string hypothesis} formulated by Takahashi~\cite{takahashi1970many,takashi1971one}. In the case of SU($N$) fermions, the Pauli exclusion principle only permits bound states composed of $N$ particles with different colours.  Consequently, this translates into having complex solutions consisting of $k$-$\Lambda$ strings (combination of one spin and $N$ charge rapidities) corresponding to a bound state of length $m$ ranging from 2 to $N$, accompanied by $\Lambda_{\alpha}$ strings (involving solely spin rapidities) pertaining to bound states albeit of smaller lengths~\cite{essler2005one,takahashi1999thermodynamics}. It is essential to note that whenever possible, the system will opt to form $N$-bound states as these are more favourable energetically. As such, in both of these string solution types, the real part is always given by the limiting spin rapidity, i.e.\ the one related to the last colour participating in the bound state, because of the nested Bethe ansatz's nature (see~\cite{chetcuti2023probe,chetcuti2023persistent} for more details). 

 In contrast to its repulsive counterpart, the attractive SU($N$) Hubbard model is not exactly solvable. The constraint of having one particle per site is no longer met, meaning the system does not behave as a Lai-Sutherland anti-ferromagnet. Thus, the scattering matrix does not satisfy the Yang-Baxter equation, and the system is classified as `diffractive'. Naturally, the SU(2) case is an exception as it intrinsically fulfils the requirement due to the particles' statistics. Another difference between the repulsive and attractive regimes lies in the mapping between the continuous and lattice theories. Whilst the lattice regularization can be performed for any arbitrary interactions, subtleties arise for the attractive case due to the lattice spacing $\Delta$ introduced upon discretizing the Gaudin-Yang-Sutherland Hamiltonian~\cite{chetcuti2023probe}. In particular, problems emerge if the correlation length of the bound states is smaller than $\Delta$, which in turn eclipses its decay. On account of this, a quantitative comparison between the continuous and the lattice one with small filling fractions is not feasible unless the interaction of the latter $U$ is suitably rescaled. Although this has been investigated thoroughly for bosons~\cite{naldesi2019rise,oelkers2007ground}, it still needs to be properly analysed for fermions. This rescaling issue makes comparing numerical results with those obtained via Bethe ansatz challenging. Once more, we note that there is a workaround for the SU(2) case as one can utilize the Lieb-Wu Bethe ansatz to monitor the numerics.

\subsubsection{Bosonizing fermionic systems}\label{sec:fermilutt}

The Luttinger description relies on writing the fermionic field operator in terms of bosonic operators using a low-energy description around the Fermi momenta where the spectrum can be linearized. This enables us to treat both particle statistics by employing a common approach. In particular, to describe the fermionic statistics, one just needs to be careful with the anti-commutation of particles, including a minus sign when two fermions exchange positions (see Eq.~\eqref{eq:bosonized_fields}). 

For systems with spin, one needs to bosonize each species separately. To simplify their description, one introduces a spin and charge combination of the field operators such that $\phi_{\rho,\sigma} = (\phi_\uparrow\pm\phi_\downarrow)/\sqrt{2}$ and $\theta_{\rho,\sigma} = (\theta_\uparrow\pm\theta_\downarrow)/\sqrt{2}$. The resulting Hamiltonian can be written as:
\begin{equation}
    H = H_\rho + H_\sigma + \frac{2g_1}{(2\pi\alpha)^2}\cos(2\sqrt{2}\phi_\sigma(x)),
\end{equation}
where $\alpha$ is a short distance cut-off, and $g_1$ is the backwards scattering \cite{schulz1991correlated} with:
\begin{equation}
    H_\nu = \frac{1}{2\pi}\int dx \left[  u_\nu K_\nu(\pi\Pi)_\nu^2 + \frac{u_\nu}{K_\nu} (\partial_x\phi_\nu)^2 \right],
\end{equation}
where the fields are:
\begin{align}
    \phi_\nu(x) &= -\frac{i\pi}{L}\sum_{q\neq0}\frac{1}{q}e^{-\alpha |q|x/2 -iqx}[\nu_+(q) + \nu_-(q)] - N_nu \frac{\pi x}{L} ,\nonumber\\
    \Pi_\nu(x) &= \frac{1}{L}\sum_{q\neq0}e^{-\alpha |q|x/2 -iqx}[\nu_+(q) - \nu_-(q)] + J_nu /L ,
\end{align}
where $\rho(q)$ and $\sigma(q)$ are the Fourier components of the charge and spin density operators for the right and left movers fermions. Note that the operators $[\phi_nu(x),\Pi_mu(y)] = i\delta_{\nu,\mu}\delta(x-y)$ fulfill bosonic commutation relations. 

The study of persistent currents in the context of the Luttinger liquid theory was originally investigated by Loss \cite{loss1992parity} and later by Schmeltzer \cite{schmeltzer1993persistent}. The current is calculated by taking the derivative of the free energy with respect to the effective flux introduced in the system (as discussed in Sec.~\ref{sec:Methods}). The free energy can be obtained through the corresponding partition function of the charge part within the Luttinger description, giving access to the current as a function of interactions as well as temperature effects \cite{schmeltzer1993persistent}:
\begin{equation}
    I = \frac{\partial F}{\partial\phi} = \frac{V K 8\pi}{L} \frac{\sum\limits_{n=-\infty}^{\infty}(n+\phi)e^{-\beta 4\pi VK(n+\phi)^2/L}}{\sum\limits_{n=-\infty}^{\infty}e^{-\beta 4\pi VK(n+\phi)^2/L}},
\end{equation}
with $V$ and $K$ corresponding to the Luttinger and velocity charge parameters within the Luttinger theory, $\beta = 1/k_BT$ for a Fermi-Hubbard model of size $L$.

\subsection{`Bloch bands' and parity dependence of  persistent currents}\label{sec:bloch}

In this section,  we will discuss how the magnetic flux periodicity and patterns of persistent currents indeed arise from the general properties of the system, like local particle conservation and symmetry of the wave function.  To be specific, the features summarized in this section hold for a generic particle translational invariant interaction and in the presence of a static disorder.   

We consider a quasi-one-dimensional system confined in a toroidal-shape potential and pierced by an effective magnetic field described by Eq.~\eqref{eq:GenHam}. Here, we analyse the dependence of the energy on the magnetic flux that, in turn, determines the periodicity of the persistent current $I(\phi)$. To this end, we resort to the property that the vector potential can be gauged away from the Hamiltonian by redefining the fields as: $\psi'= e^{i \int \mathbf{A}(\mathbf{r}) \cdot d\mathbf{r}} \psi$ implying that $\psi (\mathbf{r}+L)= \psi (\mathbf{r}) e^{i \phi/\phi_0}$. As also  noted in Sect.\ref{sec:PerCurrDev}, the theory, then, is formally similar to that of electrons moving in a periodic potential in which $\phi/\phi_0$ plays the role of the crystal momentum $k$ \cite{imry2002intro,buttiker1983josephson,leggett1991dephasing}. 
Clearly,  such periodicity changes if the flux quantum $\phi_0$ changes. Remembering that the flux quantum depends on the effective mass of the flowing particles, we arrive to the conclusion that the periodicity of the persistent current can indeed convey information (encoded in the effective mass) about the nature of the quantum fluid. This scenario provides a generalization of the halved periodicity of the  $I(\phi)$ response in superconductors that is a result of the formation of Cooper pairs \cite{byers1961theoretical,onsager1961magnetic}.  

For fermionic systems, the landscape of the persistent current is parity dependent. This property was originally noted by Leggett, and can be demonstrated through a variational argument~\cite{leggett1991theorem}. Accordingly, the many-body wave function for a system of {\it spinless} fermions with $\phi\neq 0$ is assumed of the form 
\begin{equation}
\psi_v(\mathbf{r})=e^{i\chi({\mathbf{r}})} \psi_0(\mathbf{r}),
\end{equation}
in which $\psi_0(\mathbf{r})$ is the exact ground-state of the system at $\phi=0$ and $\chi$ is a completely symmetric function of its indicies satisfying the conditions that $\chi (0) = 0$ and $\chi(2\pi) = \phi/\phi_{0}$. The eigenenergy of  $\psi_v$ is
\begin{equation}
E_v=E(\phi=0)+ {{N_{p}\hbar^2}\over{2m}} \int {|\psi_0(\mathbf{r})|^2} |\nabla \chi({\mathbf{r}})|^2,
\end{equation}
The next step of the argument relies on the single-valuedness of the many-body wavefunction $\psi_0$: if one of the particles is transported along a closed loop (keeping the other particles frozen),  $\psi_0$ must not change. On the other hand, such an operation would involve the exchange of $N_{p}-1$ particles and because of the anti-symmetry of the wavefunction, $N_{p}-1$ sign changes. For even $N_{p}$, the single-valuedness of the wave function can hold only if $\psi_0$ has (at least) a further node (on top of the $N_{p}-1$ nodes coming from Pauli principle): $\psi_0(\mathbf{\overline r})=0$, where $\overline r$ defines a `non-symmetry dictated nodal surface'.  The phase $\chi({\mathbf{r}})$ can be chosen in the form of a phase twist, namely with a phase discontinuity concentrated at $\overline r$. This choice of $\chi({\mathbf{r}})$ can be made without resulting in an increase in energy (due to the aforementioned property  $\psi_0(\mathbf{\overline r})=0$). In this case, the variational energy results to be $E_v=E(\phi=0)$ and then an upper bound of the exact energy $E(\phi)$: $E(\phi=0)\ge E(\phi)$. Following a similar argument, it can be concluded that, for odd $N_{p}$, $E(\phi=\pi)\ge E(\phi)$. Eventually, Loss proved this conjecture in the Luttinger liquid theory framework through the bosonization method~\cite{loss1992parity}. By resorting to specific inequalities fulfilled by the path integral representation of the partition function of the system, lower bounds to the many-body energy were obtained by Waintal et al.~\cite{waintal2008persistent}. Combining the Leggett and Waintal results, the ground state energy is constrained as 
\begin{eqnarray}
E(\phi=0)\le  E(\phi) \le E(\phi=\pi)\, , \quad N_{p}\, odd , \\
E(\phi=\pi)\le  E(\phi) \le E(\phi=0)\, , \quad N_{p}\, even \, .
\end{eqnarray}
We note that the above results also hold at a finite temperature, replacing the ground state energy with the free energy $F$.
For fermions with spin, the Leggett arguments cannot be applied and only the lower bounds hold:
\begin{eqnarray}
F(\phi=0)\le  F(\phi) \, , \quad N_{p\uparrow}, N_{p\downarrow} \quad odd, \\
F(\phi=\pi)\le  F(\phi) \, , \quad N_{p\uparrow}, N_{p\downarrow} \quad even.
\end{eqnarray}
These parity effects also affects the fermionic current as it causes parabolas of different angular momenta $\ell$ to traverse each other at $\phi = \frac{j}{2}\phi_{0}$ or $\phi=j\phi_{0}$ for integer $j$ depending on whether $N_{p}$ is odd or even respectively. Essentially, the landscape of the current for spinless fermions indicates that they act as a diamagnet (paramagnet) for odd (even) $N_{p}$~\cite{leggett1991dephasing} in the presence of an effective magnetic flux. For electrons/two-component fermions, i.e. having a spin-up and spin-down, this amounts to saying that the current has a diamagnetic (paramagnetic) character flow in a system consisting of $N_{p} = 4\nu +2$ ($N_{p}=4\nu$) particles for $\nu\in \mathbb{Z}^{+}$ and equal population per component~\cite{kusmartsev1994strong,waintal2008persistent}. Systems with a larger number of components will be addressed in Sec.~\ref{sec:ResFerm}. For bosons, there is no parity effect as the current is always diamagnetic due to the lack of Pauli exclusion principle. Another noteworthy feature is that the circulation of the particles, be they moving clockwise or anti-clockwise, is captured through the orientation of the persistent current on account of its parity.

\subsection{Theory for read-out persistent currents in cold atoms systems }\label{sec:Interference}

The persistent current of ultracold atoms is intrinsically related to the particles’ momenta in the condensate. Monitoring the persistent current typically involves measurement of the momentum distribution or interference that highlights the momentum occupation of the particles. From the experimental point of view, this can be done in different ways. 

Recent studies have introduced a minimally destructive method to detect persistent currents in a ring-shaped Bose-Einstein condensate (BEC) by coupling it to an optical cavity driven by light carrying orbital angular momentum. This setup allows for real-time, in-situ measurement of the condensate's rotation by analyzing the cavity transmission spectra, which are sensitive to both the magnitude and sign of the winding number. The technique has been demonstrated to detect not only persistent currents but also soliton dynamics within the condensate~\cite{pradhan2023cavity, pradhan2024cavity,pradhan2024detection}. Additionally, an approach involving coupling a ring-shaped condensate to a straight waveguide was also considered as a way to investigate currents in these systems. By observing the dynamics of atoms transitioning between the ring and the guide, one can infer the winding number of the condensate through the specific population transfer in the guide~\cite{safaei2019monitoring}. Further techniques monitor the precession or Doppler shift of collective acoustic modes in the BEC. Preparing excited counter-propagating sound waves, under the presence of a rotating condensate, leads to a precession of the density modulation that can be monitored by in-situ measurements~\cite{kumar2016minimally,marti2015collective}. %
%
%
%
%
Here we focus on the most common techniques implemented experimentally in the past years, time-of-flight (TOF) and interference dynamics after expansion. 

Already in the early experiments of ultracold atoms, the momentum distribution provided us with one of the key measurements of the condensation of particles \cite{anderson1995observation,davis1995bose}. Time-of-flight measurements are performed by releasing the atoms from their confinement and letting them fall under the action of gravity. In general, it is assumed that particles do not interact due to their fast expansion, and therefore, this process is modelled as a free particle kinetic expansion. In most of the experiments, the TOF image is achieved after $10-20$ ms releasing time. Two predominant measurements exist (i) absorption imaging, for which a laser beam probes the atomic cloud while falling and (ii) fluorescence measurements of the atoms after they are excited by the resonant beam. Both of these techniques are destructive measurements, although other approaches exist~\cite{Lettobservation1988, westbrooklocalization1990, Kohnsonline1993, courtoisrecoil1994}; these are typically quite involved and not as reliable as TOF-based measurements.In the following, we provide the theoretical analysis of the experimentally realizable protocols for TOF, also known as the homodyne protocol and self-heterodyne protocols. The specific features emerging in bosonic and fermionic persistent currents will be discussed in Secs.~\ref{sec:PerCurrBose} and~\ref{sec:PerCurrFerm}.

The theoretical approach to TOF accounts to compute the momentum distribution of the system at the instant of time in which the trap is open $t=0$~\cite{read2003free}. The connection between the initial momentum distribution of the particles in the trap and the long-time TOF momentum distribution in position space can be understood in analogy of the Fraunhofer diffraction of light. Here, we sketch such a connection in the lattice (see \cite{readfree2003} for the continuous case). The one-body correlator is defined as $n(\textbf{r},\textbf{r}',t) \equiv  \langle \Psi^{\dagger}(\textbf{r},t)\Psi (\textbf{r}',t) \rangle$ with $\textbf{r}$ being the coordinate space, and $\Psi^{\dagger}(\textbf{r})$ and $\Psi (\textbf{r})$ are the creation and annihilation operators. 

The momentum distribution can then be calculated by taking its Fourier transform,
\begin{equation}\label{eq:appx2}
    \langle n(\textbf{k})\rangle = \int e^{\imath \textbf{k}(\textbf{r}-\textbf{r}')} \langle \Psi^{\dagger}(\textbf{r})\Psi (\textbf{r}') \rangle \textrm{d}\textbf{r}\textrm{d}\textbf{r}',
\end{equation}
where $\textbf{k}$ is the momentum. One can verify that $\lim_{t\rightarrow\infty}\langle n(\textbf{r},\textbf{r}',t) \rangle \approx \langle n(\textbf{k})\rangle$, by taking the limit $t\rightarrow\infty$ of Equation~\eqref{eq:wann} and performing a Taylor expansion. In a mean-field description the interference pattern can be understood in terms of the Fourier expansion of the order parameter. For atoms in a ring-shaped potential, focused in the $x$-$y$ plane, one can write the wavefunction in polar coordinates $\mathbf{\Psi}_\ell(r,\theta)=\sqrt{n(r)}\exp(i\ell\theta)$. Its Fourier transform is related to the Bessel function of order $\ell$ as ${\mathbf{\Psi}_\ell(k,\theta_k)\approx\exp(i\ell\theta_k))J_\ell(kR)}$ for rings with a ring radius $R$ such that their ring width, $\omega$, is $R\ll\omega$ \cite{berman2004irregular}. The momentum distribution will be obtained through the density of particles measured for a time-of-flight expansion $t_{TOF}>mR^2/\hbar$, and is characterized by a hole that grows as the angular momentum $\ell$ increases~\cite{allman2023equilibrium}.

\begin{figure}
    \centering
    \includegraphics[width=0.32\linewidth]{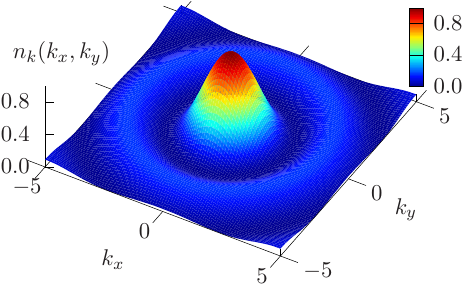}%
    \includegraphics[width=0.32\linewidth]{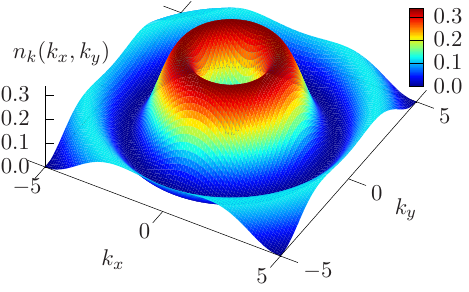}%
    \includegraphics[width=0.32\linewidth]{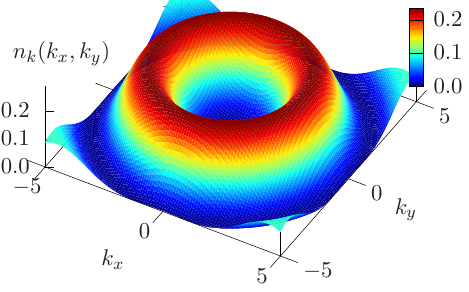}
    \put(-460,95){(\textbf{a})}
    \put(-300,95){(\textbf{b})}
    \put(-145,95){(\textbf{c})}
    
    \includegraphics[width=0.32\linewidth]{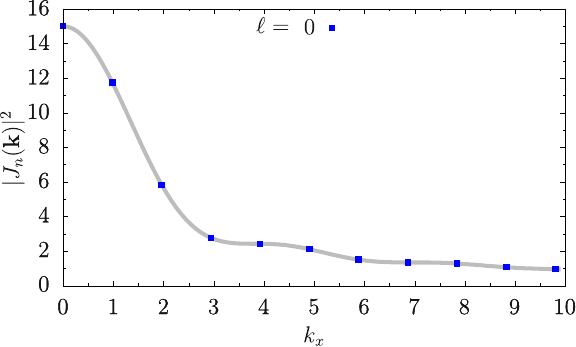}%
    \includegraphics[width=0.32\linewidth]{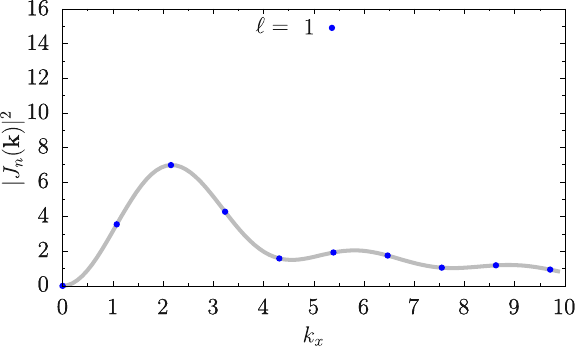}%
    \includegraphics[width=0.32\linewidth]{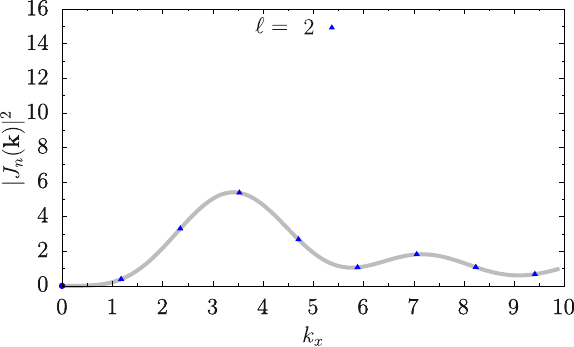} %
    \caption{Momentum distribution (top) and its cross-section (bottom) for non-interacting particles for \textbf{(a)} zero and \textbf{(b)}/\textbf{(c)} non-zero angular momentum as per Eq.~\eqref{eq:nkbessel}.  For bosonic systems, all particles occupy the same momenta and therefore, the system goes from a maximum to a hole as soon as the system acquires angular momentum. However, for fermions this is only the case when the number of particles is equal to the number of components as will be discussed in Sec.~\ref{sec:PerCurrFerm}. Figures taken from~\cite{chetcuti2022interference}.}
    \label{fig:nkbessel}
\end{figure}

For lattice systems a similar analysis can be done by expanding these operators in the single-band basis of Wannier functions $w_{j}(\textbf{r})$ where we have that $\Psi (\textbf{r}) = \sum_{j}^{L}w_{j}(\textbf{r}-\textbf{r}_{j})c_{j}$, with $w_{j}(\textbf{r}-\textbf{r}_{j})$ being the Wannier function localised at site $j$ and position coordinate space $r_j$. These Wannier functions can be further simplified considering their Taylor expansion around the lattice minima as:
\begin{equation}\label{eq:wann}
    w_{j}(\textbf{r}-\textbf{r}_{j},t) = \frac{1}{\sqrt{\pi}}\frac{\eta_{j}}{\eta_{j}^2 + \imath\omega_{0}t}\exp\bigg\{-\frac{(\textbf{r}-\textbf{r}_{j})^{2}}{2(\eta_{j}^{2} + \imath\omega_{0}t)} \bigg\},
\end{equation}
where $\eta_{j}$ provides the width and $\omega_{0} = \frac{\hbar}{m}$ with $\hbar$ and $m$ denoting Planck's constant and the particles' mass respectively.
Moreover, we note that, within the lattice description, the final momentum distribution is mainly affected by the correlations the particles had before their release from the trap, and the spatial evolution of the falling atoms is encoded on the expansion of the Wannier functions.  
For a condensate flowing along ring-shaped  circuits, TOF  displays a  characteristic shape in which the density around $\textbf{k}$ is suppressed. The TOF image (taken from the top of the expanding condensate, along the falling direction) shows a donut shape which is very different from a bell-shaped image in absence of circulation~\cite{amico2005quantum}, see Fig.\ref{fig:nkbessel}. The value of the donut radius results to change in discrete steps  corresponding to the quantization of angular momentum of the condensate~\cite{moulder2012quantized, ryu2014creation,wright2013driving,murray2013probing}.
In the non-interacting case, the momentum distribution can be calculated exactly, giving rise to a simple expression that can allow us to understand the existence of the maximum and hole of the density distribution observed in experiments. In particular, $n(\textbf{k})$ is given by \cite{pecci2022single,chetcuti2022interference}:
\begin{equation}\label{eq:nkbessel}
    \langle n(\textbf{k})\rangle \propto \frac{1}{N}\sum\limits_{\{q\}} |\mathcal{J}_{q}(\textbf{k})|^{2},
\end{equation}
where $\mathcal{J}_{q}$ is the discretized \textit{q}-th order Bessel function of the first kind, the lattice momenta $k=\frac{2\pi}{N_s} q$ with $q$ being the energy occupation number, $N_s$ is the number of sites and $\mathbf{k}$ is the momenta associated to the real-space coordinate $\mathbf{R} = \mathbf{r}-\mathbf{r}_j$. The proportionality comes from neglecting the overall envelope provided by the Wannier function that depends on the specifics of the experimental implementation of the lattice. 

In \fref{fig:nkbessel}, we show that momentum distribution \eqref{eq:nkbessel} for the momentum occupations $q=\{0,1,2\}$ corresponding to a maximum and two holes. The bottom row shows a cut on $k_x$ with the corresponding Wannier functions. We note that for fully condensed bosonic systems, particles occupy the same momentum, therefore, the transition from maximum to hole implies a change of momenta. For fermionic systems or multicomponent systems the situation is different as multiple momenta can be occupied by the particles. In the non-interacting case, the momentum distribution will correspond to the sum expressed in \eqref{eq:nkbessel}, instead of a single summand. 

In the self-heterodyne protocol, the atomic cloud containing a persistent current is co-expanded with a central cloud that provides a reference phase with which to interfere with~\cite{eckel2014interferometric,corman2014quench,mathew2015selfheterodyne}.  In experiment, the atomic density is measured in single-shot experiment consisting of thousands of particles that reveal the interference pattern between the external ring and the central cloud after $\approx1$ms of free expansion.

The interference between the two clouds can be again understood in terms of the mean-field order parameter. Considering that the two clouds are phase coherent, they will interfere depending on their relative phase, $\delta\varphi$, giving a density interference of $n = |\mathbf{\Psi}_R + \mathbf{\Psi}_C|^2 = n_R + n_C +2\sqrt{n_{R,\ell}n_C}\cos(\delta\varphi)$, with indices ${R,\, C}$ indicating the ring and center clouds respectively, where $\ell$ corresponds to the angular momentum. In the case of interference between two clouds, one wants to avoid the far-field limit as the main contribution will come from the densities and the interference term will be relatively small and thus, hard to detect. To obtain the spiral patterns observed in experiments one needs to consider both the radial and azimuthal parts of the interference. The wavefunction in polar coordinates with radius $r$ and angle $\theta$ is described as $\mathbf{\Psi}_\ell(\mathbf{r},t)=\sqrt{n(r,t)}\exp(i\varphi_\ell(r,t))\exp(i\ell\theta)$. Here, we have the ring wavefunction, $\mathbf{\Psi}_{R,\ell}$ and the center cloud, $\mathbf{\Psi}_{c}$, which combined gives the interfered density 
\begin{equation}\label{eq:meanphase}
|\mathbf{\Psi}_{R,\ell} + \mathbf{\Psi}_C| = n_{R,\ell} + n_C +2\sqrt{n_{R,\ell}n_C}\cos(\delta\varphi+\ell\theta)
\end{equation}
where the relative phase $\delta\varphi=\varphi_{R,\ell}(r,t)-\varphi_{C}(r,t)$ depends on radius $r$~\cite{eckel2014interferometric,mathew2015selfheterodyne,roscilde2016quantum}. For a fixed $r$, along the azimuthal direction $\theta$ we observe $\ell$ maxima and, as the radius $r$  changes, the position of the maxima shifts, producing a spiral pattern in the density of the expanded cloud~\cite{allman2023equilibrium}. By counting the number of spirals of the interference pattern, one can readout the phase winding $\ell$ from the experimental density images~\cite{eckel2014interferometric}.

However, the previous explanation for the spiral pattern uses a single-particle description. This logic is valid for non-interacting or mean-field description of quantum systems, yet fails for highly entangled many-body systems, which require a many-body description of the wavefunction~\cite{haug2018readout}. For example, the single-particle model cannot describe entangled NOON states, which is a superposition of all particles having either angular momentum $\ell=0$ or $\ell=1$. Such entangled states can naturally appear as the ground-state of interacting Hubbard Hamiltonians with flux. 

To study the emergence of spiral patterns in the many-body correlated regime, we resort to a second quantization approach. Accordingly, observables are obtained by ensemble averaging, implying that the system is subject to many experimental repetitions~\cite{lewenstein2012ultracold}. 
Let us give the general idea first: In the many-body picture beyond mean-field, the initial state is the Fock-state $|\phi\rangle = |\phi\rangle_{R}\otimes|\phi\rangle_{C}$, where the ring and reference clouds are tensor products, with undefined relative phase, and well defined atom number in ring and reference. During time-of-flight, both clouds are freely expanded in space and start to overlap.  The relative phase only emerges during measurement, where the positions of the atoms is recorded. As the wavefunctions of both clouds overlap, there is uncertainty whether the measured atom is from the reference or ring clouds, thus gaining uncertainty in atom number and certainty in relative phase. Most notably, one can see that a relative phase offset emerges during the measurement, which is randomly chosen in every experimental run. Now, if one averages the density over many experiments, the interference pattern of the spiral is washed away as different {\it experiments} and thus phase offsets are uncorrelated to each other.

Thus, observables that have full knowledge of the particles' origin from either center or ring cannot reveal any meaningful interference pattern on average. For example, single-body observables cannot reveal the spiral patterns when averaging over many experimental runs~\cite{haug2018readout}. In order to observe phase-related effects, we need an observable that does not fully account for the particle's origin and therefore the uncertainty of the phase is reduced. We resort to a two-body correlator
\begin{equation}\label{eq:ax1}
    G(\textbf{r},\textbf{r}',t) = \sum\limits_{\alpha,\beta}^{N}\langle n_{\alpha}(\textbf{r},t)n_{\beta}(\textbf{r}',t)\rangle ,
\end{equation}
with the density operator being defined as $n(\textbf{r},t) = \psi^{\dagger} (\textbf{r},t)\psi (\textbf{r},t)$ and $\psi^{\dagger} = (\psi^{\dagger}_{R} + \psi^{\dagger}_{C})$ being the field operator of the whole system of the ring, $R$ and the center $C$ respectively. Initially, the ring and center are decoupled and can therefore be written as $|\phi\rangle = |\phi\rangle_{R}\otimes|\phi\rangle_{C}$. However, from this correlator the terms corresponding to interference between the two clouds are only those having cross-terms between the ring and the center. Therefore, to obtain the qualitative behavior of the interference patterns one only needs to calculate:
\begin{equation}\label{eq:interx}
    G_{R,C} = \sum\limits_{\alpha,\beta}^{N}\sum\limits_{j,l=1}^{L}I_{jl}(\textbf{r},\textbf{r}',t)\big[ N_{0} (\delta_{jl} - \langle\phi_{R}|c_{l,\alpha}^{\dagger}c_{j,\alpha} |\phi_{R}\rangle ) + (1-N_{0})\langle\phi_{R}|c_{l,\alpha}^{\dagger}c_{j,\alpha} |\phi_{R}\rangle  \big ], 
\end{equation}
where $I_{jl}(\textbf{r},\textbf{r}',t) = w_{c}(\textbf{r}',t) w_{c}^{*}(\textbf{r},t) w_{l}^{*}(\textbf{r}'-\textbf{r}_{l}',t) w_{j}(\textbf{r}-\textbf{r}_{j},t)$ and $N_{0} = \langle\phi_{C}|c_{0,\beta}^{\dagger}c_{0,\beta} |\phi_{C}\rangle$ defines the expectation value of the atom number operator for the central cloud. 
\begin{figure}
    \centering
    \includegraphics[width=\linewidth]{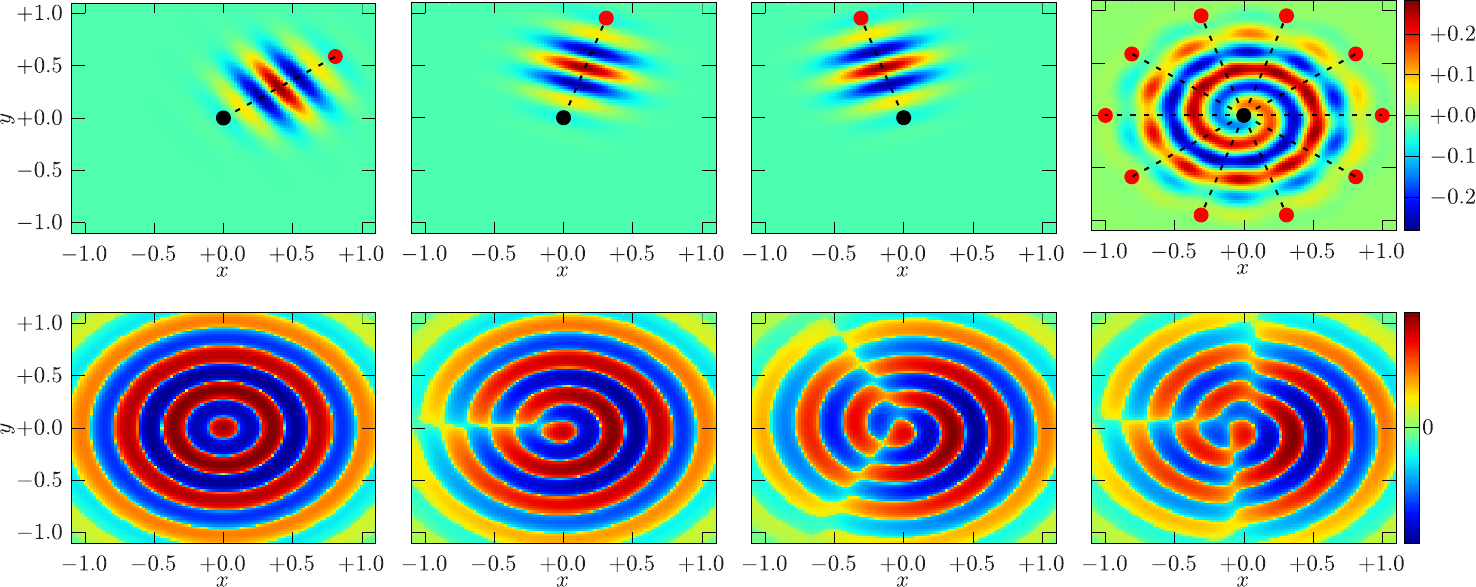}
    \put(-462,190){(\textbf{a})}
    \put(-350,190){(\textbf{b})}
    \put(-240,190){(\textbf{c})}
    \put(-130,190){(\textbf{d})}
    \put(-462,93){(\textbf{e})}
    \put(-350,93){(\textbf{f})}
    \put(-240,93){(\textbf{g})}
    \put(-130,93){(\textbf{h})}
    \caption{Spirals for non-interacting particles. The interference between the particles at each site and the center produce interference fringes (\textbf{a})-(\textbf{c}) that once joined together \textbf{(d)} as in \eqref{eq:correlation_ring_center}, form a spiral interference pattern that depends on the angular momenta ($\ell=1$ for (\textbf{a})-(\textbf{d})). Lower row shows the cumulative behavior of the interference pattern for multiple occupation of angular momenta $\ell=\{0,\,0+1,\,0+1+2,\,0+1+2+3\}$ corresponding to (\textbf{e})-(\textbf{h}), which are known as dislocations \cite{pecci2021probing,chetcuti2022interference}. }
    \label{fig:all_spirals}
\end{figure}
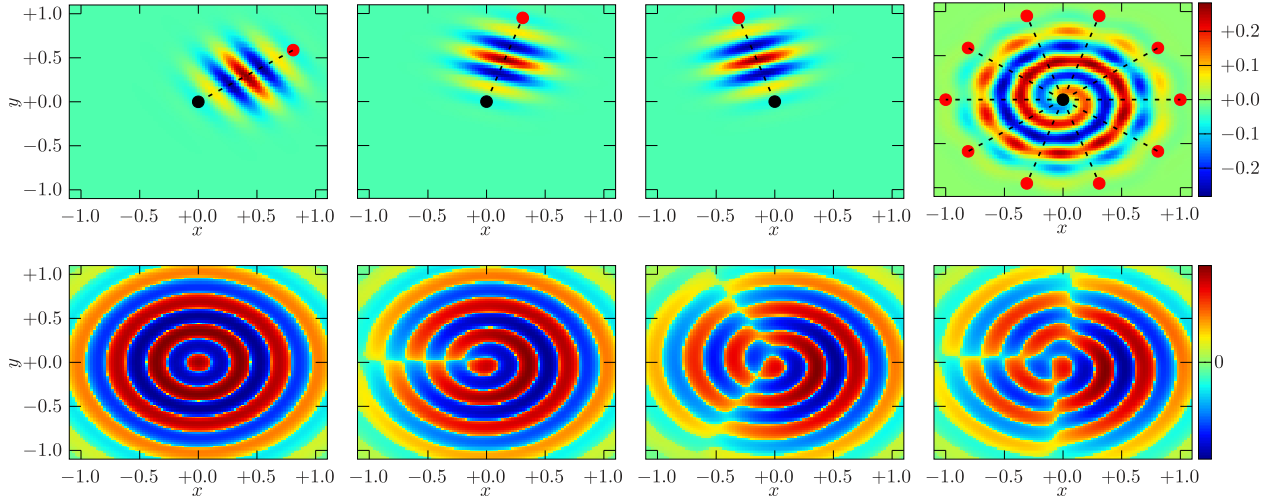
Similarly as in the case of TOF expansion of the atoms in a ring lattice, we can also describe analytically the interference between the central and external clouds for non-interacting particles. In this situation, the correlation between particles are simply $\langle c_{l}^{\dagger}c_{j}\rangle = \frac{1}{L}\sum_{\{n\}}e^{-\frac{2\imath\pi}{L}n(l-j)}$ with $n$ being the index for momentum modes: 
\begin{equation}\label{eq:correlation_ring_center}
    G_{R,C}(\mathbf{r},\mathbf{r}',t) = \frac{1}{L}\sum\limits_{\{n\}}I_{n}(\mathbf{r})I_{n}^{*}(\mathbf{r'}), 
\end{equation}
with 
\begin{equation}\label{eq:I_correlation_ring_center}
    I_{n}(\mathbf{r}) = |\mathcal{A}(\tau)|^{2}\exp\bigg\{-\frac{(r^{2})\eta_{j}^{2}}{2b(\tau)} \bigg\}\exp\bigg\{-\frac{\imath\tau(r^{2})}{2b(\tau)} \bigg\}\sum\limits_{j=1}^{L}\exp\bigg\{-\frac{(\mathbf{r}-\mathbf{r}_{j}^{2})\eta_{j}^{2}}{2b(\tau)}\bigg\}\exp\bigg\{\frac{\imath\tau(\mathbf{r}-\mathbf{r}_{j}^{2})}{2b(\tau)} -\frac{2\imath\pi}{L}nj \bigg\},
\end{equation}
and $\mathcal{A}(\tau) = \frac{1}{\sqrt{\pi}}\frac{\eta_{j}(\eta_{j}^{2}-\imath\tau)}{b(\tau)}$, $b(\tau) = \eta_{j}^{4} + \tau^{2} $ and $\tau = \omega_{0}t$ (using Eq.~\eqref{eq:wann}).

Fig.~\ref{fig:all_spirals} displays the interference pattern obtained for the interference between a single particle in the center and one in the ring at given lattice site, as well as the total $G_{R,C}(\mathbf{r},\mathbf{r}',t)$ with $\mathbf{r}'=(R,0)$. In the same figure, we also show the interference pattern in the form of spirals as well as the dislocation patterns obtained when multiple momenta are occupied, obtained through Eq.~\eqref{eq:I_correlation_ring_center}. Similar patterns are visible for interacting system featuring entangled superpositions of momenta~\cite{haug2018readout}.

Both homodyne and self-heterodyne interference protocols are simplest to implement and interpret when the interactions between the atoms have a small effect on the momentum distribution in time-of-flight, such that the evolution is well-described as a ballistic motion of free particles. In weakly interacting Bose gases of alkali atoms such as Rb, this condition is met for moderate values of the density~\cite{stamper1998optical, ernst1998free,kagan1996evolution, castin1996bose, dalfovo1997frequency}. However, the aspect ratio of the system in-situ must generally be taken into account when developing a time-of-flight expansion protocol. Tight vertical confinement leads to relatively more rapid vertical expansion, for example, and this can cause the cloud to expand beyond the imaging depth of the field on a timescale that is short compared to that of the radial expansion and interference ~\cite{ryu2014creation}.  

Interference protocols can be more challenging to implement in systems where interactions are much stronger, such as fermionic $^6$Li tuned near the Feshbach resonance. In this case, the interaction-driven expansion is usually not negligible, even to the point that depletion of the pair condensate can substantially reduce the contrast of the interference. All the persistent current experiments conducted in this regime have so far employed the same type of interaction ramp used in most fermionic experiments~\cite{ketterle2008making, menotti2002expansion}, where the scattering length is rapidly ramped to a smaller value in the BEC regime before releasing the atoms into time-of-flight expansion. There is a second purpose to doing this in experiments where pair superfluids and pair correlations are of interest. In the unitary and BCS regimes, the pairs are weakly bound and fragile; if released directly in time-of-flight they break during the expansion, and information about their motion before release is lost. A suitably engineered ramp to the BEC limit converts them into stable molecular pairs, without significantly altering their center-of-mass momentum.

When an interaction ramp is necessary, it must be fast enough to minimize losses due to three body collisions that become more severe in the molecular BEC limit, but slow enough not to drive unwanted excitations of the gas by a sudden change in the chemical potential. The optimum rate thus depends on the detailed spatial shape of the ring confinement (flatter versus tighter,  thicker  versus thinner annuli), and on  the nature of the low-lying excitations in the gas (as for example,  presence of bound states). An in-depth understanding of the time scales of the interaction and dynamics occurring under the conditions of a particular experiment is generally required~\cite{cai2022persistent, del_pace2022imprinting, allman2023quench-induced, xhani2023decay}.

Some experimental and theoretical efforts have been made to try to develop indirect and potentially less-destructive read-out schemes for persistent currents. There has been some experimental success in detecting the presence of a persistent current in a ring by observing the precession of long-wavelength acoustic excitations~ \cite{marti2015collective, kumar2016minimally}. Under the conditions of these experiments, strong damping of the acoustic modes and a low signal-to-noise ratio resulted in significant limitations on the sensitivity of this class of techniques. Another interesting line of inquiry has been that of detecting currents around a ring via coupling to an adjacent rectilinear waveguide~\cite{safaei2019monitoring} or to the modes of an optical cavity~\cite{pradhan2023cavity, pradhan2024cavity}. While an experimental implementation of these proposals would be challenging, we are hopeful that they may still be realized in the near future.

\section{Persistent currents in bosonic gases}\label{sec:PerCurrBose}
 
Persistent currents in toroidal-shaped BECs have been realized in several important experiments and thoroughly analyzed theoretically. Typically, the persistent flow is induced through phase imprinting, Raman transitions or stirring protocols, as discussed in Sec.~\ref{sec:Methods}. The latter employs a moving laser field acting on the matter wave in the form of a wide rectangular barrier. In this case, the superfluid density in correspondence with the barrier is suppressed and, for a suitable choice of physical parameters, it can be vanishing. In these conditions, persistent currents can arise since phase slips can occur with a negligible energy cost~\cite{piazza2009vortex}. It is important to remark that such a process is of a dissipative nature for the BEC, and can cause the formation of vortices in the bulk of the superfluid. Most of the experiments are carried out in spatially closed circuits with a finite radial width. While general properties of such a system can be understood through $1$D models,  vortices and phase slips dynamics of persistent currents in toroidal BECs need to be modeled by higher dimensional theories. On the practical side, such an analysis has been accomplished through GPE-like approaches that, however, cannot capture strong correlation effects. For the latter effects $1$D models remain a very valuable tool of investigation.  

\subsection{Experimental results}\label{sec:ExpResBose}

It is interesting to note in retrospect that many important ingredients for realizing a persistent current  in a BEC were incidentally present in the original experimental apparatus at MIT, where a cloud of sodium atoms was pierced by a blue-detuned ``plug'' beam to prevent spin-flip losses at the center of a magnetic quadrupole trap~\cite{davis1995bose-einstein}. In such a trap, it is possible for the BEC to be multiply connected when the chemical potential is high enough. If a quantized current ever did happen to form spontaneously in one of those early condensates, it seems it was not recognized at the time. The other trap types that were most frequently employed in early quantum gas experiments (e.g. Ioffe-Pritchard, time-orbiting potential, single-beam optical optical traps) naturally have a simply-connected geometry and were not conducive to the study of persistent currents. In those early years considerable effort was devoted to studying the basic phenomena of superfluidity and superfluid vortices in simply-connected traps, and the first generation of dedicated ring-trap experiments would not appear for nearly a decade. In 2005, a BEC was created in a millimeter-scale time-orbiting ring trap at Stanford~\cite{gupta2005circularwaveguide} and this was followed soon after by a 10 cm diameter ring-Ioffe trap at Strathclyde~\cite{arnold2006ring}.  These experiments were groundbreaking, but their large size and non-uniformity made it impossible to create a BEC that extended all the way around them. In the next few years, numerous groups worked on methods for creating smoother (and smaller) ring traps for ultracold atoms using time-averaged magnetic fields, optical fields, and combinations of the two~\cite{morizot2006ring, lesanovsky2007time, olson2007cold-atom, ryu2007observation, heathcote2008ring, schnelle2008versatile, henderson2009experimental, bruce2011smooth, ramanathan2011superflow, sherlock2011time, moulder2012quantized, ryu2013experimental, navez2016matter, bell2016bose-einstein}. For a more comprehensive review of methods for creating multiply-connected traps for ultracold atoms we refer the reader to Ref.~\cite{amico2021roadmap}; in what follows we will be focused primarily on experiments that achieved the conditions necessary for studying persistent currents.

The key technical challenges that must be overcome in an experiment focused on persistent currents is to ensure that the extended minimum of the ring trap is sufficiently smooth, and that enough atoms are trapped and cooled to temperatures sufficient for an unbroken superfluid to form around the entire circumference. One also needs a reliable means of creating and detecting the currents, which is also non-trivial. These tasks are generally easier to achieve with a system of weakly-interacting bosons than with fermionic atoms, as will be described in Sec.~\ref{sec:Interference}. So far, persistent currents have been realized in BECs of only a few alkali-metal atomic species such as (bosonic) $^{23}$Na, $^{87}$Rb, and in (fermionic) $^6$Li. While there is no fundamental obstacle to adapting the experimental techniques used with these atomic species to studying persistent currents in quantum gases of alkaline-earth atoms, or atomic species which form dipolar condensates, no experiments with these other atomic species have been reported yet. In this section, we will review the results of experiments involving persistent currents in weakly-interacting BECs of alkali-metal atoms, and in Sec.~\ref{sec:ExpResFerm} we will review more recent developments with persistent currents in rings of fermionic atoms.

\begin{figure}[h!]
   \centering
   \includegraphics[width=0.8\textwidth]{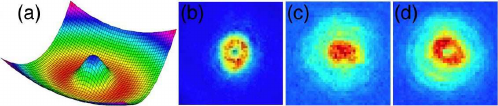}
   \caption{\textit{Experimental images for the detection protocol of the NIST Gaithersburg group.} \textbf{(a)} Image of the toroidal trap, which is a combination of the TOP magnetic trap and the plug beam. \textbf{(b)} \textit{In-situ} image of the BEC in the trap. \textbf{(c)} and \textbf{(d)} correspond to TOF images of the BEC after release from the trap with a non-circulating and circulating current state respectively. Taken from\cite{ryu2007observation}.}
   \label{fig:ryu2007}
\end{figure}

The first confirmed detection of a persistent flow in a quantum gas was performed at NIST Gaithersburg in an experimental system with some qualitative similarities to the original setup at MIT: a blue-detuned ``plug'' beam piercing a sodium BEC in a time-orbiting potential (TOP) magnetic trap, where the plug could be removed without spin-flip losses occurring at the center~\cite{andersen2006quantized}. The most important new development in that experiment was a method for coherently transferring quantized orbital angular momentum from laser fields to the condensate, as described in Sect.~\ref{sec:ExpMethods}. Consistently with the prediction in \cite{amico2005quantum},  a  vortex-like ``hole'' appearing in the density profile after it was released into TOF expansion was the signature of quantized circulation around the annular condensate -- Fig.~\ref{fig:ryu2007}. The current decayed in around 0.5 s if the ``plug'' beam was removed, but was observed to be stable as long as there was an unbroken BEC along a path around the plug beam ($\approx$ 10 s). Another notable aspect of this experiment was the first creation of a doubly charged vortex, and the observation that it decayed into two singly-charged vortices when the plug beam was removed~\cite{ryu2007observation}. This experiment was followed soon after by investigating the spontaneous appearance of vortices as a Bose gas is cooled through the superfluid phase transition~\cite{weiler2008spontaneous}. In particular, the probability of observing a vortex was found to be increased when the BEC was deformed into a wide annulus by an optical plug beam. 

Because of difficulties like compensating for drift in the position of the magnetic trap relative to the plug beam, experiments  switched to using all-optical ring trap techniques that had been under conceptual and technical development for some time~\cite{wright2000toroidal, brand2001generating, franke-arnold2007optical, olson2007cold-atom, schnelle2008versatile,gauthier2021chapter,turpin2015blue}, and that had been leading to successful implementations employing ``painted'' potentials~\cite{henderson2009experimental} and spatial light modulators~\cite{bruce2011smooth}. The laser configuration most commonly used to generate an optical ring trap includes a horizontally propagating laser field that mainly provides the vertical confinement, and a patterned vertically propagating beam that provides radial confinement leading to a ring potential or other desired configuration. The experiment at NIST~\cite{wright2000toroidal} was overhauled to implement an all red-detuned ring trap, with a single horizontal ``sheet'' beam, and a vertically propagating Laguerre-Gauss beam providing radial confinement. The trap was sufficiently smooth and stable that persistent currents were observed to survive for up to 40 seconds, appearing to be limited only by losses due to collisions with background gas molecules~\cite{ramanathan2011superflow}. In these experiments, the persistent current was created using a different two-photon technique than the Bragg method used earlier. Co-propagating beams of differing OAM were used to drive a Raman transition between different magnetic sublevels, instead of using an intermediate state of large linear momentum. This approach required only a single pulse-pair instead of two pairs, and resulted in less spurious excitation of the condensate by beam mode imperfections. When a superfluid ``weak link'' created by a repulsive blue-detuned barrier beam was introduced, the current was observed to decay when the velocity through the barrier approached the estimated local sound speed. 

These results were rapidly followed by two experiments with a ring-shaped $^{87}$Rb BEC at Cambridge, where a similar Raman technique was used to imprint a phase winding on the condensate~\cite{moulder2012quantized}. A Laguerre-Gaussian beam was used to provide radial confinement, and was also used as one of the Raman beams. Persistent currents of $\ell=3$ were observed to be stable for up to one minute, and currents with up to $\ell=10$ were realized. The rate of current decay via spontaneous phase slips was also observed to depend significantly on the azimuthal uniformity of the trap potential. The observed current decay showed good agreement with the maximum current that can be sustained in a 2D ring trap predicted in~\cite{dubessy2012critical} through Bogoliubov analysis. These experiments are particularly notable in that they also permitted the creation of persistent currents in a spin mixture of the $|F=1,m_F=0\rangle$ and $|F=1,m_F=1\rangle$ magnetic sublevels~\cite{beattie2013spinor}. They observed the existence of a critical population imbalance of around 0.6 for stability of the superflow, and were able to validate some of the earlier predictions made for such systems~\cite{smyrnakis2009mixtures, bargi2010persistent, anoshkin2013persistent}, but much remains to be investigated in this area. 
\begin{figure}[h!]
    \centering
    \includegraphics[width = 0.7\textwidth]{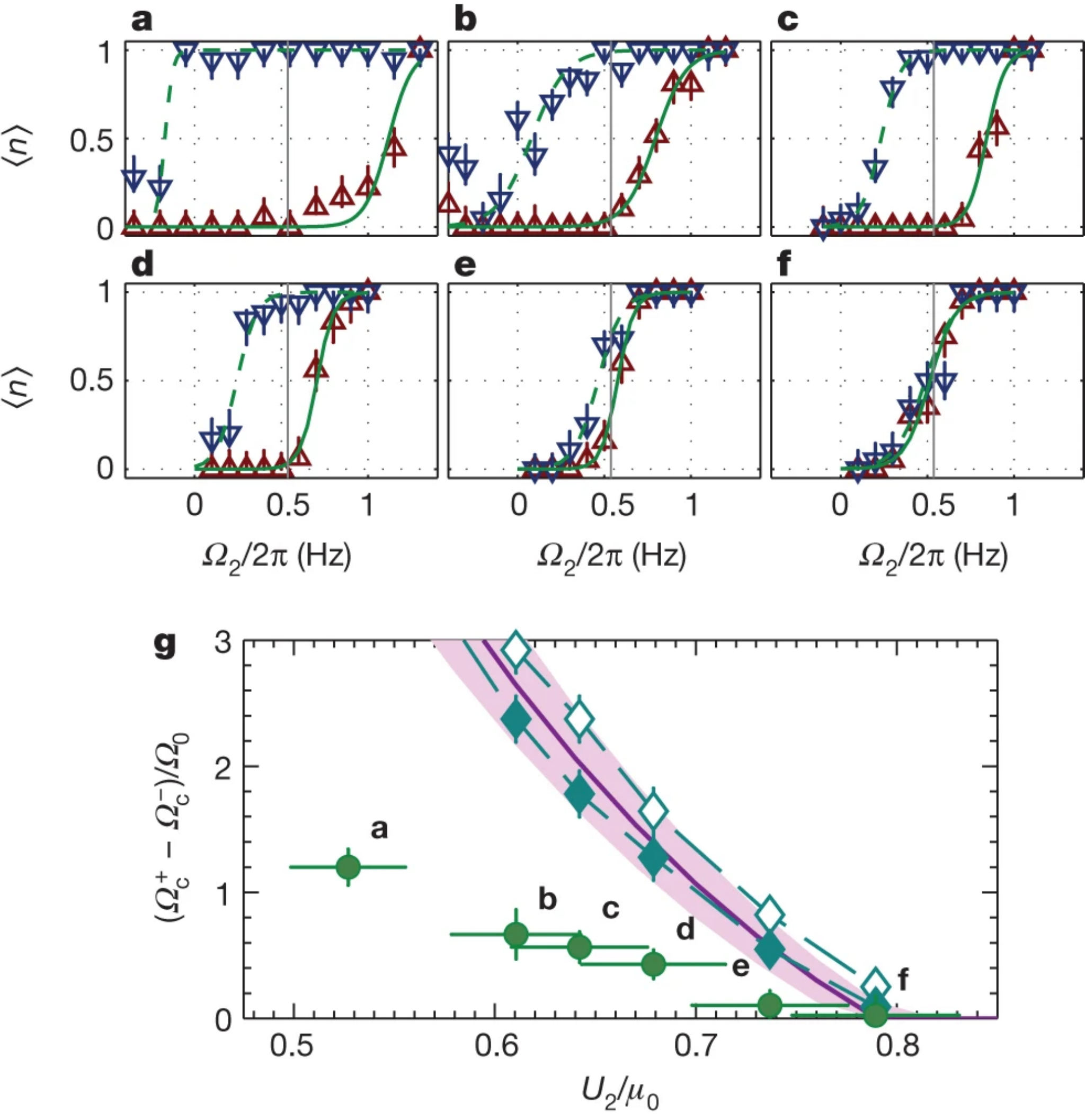}
    \caption{\textbf{(a)}-\textbf{(f)} Hysteresis loops with sigmoid fits. The red triangle (blue inverted triangles) show the winding number $\ell$ when starting with $\ell=0$ ($\ell=1$). \textbf{(g)} Hysteresis loop size vs. barrier height $U_2$. The green circles show the experimental data. The magenta line is the prediction of an effective 1D hydrodynamic model. The open (filled) cyan diamonds are the results of GPE simulation with dissipation $\Lambda=0$ ($\Lambda=0.01$). Figure from Ref.~\cite{eckel2014hysteresis}.}
    \label{fig:hysteresis}
\end{figure}

Another important experimental development was the introduction of movable barrier potentials that could be used to drive currents around a ring and/or probe the circulation state of the superfluid in the ring. As described in Sec.~\ref{sec:ColdAtomsMotion}, if one transforms to a rotating frame where the barrier potential is stationary,  there is an effective gauge potential and a non-zero ``circulation flux'' threading through the ring. It was shown to be possible to deterministically and reversibly change the circulation state by driving phase slips with a moving barrier~\cite{wright2013driving, wright2013threshold}, and that the response of the system to a cyclic change of the rotation speed became increasingly hysteretic as the critical current of the moving weak link was increased~\cite{eckel2014hysteresis} - see also Sec.~\ref{VortexDynamicsBose}. The hysteresis experiment was conducted with a ring trap employing a blue detuned ``dark-ring'' potential, instead of a red-detuned ``bright-ring'', which was found to have some advantages for creating a more azimuthally uniform ring in the presence of optical imperfections --see Fig.~\ref{fig:hysteresis}. Continuing experiments at the University of Arizona investigated the formation and evolution of turbulent superfluid states involving large numbers of vortices ($\ell>20$) generated by stirring of a wide, flat annular BEC. They found that the system evolved dissipatively toward a persistent current state with some of the vortices being pinned by the central potential, and others leaving the system, and compared their experimental data against numerical simulations showing evidence of an inverse energy cascade consistent with expectations for 2D quantum turbulence~\cite{neely2013turbulence}.
\begin{figure}[h!]
    \centering
\includegraphics[width=0.7\linewidth]{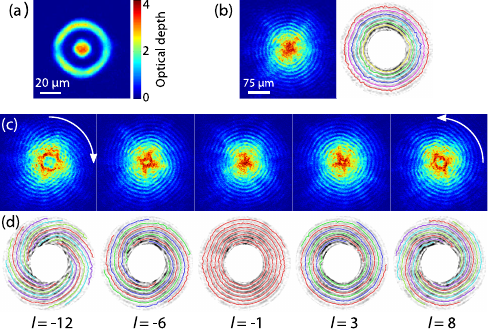}
    \caption{\textit{Experimental images for the interferometeric protocol of the NIST group.}\textbf{(a)} \textit{In-situ} image of the ring and disc BECs. \textbf{(b)} Interferogram of the ring BEC with no current after 15ms of TOF. Interferograms for circulating current states of different winding numbers are displayed in \textbf{(c)} with the images in panel \textbf{(d)} corresponding to traces in the interference fringes as a guide to the eye.    
    Figure from Ref.~\cite{eckel2014hysteresis}.}
    \label{fig:eckel_spiral}
\end{figure}

Imaging objectives for ultracold atoms experiments require high resolution and large numerical aperture; often the imaging system needs to be customized to make it compatible with the actual spatial configuration of the experimental apparatus. Indeed, important progress was made in experiments with an apparatus at Los Alamos, which had substantially higher optical resolution than the experiments at NIST. This allowed for the creation of features with a width similar to the condensate healing length in a ``painted'' ring potential created by 2D scanning of a tightly focused beam~\cite{ryu2013experimental}. Narrow barriers are necessary for realizing tunnel junctions with coherent dynamics; the flow through longer hydrodynamic weak links in the NIST experiments was found to be more strongly damped and resistive in character when the critical current was exceeded~\cite{jendrzejewski2014resistive}. In those experiments the full-width of the weak link was 12 $\mu$m, much larger than the healing length of 0.3 $\mu$m, while the barrier width in the Los Alamos experiments could be made as small as 2 $\mu$m. This was close enough to the healing length that they were ultimately able to observe evidence of interference between the currents through two movable junctions in a configuration functionally equivalent to a DC SQUID, in a landmark proof-of-concept of the idea of superfluid ``Atomtronics''~\cite{ryu2020quantum}. 

Experimental progress was driven by the introduction of more sophisticated interferometric techniques for obtaining information about the state of a superfluid ring. The main technique used to detect currents in early ring BEC experiments was to look for vortex-like density features in the cloud after ballistic expansion. There were two experiments that independently and concurrently demonstrated that it was possible to measure the gradient of the phase around a superfluid ring by interfering it with a second disk-shaped BEC in the center of the ring -- see Fig.~\ref{fig:eckel_spiral}. At NIST, the ``target'' trap technique was initially used to measure the current-phase relation of a superfluid weak link in the hydrodynamic flow regime, as the height of the barrier was varied~\cite{eckel2014interferometric}. 
Clearly, resolving the arms in the spiral interferogram tolerates only small optics misalignments. This was followed by a more detailed analysis of and validation of this approach to self-heterodyne interferometry, comparing numerical simulations of time-of-flight expansion against experimental data, for the specific case of a quantized supercurrent flowing around a ring with a constriction~\cite{mathew2015selfheterodyne}. 

\begin{figure}[h!]
    \centering
\includegraphics{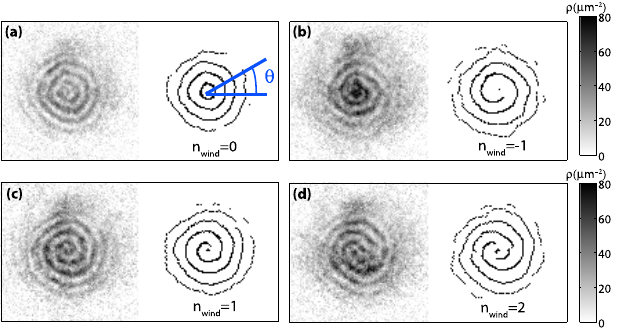}
    \caption{\textit{Experimental images for the interferometeric protocol of the ENS group.} \textbf{(a)} Interference pattern for a non-circulating current state. \textbf{(b)}, \textbf{(c)} and \textbf{(d)} correspond to interference patterns where the system has a phase winding of $+2\pi$, $-2\pi$ and $+4\pi$.  
    Figure from Ref.~\cite{eckel2014hysteresis}.}
    \label{fig:corman_spiral}
\end{figure}

Researchers at ENS independently developed the same interferometric technique to measure the circulation around a ring BEC after a rapid quench of the system through the normal-to-superfluid phase transition~\cite{corman2014quench} (see Fig.~\ref{fig:corman_spiral}). The distribution of spontaneously formed persistent currents was found to depend on the quench rate with a power-law scaling that was in good general agreement with predictions of Kibble-Zurek theory. They also showed that it is possible to extract the azimuthally varying phase profile and the angular correlation function of the ring from the interference pattern. The group at ENS later conducted a very closely related experiment investigating the spontaneous appearance of quantized currents when a chain of $N$ uncorrelated condensates is connected to form a ring. They observed that the rms width of the winding number distribution increased with $\sqrt{N}$ as expected, and characterized the time scale for phase defect relaxation in the system~\cite{aidelsburger2017relaxation}. Spontaneous current formation in a ring of fermionic superfluid after a quench has also been studied~\cite{allman2023quench-induced}, but that work will be described in Sec.~\ref{sec:PerCurrFerm}.

The propagation of phononic and other collective excitations around a ring BEC has been the subject of several experimental studies to date. The first of these was a realization of a collective excitation interferometer at Stanford, where phase imprinting was used to excite acoustic standing waves of order up to $m=7$ in a ring BEC. The modes were expected to precess in the presence of rotation, but the potential sensitivity was found to be substantially limited by rapid acoustic damping and by noise.~\cite{marti2015collective}.  A conceptually similar experiment at NIST was later able to detect the precession of long-wavelength acoustic standing waves in a ring in the presence of a persistent current and use the effect to identify the current state with a confidence of around 90\%~\cite{kumar2016minimally}. Dynamically changing the radius of a ring-shaped BEC was also found to shift the wavelength of acoustic excitations in a manner that is analogous to the redshift caused by cosmological inflation, with creation of azimuthal phonons leading to stochastic formation of persistent currents at late times~\cite{eckel2018rapidly}.

The metastability of persistent currents arises because of an energy barrier between current states that is large enough to make decay improbable, and an understanding of the mechanisms leading to decay in various limits is important. In the conditions typical of most experiments with ultracold atoms, the system temperature is high enough that thermal fluctuations cannot be neglected, and efforts to explain experimentally observed critical velocities and decay rates have had mixed results. Qualitatively, decay rates have been shown to increase with temperature and with the strength of a potential barrier applied to a superfluid ring~\cite{kumar2017temperature}, but simple models of decay dynamics consistently predict much larger critical velocities and current lifetimes than are observed in experiment. Models that attempt to more realistically account for finite temperature effects tend to be in better agreement with experimental observations~\cite{mathey2014decay, snizhko2016stochastic, kunimi2019decay, mehdi2021superflow} but more experimental and theoretical work will be required to better understand the origin of the remaining discrepancies, especially at high temperatures and under experimental conditions where atom loss and non-equilibrium effects may play a role.

Creation and stability of higher-order current states around a ring BEC were addressed qualitatively in early experiments at NIST~\cite{wright2013threshold}, Cambridge~\cite{moulder2012quantized}, Los Alamos~\cite{ryu2014creation}. The group at the University of Arizona conducted the first systematic investigations of how the maximum stable winding number changed as a central pinning potential was varied~\cite{law2014dynamic}. This work was motivated in part by difficulties controlling the relative position of the magnetic and optical traps, but with improvements in control and stability the same group was later able to generate stable currents with substantially higher winding numbers, up to $\ell=11$, and observe the dynamical breakup of giant vortices into a cluster of singly quantized vortices when the pinning potential was weakened~\cite{wilson2022generation}

The notable success realizing stable high-order currents in the experiments at Arizona was in part due to the smoothness of their optically plugged magnetic trap. One substantial advantage of time-averaged and RF dressed magnetic traps in general~\cite{morizot2006ring, lesanovsky2007time, heathcote2008ring, schnelle2008versatile, henderson2009experimental, sherlock2011time, navez2016matter} is that they are intrinsically much smoother than optical traps. There has been a string of remarkable recent results leveraging this advantage in experiments using magnetic and hybrid magnetic/optical traps~\cite{herve2021versatile} capable of realizing unusual geometries like rings and bubbles~\cite{dubessy2024perspective} with sizes ranging up to the millimeter scale. One breakthrough experiment conducted at FORTH in Heraklion was able to accelerate a BEC up to supersonic speeds (16$\times$ their speed of sound) around such a ring using a ``bang-bang'' scheme\footnote{A ``bang-bang" scheme is a protocol derived from optimal control, characterized by abrupt discrete switching between two states or extreme values~\cite{chen2014optimal}. A common example of such a scheme is the discrete Heaviside function. In this particular experiment, the system experiences rapid pulses that act as strong discrete kicks to attain the desired state.}, with no evidence of heating as it continued around the ring for 15~seconds over a total travel distance of 40~cm~\cite{pandey2019hyper}. Concurrently, experiments at LPL in Paris were carried out where angular momentum selective evaporation of a rotating cloud was used to spin it up to supersonic speeds where it formed a narrow dynamically stabilized annulus 50~$\mu$m in diameter and continued to rotate for more than one minute as shown in~\cite{guo2020supersonic}. The potential of such smooth rf-dressed potentials and waveguides for guided matter-wave interferometer is an important area of continuing investigation.

\subsection{Theoretical understanding - persistent currents in one-dimensional rings }\label{sec:ResBose}

Consider now the idealized case of a ring trap of radius $R=L/2\pi$ with very tight transverse confinement, such that the particles have essentially a one-dimensional motion. We first take the case of $N$ bosons on a uniform ring at zero temperature, subjected to an artificial gauge field $\phi$. In this case, the system is described by the Lieb-Liniger Hamiltonian Eq.~\eqref{eq:LLHam} with periodic boundary conditions. The persistent current can be readily calculated in several cases from the identity relating it to the ground-state energy (\ref{eq:ZeroTempCurrDef}). 

In the case of non-interacting bosons, i.e.\ upon setting $g=0$ all particles form a Bose-Einstein condensate, and they all occupy the same single-particle energy level. In this case, the total energy is given by 
\begin{equation}
    E(\phi)=\frac{N \hbar^2}{2 m R^2} \left(\ell-\frac \phi \phi_0 \right)^2,
\end{equation}
where $\ell$ is an integer chosen such that the energy takes its minimal value, hence, specifically, $\ell=0$ for $-1/2\leq\phi/\phi_0 \leq 1/2$, $\ell=1$ for $1/2\leq \phi/\phi_0 \leq 3/2$ and so on. The energy as a function of flux is piece-wise parabolic. The integer $\ell$ indicates the angular momentum per particle and labels each parabola. 

The opposite limiting case is given by the Tonks-Girardeau (TG) limit $g\rightarrow\infty$. Also in this case the persistent current can be calculated exactly.  For this purpose, we use the Bose-Fermi mapping~\cite{girardeau1960relationship}, which states that
the many-body wavefunction of TG bosons is given by 
\begin{equation}
\Psi_B(x_1,...,x_N)=\prod_{1\le j<l\le N} {\rm sign} (x_j-x_l) \; \Psi_F(x_1,...x_N),
\end{equation}
where, on a ring, the free fermionic wavefunction is 
$\Psi_F=\frac{1}{\sqrt{N!}} \left. \det[e^{i k_j x_l}] \right|_{j,l=1...N}$.
The wavevectors $k_j$ entering in $\Psi_F$ above depend on the value of the applied gauge field. In the TG case, as well as in the general case of arbitrary interactions in the Lieb-Liniger model, the ground-state energy is given by $E_{GS}=\frac{\hbar^2}{2m}\sum_{j=1}^N \left[k_j-(2 \pi/L) (\phi/\phi_0)\right]^2$.
One can readily check by substitution that for 
 $-1/2<\phi/\phi_0 <1/2$ the wavevectors yielding the lowest (ground-state) energy for odd number $N$ of bosons are given by 
 \begin{equation}
 k_j=\frac{2 \pi}{L}\left\{ -\frac{(N-1)}{2}, ...., 0, ..., \frac{(N-1)}{2} \right\} .
 \label{eq:TGrapidities}
 \end{equation}
In fact for an {\em odd} boson number, the $\prod \text{sgn} (x_j-x_l)$ mapping function needed to map fermions onto bosons is periodic, hence the free-fermion wavefunction must satisfy periodic boundary conditions. For an {\em even} number of bosons, instead, the mapping function is antiperiodic. Hence, in order to have a bosonic wavefunction which is periodic on the ring, we need to employ an {\em anti-periodic} fermionic wavefunction. For  $-1/2<\phi/\phi_0 <1/2$ the bosonic ground state is obtained by choosing 
\begin{equation}
 k_j=\frac{2\pi}{L}\left\{ -\frac{(N-1)}{2},  -\frac{(N-3)}{2},  ......, \frac{(N-3)}{2},\frac{(N-1)}{2} \right\} .
 \end{equation}
This different choice in the odd/even case  agrees with the general rule for choosing the $I_j$ entering the Bethe-Ansatz equations (\ref{eq:BArapidities}),  and provides a physical interpretation to it.
The above reasoning can be extended to arbitrary values of $\phi/\phi_0$. In each sector $-1/2+\ell <\phi/\phi_0 <1/2+\ell$, the ground-state wavevectors correspond to shifting the wavevectors given in (\ref{eq:TGrapidities}) each by $2 \pi \ell/L$. In fact, such a state corresponds to shifting the whole Fermi sphere, by giving the value $2\pi N \ell/L$ to its center-of-mass momentum. Such states are also known as yrast states in nuclear physics \cite{kanamoto2010metastable}.
In both the odd and even $N$ case, the ground-state energy of the TG gas on a ring is given by a set of parabolas centered at $\phi/\phi_0=0$ and with periodicity 1. Hence it is remarkable that, although there is a different set of quantum numbers for even or odd numbers of bosons, there is no parity effect. This is strikingly different from the case of a two component Fermi gas discussed in Sec.~\ref{sec:PerCurrFerm} below.

We next proceed in extending the above results to arbitrary repulsive interactions. In this case, to obtain the ground-state energy we use the Bethe Ansatz solution Eq.~\eqref{eq:BArapidities}. The Bethe rapidities are all real. For given interval $ \ell -1/2 \le  \phi/\phi_0 \le  \ell+1/2$ the ground state is obtained by choosing the quantum numbers  $I_j = - (N -1)/2 + j + \ell$. In a homogeneous ring, center-of-mass and relative coordinate are decoupled, and the artificial gauge fields couple only to the center-of-mass momentum $P_{cm}=\hbar \sum_{j=1}^N k_j$:
\begin{equation}
    E_{GS}=\frac{1}{2 N m}(P_{cm} -N m \phi R^2)^2+ E_{int},
\end{equation}
where $E_{int}$ is the internal energy associated with the relative motion and depends on interactions. Using that  $\arctan[(k_j-k_l)/c]$ entering Eq.~(\ref{eq:BArapidities}) is an odd function of its argument, one readily finds  $P_{cm}=  \hbar  N \ell/R$.
The ground-state energy is hence given by 
\begin{equation}
      E(\phi)=\frac{N \hbar^2}{2 m R^2} \left(\ell-\frac \phi \phi_0 \right)^2 + E_{int},
      \label{eq:EgsBArep}
\end{equation}
embodying the logic of Leggett's theorem: for any interaction strength $g\ge0$, the ground-state energy is a periodic function of $\phi/\phi_0$ with periodicity 1. The energy landscape is the same for any value of the interactions apart from an interaction-dependent constant shift.

\begin{figure}[h!]
\centering\includegraphics[width=\linewidth]{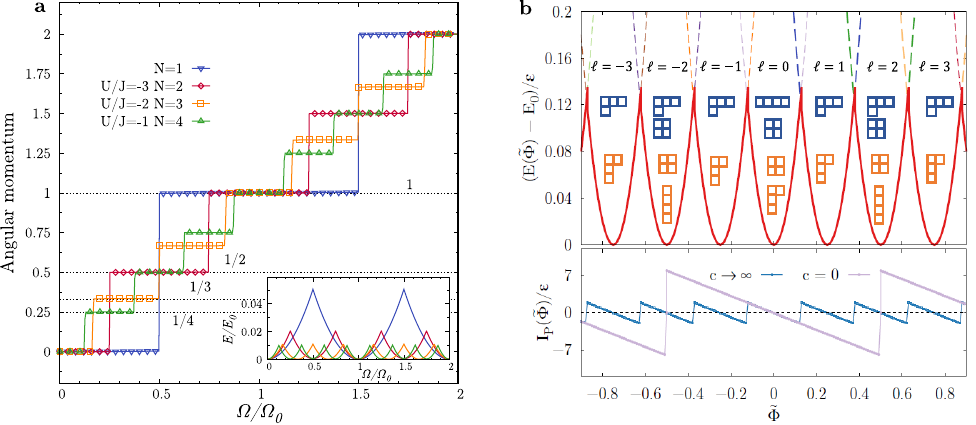}
    \caption{\textbf{(a)} Average angular momentum per particle (main) and ground-state energy $E$ (inset) for bosons with attractive interactions $U$ on a lattice ring as a function of artificial gauge field $\Omega$, from numerical exact diagonalization calculations at varying particle number $N$, for chosen values of interaction strength as indicated in the figure. \textbf{(b)} Ground-state energy (top) and persistent current (bottom) for SU(2)  bosons on a lattice ring with strong repulsive interactions $c$ as a function of the dimensionless synthetic gauge field $\tilde{\Phi}$, from Bethe Ansatz for 4 particles.  The angular momentum $\ell$ and the Young tableau reflecting the symmetry of the ground-state for each piece-wise parabola are shown for bosons (blue) and fermions (orange). Figures from \cite{naldesi2022enhancing,pecci2023persistent}. }
\label{fig:frac_bos}
    \end{figure}

The situation is remarkably different in the case of attractive boson-boson interactions. In such a case the Bethe rapidities are complex,
and the ground state is a many-body bound state made of $N$ bosons. In the thermodynamic limit $|c|,L\rightarrow \infty$ the string hypothesis yields $k_j=k_0 - i (N +1- 2j )c/2$, with $j=1...N$. This state is the quantum analogue of a bright soliton. 
In this case the ground-state energy is given by 
\begin{equation}
   E(\phi)=\frac{ \hbar^2}{2 m N R^2} \left(\ell-N \frac \phi \phi_0 \right)^2 + E_{int}.
      \label{eq:EgsBAattr}  
\end{equation}
Notice that here the periodicity of the ground-state energy as a function of the flux is reduced by a factor $1/N$ where $N$ is the number of bosons~\cite{naldesi2022enhancing}. The center-of-mass of the many-body bound state can take all integer multiples of $\hbar /R$, and the energy landscape displays $N$ parabolas in the interval  $ \ell -1/2 \le  \phi/\phi_0 \le  \ell+1/2$ -- Fig.~\ref{fig:frac_bos}\textbf{(a)}. Correspondingly, the persistent currents have a much smaller period $\phi_0/N$. Going back to the labframe,  this corresponds to a step-like increase of angular momentum as a function of $\phi$ with $N$ intermediate steps between each integer value. 

This indicates that angular momentum per particle can take fractional values. The {\em angular momentum fractionalization} \cite{naldesi2022enhancing} is a truly quantum effect arising from the change of nature of the bosonic ground-state with respect to the case of repulsive interactions.  We stress that -- due to the presence of the bound states -- Leggett's theorem in its simplest form does not hold in this case. Rather, it can be reformulated in terms of the elementary constituents of the attracting Bose fluid, i.e.\ the many-body bound states, corresponding to bosons of mass $Nm$ and effective flux quantum $\phi_{obs}=\phi_0/N$.
We shall see below that a similar feature occurs  -- for similar reasons -- for paired Fermi gases.

It is also worth briefly mentioning that  angular momentum fractionalization is found also in  multicomponent Bose-Bose mixtures with repulsive interactions~\cite{pecci2023persistent}. In this case the fractionalization is due to the spin degrees of freedom of the ground state of the fluid, and the emergence of additional parabolas is due to the change of the symmetry of the ground state under application of the gauge field $\phi$. In essence, the result stems from the Bethe Ansatz equations for the mixture at strong coupling, where spin and orbital degrees of freedom decouple. In this limit, the Bethe equations read  \cite{oelkers2006bethe}
\begin{equation}
\begin{cases}
L k_j = 2\pi \Bigl(\mathcal{I}_j - \frac{1}{N} \sum_{m=1}^M \mathcal{J}_m\Bigr)\\
2N \arctan (\lambda_m) = 2\pi  \mathcal{J}_m + \sum_{n=1}^M 2 \arctan (\lambda_m - \lambda_n).
\label{bethe_eqs_infinite}
\end{cases}
\end{equation}
where $k_j$ are the charge rapidities with quantum numbers $\mathcal{I}_j$ and $ \lambda_m$ the spin rapidities with quantum numbers $\mathcal{J}_m$. One sees that the spin quantum numbers enter in the definition of the charge rapidities, leading to the formation of additional intermediate parabolas in the energy landscape and correspondingly, fractionalized steps in angular momentum -- Fig.~\ref{fig:frac_bos}\textbf{(b)}.  For example, for the ground state of $N=2+2$ bosons, one has $\mathcal{I}_j = \{-\frac{3}{2}, -\frac{1}{2},\frac{1}{2},\frac{3}{2}\}$ and  $\mathcal{J}_1 + \mathcal{J}_2 = 0,1,2,3$, leading to $1/4$ fractionalization.

The addition of a localized barrier potential on a ring breaks rotational symmetry and allows to mix angular momentum states. This leads to opening of gaps in the energy landscape in correspondence to the crossing points. The shape of the angular momentum steps gets smeared, and the persistent current amplitude decreases as compared to the clean ring case.  In fact, the persistent current amplitude displays a non-monotonous behaviour as a function of the interaction strength, for any value of the barrier strength \cite{cominotti2014optimal}. This is due to a competition of different effects. At weak interactions, mean-field interactions screen the barrier, which results in its strength effectively smaller than its nominal value. As a result, the persistent amplitude increases with interaction strength. Mean-field description breaks down at intermediate interactions, where quantum fluctuations start to play an important role. In this regime, the system can be suitably described by the Luttinger liquid theory. The fluid is described at low-energy by a harmonic-fluid model, recalling Eq.(\ref{eq:BoseLuttingerHam})
\begin{equation}
  \mathcal{H}_{0}=\frac{\hbar v}{2\pi}\!\!\int_{0}^{L}\!\!\!\mbox{d}x\bigg[K\!\!\left(\partial_{x}\phi(x)\!-\!\frac{2\pi}{L}\Omega \right)^2\!\!\!+\!\frac{1}{K}(\partial_{x}\theta(x))^2 \bigg],
  \label{eq:llhamiltonian}
\end{equation}
where $K$ is the Luttinger parameter, and $v$ the sound velocity, and $\phi(x)$, $\partial_{x}\theta(x)$ are the phase and density fluctuations in the fluid, satisfying the commutation relations $[\phi(x),\partial_{x}\theta(x)/\pi]=i \delta(x-x')$. Within the Luttinger liquid formalism for bosons, a weak delta barrier potential gives rise to the lowest order to the backscattering term $\mathcal{H}_{b}\sim 2 U_{0}n_{0}\cos[2\theta(0)]$. By averaging over the quantum fluctuations of the density, $ U_{\text{eff}}=U_{0}n_{0}\langle\cos[2\theta(0)]\rangle$ one obtains $U_{\rm eff}=U_0 (d/L)^K$, where $d$ is a short-distance cut-off length. Hence, quantum fluctuations renormalize the barrier strength. As at increasing interactions, the quantum fluctuations of the phase increase, and those of the density decrease, the barrier is less and less renormalized at increasing interactions. Combination of the two regimes of weak and large interactions yields a non-monotonous behaviour of the effective barrier strength, and hence of the persistent current amplitude. In the case of attractive interactions, the dispersion of collective excitations is not linear but quadratic. Hence, it is not possible to bosonize the fluid. An exact diagonalization study shows that the barrier has a non-trivial behaviour in terms of the ratio of interaction to barrier strength, with a universal scaling emerging  \cite{polo2022quantum}.

The decay and oscillations of supercurrents in the presence of a barrier after an initial phase imprint have been studied in~\cite{polo2019oscillations}. In this work, both thermally driven incoherent phase slips as well as quantum coherent phase slips were observed depending on the interaction regime. 

A quasi-one-dimensional geometry is realized by coupling two one-dimensional rings. If the rings are stacked on top of each other or concentric, particles can tunnel uniformly among each ring. This corresponds to a mesoscopic ladder geometry with periodic boundary conditions and allows to realize both Meissner phase, corresponding to chiral currents, with negligible inter-ring tunneling and the vortex phase, corresponding to local circulating flows among the two rings~\cite{haug2018mesoscopic,victorin2018bosonic}, where the advantage of the ring geometry is that it is possible to probe the current states by established readout methods (see Sec.~\ref{sec:Interference} for readout).

Another possibility of ring coupling is to consider two lattice rings side-by-side, where a single site of a ring is tunnel coupled to a single site of the second ring. This geometry is particularly suited for demonstrating quantum coherent phase slips by following the winding number transfer among the two rings~\cite{perezobiol2022coherent}.

\subsection{Persistent currents in weakly interacting three-dimensional ring traps}
While the one-dimensional analysis can capture relevant traits of the physics of the persistent currents, important features of the phenomenon are distinctive of higher dimensions - see Sec.~\ref{sec:ExpResBose}. 
In this case, the theoretical analysis can be conducted through mean-field methods. In particular, the time-dependent Gross-Pitaevskii equation (TDGPE, Eq.~\eqref{eq:GPE}) has turned out to provide an excellent approximation to study the system in weak interaction regimes. Typically, the initial condition for the TDGPE is the equilibrium condition in the initial trap and is obtained by first solving the time-independent Gross-Pitaevskii equation, e.g.~by evolution in imaginary times. Then, a specific dynamical protocol is implemented and the condensate density $\rho(\mathbf{r},t)$ and phase $\phi(\mathbf{r},t)$ are obtained from the condensate wavefunction by setting $\Psi(\mathbf{r},t)=\sqrt{\rho(\mathbf{r},t)}e^{i \phi(\mathbf{r},t)}$. Similarly, the momentum distribution is obtained by   $n(\mathbf{k},t)=|{\tilde \Psi}(\mathbf{k},t)|^2$, where ${\tilde \Psi}(\mathbf{k},t)$ is the spatial Fourier transform of the condensate wavefunction.

The TDGPE has been used e.g. in Ref.~\cite{murray2013probing} for simulating the experiment probing the persistent current states by time-of-flight expansion. The circulating state was included in the simulation by imprinting a phase $e^{i \ell \phi}$ with integer $\ell$ values, and then the TDGPE was used to describe the expansion protocol. At the end of the expansion, an annulus forms in the TOF images, with a radius that depends on the imprinted circulation. A very good agreement between the numerical simulations and the experimental data was found, showing the accuracy of the TDGPE for this case.

The TDGPE allows also to study the way to prepare states with given angular momentum by suitable phase imprinting. Thanks to the accurate modeling scheme, the shape of the intensity pattern can be designed and optimized in order to account for finite experimental resolution~\cite{kumar2018producing}.

There are indeed some cases where the GPE does not provide a sufficiently accurate description of the experiment. This is the case when fluctuations, typically of thermal origin, enter into  play  in the experiment and need to be included in the simulations. In this regime, a useful approach is provided by the Truncated Wigner Approximation as implemented e.g. in Refs.~\cite{mathey2014decay,mathey2016realizing}. In essence, this classical field approach includes to lowest order thermal and quantum fluctuations beyond the Gross-Pitaevskii equation. Starting from a 3D Bose-Hubbard Hamiltonian, the method performs a classical field approximation on each site. The fields are propagated by classical equations of motion, with initial conditions generated by sampling from the Wigner distribution of a weakly interacting Bose gas in the Bogoliubov approximation at finite temperature. The resulting classical trajectories are finally averaged over the initial conditions for the calculation of observables. This method has been used  to perform theoretical investigations of the decay of supercurrents~\cite{mathey2014decay} and to design a protocol to realize an atomtronic SQUID~\cite{mathey2016realizing}. Other approaches investigating both formation and decay of persistent currents, including stochastic projected GPE, stochastic GPE, stochastic Landau-Ginzburg, Bogoliubov and others \cite{montgomery2010spontaneous,corman2014quench,allman2023quench-induced,piazza2009vortex,dubessy2012critical,mathey2014decay,jendrzejewski2014resistive,snizhko2016stochastic,kumar2017temperature,polo2019oscillations,kunimi2019decay,mehdi2021superflow,perez2020current}.

The coherent tunneling of current states has also been investigated in different geometric scenarios. One can consider how currents can be transported between rings in concentric, sided-coupled, and transverse configurations. For a concentric configuration and sufficiently large rings, currents can be moved from the inner to outer ring using adiabatic passage techniques~\cite{polo2016transport}, allowing for efficient transport and control of the current. Another configuration that has received a lot of attention is that of sided-coupled ring potentials. From a geometric perspective, this geometry brings some interesting new features, as the breaking of the cylindrical symmetry leads to complex Josephson tunneling amplitudes for triangular configurations~\cite{polo2016geometrically}. These adiabatic techniques have also been proposed for the dynamical generation of angular momentum states via spatial adiabatic passage~\cite{menchonenrich2014tunneling}. Sided-coupled rings have also been investigated when they have a physical overlap and the currents move from the left to the right ring following a fluid description, provided by models such as the GPE, while keeping the quantization restrictions of the circulation~\cite{bland2020persistent,bland2022persistent}. Stacked configurations have also been investigated both in the continuum~\cite{lesanovsky2007spontaneous,brand2010sign,oliinyk2019symmetry,oliinyk2019tunneling,nicolau2020orbital,bazhan2022generation} and in lattices~\cite{amico2014superfluid,escriva2019tunneling} investigating the Josephson dynamics between the rings and how currents can spontaneously emerge in these configurations. More complex scenarios have also been investigated and proposed in regard to the artificial magnetic field acting in bosonic ring geometries. When an inhomogeneous artificial gauge field is applied, the behavior of the currents is strongly affected by spatial dependence of the field, affecting their quantization properties, introducing a linear dependence of the plateaus versus the strength of the applied field~\cite{hejazi2022formation}.

\subsection{Vortex dynamics and phase slips in toroidal BEC}\label{VortexDynamicsBose}

\begin{figure}[h!]
    \centering
    \includegraphics[width = 0.7\textwidth]{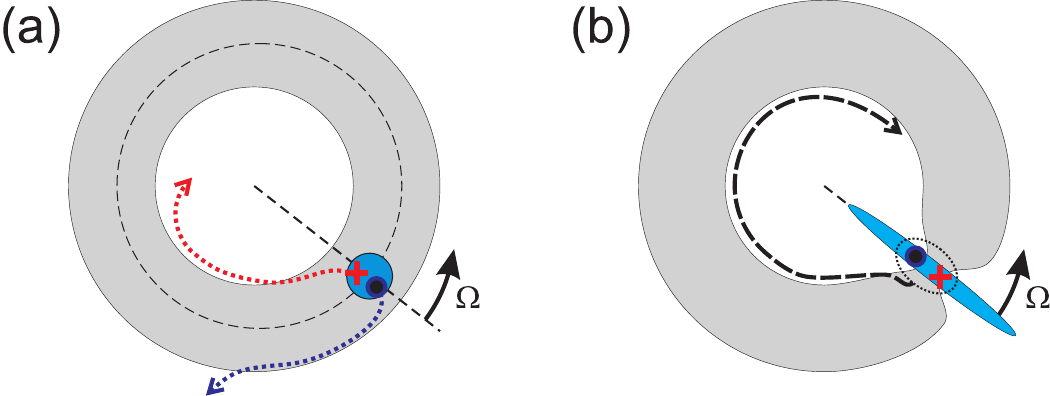}
    \caption{ Schematics of two different scenarios of the persistent current generation in toroidal BEC by rotating blue-detuned laser beams: \textbf{(a)} For small stirrers a vortex-antivortex pair is nucleated near the center of the rotating barrier in the bulk of the condensate. Then the pair undergoes a breakdown and the antivortex (blue circle) moves spirally to the external surface of the condensate and finally decays into elementary excitations, while the vortex (red cross) becomes pinned in the central hole of the annulus. \textbf{(b)} For phase slips driven by a rotating weak link, the vortex line from the external periphery and the anti-vortex line from the internal region approach each other and create a vortex-antivortex dipole. This coupled pair of vortices circles clockwise inside the central hole of the toroidal condensate. Figure from Ref.~\cite{yakimenko2014generation}.}
    \label{fig:vortexdynamics}
\end{figure}

The persistent flow is indeed started by quantum phase slips of the superfluid order parameter $\psi(r)=\sqrt{\rho(r)} e^{i\phi(r)}$ occurring in the proximity of the region where the superfluid is stirred and accompanied by production and motion of vortices. Quantum phase slips and vortices determine quantized winding numbers of the superfluid flow and its decay\cite{piazza2009vortex,kumar2017temperature,jendrzejewski2014resistive,dubessy2012critical,wright2013threshold,eckel2014hysteresis}. Vortices are generated as a result of a non-uniform backflow along the cross-section of the torus; the backflow itself occurs as the superfluid tends to fill the matter's low-density region created by the stirring~\cite{mehdi2021superflow}. A semi-classical picture, obtained by equating the velocity of the flow and the critical sound speed inside the barrier, gives an accurate estimate of the conditions for vortex excitations to enter the ring~\cite{piazza2011instability}. By increasing the height of the potential barrier set by the stirring laser, the backflow can be reduced, affecting in turn vortex and phase slip dynamics. The quantitative study of such barrier height dependence was carried out by Yakimenko's group in a series of  works based on  3D and 2D GPE~\cite{yakimenko2015vortices,yakimenko2015vortex,yakimenko2014generation} --see Fig.~\ref{fig:vortexdynamics}. The analysis indicates that,  for moderate barrier heights, a vortex-antivortex pair is in the immediate spatial position of the stirring laser beam; then, above a critical angular velocity of the stirring, the vortex-antivortex pair unbinds,  with the vortices propagating in the condensate at nearly the speed of sound; finally, the antivortex decays as result of dissipation and the vortex is effectively pinned at the center of the ring and sets the persistent flow. For a stronger barrier,  the vortex-antivortex pair is formed again close to the stirring potential, but {\it the whole pair} moves around the internal side of the torus, rotating in the opposite direction to the stirring; in this rotation, vortices transfer their winding number to the condensate that is eventually set into motion.\footnote{The effects caused by a red-detuned attractive stirring potential were also analysed~\cite{weimer2015critical}.  In this case, theoretical simulations predict that the superfluid is characterized by a specific phononic decay with suppression of vortex-antivortex pairs nucleation~\cite{singh2016probing}.}
 
Phase slip dynamics was demonstrated to provide the underlining mechanism to set a persistent current also in non-ring geometries as racetracks confinements~\cite{eller2020producing}. For stirring barrier strengths above a critical value,  transitions from curved to straight racetrack parts are characterized by a circulation that results in an increase of one unit; in the transitions from straight to curved parts, instead the circulation results in  a decrease by one unit. The analysis theoretically demonstrated that, playing with stirring barrier height, stir speed and racetrack geometry, quantized smooth atomic flows can be engineered on–demand in such systems. The interplay between curvature and atom-atom interactions in elliptic systems was explored in Refs.~\cite{nikolaieva2023engineering,tononi2024quantum}. 

In one-dimensional systems, the phenomenon of phase slips is no longer associated with vortices but with other excitations such as gray solitons. Different dynamical regimes have been evidenced. Indeed, currents can result to be self-trapped, oscillating or decaying when a defect is present in the system. These dynamics can be explained in terms of a dual Josephson model involving phase slips of thermal or quantum nature~\cite{polo2019oscillations}. In particular, for weak interactions phase slips are triggered by the formation of sound waves and thermally activated dark solitons, whereas for strong interactions they appear due to a degradation of the phase by higher modes at large barrier strength or by thermal fluctuations at finite temperature. 

\begin{figure}
    \centering
\includegraphics{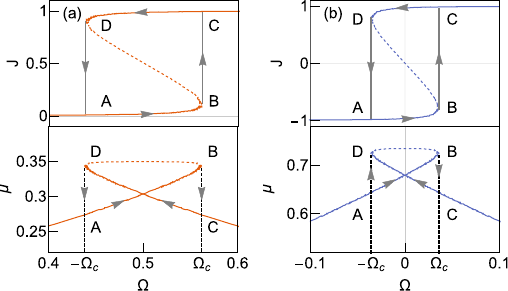}
    \caption{Hysteresis cycles in terms of currents and stirring velocities (top plots) and corresponding energy diagrams in form of swallowtails (bottom plots). \textbf{(a)} Stirring of the ground state up to $\Omega_\text{c}$, A$\rightarrow$B, transition to current $J=1$ through the gray line (vertical arrow), B$\rightarrow$C, and stirring in opposite direction, C$\rightarrow$D, where the condensate decays to the original state, D$\rightarrow$A. \textbf{(b)} When a link is set in a vortex state, two solitons are produced, but the hysteresis paths are analogous to the ground state case. In this cycle, the paths are cut short at much smaller critical velocities.  Figure from Ref.~\cite{perez2020current}.}
    \label{fig:perez}
\end{figure}

As was discussed in Sec.~\ref{sec:ExpResBose}, the persistent flow is characterized by a hysteretic behavior as shown in Ref.~\cite{eckel2014hysteresis} and shown in Fig.~\ref{fig:hysteresis}: when a rotating flow is made to decrease,  the transition to lower winding numbers occurs at stirring frequencies smaller than the ones that were imparted to achieve the original rotation one started from~\cite{eckel2014hysteresis}. The hysteresis results in a decrease as the barrier strength increases. While the quantitative understanding of such a phenomenon is still missing, there is a consensus in the community that hysteresis occurs as a result of the irreversibility encoded in vortex decay~\cite{snizhko2016stochastic,yakimenko2014generation,perez2020bose}. In Ref.~\cite{eckel2014hysteresis}, hysteresis among different branches is observed and proposed as a method to control an atomtronics device. In the mean-field regime, one can also investigate how hysteresis can be understood when following the ground state solution of a stirred BEC. In particular, in \cite{perez2020bose,perez2020current}, an exact analysis through GPE was shown to produce unstable solutions forming loops in the energy landscape of the system. As shown in Fig.~\ref{fig:perez}, this leads to the facilitation of phase slips to access more stable states as the velocity of the barrier is increased above a critical velocity. Motivated by the implementations of the atomic counterpart of the dc-SQUID~\cite{ryu2013experimental,ryu2020quantum}, two moving weak links on the ring in opposite directions have been studied~\cite{jendrzejewski2014resistive}.  For such a configuration, vortex formation and phase slip dynamics explaining the current-phase relation was studied~\cite{singh2016probing,eckel2014interferometric}. Bosonic ring condensates with radii of a few hundred microns have been considered, therefore demonstrating the remarkable progress achieved in the coherent manipulation of condensates that can be seen with the naked eye. 

Even though the current status of the experimental techniques allows to image vortices~\cite{reeves2022turbulent,kwon2021sound}, the effects of the phase slips and vortices in the read-out was not analysed in detail yet.  
Indeed,  spiral interferograms obtained by the heterodyne dynamics protocols (see Secs.~\ref{sec:Interference} and~\ref{sec:ExpResBose}) are sensitive to phase fluctuations. In particular, a specific discontinuity in the azimuthal dependence of the read-out density  (dislocations)  can be noticed to reflect phase slips ~\cite{corman2014quench,roscilde2016quantum}. For correlated systems on lattices, the phase information can be achieved also by studying noise correlations in the expanding density~\cite{haug2018readout}. 

\section{Persistent currents in fermionic gases}\label{sec:PerCurrFerm}

Even though a snapshot of a persistent current of an ultracold interacting fermionic gas trapped in a ring-shape optical potential and pierced by a synthetic field  was achieved long ago~\cite{amico2005quantum}, an in-depth analysis of the problem has been carried out only recently. In this section, we summarize the bulk of knowledge acquired so far in the subject. Clearly, the existence of a Fermi energy, the BCS-BEC crossover~\cite{chen2005bcs,zwerger2011bcs,giorgini2008theory,strinati2018BCS-BEC}, together with the different non-classical many-body states  characterizing the system are just some of the aspects making the persistent currents of fermionic ultracold atoms a very relevant problem to analyse, both for basic and applied research in quantum technology. Important breakthroughs in preparing, trapping, coherently manipulating  and  imaging fermionic matter-waves  have been accomplished in the experiments only recently. 

\subsection{BCS-BEC cross-over}\label{sec:FermCross}

A mixture of weakly attractive two-component Fermi gases forms pairs 
 at opposite momenta around the Fermi sphere. This is the renowned Bardeen-Cooper-Schrieffer (BCS) state, which has been introduced to explain superconductivity. By tuning to stronger values the attraction among the fermions, the Cooper pairs modify onto smaller and finally close to point-like bosons, which can undergo Bose-Einstein condensation of such composite bosons. This is the celebrated BCS-BEC crossover, which  plays an important role in different domains, ranging from nuclear to condensed matter physics. Ultracold atomic gases provide a unique platform to explore the BCS-BEC crossover as the interactions are tunable. This allows in particular to test different theoretical approximations. In the BCS regime, the correlation length of the bound pairs is large compared with the typical interparticle distance; in the BEC regime, instead, the pairs are tightly bound in the real space and the pair correlation length is much smaller than the distance between the particles. At intermediate values of the  BCS-BEC crossover, one reaches a regime of infinitely attractive interactions, or unitary limit, where the gas properties are fixed just by the density and temperature of the gas. 

\subsection{Experimental results}\label{sec:ExpResFerm}
Following the experimental successes with ring-shaped condensates of weakly-interacting bosonic atoms, there remained the challenge of how to adapt those tools and techniques to work with pair condensates of fermionic atoms. Beyond the primary motivation of enabling the study of fermionic pair superfluids, there was also the promise of being able to study superfluid rings with tunable interactions for the first time, and study superfluidity across the BEC-BCS crossover. There were numerous technical challenges to be overcome to make this possible, however. Achieving deep degeneracy with ultracold fermionic atoms is comparatively more difficult than with bosonic atoms, and fermionic systems are extremely susceptible to heating caused by losses. Condensate fraction is invariably lower in fermionic systems, which reduces the fringe contrast when matter-wave interference techniques are used. Creating a fermionic pair superfluid also requires working with spin mixtures and considering state-dependent interaction effects between atoms and with external fields. Avoiding inelastic collisions and managing 3-body losses is a larger challenge, particularly in the molecular BEC limit. Techniques that rely on matter-wave interference in time-of-flight are made more complicated by higher chemical potential and interaction energy, and the fact that when interactions are tuned to the attractive (BCS) limit, the pairs break during ballistic expansion and information about pair coherence is lost. Uniformity and agile control of large magnetic fields needed to tune interactions using Feshbach resonances is also critical, given the stringent flatness requirements for a ring potential. Programs to overcome these obstacles were undertaken in parallel at Dartmouth and LENS, with the first results appearing in 2022~\cite{cai2022persistent,del_pace2022imprinting}. Both of these groups chose to work with $^6$Li, which has an unusually broad Feshbach resonance and much lower losses in the molecular limit than other fermionic species such as $^{40}$K.

\begin{figure}[h!]
    \centering
    \includegraphics[width=0.6\textwidth]{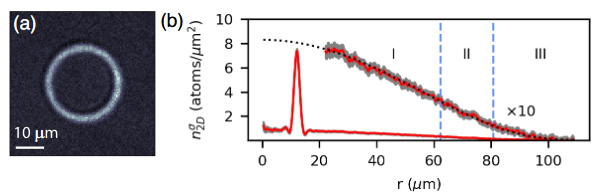}
    \caption{(a) Absorption image of an equal
spin mixture of lithium atoms in a ring-dimple trap, where there is a ring-shaped region of high chemical potential embedded in a broad halo of lower chemical potential. (b) Experimental radial column density. The black dotted line is the expected density profile for 
an ideal Fermi gas in the halo region. In region I, the 
system is 3D degenerate, II is quasi-2D degenerate, and III is quasi-2D thermal. The temperature of the system was inferred by fitting to the profile of the halo. Figure from Ref.~\cite{cai2022persistent}.}
    \label{fig:DartmouthFermion1}
\end{figure}

The group at Dartmouth used an all red-detuned optical ring-dimple trap with an average ring radius of 12 $\mu m$ and a radial half-width of $1/e^2$ 2.2 $\mu$m (See Fig.~\ref{fig:DartmouthFermion1}). The ring-pattern beam was initially generated using an axicon-based mode converter, but this was later replaced with a digital micromirror device to allow greater experimental flexibility. Evaporating near resonance with an equal spin mixture of atoms in the lowest two hyperfine states, they were able to reach temperatures well below the pair-superfluid phase transition for a population of a few thousand atoms localized to the ring-dimple region. They were then able to reliably initialize the superfluid in a chosen current state using a focused blue-detuned stirring beam~\cite{cai2022persistent}. Currents flowing around this single ring were detected using a time-of-flight self-interference technique, after ramping interactions to the molecular BEC limit to prevent pair fragmentation. The observed current lifetimes exceeded 10 seconds in the strongly-interacting limit, and temperatures were low enough for the currents to be robust as interactions were tuned to the BCS limit, as far as $1/k_F a=-1$. The Dartmouth group's ability to work at very low temperatures over comparatively long time scales was later determined to be a serendipitous feature of the all-red-detuned ring-dimple trap used in the experiments, which was found to significantly reduce the heating caused by background gas losses~\cite{allman2023heating}. The rate of current decay increased further into the BCS limit, as the critical temperature reached the system temperature. Interestingly, driving the system into the normal phase did not cause a rapid decrease in the current around the ring; the supercurrents reappeared with a probability of around 0.3 after a slow interaction ramp back to the superfluid regime, indicating weak damping of the flow even in the normal phase.

\begin{figure}[h!]
    \centering
    \includegraphics[width=0.6\textwidth]{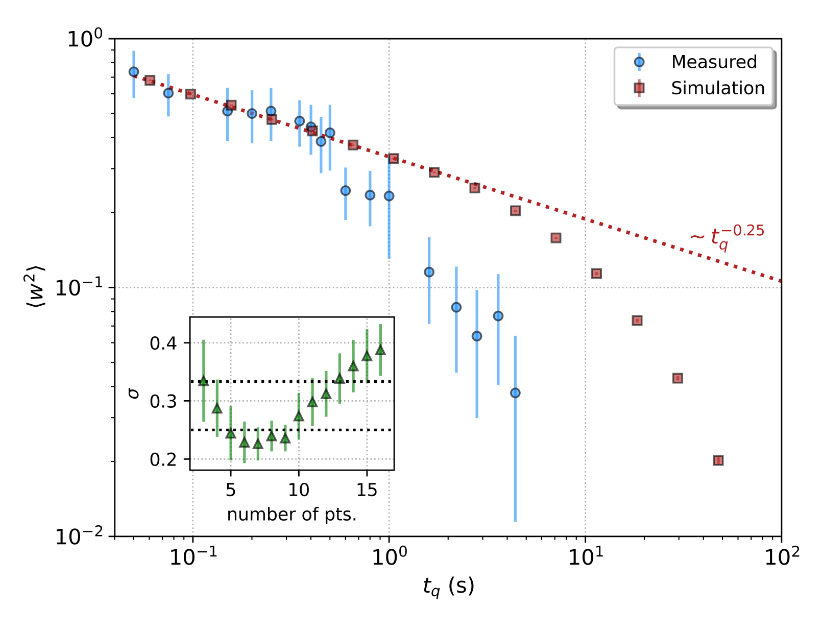}
    \caption{Measured mean-square winding number versus ramp duration (blue circles).  The straight red dotted line shows a power law with exponent 0.25. The inset shows the power law fit-extracted exponents $\sigma$ obtained from fits to various numbers of fastest-quench data points. The dotted lines show the mean-field and F-model predictions $\sigma = 1/4$ and 1/3, respectively. Figure from Ref.~\cite{allman2023quench-induced}.}
    \label{fig:dan_quench}
\end{figure}

Faster interaction-driven ramps across the normal-to-superfluid phase transition were observed to result in spontaneous current formation, even for a non-rotating normal ring. This motivated the Dartmouth group to measure the rate of spontaneous current formation as a function of the quench rate, and compare the results against predictions obtained from Kibble-Zurek theory as shown in Ref.~\cite{allman2023quench-induced} and Fig.~\ref{fig:dan_quench} (see also Sec.~\ref{sec:ExpResBose}). In these experiments they drove thermal quenches as fast as 50 ms without substantial change in the peak density and chemical potential of the ring by using a hybrid quench protocol where the interaction strength and trap depth were ramped simultaneously. The trap geometry was a concentric double-ring, with one ring treated as a reference ring for interferometric detection of the current state in the other ring. For the fastest quench rates, the freeze-out time was reached before beyond-mean-field effects become significant and the spontaneous current formation rate matched well with predictions of mean-field Kibble-Zurek theory. For slower quenches, the rate of current production declined more rapidly. This was attributed to finite size effects, with the correlation length becoming comparable to the ring circumference for sufficiently slow quench rates.

\begin{figure}[h!]
    \centering
    \includegraphics[width=0.8\textwidth]{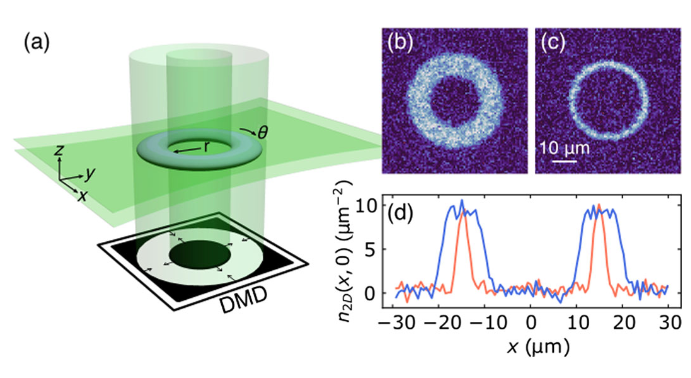}
    \caption{Experimental demonstration of rings with ultracold fermionic superfluids. \textbf{(a)} Sketch of the experimental setup of blue-detuned beam for confinement in $z$-axis, and DMD-made repulsive potential in the $x$-$y$ plane. \textbf{(b)}-\textbf{(c)} In situ images of a unitary Fermi superfluid confined in two ring geometries with different radial thickness. (d) Density cut along the diameter of the rings in \textbf{(b)} and \textbf{(c)} are shown as blue and orange lines. Figure from Ref.~\cite{del_pace2022imprinting}.}
    \label{fig:LENS_1}
\end{figure}

The work undertaken in parallel by the group at LENS used an all-blue-detuned trap configuration, with a DMD-generated pattern beam allowing them to create rings of varying radius and aspect ratio, as shown in Fig.~\ref{fig:LENS_1}. In their earliest experiments, they mainly used a ring of typical inner (outer) radius of 10 (20) $\mu$m, with a disk-shaped central region in a ``target'' trap that allowed detection of the current in the ring by matter-wave interference~\cite{del_pace2022imprinting}, as had previously been demonstrated with BECs~\cite{eckel2014interferometric}. One of the most notable aspects of this work is that it was the first use of gradient phase imprinting to create currents in a ring superfluid, a technique which we discussed in Sec.~\ref{sec:ExpMethods}. They were able to create low current states with excellent fidelity, and in their system the $\ell=1$ state was observed to be stable for around 3 seconds in the unitary limit. Winding numbers up to $\ell=9$ were created by up to 5 repetitions of the phase imprinting protocol, and the maximum stable winding number was found to vary as the interactions were tuned across the Feshbach resonance, with a maximum occurring in the unitary regime.

\begin{figure}[h!]
    \centering
    \includegraphics[width=0.6\textwidth]{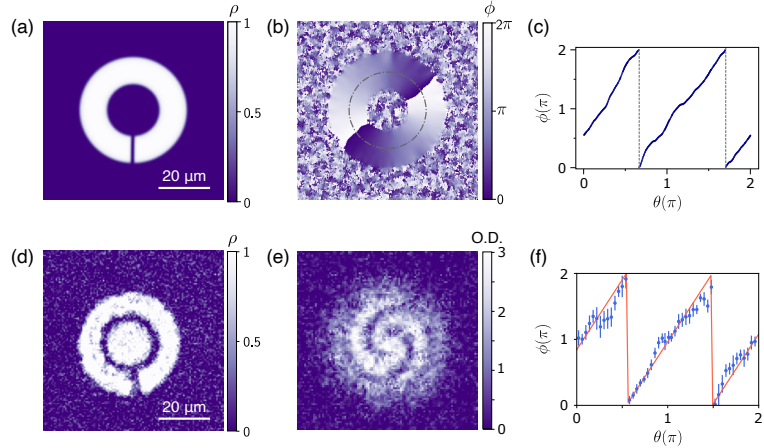}
    \caption{Comparison between numerical and experimental observables. \textbf{(a)} Numerical density profile extracted immediately after a realistic phase imprinting. \textbf{(b)} Numerical superfluid phase in the xy-plane 50 ms after the imprinting. \textbf{(c)} Azimuthal profile of the numerical phase profile as in \textbf{(b)} along the circle marked by the dash-dotted gray line. (d) Experimental in situ density profile of the superfluid ring immediately after the imprinting. \textbf{(e)} Experimental interferograms obtained 50 ms after the same imprinting as in \textbf{(d)}. \textbf{(f)} Azimuthal profile of the phase difference measured from the interferograms of \textbf{(e)}. Figure from Ref.~\cite{xhani2023decay}.}
    \label{fig:LENS_2b}
\end{figure}

In a follow-up study which compared experimental data in the BEC limit against numerical modeling using the Gross-Pitaevskii equation as shown in Fig.~\ref{fig:LENS_2b}, the group at LENS was able to characterize the effects of imperfections in the potential both on the optical gradient imprinting procedure and on the subsequent dynamics and decay of superflow~\cite{xhani2023decay}. Optical resolution limits make it impossible to realize the ideal gradient profile, and this leads to the inevitable formation of density excitations in the vicinity of the localized reverse-gradient part of the optical potential. They observed that there was a maximum achievable winding number using gradient phase imprinting, and that this was principally due to the  formation and evolution of these density excitations, which lead to the entrance of vortices into the inner edge of the superfluid ring that subsequently migrate to the outer edge of the ring and decrease the circulation in the system. The BCS and unitarity regimes were theoretically investigated in~\cite{pisani2024critical}, where the maximum winding number was explained in terms of the Landau critical velocity at which the intrinsic critical current of the superfluid is reached and spontaneous decay of the superflow begins.

The LENS group investigated the dynamics of vortex nucleation in further detail by introducing a localized defect in the ring potential. In their earliest work they found that a weak, localized defect ($V_0\leq 0.2 E_F$, $D= 1.6$ $\mu$m) in a wide ring had no effect for currents below a critical winding number of $\ell\approx5$. In their later work they created potential bumps in the middle of the flow channel with height up to twice the chemical potential. As they increased the winding number/current they found that the flow velocity between the obstacle and the inner edge of the ring would eventually exceed the local speed of sound, leading to nucleation of a vortex that then enters the bulk of the superfluid in the ring. A higher potential barrier in the defect was found to lead to a lower threshold for the nucleation of a vortex and subsequent decrease in the circulation around the ring. 

\begin{figure}[h!]
    \centering
    \includegraphics[width=0.8\textwidth]{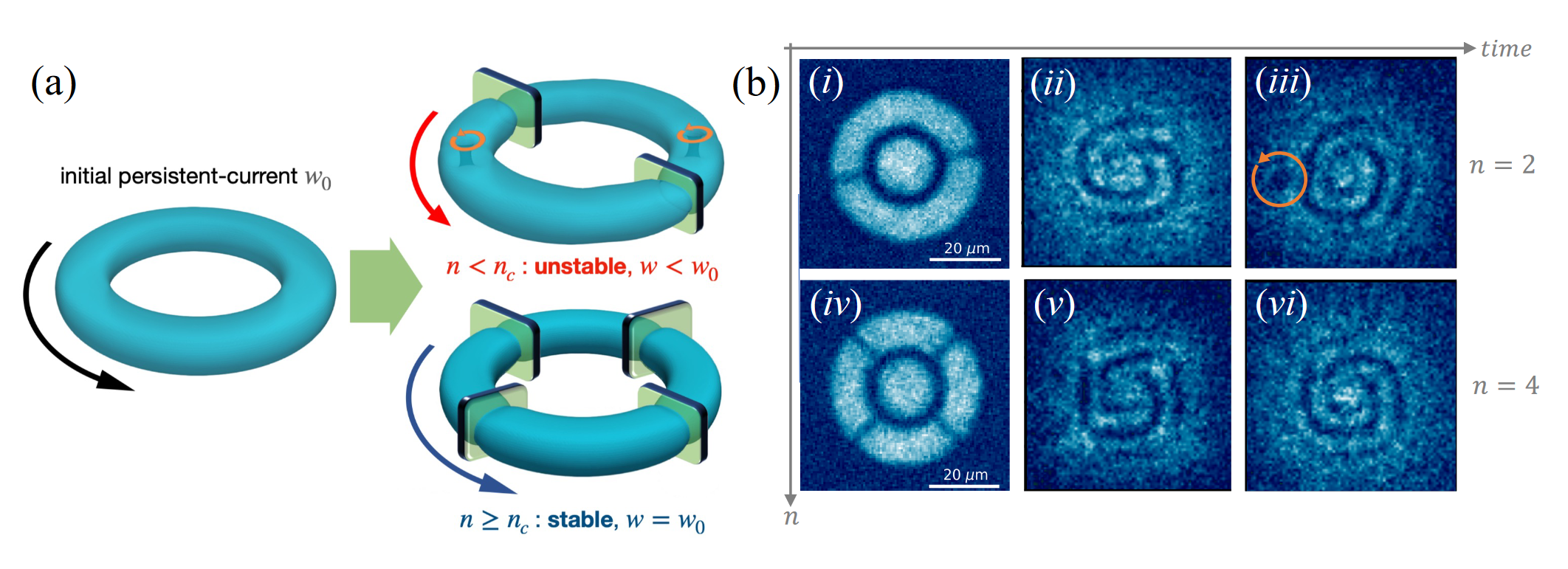}
    \caption{\textbf{(a)} After preparing an initial persistent current state with circulation $w_0$, the $n$ junctions are ramped up. The 3D density plots are isosurfaces obtained from 3D GPE numerical simulations of the experimental setup for the molecular BEC regime. If $n<n_c$ is below a critical value, the initial current is dissipated via the nucleation of vortices. Conversely, if $n\geq n_c$, the system remains stable with $w = w_0$ (lower right plot). \textbf{(b)} Examples of single-shot experimental in-situ images and interferograms obtained for $w_0 = 2$ and for the same number of junctions $n$ as in \textbf{(a)}: n = 2 (unstable configuration), at $t = 0$ (i), $t = 1$ms (ii) and $t = 7$ms (iii); and $n = 4$ (stable configuration) for $t = 0$ms (iv), $t = 1$ms (v) and $t = 20$ms (vi). In the case (iii), the circulation has decayed ($w(t) < w_0$) and the vortex emission is identified by the single spiral arm and the presence of a vortex. Figure from Ref.~\cite{pezze2023stabilizing}.}
    \label{fig:LENS_3}
\end{figure}

Stabilizing the current around a superfluid circuit against such decays is an important experimental objective, and the LENS group recently presented important results as shown in Fig.~\ref{fig:LENS_3}. They demonstrated that the maximum stable winding number rises as one inserts an increasing number of tunnel junctions in a 1D array around a superfluid ring~\cite{pezze2023stabilizing}. While this result seems at first paradoxical when compared to the effect of inserting a point defect, there is an essential difference in the way the flow around the ring is altered. While there is an increase in the flow velocity at any junction relative to the velocity of flow the bulk, if the winding number is held fixed, the velocity increase at each junction is reduced as the number of junctions is increased. This effectively stabilizes the flow against nucleation of vortices that would otherwise occur if the phase drop across any junction approaches $\pi$. While this experiment was conducted in a molecular BEC, the results have potentially important implications for any system supporting metastable superflow, and where Josephson junctions may be employed.

\subsection{Theoretical Understanding}\label{sec:ResFerm}

\begin{figure}[h!]
    \centering
    \includegraphics[width=0.6\textwidth]{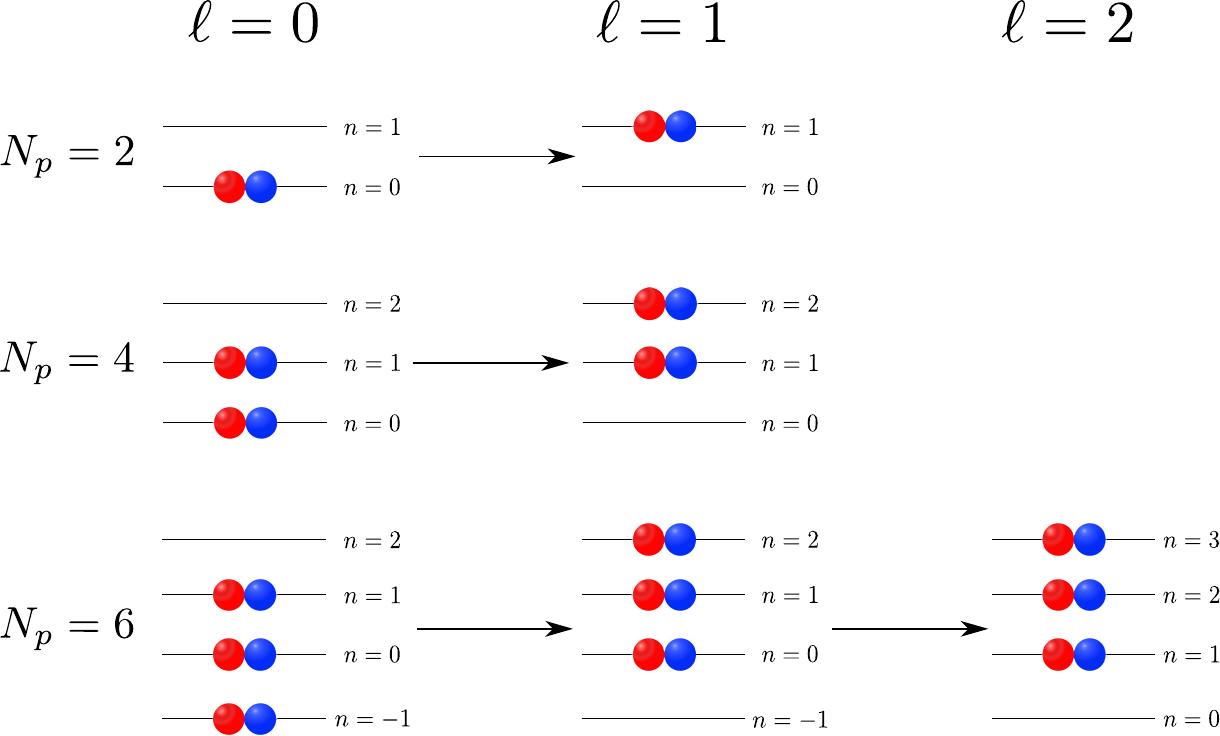}
    \caption{Schematic representation of the energy level occupations for non-interacting $N_{p}$ fermions with SU(2) symmetry and their behaviour with increasing angular momentum per particle $\ell$. To counteract the increase in $\ell$, the Fermi sphere displacement acquires a shift. Naturally, due to the Pauli exclusion, such a displacement for larger $N_{p}$ is mediated by going to higher $\ell$. Image reprinted from~\cite{chetcuti2022interference}.}
    \label{fig:levelsFerm}
\end{figure}

 Still in its infancy, the theoretical study of persistent currents  of ultracold matter in fermions is currently restricted to one-dimensional rings. At zero temperature and in the absence of interactions, the total energy of a spinless fermionic system is given by 
\begin{equation}\label{eq:FermEnergy}
    E(\phi) = \sum\limits_{\{n\}}\frac{\hbar^{2}}{2m}\bigg[\frac{2\pi}{L}\bigg(n-\frac{\phi}{\phi_{0}}\bigg)\bigg]^{2},
\end{equation}
with $k_{n}=\frac{2\pi}{L}(n-\phi/\phi_{0})$ being the associated momentum of the eigenstates and $\{n\}$ being the set of quantum numbers corresponding to the energy levels occupied by the particles as shown in Fig.~\ref{fig:levelsFerm}.\footnote{The set of quantum numbers $\{n\}$ is inherently related to the charge quantum number $I_{j}$ parameterizing the Bethe equations with $I_{j}=n$ or $I_{j}=n+\frac{1}{2}$ with $j=1,\hdots, N_{p}$. By construction, the sum of these charge quantum numbers is integer and is related to the angular momentum per particle as $\ell = N\frac{\sum_{j} I_{j}}{N_{p}}$.} In contrast with bosons, only $N$ fermions can occupy a single-particle energy level due to the particle statistics. The resulting momentum distribution, centered compactly around zero to minimize the energy, is symmetric or asymmetric depending on whether the number of particles is $N_{p}=(2m+1)N$ or $N_{p}=(2m)N$ for integer $m$, i.e.\ $\frac{N_{p}}{N}$ is odd or even respectively, giving rise to the parity effect: a diamagnetic or paramagnetic current respectively --Fig.~\ref{fig:levelsFerm}. Such a phenomenon is the straightforward generalization of the facets of the Leggett theorem discussed in Sec.~\ref{sec:bloch} to $N$-component fermions~\cite{waintal2008persistent,chetcuti2021persistent}. As the effective magnetic flux threading the system increases, the Fermi sphere is displaced so as to counteract it and minimize the energy as per Eq.~\eqref{eq:FermEnergy}~\cite{pecci2022single,chetcuti2022interference}. 

\begin{figure}[h!]
    \centering
    \includegraphics[width=0.37\textwidth]{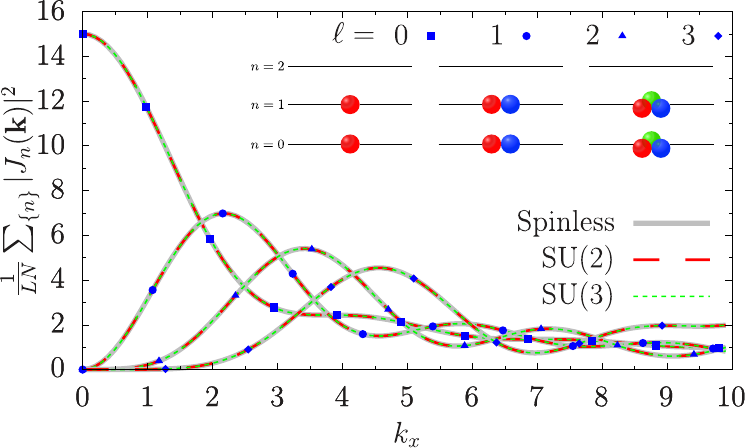}
    \includegraphics[width=0.58\textwidth]{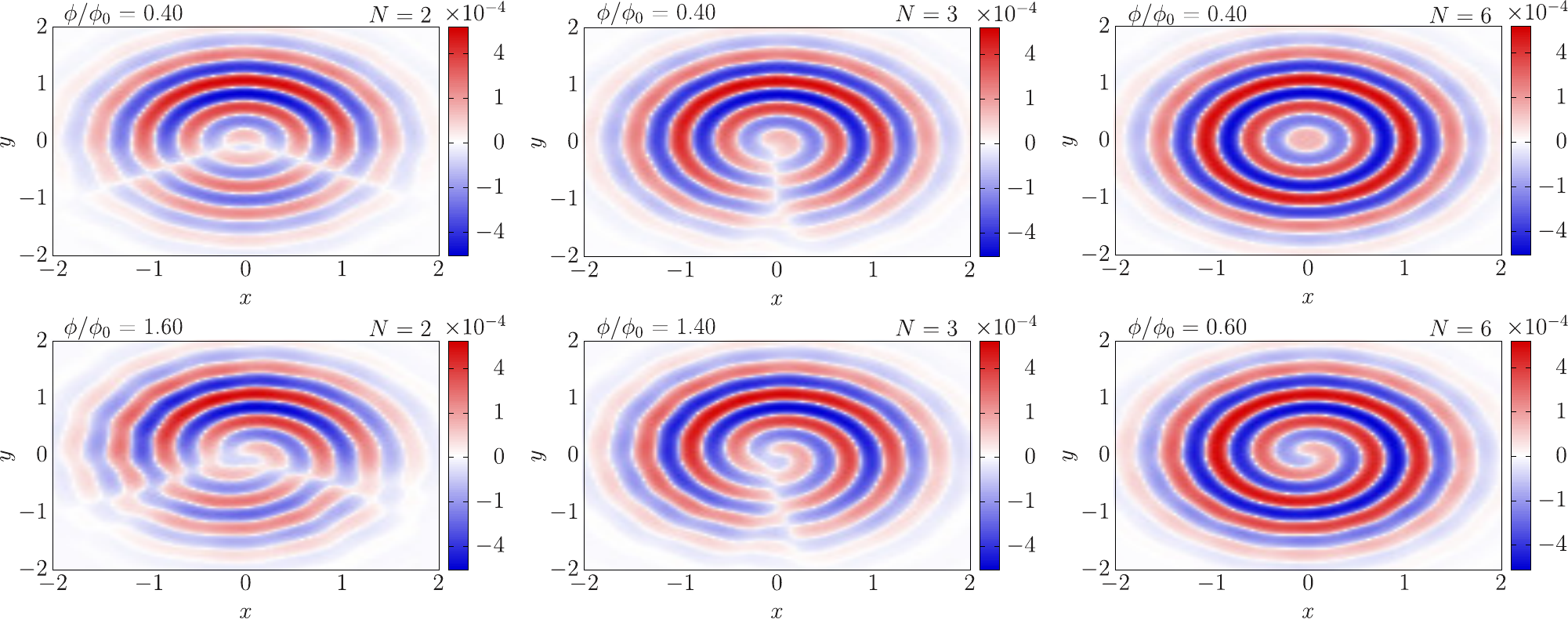}
    \put(-460,105){(\textbf{a})}
    \put(-282,105){(\textbf{b})}
    \caption{Interference dynamics for weakly interacting SU($N$) fermions. (\textbf{a}) \textit{Homodyne protocol}: Momentum distribution $\sum_{\{n\}}|J_{n}(\mathbf{k})|^{2}$ re-scaled by the number of components $N$ as defined in Eq.~\eqref{eq:nkbessel} with $n$ corresponding to level occupation of the particles. At angular momentum per particle $\ell=0$, the momentum distribution is peaked at the origin. Nevertheless, for $\ell>0$ there is a collapse, since the Fermi sphere is displaced as in Fig.~\ref{fig:levelsFerm} causing a characteristic hole to appear. The figure depicts three cases of equal number of particles per components: spinless, SU(2) and SU(3). \textit{Self-heterodyne protocol}: Ring-centre interference $G_{R,C}$ for $N_{p}=6$ particles as a function of $N$ and $\ell$ at short time expansion $t=0.025$. Just as in its homodyne counterpart, the appearance of spirals necessitates that the Fermi sphere is displaced by half. For a fixed number of particles ($N_{p}=6$ in this case), the spirals emerge at lower $\ell$ (earlier value of the flux $\phi$) on increasing $N$. Additionally, dislocations (radially segmenting white lines) characterize the interferograms of different $N$. The colour bar is non-linear by setting $\mathrm{sgn}(G_{R,C})\sqrt{|G_{R,C}|}$. Images taken from~\cite{chetcuti2022interference}.}
    \label{fig:FreeMom}
\end{figure}

While persistent currents of interacting fermions and their read-out through TOF expansion were theoretically drawn in~\cite{amico2005quantum},  in-depth studies were carried out only recently~\cite{pecci2022coherence,chetcuti2022interference}. The Fermi sphere effect is a particularly important to understand the intereference fringes.  Specifically, for bosons the characteristic hole and spirals in the homodyne and self-heterodyne protocols emerge when the system acquires angular momentum $\ell$. However, such is not the case for fermions. Recalling Eq.~\eqref{eq:nkbessel}, we have that the momentum distribution is a sum of discrete $n$-th order Bessel functions running over $\{n\}$, with all orders being zero valued at $\mathbf{k}=0$ barring the zeroth order. 

On account of this, the momentum distribution will only collapse at $\mathbf{k}=0$ when there are no particles occupying the $n=0$ such that the zeroth order Bessel function is excluded from the summand --Fig.~\ref{fig:FreeMom}(\textbf{a}). The angular momentum required for this displacement to occur is given by the ceiling function $\lceil \frac{N_{p}}{2N}\rceil$. Accordingly, the momentum distribution of systems having an equal and commensurate $W=\frac{N_{p}}{N}$ are equivalent to each other reflecting their similar features such as the parity effect mentioned previously. Although this also holds for the self-heterodyne interferograms, the presence of dislocations (radially segmenting white lines) arising due to the multiple momenta contributions associated to the specific particle distribution (see Figs.~\ref{fig:all_spirals} and~\ref{fig:FreeMom}(\textbf{b})), allows one to distinguish between systems with equal and commensurate $W$. 

\begin{figure}[h!]
    \centering
    \includegraphics[width=0.5\textwidth]{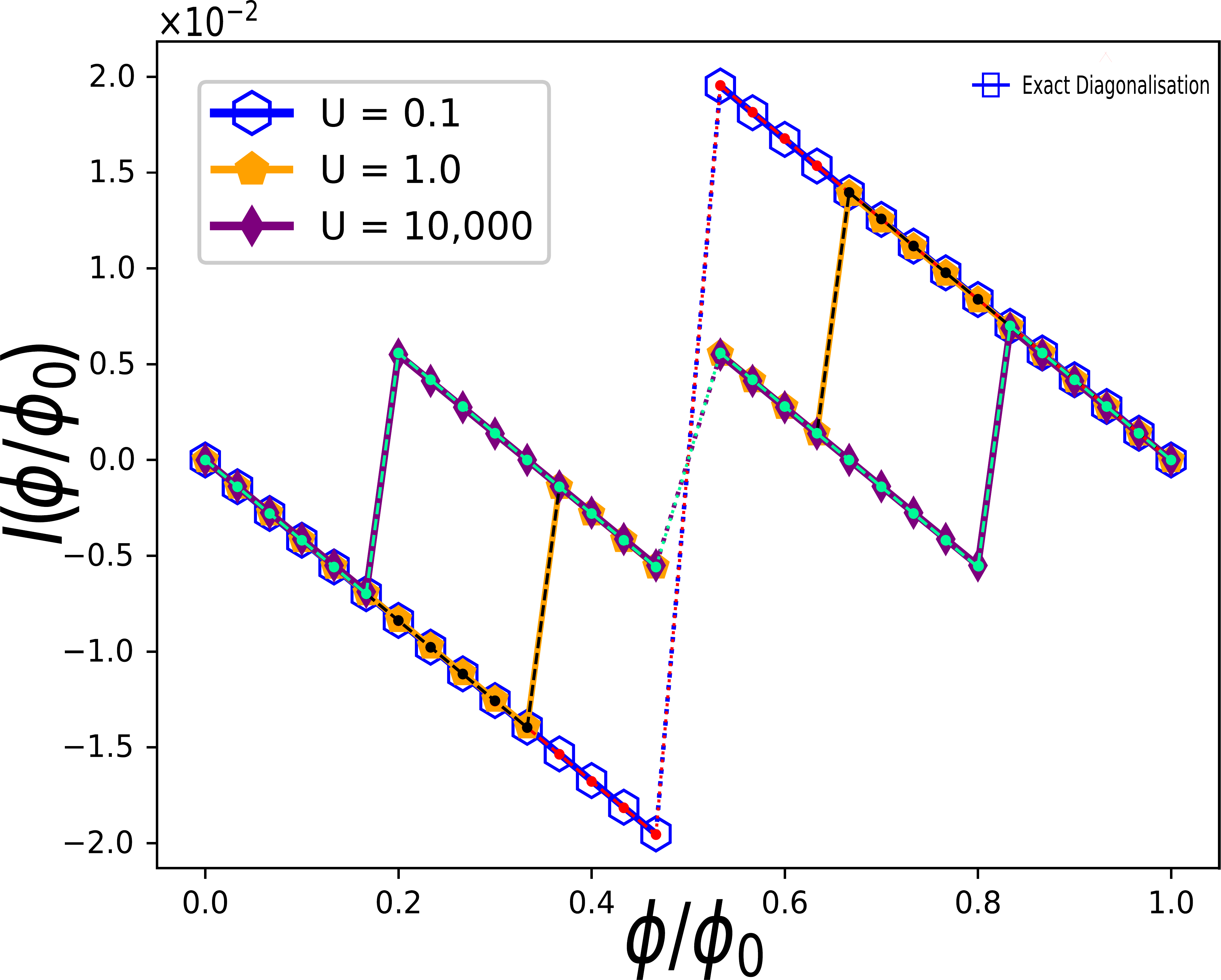}
    \hspace*{4mm}
    \includegraphics[width=0.45\textwidth]{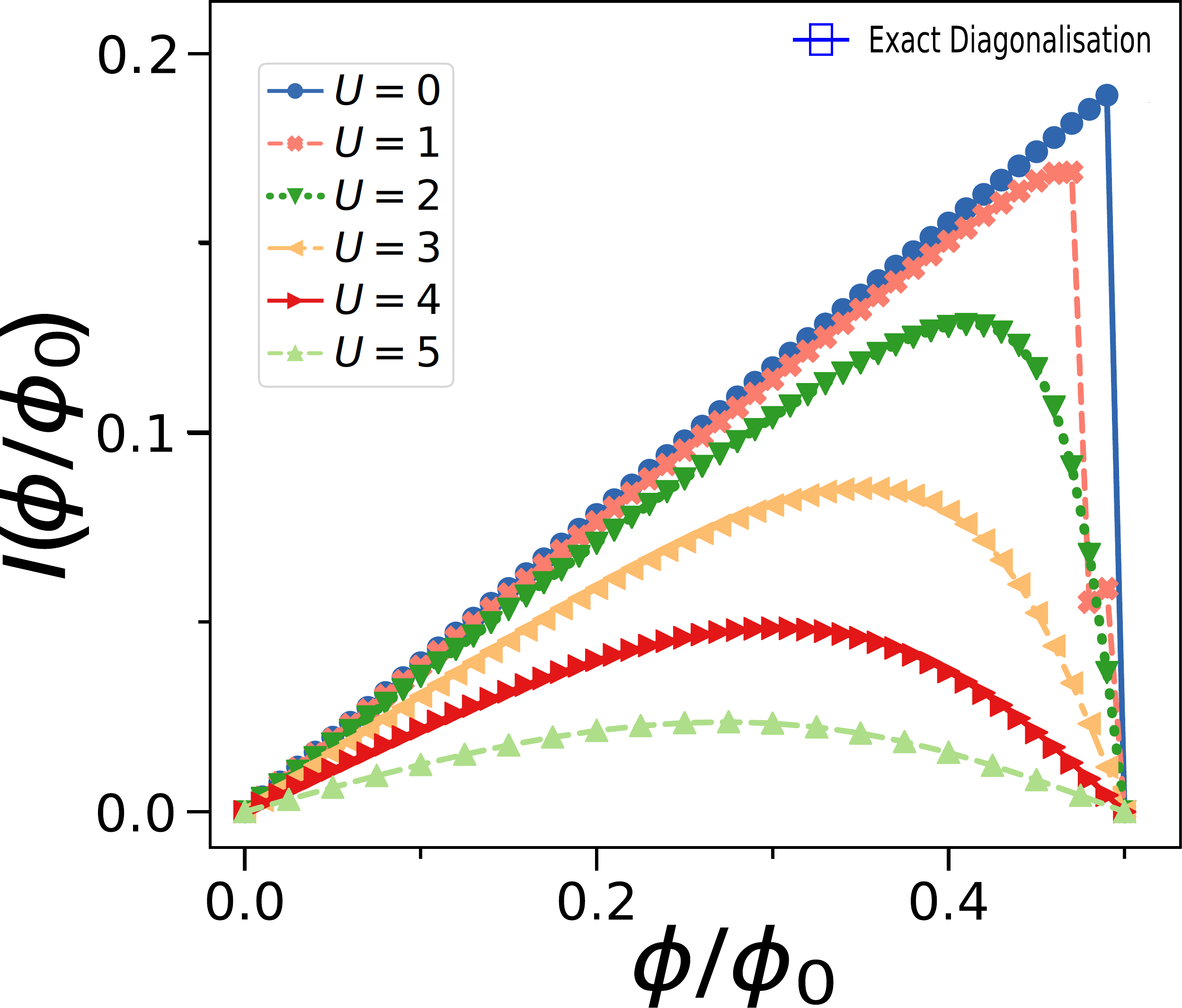}
    \put(-455,180){(\textbf{a})}
    \put(-200,180){(\textbf{b})}
    \caption{Persistent current $I(\phi)$ of SU($N$) symmetric fermions with repulsive interactions $U$ as a function of the synthetic gauge field $\phi$ for different filling fractions. (\textbf{a}) \textit{Incommensurate regime}: Current exhibits a reduced periodicity when going to strong $U$. The quantities $\phi_{d}$ and $\phi_{s}$ correspond to the degeneracy point and flux at which spinons are generated respectively. (\textbf{b}) \textit{Mott regime}: Persistent current as a function of $U$. Changes in its profile (magnitude and shape) capture the onset to the Mott phase transition.  Figure taken from~\cite{chetcuti2021persistent}.}
    \label{fig:PerrFermRep}  
\end{figure}
Similar to the bosonic case, fermionic matter-wave currents also experience a reduced periodicity, dependent on the nature of interactions the atoms are subject to, reflecting the angular momentum fractionalization of the system. Firstly, we turn our attention to strongly repulsive fermions in the incommensurate regime $\nu =\frac{N_{p}}{N_{s}}\ll 1$ modeled by the $\mathcal{H}_{GYS}$ in Eq.~\eqref{eq:GYSHam}, where the periodicity was found to be reduced by $1/N_{p}$ for two-component fermions~\cite{kusmartsev1994strong,yu1992persistent} and later extended to $N$-component fermions~\cite{chetcuti2021persistent} --Fig.~\ref{fig:PerrFermRep}(\textbf{a}). Although this fractionalization is qualitatively similar to what occurs in attractive bosonic systems, the origin of the fractionalization is completely different. From Bethe ansatz analysis, the persistent current can be analytically calculated in the limit of $U\rightarrow\infty$ as
\begin{equation}\label{eq:FermCurrInfU}
    I(\phi) = -2\bigg(\frac{2\pi}{L}\bigg)^{2}\sum\limits_{j}^{N_{p}}\bigg[I_{j}+\frac{X}{N_{p}}+\phi\bigg],
\end{equation}
where $I_{j}$ and $X=\sum_{j}^{N-1}\sum_{\alpha}J_{\alpha_{j}}$ are the charge and spin quantum numbers parameterizing the Bethe ansatz equations (see Sec.~\ref{sec:BA}), which label the entire spectrum. To minimize the energy, spinon excitations are created in the ground-states (acquired by altering $X$) to counterbalance the increasing synthetic gauge field threading the system --Fig.~\ref{fig:PerrFermRep}(\textbf{a}). Due to being quantized in nature, these excitations can only partially compensate the external flux, a continuous quantity, resulting in a current with $1/N_{p}$ periodic oscillations. An additional effect stemming from these energy level crossings between spinon and non-spinon states is the washing out of the parity effect observed for free fermions, with the persistent current being diamagnetic in nature for all cases. The reduced periodicity can be readily seen from Eq.~\eqref{eq:FermCurrInfU}, which reflects that the spin correlations in the system give rise to an effective $N_{p}$-body bound state in a sort of `attraction from repulsion', that albeit following a different route  bears qualitative similarities to the fractionalization occurring in electrons with pairing force interactions and the bright quantum soliton formation discussed in Sec.~\ref{sec:ResBose}. Remarkably, the observed behaviour occurs due to the inherent coupling between the matter and spin degrees of freedom irrespective of the particle statistics. For instance, single-component bosons with repulsive interactions do not exhibit fractionalization, which can be straightforwardly inferred from the fact that in the TG limit they map to spinless fermions. Using similar logic, multicomponent bosons experience fractionalization as in their fermionic counterparts due to the introduction of extra internal degrees of freedom~\cite{pecci2023persistent}. 

 Another interesting regime for persistent currents of repulsive fermions is the commensurate one for integer fillings $\nu =1$ of one particle per site governed by the Lai-Sutherland Hamiltonian. For $N>2$, the system becomes a Mott phase insulator on increasing interactions, as the particles need to pay an energy penalty to move around, thereby constraining their motion~\cite{cazalilla2014ultracold}. Seeing as the hopping of the fermions is virtual, the low-energy physics of the system is captured by the anti-ferromagnetic Heisenberg model, obtained through second-order degenerate perturbation theory~\cite{capponi2016phases}. Such behaviour is inherently reflected in the persistent current through its profile and magnitude as depicted in Fig.~\ref{fig:PerrFermRep}(\textbf{b}): (i) the characteristic saw-tooth shape the current exhibits in the metallic-like phase at small interactions, is eventually smoothened out on going to the Mott phase at stronger interactions; (ii) the maximum amplitude of the persistent current decreases reflecting its exponential suppression due to the restricted motion of the particles. The Mott phase transition is characterized by the opening of a spectral gap in the charge sector, which for $N>2$ occurs at a finite value of the interaction in contrast to the $N=2$ case where such a transition is absent as the gap is open for any repulsive interaction~\cite{lieb1968absence}. The presence of this gap hinders any energy level crossings between the no spinon ground-state and spinon states -- see insets of Fig.~\ref{fig:PerrFermRep}(\textbf{b}). As a result, there is no fractionalization and the parity effect is still observed at large repulsive interactions. 
\begin{figure}[h!]
    \centering
    \includegraphics[width=0.5\textwidth]{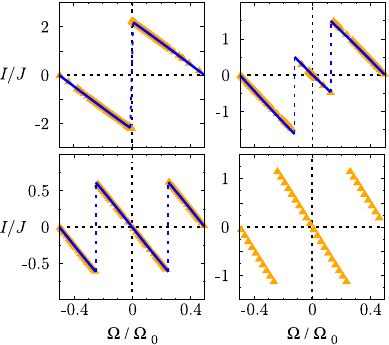}
    \hspace*{4mm}
    \includegraphics[width=0.39\textwidth]{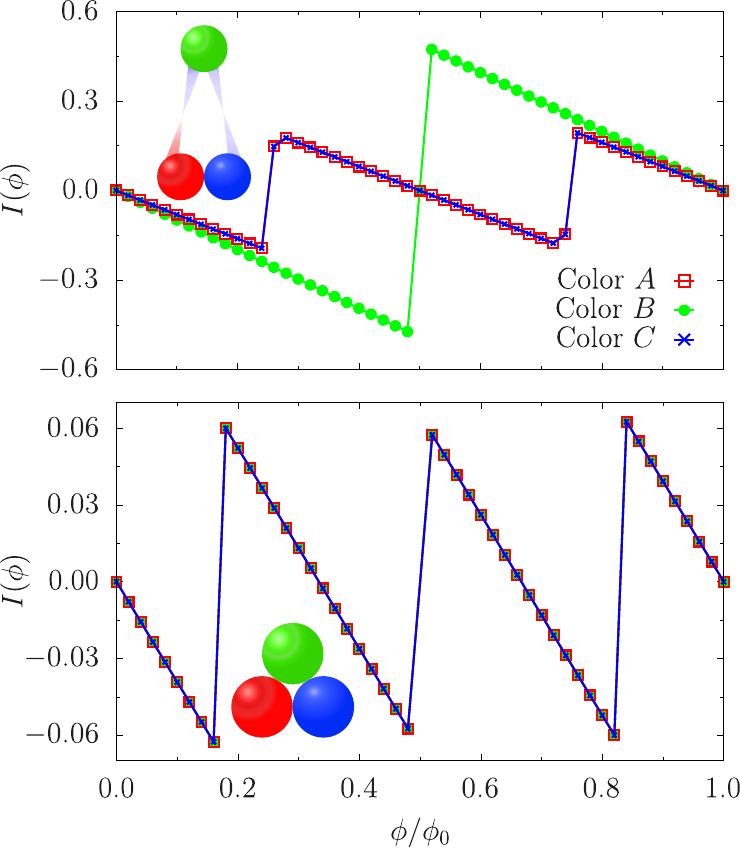}
    \put(-435,190){(\textbf{a})}
    \put(-185,190){(\textbf{b})}
    \caption{Persistent current $I(\phi)$ of attractively interacting $N$-component fermions against the effective magnetic flux $\phi$. (\textbf{a}) \textit{Monitoring the BCS-BEC crossover}. Panels depict the current acquiring a halved periodicity with increasing attractive interactions and the washing out of the parity effect as is crosses from the BCS to BEC side. Figure adapted from~\cite{pecci2021probing}.  (\textbf{b}) \textit{Probe for bound states of three-component fermions}. Top (bottom) panel shows the persistent current for a colour superfluid (trion) configuration.  The reduced periodicity exhibited by the current reflects the number of particles forming the bound state. Figure reprinted from~\cite{chetcuti2023probe}.  }
    \label{fig:PerrFermAtt}
\end{figure}

The qualitative and quantitative changes of the persistent current make it a suitable probe to distinguish between the various bound states and to map out the system's phase diagram~\cite{chetcuti2021persistent,consiglio2022variational}. The current exhibits critical behaviour during a quantum phase transition despite its mesoscopic nature. For the case in question, the onset to the Mott phase transition in SU($N>2$) Hubbard models was demonstrated to be captured with the persistent current through finite-size scaling analysis~\cite{chetcuti2021persistent}. Studies of Mott phase transitions with currents were also extended to multi-orbital SU($N$) Hubbard models to construct a conductivity-based phase diagram~\cite{richaud2021interaction,richaud2022mimicking,ferraretto2023enhancement}. In contrast with the single-band model, it was shown that the current's suppression with interaction is not always the case. Instead, it thrives in the deep insulating state, and in certain regions known as Hund's metal region, it is also enhanced as the hopping processes across the different orbitals facilitate the fermions' motion even at strong interactions. 

\begin{figure}[h!]
    \centering
    \includegraphics[width=0.49\textwidth]{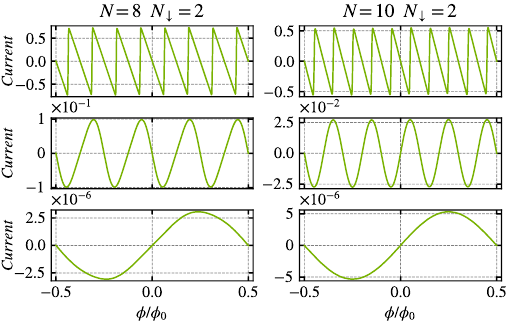}
    \hspace*{4mm}
    \includegraphics[width=0.38\textwidth]{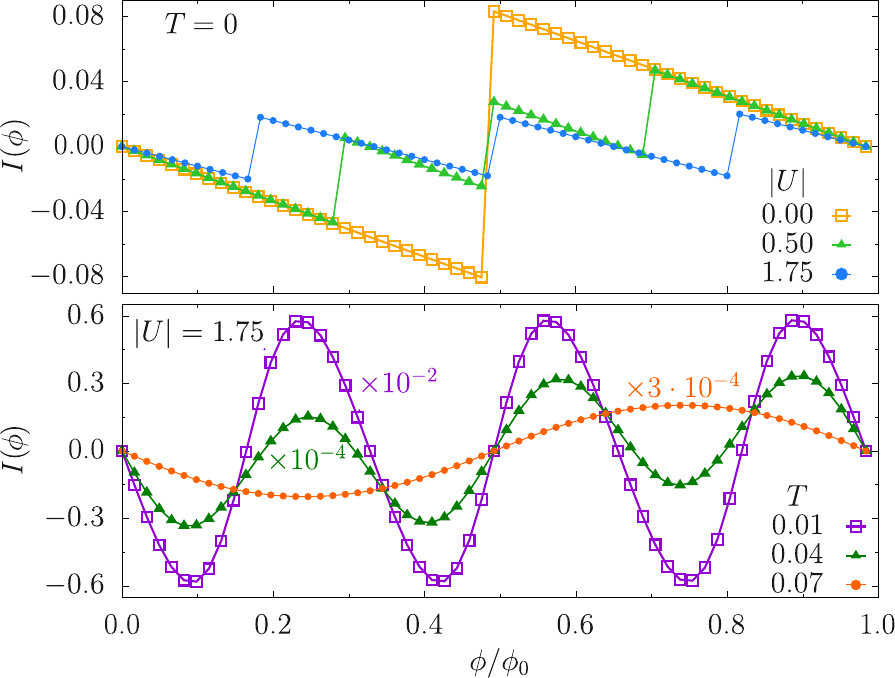}
    \put(-445,120){(\textbf{a})}
    \put(-190,120){(\textbf{b})}
    \caption{Dependence of the persistent current $I(\phi)$ on the interplay between temperature $T$ and interaction $U$. (\textbf{a}) Persistent current of a two-component fermionic system with strong repulsive interaction of $U=100$ with $N_{p}/N_{s}=0.1$ with increasing $T$ going down the panels. Figure adapted from~\cite{patu2022temperature}.  (\textbf{b}) Persistent current of a three-component fermionic system with strong attractive interactions at zero and finite temperature in the top and bottom panels respectively. Figure reprinted from~\cite{chetcuti2023probe}. Qualitatively, the effect of interaction and temperature oppose each other, in that they cause the system to fractionalize and de-fractionalize, which manifests in the current's period.}
    \label{fig:PerrFermAttT}
\end{figure}

Fractionalization is also present for attractively interacting fermionic systems and stems from forming many-body bound states. While the mechanism that facilitates this phenomenon is similar to bosons, the different particle statistics generate a different outcome in that for fermions, the Pauli exclusion principle restricts the formation of bound states. The current's reduced periodicity dictated the bound state's effective mass, solely dependent on the number of particles constituting the bound states, irrespective of the total number of particles present in the system is given by $1/r$ with $r$ being the number of bound particles -- Fig.~\ref{fig:PerrFermAtt}. Relying on this feature, the persistent current has been utilized to probe the celebrated BCS-BEC crossover in ultracold platforms~\cite{pecci2021probing} and distinguish between the different types of bound states of SU(3) fermions~\cite{chetcuti2023probe}. The fractionalized nature of the system can be inferred through the Bethe ansatz analysis of the GYS model in the limit of $L|U|\gg 1$~\cite{chetcuti2023probe}. For strong attractive interactions and equal number of particles per colour, the Bethe equations simplify and give an exact expression for the persistent current 
\begin{equation}\label{eq:FermCurrInfUa}
    I(\phi) = -2N\bigg(\frac{2\pi}{L}\bigg)^{2}\sum\limits^{n_{N}}_{a=1}\bigg[\frac{K_{a}}{N}+\phi\bigg],
\end{equation}
where $n_{N}$ is the number of maximally bound fermions for a given $N$-component system and $K_{a}$ being the corresponding quantum numbers.  Essentially, the above equation behaves similarly to Eq.~\eqref{eq:FermCurrInfU} for repulsive fermions, where an increase in the effective flux necessitating a shift in the quantum numbers to counteract them and decrease the total energy resulting in $1/N$ oscillations. Despite sharing various qualitative features, it is worth remarking that Eq.~\eqref{eq:FermCurrInfUa} highlights the distinction between the two regimes. Whilst for repulsive fermions the generation of a single spin excitation, shields the whole system from the flux piercing the system, such is not the case for the attracting case. From Eq.~\eqref{eq:FermCurrInfUa}, it is clear that as $U\rightarrow -\infty$, the Bethe equations decouple from one another and that each quantum number $K_{a}$ is associated to a given $N$-body bound state. When it comes to small interactions, it was demonstrated that for SU(2) fermions bound states are formed at arbitrarily small interactions~\cite{pecci2021probing}. By solving the Bethe ansatz equations in the product form, one obtains that the charge rapidities $k_{j}$ admit complex values for any $U<0$. It must be stressed that such a feature is parity dependent as for $4n$ ($4n+2$) systems the rapidities are complex when they correspond to the unfractionalized parabolas centered around $\phi =0.5$ ($\phi =0$). On the other hand, the rapidities associated to the fractionalized parabolas acquire their complex nature at finite value of the interactions when $|U|>|U_{c}|$ with $U_{c}$ being the critical interaction dependent on the gas' density. Such a result was later extended for SU(3) fermions in the three-particle sector demonstrating that a three-body bound state is formed immediately from the free system instead of sequentially, i.e.\ going through the intermediate step of being a two-body bound state~\cite{chetcuti2023probe}. Another interesting aspect of persistent currents in attractive systems is the parity effect, which is not washed out due to fractionalization in contrast to the repulsive case. Rather it would be more apt to say that whether the parity effect is retained or lost, is rooted in the number of particles composing the bound states, be it odd or even corresponding to an anti-symmetric and symmetric wavefunction~\cite{chetcuti2023probe}. Indeed, the $N$-body bound state of multicomponent fermionic systems can be effectively treated as a composite boson or fermion modeled by super Tonks-Girardeau or super Fermi Tonks-Girardeau gas for even and odd $N$ respectively~\cite{yin2011effective}. This trend in the parity effect was demonstrated to hold for systems with up to $N=5$. The persistent current's parity is another feature that re-affirms its suitability as a diagnostic tool as was found for SU(2) fermions when studying the BCS-BEC crossover~\cite{pecci2021probing}. 
\begin{figure}[h!]
    \centering
    \includegraphics[width=0.49\textwidth]{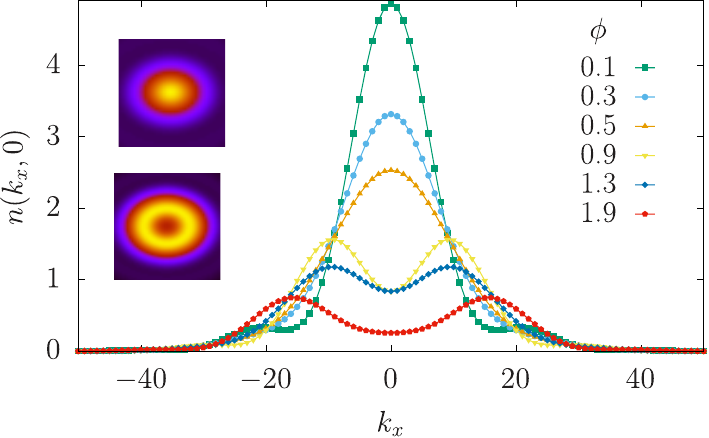}
    \includegraphics[width=0.49\textwidth]{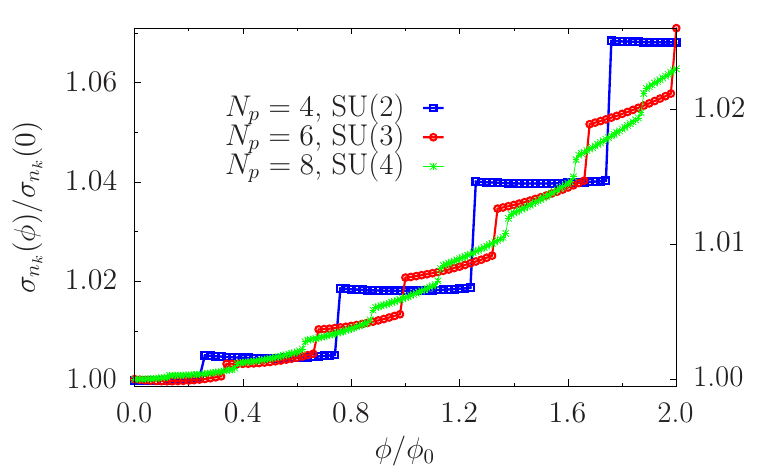}
    \put(-465,130){(\textbf{a})}
    \put(-220,130){(\textbf{b})}
    \caption{Homodyne protocol for strongly interacting SU($N$) fermionic systems. (\textbf{a}) Momentum distribution cross-section $n(k_{x},0)$ for $N_{p}=3$ particles with large repulsive interactions $U=10,000$ as a function of the effective flux $\phi$. The initially peaked momentum distribution collapses at $k_{x}=0$ once the additional threshold constraint $(N_{p}-1)/2N_{p}$ enforced by the fractionalization is cleared.  Insets correspond to the full momentum distribution. (\textbf{b}) Variance of the width of the momentum distribution $\sigma_{n_{k}}(\phi)$ for strongly attractive fermions. The appearance of discrete steps per flux quantum $\phi_{0}$ corresponds to the number of particles composing the bound state and ($N$ in the case shown), in turn, reflects the angular momentum fractionalization. Figure taken from~\cite{chetcuti2022interference}.}
    \label{fig:MomStrongU}
\end{figure}

In the presence of finite temperature, changes in the persistent current periodicity stem from the interplay between interactions and thermal fluctuations. Employing a thermodynamic Bethe ansatz approach for repulsively interacting two-component fermions, it was demonstrated that apart from having its sawtooth shape smoothened out, the persistent current's amplitude is also exponentially reduced with the decay rate dependent on the system's parameters~\cite{patu2022temperature}. The suppression of the current could be attributed to temperature activated transitions between states with different angular momenta. Alternatively, the persistent current contribution from the populated levels are equal in magnitude but with different directionalities thereby cancelling each other out.  A remarkable effect of finite temperature is that the single-particle frequency of the persistent current is re-instated, imitating, in a way, a reduction in the interaction strength -- Fig.~\ref{fig:PerrFermAttT}. This `de-fractionalization' phenomenon was also demonstrated to hold for $N$-component fermions with attractive interactions~\cite{chetcuti2023probe}. The mechanism behind this is that the contributing excited states have different frequencies that modify the current's periodicity, with the general trend being that the system becomes more robust to thermal fluctuations with increasing interactions. Nevertheless, for strong attractive interactions the system is more susceptible to temperature. It turns out that the bound states sub-band, which is the only contributor to the current in this regime, flattens out due to a decrease in the level spacing between the states constituting it. Consequently, at small temperatures the system exists as a `gas of bound states' altering the persistent current periodicity in the same manner discussed previously.
\begin{figure}[h!]
    \centering
    \includegraphics[width=0.7\textwidth]{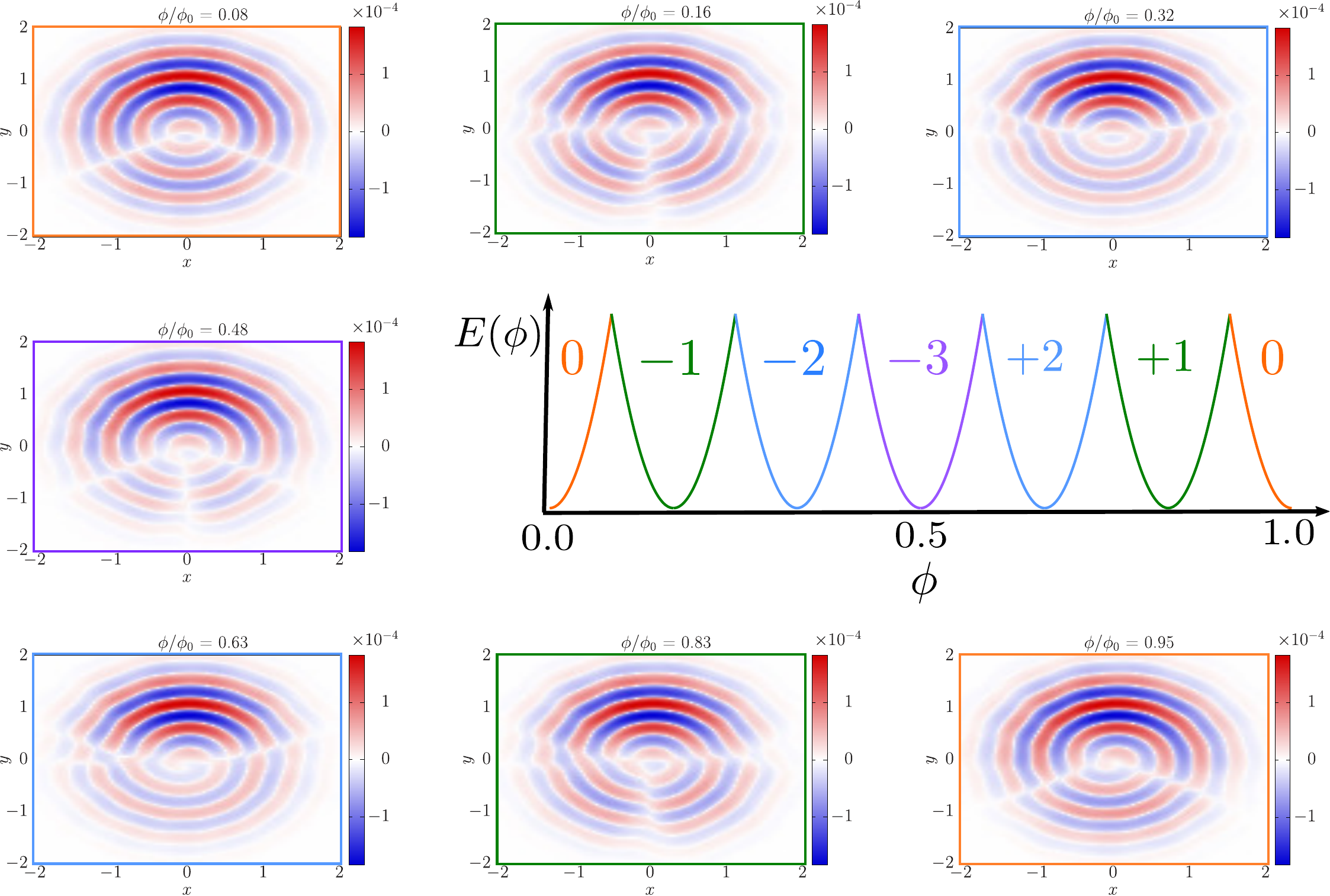}
    \caption{Ring-centre interference $G_{R,C}$ as a function of the angular momentum quantization per particle $\ell$ for $N_{p}=6$ strongly repulsive two-component fermions at short time expansions $t=0.025$. The schematic in the middle right shows the fractionalized energy parabolas for the system in question and the corresponding spin quantum numbers $X$ parameterizing the Bethe equations. Each parabola is characterized by different fractional values of $\ell$, which result to visualized in the interferograms through the manifested dislocations by monitoring both their number and orientation. The colour bar is non-linear by setting $\mathrm{sgn}(G_{R,C})\sqrt{|G_{R,C}|}$. Figure adapted from~\cite{chetcuti2022interference}.}
    \label{fig:IntStrong}
\end{figure}

To round things off, we consider the read-out of persistent currents in the strongly interacting regimes through interferometric means.\footnote{From the theoretical standpoint, the calculation of interference patterns for strongly interacting systems is limited to small particle numbers and system sizes. Unlike the free fermion regimes, there is no exact expression for the correlators in this regime and these systems are intractable by numerical methods such as exact diagonalization and DMRG for large particle numbers and system sizes. Nevertheless for the strongly repulsive interacting regime, it was shown that by employing a Bethe ansatz scheme to split the SU($N$) Hubbard model into the spinless fermionic and SU($N$) Heisenberg models, interference patterns for systems well beyond the current state-of-the-art numerical methods can be obtained~\cite{osterloh2023exact}.} In the homodyne protocol, the fractionalization in the system generates a momentum distribution depression, i.e.\ a non-zero local minimum in the origin, instead of the characteristic hole. Furthermore, the appearance of this depression occurs at larger flux values compared to the zero and weak interaction picture. On top of the Fermi sphere effect explained previously, there is an additional delay of $\frac{N_{p}-1}{2N_{p}}$ ($\frac{N-1}{2N}$) in strongly repulsive (attractive) regimes, which is a signature of the fractionalization --Fig.~\ref{fig:MomStrongU}(\textbf{a}). The visibility of the depression for attracting fermions is significantly diminished due to the reduced coherence associated with the bound state formation. Nevertheless, information can be acquired indirectly from the variance of the width of the momentum distribution, with the appearance of discrete steps whose number in a given flux quantum correspond to the angular momentum fractionalization --Fig.~\ref{fig:MomStrongU}(\textbf{b}). For the self-heterodyne protocol, the interferograms also embody the features of the fractionalized with the emergence of the spiral corresponding with the depression in the momentum distribution. Aside from being able to track the angular momentum quantization by spiral's specific delay, it can also be monitored through the change in the number and orientation of the dislocations. At variance with free fermions, the dislocations in the strong interaction regime are flux dependent being modified by the presence of spin excitations giving a visual cue about the angular momentum fractionalization --Fig.~\ref{fig:IntStrong}. In the case of attractive interactions, self-heterodyne interference patterns are not a good figure of merit for SU($N>2$) systems since the features of $N$-body bound are not captured by the density-density correlators used to generate these interferograms. 

\section{Applications for quantum technologies \& Atomtronics}\label{sec:atomtronics}

Persistent currents in superconducting rings set the basis for magnetic field high precision sensing; at the same time SQUIDs have moved macroscopic quantum tunneling and macroscopic phase coherence from being `esoteric'  to the realm of experimentally measurable quantities. Such notions defined the persistent-current qubits~\cite{mooij1999josephson,chiorescu2003coherent}.  With a similar logic, persistent currents of ultracold atoms can set the basis for high precision rotation sensing. The atomic counterpart of SQUIDs have been experimentally realized and show promise  to explore fundamental quantum science in malleable coherent platforms; indeed neutral persistent current qubits have been conceived.  Playing with quantum fluids of different nature as allowed by cold atoms technology,  different types of entangled states can be engineered.

\subsection{Rotation sensors }\label{sec:RotSens}

Quantum sensing is one of the most direct applications of matter-wave systems in quantum technologies. It leverages the unique quantum properties of the system to detect physical changes in its surroundings with high precision. The advantages of matter-wave systems rely on properties such as correlations, long coherence and quantization, which lead to remarkably high sensitivity to environmental changes. While this technology is still in its early stages of development, it holds great promise for various fields, such as navigation, seismology, and earth observation \cite{travagnin2020cold,geiger2020high}. Some examples have already been implemented successfully, for instance, gravimeters based on ultracold atom technology \cite{menoret2018gravity}. Further advantage is expected when utilizing quantum correlations for which the sensitivity of a quantum system can surpass that of its individual constituent elements, reaching sensitivities beyond the standard quantum limit. 

\begin{figure}[h!]
    \centering
    \includegraphics[width=0.8\linewidth]{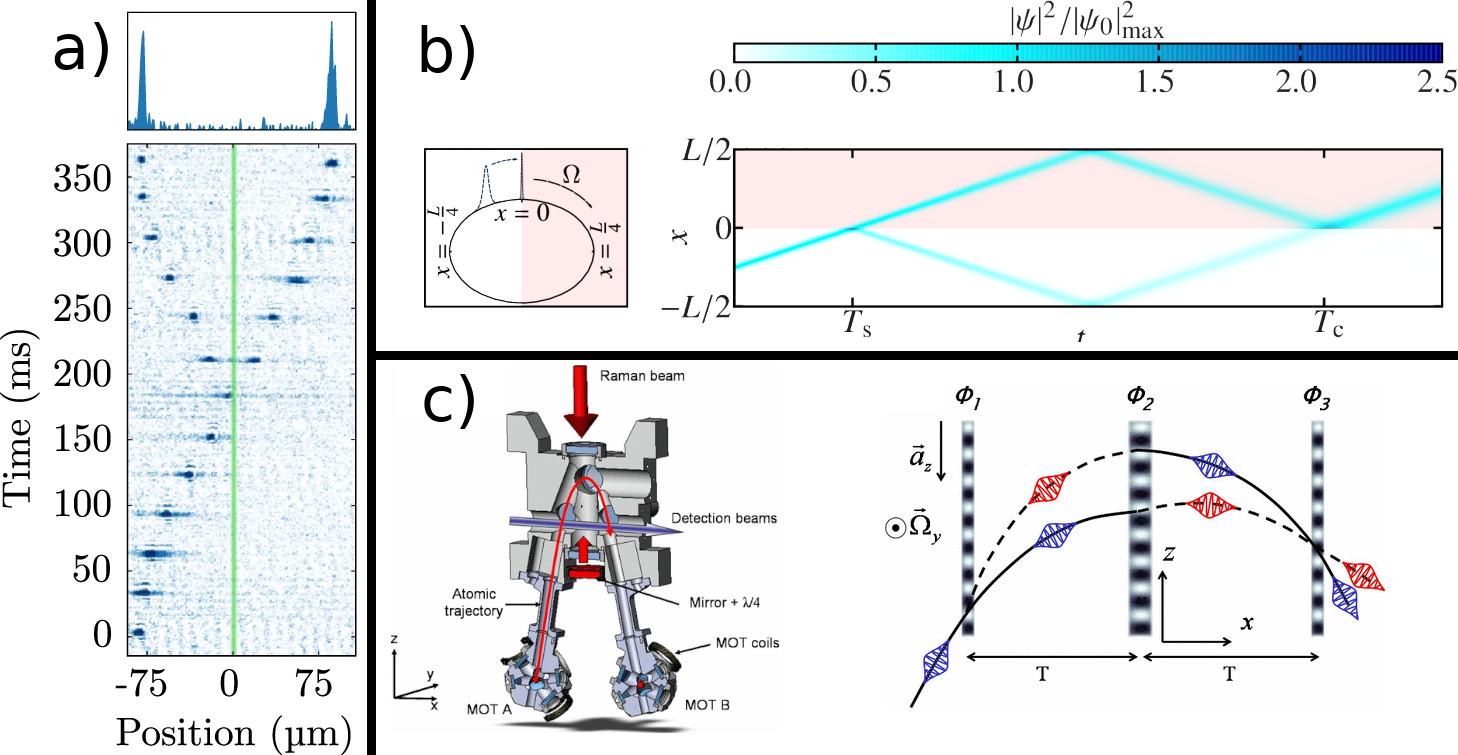}
    \caption{\textbf{(a)} Example sequence of a soliton being split by the narrow barrier into two solitons of approximately equal atom number. \textbf{(b)} Stages of Sagnac interferometry. An incoming soliton splits at time $T_{s}$ on a barrier into two solitons of equal amplitude and opposite velocity. After circumnavigating the ring trap, at time $T_{c}$ the solitons recombine either at the same barrier. The resulting phase difference, incorporating the Sagnac phase due to the rotating reference frame, is read out via the population difference in the final output products within the positive (shaded) and negative domains. \textbf{(c)} Scheme of a cold-atom gyroscope based on a $\pi/2-\pi-\pi/2$ interferometer in a vertical Raman configuration. The interferometer is sensitive to the vertical acceleration and the rotation in the $y$ axis. Figures adapted from \cite{wales2020splitting,helm2015sagnac,gauguet2009characterization}.}
    \label{fig:splitting_soliton}
\end{figure}

In ring geometries, besides persistent currents, the Sagnac effect  is very relevant to rotation sensing~\cite{barrett2014sagnac}.
The Sagnac effect is based on the resulting interference between two waves that follow a closed guided path in opposite directions and are recombined after its evolution Fig.~\ref{fig:splitting_soliton}\textbf{(b)}. These waves acquire a relative phase depending on the angular velocity that the device is subjected to. In ultracold atoms we deal with matter-waves, therefore the interference fringes will be associated to the wave character of the atoms, which is given by $\phi_S = 4\pi m\mathbf{\mathcal{A}}\mathbf{\Omega}/h$, where $\mathbf{\mathcal{A}}$ corresponds to the enclosed area and $\mathbf{\Omega}$ to the rotational vector~\cite{amico2022atomtronic}. 
To take advantage of this phenomenon, matter-wave devices typically employ an interferometric scheme that requires a splitting mechanism that is used for both creating two counter-propagating matter-waves and for the recombination process. The latter converts the phase difference into a population imbalance through the quantum interference between the two matter-waves. A natural way to implement this splitting in ultracold atoms is the use of potential barriers that can separate spatially propagating wavepackets into two -- Fig.~\ref{fig:splitting_soliton}\textbf{(a)}. One important approach that has been followed is the use of attractively interacting self-focusing matter-waves where the dispersion of the wave-packet is compensated by the interactions forming the so-called bright solitons. Within mean-field, the splitting through thin barriers has been investigated, accounting for different phenomena occurring during the splitting process~\cite{mcdonald2014bright,wales2020splitting,polo2013soliton,helm2014splitting,helm2012bright,naldesi2023massive}. From a many-body perspective, one can also consider taking advantage of quantum correlations to enhance the sensitivity by creating superpositions of matter-waves~\cite{streltsov2009efficient}. This contra-propagating wavepackets along with the Sagnac phase acquired in the system \cite{polo2013soliton}, can be used to develop a sensor whose interference depends on the external rotation \cite{durfee20006long,barrett2014sagnac,travagnin2020cold,helm2015sagnac}. In the last two decades owing to the experimental advances in cold atoms technology, matter-wave interferometric setups based on the Sagnac effect  have been implemented experimentally~\cite{wu2007demonstration,gauguet2009characterization,burke2009scalable,krzyzanowska2023matter,moan2020quantum,navez2016matter,stevenson2015sagnac,beydier2024guided} (e.g. Fig.~\ref{fig:splitting_soliton}\textbf{(c)}).

\begin{figure}[h!]
    \centering\includegraphics[width=0.7\linewidth]{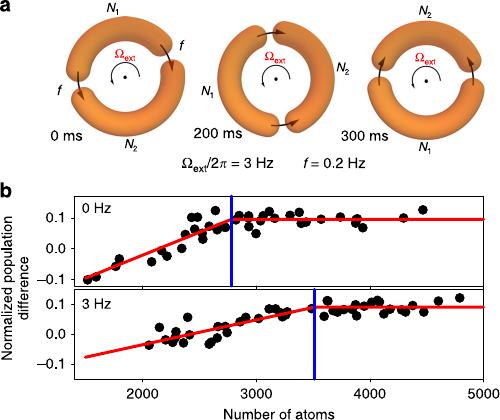}
    \caption{\textbf{(a)} Schematic of a double junction atomtronic SQUID with various rotation rates $\Omega_{ext}$ with $f$ being the frequency for the current bias generation and $N_{1},N_{2}$ corresponding to the number of atoms in each half of the ring. Arrows represent the movement of the junctions. \textbf{(b)} Change in normalized population difference $z$ as a function of the atom number at a rotation frequency of $0$ Hz and 3 Hz. The red line corresponds to the optimal fit whilst the blue line indicates the critical atom number and the point of transition. Figures adapted from \cite{ryu2020quantum}.}
    \label{fig:aquid}
\end{figure}

A more direct use of persistent currents for rotation measurements can also be found. Taking advantage of the large control in experiments, a differential measurement of rotation around the critical rotation velocity at which a persistent current is created can be used to measure changes on rotational frequencies \cite{naldesi2022enhancing}. This, however, can have some limitations due to the hysteresis phenomenon~\cite{eckel2014hysteresis} occurring around this critical point as well as unwanted phase slippage of coherent or incoherent nature \cite{polo2019oscillations,klejdja2023decay}. Nonetheless, the use of attractive interactions can give an edge on these measurements, where the angular momentum is fractionalized and therefore a current is created at much smaller changes of rotation, offering possibilities for creating absolute measures of rotation~\cite{naldesi2022enhancing,polo2022quantum}.
Another approach for persistent current-based sensors can have been investigated using the analogue of the superconducting interference devices (SQUIDs), known as atomic SQUID or AQUID~\cite{ryu2020quantum,kiehn2022implementation} (see Fig.~\ref{fig:aquid}). This can be exploited to create rotation sensors with entanglement-enhanced precision~\cite{ragole2016interacting}. The superposition of orbital angular momentum states can also be used in a different manner to devise a rotation sensor. In a superposition of two orbital angular momentum states, a density minima appears due to the interference between the two counter-propagating modes. By precisely controlling the imbalance between the two states, one can sense rotations through the change on the frequency at which the nodal point of the density moves \cite{pelegri2018quantum}. Double ring configurations have also been considered, in particular, sided-coupled rings are sensitive to external rotations \cite{perezobiol2022coherent} and can be utilized to build rotation and acceleration sensors \cite{adeniji2024double,chaika2024acceleration}

Rotation sensors can also be constructed by studying the current going through the ring~\cite{lau2023atomtronic}. In such a setup, leads are attached to the ring. One of the lead acts as a source, inducing a current through the ring into the other leads, called drains. Depending on the applied flux, the current through the ring is modified. Specifically, in a three-lead setup, it was shown that depending on the applied flux the current flows into either the first or second drain lead. From this flux-dependent non-reciprocal current, one could design an atomtronic flux sensor. An advantage of this setup is that one does not need to monitor the current within the ring, but it suffices to measure the atomic density in the drains in order to read out the rotation. 

\subsection{Current-based quantum simulators and analogues}\label{sec:CurrSims}

 As discussed previously in this manuscript, persistent current represents one of the most fundamental instances of matter-wave transport. As such, using the persistent current to simulate and understand transport on other quantum systems is a very promising endeavour.

The realization of fermionic persistent currents is one example that can provide us with a toolbox to simulate strongly interactive matter in unexplored regimes. The unitary Fermi limit is characterized by its divergent scattering length, leaving only a single intrinsic length scale for the system, the inter-particle distance. In this regime, the thermodynamic behaviour of the system becomes independent of the underlying microscopic interactions, and only macroscopic quantities become relevant, allowing us to explore the universal behaviour of strongly correlated fermionic systems \cite{makotyn2014universal,hu2007universal,nascimbne2010exploring,ho2004universal}. 

\begin{figure}[h!]
    \centering
    \includegraphics{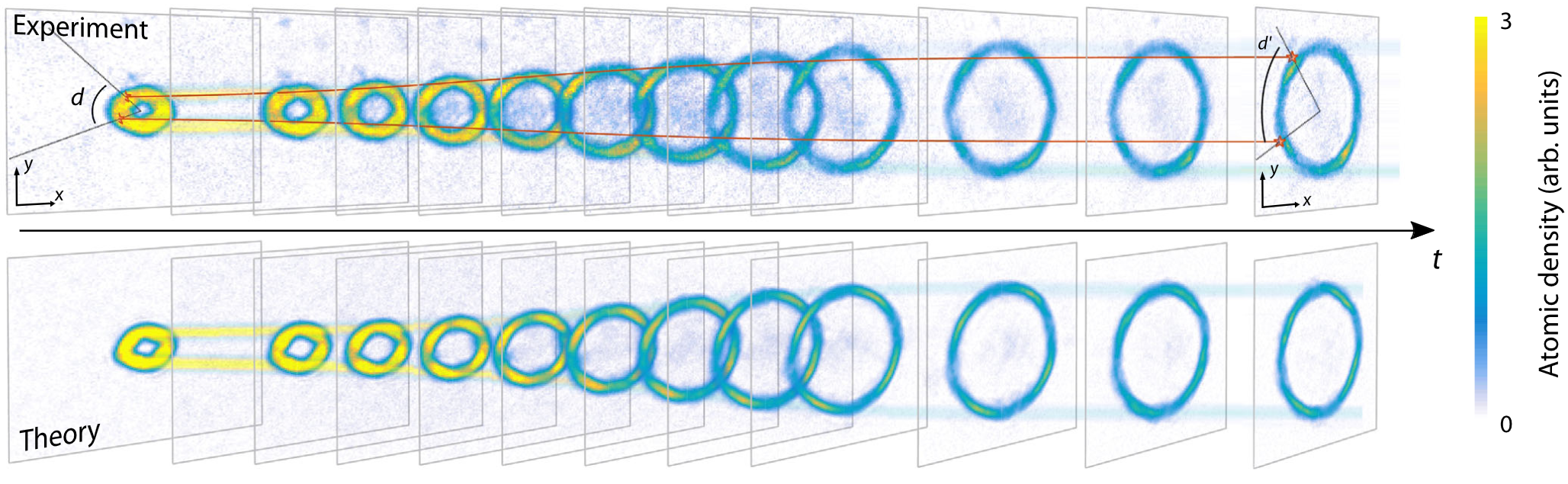}
    \caption{Study of the dynamics of a supersonically expanding, ring-shaped Bose-Einstein condensate displaying the expansion redshifts long-wavelength excitations, reminiscent of an expanding universe. Image taken from \cite{eckel2018rapidly}.}
    \label{fig:holey}
\end{figure}

Bose gases provide an excellent framework to investigate cosmology and astrophysics in a controlled environment by taking advantage of the metric followed by phonons when propagating in a BEC \cite{barcelo2003probing}.   
Examples include the study of black holes and the event horizon \cite{giovanazzi2005hawking,nguyen2015acoustic} or Hawking radiation \cite{steinhauer2016observation,belgiorno2010hawking}, both typically relying on the aforementioned phonon excitations in the bulk of a BEC. Ring geometries have also been used to investigate cosmological questions; for instance, the expansion and contraction of the universe in cold atom platforms was studied in \cite{banik2022accurate} where they considered the excitations produced in the condensate, which include phonons, topological excitations such as vortices and solitons as well as the spontaneous generation of persistent currents \cite{eckel2018rapidly} (see Fig.~\ref{fig:holey}).  

Simulating phenomena of high energy physics is also now possible to do in ultracold experiments \cite{cao2011universal} thanks to the recent developments in the field of multicomponent fermions, where one can achieve highly symmetric interacting fermions, the so-called SU($N$) fermions~\cite{scazza2014observation,gorshkov2010two,cazalilla2009ultracold}. The systems present multiple phases of matter, where particles form bound states structured as pairs, pairs plus free particles, and trions \cite{guan2013fermi}. These phases are analogues of the colour superfluid and baryonic phases found in quantum chromodynamics (QCD) and have been investigated in multiple works \cite{rapp2007color,ozawa2010population,He2006superfluidity}. In the context of rings and persistent currents, it was shown that one could probe these phases by analyzing the change of periodicity of persistent current versus temperature, finding phenomenology analogue to the deconfinement in QCD \cite{chetcuti2023probe}.

In this review, we discussed the use of lattice systems to investigate persistent current in ultracold atoms using a full many-body description in Sec.~\ref{sec:BoseHubb}. Similarly, lattices are also used to investigate gauge theories for fundamental science. The study of lattice gauge theories has brought much attention to the ultracold atom community 
\cite{wiese2013ultracold,zohar2016quantum,schweizer2019floquet,aidelsburger2021cold}. These studies also open new ideas within the field of cold atoms, where now we can also ask questions related to the coherence properties of analogues of mesons \cite{domanti2023coherence} or other particles that appear in gauge theories.
In addition, lattice systems have also provided a playground for simulating and understanding solid-state physics with Hubbard-like Hamiltonians. Recently, the access to SU($N$) multicomponent fermions have open the possibility to explore currents in synthetic heterostructure by considering the internal degrees of freedom as extra dimension~\cite{ferraretto2023enhancement} opening new pathways for simulating new solid-state devices \cite{altman2021quantum}.

\subsection{Quantum transport}\label{sec:QT}
Persistent currents can display a highly interesting interplay with dynamical systems. While persistent currents by themselves are time-independent eigenstates, by subjecting them to external perturbations a rich physics can emerge. In particular, consider a system supporting persistent currents, which is subject to an additional non-equilibrium flow of particles~\cite{chien2015quantum}.  In condensed matter systems, such systems have been studied in the context of SQUIDs with external leads attached to the device, which induce an electronic current through the SQUID~\cite{fagaly2006superconducting}. The transport of electrons through the SQUID is strongly affected by the persistent current, allowing one to read out magnetic fields threading the SQUID.

Similarly, cold atoms pose a natural platform to study the interaction of transport and persistent currents. In particular, cold atoms can study quantum transport in a highly controlled and versatile way, allowing to probe regimes of transport not possible in other platforms~\cite{chien2015quantum,krinner2017two,loiko2014coherent}. For example, transport through a quantum ring depending on flux and persistent currents can be studied in novel ways using cold atoms~\cite{haug2019aharonov,haug2019andreev,haug2021quantum,haug2019topological,lau2023atomtronic}. In such a setup, a quantum ring threaded by flux $\Omega$ has two or more leads are attached, where the quantum transport through the ring is facilitated by the leads. One lead, commonly dubbed source, has a higher chemical potential, i.e.\ more atoms than the ring and other leads. Thus, atoms will start flowing from the source into the ring. The other leads, dubbed drain, have a lower chemical potential and thus the atoms in the ring start flowing into the drain leads. Now, the atoms propagating through the ring interact with the flux, which induces a path-dependent phase shift which can affect the transported current.

The interplay between flux and quantum transport depends on the interaction strength between the atoms, as well as whether the atoms are fermions or bosons. For interacting bosons, the Aharonov-Bohm effect is washed out and the flux has little control over the current~\cite{haug2019aharonov,haug2019andreev}. The transport dynamics depends strongly on the ring-lead coupling, with weakly coupled rings showing oscillatory dynamics between source and lead which are modulated by the flux~\cite{haug2019aharonov}, while for strong coupling the Aharonov-Bohm effect is quickly washed out. In contrast, for non-interacting fermionic systems, the Aharonov-Bohm effect is present and the flux strongly affects the current. Curiously, we find that the dynamics of non-interacting fermions and hard-core bosons shows very different behaviors in a ring system with leads, while without leads the dynamics is exactly identically as the both species can be mapped onto each other. 

Such a flux dependence is also observed for topological pumping. Here, the system Hamiltonian is adabiatically modulated in time such that topological features of the Hamiltonian induce a transport which is robust against perturbations and disorder~\cite{thouless1983quantization}.
When atoms are transported through a ring with flux via topological pumping, the flux induces a non-trivial phase shift on the dynamics~\cite{haug2019topological}. This can create highly entangled states depending on the flux as well as modify the transmission and reflection coefficient depending on the flux and geometry of the ring.

One can also apply the interaction of flux and transport to engineer novel devices.In Ref.~\cite{lau2023atomtronic}, transport through the ring system with three leads was studied.  By tuning the flux, the transport can be controlled to flow from the source into either the first or second drain lead, realizing a non-reciprocal switch to control the flow of atoms. This behavior is found consistently for zero, weak and strong interactions, as well as for lattice systems and continuous potentials with finite width simulated via GPE equations. 

A different dynamics is found when the transport constitutes of low-energy excitations. Here, the system is prepared in the ground state, and a density wave is induced by slightly quenching the source lead. The incoming density wave is then either reflected or transmitted at the ring, where the transmission and reflection strongly depends on the flux. For zero flux, the density wave is transmitted with an amplitude of $4/3$ and a negative Andreev reflection, while for half flux quantum the density wave is not transmitted, but instead one observes a positive reflection~\cite{lau2023atomtronic}.

Finally, the flux can also be changed periodically in time in such ring-leads systems~\cite{lau2023atomtronic}. The time-dependence leads to a time-dependent Aharonov-Bohm effect, which modulates the transported current and the persistent current within the ring. In particular, a constant incoming current from the source is sent through a ring with linearly increasing flux. This transforms the current into a sinusoidal oscillating current in the drain, similar to a constant to alternating current transformer known from electronics.  Surprisingly, a linear increase in flux results in sinusoidal current oscillations, which is the result of the periodicity of the quantized flux. In particular, the oscillation time is directly a function of the flux quantization.

Recently, quantum transport through a flux ring has been realized in experiment~\cite{gou2020tunable}. Experimentally, the lattice is realized in momentum space with non-interacting cold atoms subject to particle loss. The lattice sites are coupled via two-photon Bragg transitions induced by laser fields of different frequencies, whereas a coupling term with variable flux for the ring is induced by a four-photon process. The study find that transport through the ring coupled to bosonic condensate, which act as reservoir for atoms, is strongly modified by the flux.

\subsection{Persistent current-based qubits and quantum phase slips}\label{sec:Qubits}
Qubits are the building blocks of quantum computers. From a fundamental stand-point, qubits are a construct that represents a physical systems whose dynamics is effectively described by a two-dimensional Hilbert space, i.e.\ a two-level or two-state quantum system. Some systems can naturally be described by qubits e.g. the spin in electrons or the polarization of light. Others require an effective description such as superconducting circuits, Rydberg atoms or ion-based qubits. As an example, in superconducting circuits the dynamics of the electrons and Cooper pairs in the system can be be described by an order parameter that resides in a two-dimensional Hilbert space, which is used as a computational qubit. In all cases, the effective qubits representation is limited by some factors such as cross-talk to other quantum states or the coherence-time and life-time.

\begin{figure}[h!]
    \centering\includegraphics[width=0.8\linewidth]{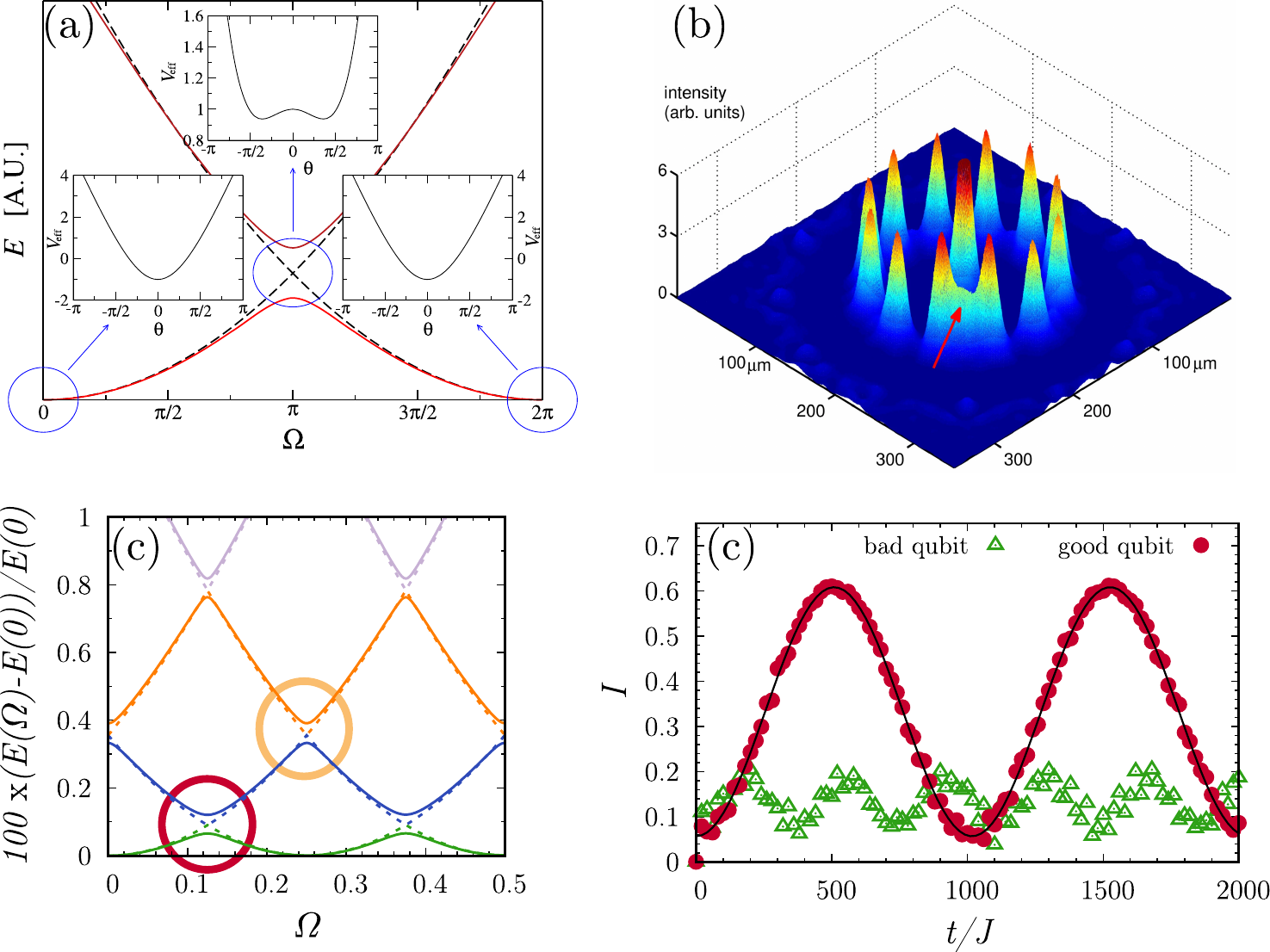}
    \caption{\textbf{(a)} Sketch of the effective two-level system/qubit formed due to the energy splitting of the ground-state (black dotted lines) by a barrier. Insets corresponds to the qualitative form of the effective potential $V_{\mathrm{eff}}$ at effective flux $\Omega = 0,\pi,2\pi$. \textbf{(b)} Experimental realization of a ring-lattice potential with an adjustable weak link (red arrow). Measured intensity distribution with an azimuthal lattice spacing of 28 $\mu$m and a ring radius of 88 $\mu$m. The size of the structure is scalable and a lower limit is imposed by the diffraction limit of the focusing optics. \textbf{(c)} Energy spectrum for $N_p=4$ attractively interacting bosons where a bound state has been formed and a single-site barrier is present with $\lambda=0.01$. We observe emerging avoided crossing forming new possible qubit states at lower rotational frequencies. Right panel corresponds to the current oscillations of a superposition of currents forming a qubit-like state created by quenching from $\Omega_i/\Omega_0 = 0$ to $\Omega_f/\Omega_0 = 0.125$ while the bad one is created by quenching to $\Omega_f/\Omega_0 = 0.175$. Figure adapted from~\cite{aghamalyan2015coherent,amico2014superfluid,polo2022quantum}. }
    \label{fig:aquid1}
\end{figure}

Persistent currents can also be regarded as a two-level quantum system. In this review we have shown that a quantum gas subject to an artificial magnetic field presents energy level crossing around integer (or fractional) values of the flux quantum. These degeneracies can be lifted by introducing an element that breaks the rotational symmetry of the system, creating avoided level crossing between different angular momentum states --Fig.~\ref{fig:aquid1}\textbf{(a)}-\textbf{(b)}. The two states involve persistent currents that, as we have discussed, possess some particular properties that make them good candidates for creating qubits such as robustness, being long-lived and having long coherent life-times due to the low temperatures. Moreover, the high control of ultracold atoms makes then ideal to study fundamental aspects of qubit systems such as coupling to external systems and simulated environments in a more controlled manner. Therefore, current-based qubits could also be used as simulation of other systems to improve quantum technologies.

From a theoretical point of view, two-level systems have been investigated both in continuum \cite{hallwood2010robust,nunnenkamp2011superposition} and lattice systems \cite{amico2014superfluid}. Important elements have been discussed such as the scaling of the gap, or possible implementations of gates. The high controllability also brings a large parameter space to explore for instance, the amplitude of persistent currents in presence of a barrier~\cite{cominotti2014optimal} (being one of the main mechanisms to couple different angular momentum states) or studies of AQUID with attractive interactions~\cite{polo2022quantum} (see Fig.~\ref{fig:aquid1}\textbf{(c)}-\textbf{(d)}). Important issues on dissipation have also been explored in rings with weak impurities \cite{polo2018damping,polo2019oscillations}. Other proposals, inspired by the the analogous circuit found in electronics (fluxonioums), include three weak links \cite{aghamalyan2016atomtronic} where the links facilitating the function of the system as a qubit and enlarging the parameter space of the two-level effective dynamics. 

Qubit-like states consisting of persistent currents strongly rely on the control one can have on those states. Coherent control of current states with present experimental techniques remains a challenge, as they typically involve using barriers or defects to control or imprint specific winding numbers. The change from one persistent current to another is known as ``phase slips'' \cite{wright2013driving}. Many investigations on incoherent phase slips exists, which account for the main process for which the current dissipates. Coherently controlling and understanding this process can allow for the engineering of states~\cite{perezobiol2022coherent}.

\subsection{Rydberg atomtronics}\label{sec:Rydberg}
Rydberg states are highly excited electronic states of atoms. When cooled down to micro Kelvin temperatures, these excitations can be controlled through optical or magnetic fields~\cite{low2012experimental}. Due to the large dipole moment, Rydberg excitations strongly interact between different atoms, allowing for fast and long-range interactions between atoms. In particular, excitations can move between atoms several orders of magnitude faster than the movement of the underlying atoms~\cite{browaeys2020many}. Recent works established the  paradigm of Rydberg Atomtronics: Here, the flow of excitations between Rydberg states of different atoms is controlled to create meaningful devices~\cite{perciavalle2022controlled}.

The directional flow of excitations in Rydberg atoms in a lattice ring was studied in Ref.~\cite{perciavalle2022controlled}. In this work, Rydberg atoms are arranged in a ring configuration, which are coupled via dipole-dipole interaction. The Hamiltonian is given by
\begin{equation}
\mathcal{H}=\frac{\Omega}{2}\sum_{j} \big(\sigma_{j}^{+} + {\rm H.c.} \big) +\frac{\Delta}{2}\sum_{j}\sigma_{j}^{z}+ \sum_{k<j}J_{kj} \big( \sigma_{k}^{+}\sigma_{j}^{-} + {\rm H.c.} \big)\,,
\label{eq:Ham_rydberg}
\end{equation}
where $\Omega$ is the Rabi driving between the two Rydberg states $\ket{\uparrow}$ and $\ket{\downarrow}$, $\Delta$ the frequency detuning of the driving and $J_{kj} = 2 C_{3}/D_{kj}^{3}$ describes the isotropic dipole-dipole interactions between atoms, $D_{kj}$ the distance between atoms $k$ and $j$, and $C_3$ the dipole interaction strength. Eq.~\eqref{eq:Ham_rydberg} is time-reversal symmetric and thus by itself cannot support a directional current. To induce a current, a non-trivial phase is encoded into the Hamiltonian which breaks time reversal symmetry. This is achieved by inducing the Rabi driving $\Omega$ via a Laguerre-Gauss beam with nontrivial phase winding $\theta$, as well as an additional laser field and microwave field which couple two intermediate Rydberg states $\ket{P}$ and $\ket{D}$. As a result, the effective Rabi coupling terms between the $\ket{\uparrow}$ and $\ket{\downarrow}$ now becomes $\sigma_{j}^{\pm} \longrightarrow\sigma_{j}^{\pm}e^{\mp i\theta j}$.
In  the rotating frame, the effective Hamiltonian can then be written as
\begin{equation}
\mathcal{H} = \frac{1}{2}
\sum_{j} \Big( \Omega\sigma_{j}^{x} + \Delta\sigma_{j}^{z} \Big)+ \sum_{k<j}J_{kj} \big( e^{i\theta(j-k)}\sigma_{k}^{+}\sigma_{j}^{-} + {\rm H.c.}  \big),
\label{eq:Ham_rotated}
\end{equation}
which now breaks time-reversal symmetry along the inter-atom hopping terms. Note that flux in Rydberg atoms can be introduced also in different manners. Using the directional character of dipole-dipole interaction in Rydberg atoms, one can create a non-trivial flux by arranging atoms in a hexagonal lattice~\cite{lienhard2020realization}. Alternatively, suitable laser drivings with an additional magnetic field can introduce non-trivial gauge fields in the coupling between Rydberg atoms~\cite{wu2022manipulating,li2022coherent,han2024tuning}.

Ref.~\cite{perciavalle2022controlled} then proceeds to study the density of excitations
\begin{equation}
n_j(t) = \langle \psi(t)| \frac{1}{2} (1 + \sigma^z_j)|\psi(t) \rangle,
\end{equation}
and the current of excitations for nearest-neighbors
\begin{equation}
\mathcal{I}_{j}^{(nn)}(t) = -i J_{nn} \langle \psi(t)| \big( e^{i\theta}\sigma_{j}^{+}\sigma_{j+1}^{-} - {\rm H.c.} \big) |\psi(t) \rangle.
  \label{eq:currentNN}
\end{equation}
The system is initially prepared with one excitation in the $\ket{\uparrow}$ state, while all other atoms are in the $\ket{\downarrow}$ state. Without flux $\theta=0$, the dynamics is equal in either directions. However, with flux $\theta\ne0$ excitations favor propagation in one direction, generating a chiral current. This effect persists even with dephasing present, making it robust even in an experimental setting. Furthermore, the authors study the time-averaged nearest-neighbor current of the dynamics and find that it remarkably matches the current of  the ground state of the system. It is difficult to prepare ground states in Rydberg atoms, thus studying the persistent current directly can be a challenge. However, this finding shows that the dynamical motion shares a curious relationship with the ground state current, allowing for potential study of ground state persistent currents via the dynamics of the system.  

Currents can also be induced by adding leads to the ring such as in Ref.~\cite{perciavalle2023coherent}. This setup mirrors the cold atom transport studied in Sec.~\ref{sec:QT}. As Rydberg atoms have long-range interaction, all sites of the ring are coupled to the leads with a coupling strength which depends on the respective distance. This system shows intricate source-drain oscillations which depend on flux and coupling strength~\cite{perciavalle2023coherent}.

In Ref.~\cite{perciavalle2024quantum}, persistent currents in Rydberg atoms are created by time-dependent control of the detuning $\Delta_j(t)$ of each atom $j$. Using optimal control technique GRAPE~\cite{khaneja2005optimal}, one can find the driving protocol that creates the current state in short amount of time. Ref.~\cite{perciavalle2024quantum} engineers current eigenstates of the bare Hamiltonian with fixed winding number. Moreover, they create superposition states of different winding numbers which evolve in time under the bare Hamiltonian and can show directional flow of the excitation density.

Experimentally, Rydberg states with persistent currents have been generated in Ref.~\cite{bornet2024enhancing}. Here, a non-trivial phase winding has been imprinted on three Rydberg atoms arranged in a triangular configuration, where the atoms are subject to long-range dipolar interactions. The fidelity of the state and the phase winding was read-out via tomography.

Atomtronic currents can also be engineered using only the dynamics between the atomic ground state and a Rydberg state~\cite{kitson2023rydberg}. Notably, in this setup only the excited Rydberg states interact via van-der-Waals forces, however, no direct transfer of excitation occurs between different atoms as there are no hopping terms of the form $\sigma_{k}^{+}\sigma_{j}^{-}$. While the atoms cannot move between atoms themselves, one can create indirect propagation by the facilitation mechanism~\cite{morsch2018dissipative}. Here, one applies a constant driving term which can excite atoms, however, also applies a large detuning to suppress the generation of excitations. Only when an excited Rydberg state is at a particular distance, then  an effective detuning is induced, which allows the creation of new excitations. Now, one can engineer a directional propagation of excitations by careful engineering of the distances and detunings of Rydberg atoms~\cite{kitson2023rydberg,valencia2023rydberg}.
Ref.~\cite{kitson2023rydberg} used this effect to create directional control over the transport of excitations both for Rydberg atoms in lattices as well as three-dimensional gases. Further, one can create various atomtronic devices such as diodes, transistors as well as universal logic gate sets.

\section{Remarks \& Outlook }\label{sec:Conc}

Persistent currents are connected to  important problems in quantum science. The exceptional control and flexibility afforded by ultracold atoms makes them a powerful platform for investigating a wide range of interesting phenomena. We have reviewed how persistent currents arise as a response to an effective gauge field, and discussed how this response is expected to vary under different physical conditions, from scenarios as interacting bosonic systems with tunable interactions  and the BEC-BCS crossover in fermionic systems~\cite{chen2005bcs, zwerger2011bcs, giorgini2008theory, strinati2018BCS-BEC, del_pace2022imprinting, cai2022persistent, pecci2021probing}, to more exotic situations such as supersolids~\cite{Leonard2017, Li2017, chomaz2022dipolar, Li2013, Tanzi2019a, Geier2023}, systems with quantum impurities~\cite{windt2024fermionic, hewitt2024controlling, knap2012time, spethmann2012dynamics}, and $N$-component fermions/bosons~\cite{bauer2023quantum, aidelsburger2021cold, halimeh2023cold, domanti2023coherence}.

In view of the mild decoherence properties together with the aforementioned  capability to control the system's conditions, atomic SQUIDs provided by toroidal quantum gases interrupted by weak links and subjected to an effective magnetic flux would be an ideal setting to observe and manipulate  macroscopic quantum tunneling \& coherence~\cite{leggett1980macroscopic, leggett1991concept}.  Rydberg atoms, providing a new concept of currents~\cite{polo2024perspective}, can define a playground to explore  coherence properties in many body quantum systems with long-range interactions and in presence of driving and dissipation~\cite{perciavalle2022controlled,perciavalle2024coherent,perciavalle2024quantum}; given the considerable interests on Rydberg atoms as platform for  quantum computation~\cite{bluvstein2024logical, evered2023high, xu2024constant}, Rydberg atomtronics can assist in the engineering of quantum gates.     

Persistent current can form the basis of quantum devices exploiting propagating matter-wave. Our first thought is towards compact quantum sensing~\cite{akatsuka2017optically, burke2009scalable,kim2022one, krzyzanowska2023matter, mcdonald201380hk, mcdonald2013optically, moan2020quantum, qi2017magnetically, wu2007demonstration, reilly2023optimal}, that is clearly relevant in many different contexts including quantum labs in microgravity~\cite{alonso2022cold, thompson2023exploring, gaaloul2022space} and inertial navigation~\cite{barrett2016inertial, abend2023technology, hensel2021inertial, richardson2020quantum, templier2022tracking}. 
In this context, based on the lesson taught by high precision photonic interferometry~\cite{vahlbruch2008observation,treps2002surpassing}, an interesting direction to pursue  would be exploring protocols employing cold atoms in non-classical states~\cite{kaubruegger2019variational,mcdonald2014bright,naldesi2023massive,polo2013soliton,helm2015sagnac}. A second direction to explore would be studying integrated atomtronics and hybrid circuits. In the simplest schemes, such circuits can be made of suitable coupled toroidal degenerate gas characterized by given winding numbers. The ability to transfer winding numbers within such structures~\cite{perezobiol2022coherent} encloses  a great potential both for basic quantum science, for example providing experimentally feasible platforms for observing coherent quantum phase slips or  macroscopic quantum coherence,  and to conceive new concepts of quantum devices as  atomic counterparts of  Rapid Single Flux Quantum Logic superconducting circuits~\cite{likharev1991rsfq}. Ultracold atoms confined within optical cavities hybrid circuits~\cite{kumar2021cavity} can open the way  for non-destructive probing of ultracold matter through persistent current,  which can be relevant both for basic investigations in quantum science~\cite{pradhan2023ring,das2024hybrid} and  introduce novel avenues for quantum technology~\cite{jing2019entanglement,yong2020entanglement}.  However, challenges persist in achieving high-performance light-matter interactions~\cite{bao2012efficient,cho2016coherent}and in the low density regimes, where quantum gas microscopy could be required~\cite{bakr2009quantum,buob2024strontium,holten2022observation}.

Clearly, the concept of persistent currents traces back to mesoscopic physics of electronic systems or superfluid Helium. As discussed in this review, because of the specific features of the ultracold atoms platform, persistent currents in ultracold matter serve to advance in new physics and draw new specific technological applications. At the same time, notorious bottlenecks and unsolved problems in other platforms can be eased from the lesson learned  for ultracold matter. For example, the bulk of knowledge we acquired so far  can impact quantum simulations through superconducting transmon chains \cite{Ustinov,solitons,stacks} or excitons-polaritons in semiconductor microcavities  \cite{carusotto2013quantum,bloch2022nonequilibrium}, where persistent currents can be also excited, in regimes that can be complementary to cold atoms systems~\cite{sanvitto2010persistent,wouters2010superfluidity,li2015stability,gallemi2018interaction,lukoshkin2018persistent,barrat2024stochastic,chirolli2024synthetic}.

\section*{Acknowledgements} 

We thank A.J. Leggett, A. Osterloh, V. Singh and the Quantum Physics division at Technology Innovation Institute (Abu Dhabi) for many useful discussions. We acknowledge V. Ahufinger,  H. Perrin, P. Pieri, A. Yakimenko for insightful comments to the manuscript. LA acknowledges the Julian Schwinger Foundation grant JSF-18-12-0011.

\bigskip

\clearpage


\begin{thebibliography}{549}
\expandafter\ifx\csname natexlab\endcsname\relax\def\natexlab#1{#1}\fi
\providecommand{\url}[1]{\texttt{#1}}
\providecommand{\href}[2]{#2}
\providecommand{\path}[1]{#1}
\providecommand{\DOIprefix}{doi:}
\providecommand{\ArXivprefix}{arXiv:}
\providecommand{\URLprefix}{URL: }
\providecommand{\Pubmedprefix}{pmid:}
\providecommand{\doi}[1]{\href{http://dx.doi.org/#1}{\path{#1}}}
\providecommand{\Pubmed}[1]{\href{pmid:#1}{\path{#1}}}
\providecommand{\bibinfo}[2]{#2}
\ifx\xfnm\relax \def\xfnm[#1]{\unskip,\space#1}\fi
\bibitem[{Onsager(1961)}]{onsager1961magnetic}
\bibinfo{author}{L.~Onsager},
\newblock \bibinfo{title}{{Magnetic Flux Through a Superconducting Ring}},
\newblock \bibinfo{journal}{Physical Review Letters} \bibinfo{volume}{7}
  (\bibinfo{year}{1961}) \bibinfo{pages}{50--50}. \URLprefix
  \url{https://link.aps.org/doi/10.1103/PhysRevLett.7.50}.
  \DOIprefix\doi{10.1103/PhysRevLett.7.50}.
\bibitem[{Byers and Yang(1961)}]{byers1961theoretical}
\bibinfo{author}{N.~Byers}, \bibinfo{author}{C.~N. Yang},
\newblock \bibinfo{title}{{Theoretical Considerations Concerning Quantized
  Magnetic Flux in Superconducting Cylinders}},
\newblock \bibinfo{journal}{Physical Review Letters} \bibinfo{volume}{7}
  (\bibinfo{year}{1961}) \bibinfo{pages}{46--49}. \URLprefix
  \url{https://link.aps.org/doi/10.1103/PhysRevLett.7.46}.
  \DOIprefix\doi{10.1103/PhysRevLett.7.46}.
\bibitem[{Bloch(1965)}]{bloch1965off}
\bibinfo{author}{F.~Bloch},
\newblock \bibinfo{title}{{Off-Diagonal Long-Range Order and Persistent
  Currents in a Hollow Cylinder}},
\newblock \bibinfo{journal}{Physical Review} \bibinfo{volume}{137}
  (\bibinfo{year}{1965}) \bibinfo{pages}{A787--A795}. \URLprefix
  \url{https://link.aps.org/doi/10.1103/PhysRev.137.A787}.
  \DOIprefix\doi{10.1103/PhysRev.137.A787}.
\bibitem[{Deaver and Fairbank(1961)}]{deaver1961experimental}
\bibinfo{author}{B.~S. Deaver}, \bibinfo{author}{W.~M. Fairbank},
\newblock \bibinfo{title}{{Experimental Evidence for Quantized Flux in
  Superconducting Cylinders}},
\newblock \bibinfo{journal}{Physical Review Letters} \bibinfo{volume}{7}
  (\bibinfo{year}{1961}) \bibinfo{pages}{43--46}. \URLprefix
  \url{https://link.aps.org/doi/10.1103/PhysRevLett.7.43}.
  \DOIprefix\doi{10.1103/PhysRevLett.7.43}.
\bibitem[{Doll and N\"abauer(1961)}]{doll1961experimental}
\bibinfo{author}{R.~Doll}, \bibinfo{author}{M.~N\"abauer},
\newblock \bibinfo{title}{{Experimental Proof of Magnetic Flux Quantization in
  a Superconducting Ring}},
\newblock \bibinfo{journal}{Physical Review Letters} \bibinfo{volume}{7}
  (\bibinfo{year}{1961}) \bibinfo{pages}{51--52}. \URLprefix
  \url{https://link.aps.org/doi/10.1103/PhysRevLett.7.51}.
  \DOIprefix\doi{10.1103/PhysRevLett.7.51}.
\bibitem[{Reppy and Depatie(1964)}]{reppy1964persistent}
\bibinfo{author}{J.~D. Reppy}, \bibinfo{author}{D.~Depatie},
\newblock \bibinfo{title}{{Persistent Currents in Superfluid Helium}},
\newblock \bibinfo{journal}{Physical Review Letters} \bibinfo{volume}{12}
  (\bibinfo{year}{1964}) \bibinfo{pages}{187--189}. \URLprefix
  \url{https://link.aps.org/doi/10.1103/PhysRevLett.12.187}.
  \DOIprefix\doi{10.1103/PhysRevLett.12.187}.
\bibitem[{Gammel et~al.(1984)Gammel, Hall, and Reppy}]{gammel1984persistent}
\bibinfo{author}{P.~L. Gammel}, \bibinfo{author}{H.~E. Hall},
  \bibinfo{author}{J.~D. Reppy},
\newblock \bibinfo{title}{{Persistent Currents in Superfluid
  $^{3}\mathrm{He}$-$B$}},
\newblock \bibinfo{journal}{Physical Review Letters} \bibinfo{volume}{52}
  (\bibinfo{year}{1984}) \bibinfo{pages}{121--124}. \URLprefix
  \url{https://link.aps.org/doi/10.1103/PhysRevLett.52.121}.
  \DOIprefix\doi{10.1103/PhysRevLett.52.121}.
\bibitem[{Pekola and Simola(1985)}]{pekola1985persistent}
\bibinfo{author}{J.~P. Pekola}, \bibinfo{author}{J.~Simola},
\newblock \bibinfo{title}{{Persistent currents in superfluid 3 He}},
\newblock \bibinfo{journal}{Journal of low temperature physics}
  \bibinfo{volume}{58} (\bibinfo{year}{1985}) \bibinfo{pages}{555--590}.
\bibitem[{Leggett(1999)}]{leggett1999superfluidity}
\bibinfo{author}{A.~J. Leggett},
\newblock \bibinfo{title}{Superfluidity},
\newblock \bibinfo{journal}{Reviews of Modern Physics} \bibinfo{volume}{71}
  (\bibinfo{year}{1999}) \bibinfo{pages}{S318}.
\bibitem[{Yang(1962)}]{yang1962concept}
\bibinfo{author}{C.~N. Yang},
\newblock \bibinfo{title}{Concept of off-diagonal long-range order and the
  quantum phases of liquid he and of superconductors},
\newblock \bibinfo{journal}{Reviews of Modern Physics} \bibinfo{volume}{34}
  (\bibinfo{year}{1962}) \bibinfo{pages}{694}.
\bibitem[{Anderson(1966)}]{anderson1966considerations}
\bibinfo{author}{P.~W. Anderson},
\newblock \bibinfo{title}{Considerations on the flow of superfluid helium},
\newblock \bibinfo{journal}{Reviews of Modern Physics} \bibinfo{volume}{38}
  (\bibinfo{year}{1966}) \bibinfo{pages}{298}.
\bibitem[{Büttiker et~al.(1983)Büttiker, Imry, and
  Landauer}]{buttiker1983josephson}
\bibinfo{author}{M.~Büttiker}, \bibinfo{author}{Y.~Imry},
  \bibinfo{author}{R.~Landauer},
\newblock \bibinfo{title}{Josephson behavior in small normal one-dimensional
  rings},
\newblock \bibinfo{journal}{Physics Letters A} \bibinfo{volume}{96}
  (\bibinfo{year}{1983}) \bibinfo{pages}{365--367}. \URLprefix
  \url{https://www.sciencedirect.com/science/article/pii/0375960183900117}.
  \DOIprefix\doi{https://doi.org/10.1016/0375-9601(83)90011-7}.
\bibitem[{B\"uttiker et~al.(1984)B\"uttiker, Imry, and
  Azbel}]{buttiker1984quantum}
\bibinfo{author}{M.~B\"uttiker}, \bibinfo{author}{Y.~Imry},
  \bibinfo{author}{M.~Y. Azbel},
\newblock \bibinfo{title}{Quantum oscillations in one-dimensional normal-metal
  rings},
\newblock \bibinfo{journal}{Physical Review A} \bibinfo{volume}{30}
  (\bibinfo{year}{1984}) \bibinfo{pages}{1982--1989}. \URLprefix
  \url{https://link.aps.org/doi/10.1103/PhysRevA.30.1982}.
  \DOIprefix\doi{10.1103/PhysRevA.30.1982}.
\bibitem[{Webb et~al.(1985)Webb, Washburn, Umbach, and
  Laibowitz}]{webb1985observation}
\bibinfo{author}{R.~A. Webb}, \bibinfo{author}{S.~Washburn},
  \bibinfo{author}{C.~P. Umbach}, \bibinfo{author}{R.~B. Laibowitz},
\newblock \bibinfo{title}{{Observation of $\frac{h}{e}$ Aharonov-Bohm
  Oscillations in Normal-Metal Rings}},
\newblock \bibinfo{journal}{Physical Review Letters} \bibinfo{volume}{54}
  (\bibinfo{year}{1985}) \bibinfo{pages}{2696--2699}. \URLprefix
  \url{https://link.aps.org/doi/10.1103/PhysRevLett.54.2696}.
  \DOIprefix\doi{10.1103/PhysRevLett.54.2696}.
\bibitem[{Saminadayar et~al.(2004)Saminadayar, Bauerle, and
  Mailly}]{saminadayar2004equilibrium}
\bibinfo{author}{L.~Saminadayar}, \bibinfo{author}{C.~Bauerle},
  \bibinfo{author}{D.~Mailly},
\newblock \bibinfo{title}{Equilibrium properties of mesoscopic quantum
  conductors},
\newblock \bibinfo{journal}{Enc. Nanosci. Nanotech} \bibinfo{volume}{4}
  (\bibinfo{year}{2004}) \bibinfo{pages}{267}.
\bibitem[{{Zvyagin} and {Krive}(1995)}]{zvyagin1995persistent}
\bibinfo{author}{A.~A. {Zvyagin}}, \bibinfo{author}{I.~V. {Krive}},
\newblock \bibinfo{title}{{Persistent currents in one-dimensional systems of
  strongly correlated electrons (review)}},
\newblock \bibinfo{journal}{Low Temperature Physics}
  \bibinfo{volume}{\textbf{21}} (\bibinfo{year}{(1995)})
  \bibinfo{pages}{533--555}.
\bibitem[{Bouchiat et~al.(1991{\natexlab{a}})Bouchiat, Montambaux, and
  Sigeti}]{bouchiat1991persistent}
\bibinfo{author}{H.~Bouchiat}, \bibinfo{author}{G.~Montambaux},
  \bibinfo{author}{D.~Sigeti},
\newblock \bibinfo{title}{Persistent currents in mesoscopic rings, conductance,
  and boundary conditions},
\newblock \bibinfo{journal}{Physical Review B} \bibinfo{volume}{44}
  (\bibinfo{year}{1991}{\natexlab{a}}) \bibinfo{pages}{1682--1691}. \URLprefix
  \url{https://link.aps.org/doi/10.1103/PhysRevB.44.1682}.
  \DOIprefix\doi{10.1103/PhysRevB.44.1682}.
\bibitem[{Bouchiat et~al.(1991{\natexlab{b}})Bouchiat, Montambaux, Levyt,
  Dolan, and Dunsmuir}]{bouchiat1991persistentb}
\bibinfo{author}{H.~Bouchiat}, \bibinfo{author}{G.~Montambaux},
  \bibinfo{author}{L.~P. Levyt}, \bibinfo{author}{G.~Dolan},
  \bibinfo{author}{J.~Dunsmuir}, \bibinfo{title}{{Persistent Currents in
  Mesoscopic Rings: Ensemble Average and Half-Flux-Quantum Periodicity}},
  \bibinfo{publisher}{Springer US}, \bibinfo{address}{Boston, MA},
  \bibinfo{year}{1991}{\natexlab{b}}, pp. \bibinfo{pages}{245--260}. \URLprefix
  \url{https://doi.org/10.1007/978-1-4899-3698-1_16}.
  \DOIprefix\doi{10.1007/978-1-4899-3698-1_16}.
\bibitem[{Szopa and Zipper(1991)}]{szopa1991persistent}
\bibinfo{author}{M.~Szopa}, \bibinfo{author}{E.~Zipper},
\newblock \bibinfo{title}{Persistent currents induced in small metallic rings},
\newblock \bibinfo{journal}{Solid State Communications} \bibinfo{volume}{77}
  (\bibinfo{year}{1991}) \bibinfo{pages}{739--743}. \URLprefix
  \url{https://www.sciencedirect.com/science/article/pii/003810989190780Y}.
  \DOIprefix\doi{https://doi.org/10.1016/0038-1098(91)90780-Y}.
\bibitem[{Riedel(1991)}]{riedel1991persistent}
\bibinfo{author}{E.~K. Riedel}, \bibinfo{title}{{Persistent Currents in
  Mesoscopic Rings}}, \bibinfo{publisher}{Springer US},
  \bibinfo{address}{Boston, MA}, \bibinfo{year}{1991}, pp.
  \bibinfo{pages}{261--276}. \URLprefix
  \url{https://doi.org/10.1007/978-1-4899-3698-1_17}.
  \DOIprefix\doi{10.1007/978-1-4899-3698-1_17}.
\bibitem[{Imry(2002)}]{imry2002intro}
\bibinfo{author}{Y.~Imry}, \bibinfo{title}{Introduction to mesoscopic physics},
  \bibinfo{publisher}{Oxford University Press on Demand},
  \bibinfo{year}{(2002)}.
\bibitem[{Ambegaokar and Eckern(1990)}]{ambegaokar1990coherence}
\bibinfo{author}{V.~Ambegaokar}, \bibinfo{author}{U.~Eckern},
\newblock \bibinfo{title}{Coherence and persistent currents in mesoscopic
  rings},
\newblock \bibinfo{journal}{Phys. Rev. Lett.} \bibinfo{volume}{65}
  (\bibinfo{year}{1990}) \bibinfo{pages}{381--384}. \URLprefix
  \url{https://link.aps.org/doi/10.1103/PhysRevLett.65.381}.
  \DOIprefix\doi{10.1103/PhysRevLett.65.381}.
\bibitem[{Aharonov and Bohm(1959)}]{aharonov1959significance}
\bibinfo{author}{Y.~Aharonov}, \bibinfo{author}{D.~Bohm},
\newblock \bibinfo{title}{Significance of electromagnetic potentials in the
  quantum theory},
\newblock \bibinfo{journal}{Physical Review} \bibinfo{volume}{115}
  (\bibinfo{year}{1959}) \bibinfo{pages}{485--491}. \URLprefix
  \url{https://link.aps.org/doi/10.1103/PhysRev.115.485}.
  \DOIprefix\doi{10.1103/PhysRev.115.485}.
\bibitem[{Gefen et~al.(1984)Gefen, Imry, and Azbel}]{gefen1984quantum}
\bibinfo{author}{Y.~Gefen}, \bibinfo{author}{Y.~Imry}, \bibinfo{author}{M.~Y.
  Azbel},
\newblock \bibinfo{title}{{Quantum Oscillations and the Aharonov-Bohm Effect
  for Parallel Resistors}},
\newblock \bibinfo{journal}{Physical Review Letters} \bibinfo{volume}{52}
  (\bibinfo{year}{1984}) \bibinfo{pages}{129--132}.
  \DOIprefix\doi{10.1103/PhysRevLett.52.129}.
\bibitem[{Mohanty(1999)}]{mohanty1999persistent}
\bibinfo{author}{P.~Mohanty},
\newblock \bibinfo{title}{Persistent current in normal metals},
\newblock \bibinfo{journal}{Annalen der Physik} \bibinfo{volume}{8}
  (\bibinfo{year}{1999}) \bibinfo{pages}{549--558}.
\bibitem[{Chakraborty and Pietil\"{a}inen(1994)}]{chakraborty1994electron}
\bibinfo{author}{T.~Chakraborty}, \bibinfo{author}{P.~Pietil\"{a}inen},
\newblock \bibinfo{title}{Electron-electron interaction and the persistent
  current in a quantum ring},
\newblock \bibinfo{journal}{Physical Review B} \bibinfo{volume}{50}
  (\bibinfo{year}{1994}) \bibinfo{pages}{8460}.
\bibitem[{Entin-Wohlman et~al.(2006)Entin-Wohlman, Imry, and
  Aharony}]{entin2006persistent}
\bibinfo{author}{O.~Entin-Wohlman}, \bibinfo{author}{Y.~Imry},
  \bibinfo{author}{A.~Aharony},
\newblock \bibinfo{title}{{Persistent Currents in Interacting Aharonov-Bohm
  Interferometers}},
\newblock in: \bibinfo{booktitle}{Theory of Quantum Transport in Metallic and
  Hybrid Nanostructures}, \bibinfo{publisher}{Springer}, \bibinfo{year}{2006},
  pp. \bibinfo{pages}{77--82}.
\bibitem[{Matveev et~al.(2002)Matveev, Larkin, and
  Glazman}]{matveev2002persistent}
\bibinfo{author}{K.~Matveev}, \bibinfo{author}{A.~Larkin},
  \bibinfo{author}{L.~Glazman},
\newblock \bibinfo{title}{Persistent current in superconducting nanorings},
\newblock \bibinfo{journal}{Physical Review Letters} \bibinfo{volume}{89}
  (\bibinfo{year}{2002}) \bibinfo{pages}{096802}.
\bibitem[{Riedel and von Oppen(1993)}]{riedel1993mesoscopic}
\bibinfo{author}{E.~K. Riedel}, \bibinfo{author}{F.~von Oppen},
\newblock \bibinfo{title}{Mesoscopic persistent current in small rings},
\newblock \bibinfo{journal}{Physical Review B} \bibinfo{volume}{47}
  (\bibinfo{year}{1993}) \bibinfo{pages}{15449}.
\bibitem[{Cheung et~al.(1989)Cheung, Riedel, and Gefen}]{cheung1989persistent}
\bibinfo{author}{H.-F. Cheung}, \bibinfo{author}{E.~K. Riedel},
  \bibinfo{author}{Y.~Gefen},
\newblock \bibinfo{title}{{Persistent Currents in Mesoscopic Rings and
  Cylinders}},
\newblock \bibinfo{journal}{Physical Review Letters} \bibinfo{volume}{62}
  (\bibinfo{year}{1989}) \bibinfo{pages}{587--590}. \URLprefix
  \url{https://link.aps.org/doi/10.1103/PhysRevLett.62.587}.
  \DOIprefix\doi{10.1103/PhysRevLett.62.587}.
\bibitem[{Kulik(2010)}]{kulik2010persistent}
\bibinfo{author}{I.~O. Kulik},
\newblock \bibinfo{title}{{Persistent currents, flux quantization, and
  magnetomotive forces in normal metals and superconductors (Review Article)}},
\newblock \bibinfo{journal}{Low Temperature Physics} \bibinfo{volume}{36}
  (\bibinfo{year}{2010}) \bibinfo{pages}{841--848}. \URLprefix
  \url{https://doi.org/10.1063/1.3514415}. \DOIprefix\doi{10.1063/1.3514415}.
  \href{http://arxiv.org/abs/https://pubs.aip.org/aip/ltp/article-pdf/36/10/841/14108057/841\_1\_online.pdf}{{\tt
  arXiv:https://pubs.aip.org/aip/ltp/article-pdf/36/10/841/14108057/841\_1\_online.pdf}}.
\bibitem[{Yerin et~al.(2021)Yerin, Gusynin, Sharapov, and
  Varlamov}]{yerin2021genesis}
\bibinfo{author}{Y.~Yerin}, \bibinfo{author}{V.~P. Gusynin},
  \bibinfo{author}{S.~G. Sharapov}, \bibinfo{author}{A.~A. Varlamov},
\newblock \bibinfo{title}{{Genesis and fading away of persistent currents in a
  Corbino disk geometry}},
\newblock \bibinfo{journal}{Physical Review B} \bibinfo{volume}{104}
  (\bibinfo{year}{2021}) \bibinfo{pages}{075415}. \URLprefix
  \url{https://link.aps.org/doi/10.1103/PhysRevB.104.075415}.
  \DOIprefix\doi{10.1103/PhysRevB.104.075415}.
\bibitem[{L\'evy et~al.(1990)L\'evy, Dolan, Dunsmuir, and
  Bouchiat}]{levy1990magnet}
\bibinfo{author}{L.~P. L\'evy}, \bibinfo{author}{G.~Dolan},
  \bibinfo{author}{J.~Dunsmuir}, \bibinfo{author}{H.~Bouchiat},
\newblock \bibinfo{title}{Magnetization of mesoscopic copper rings: Evidence
  for persistent currents},
\newblock \bibinfo{journal}{Physical Review Letters} \bibinfo{volume}{64}
  (\bibinfo{year}{1990}) \bibinfo{pages}{2074--2077}. \URLprefix
  \url{https://link.aps.org/doi/10.1103/PhysRevLett.64.2074}.
  \DOIprefix\doi{10.1103/PhysRevLett.64.2074}.
\bibitem[{Bluhm et~al.(2009)Bluhm, Koshnick, Bert, Huber, and
  Moler}]{bluhm2009persistent}
\bibinfo{author}{H.~Bluhm}, \bibinfo{author}{N.~C. Koshnick},
  \bibinfo{author}{J.~A. Bert}, \bibinfo{author}{M.~E. Huber},
  \bibinfo{author}{K.~A. Moler},
\newblock \bibinfo{title}{{Persistent Currents in Normal Metal Rings}},
\newblock \bibinfo{journal}{Physical Review Letters} \bibinfo{volume}{102}
  (\bibinfo{year}{2009}) \bibinfo{pages}{136802}. \URLprefix
  \url{https://link.aps.org/doi/10.1103/PhysRevLett.102.136802}.
  \DOIprefix\doi{10.1103/PhysRevLett.102.136802}.
\bibitem[{Bleszynski-Jayich et~al.(2009)Bleszynski-Jayich, Shanks, Peaudecerf,
  Ginossar, von Oppen, Glazman, and Harris}]{bleszynski2009persistent}
\bibinfo{author}{A.~C. Bleszynski-Jayich}, \bibinfo{author}{W.~E. Shanks},
  \bibinfo{author}{B.~Peaudecerf}, \bibinfo{author}{E.~Ginossar},
  \bibinfo{author}{F.~von Oppen}, \bibinfo{author}{L.~Glazman},
  \bibinfo{author}{J.~G.~E. Harris},
\newblock \bibinfo{title}{{Persistent Currents in Normal Metal Rings}},
\newblock \bibinfo{journal}{Science} \bibinfo{volume}{326}
  (\bibinfo{year}{2009}) \bibinfo{pages}{272--275}. \URLprefix
  \url{https://www.science.org/doi/abs/10.1126/science.1178139}.
  \DOIprefix\doi{10.1126/science.1178139}.
  \href{http://arxiv.org/abs/https://www.science.org/doi/pdf/10.1126/science.1178139}{{\tt
  arXiv:https://www.science.org/doi/pdf/10.1126/science.1178139}}.
\bibitem[{Anderson et~al.(1995)Anderson, Ensher, Matthews, Wieman, and
  Cornell}]{anderson1995observation}
\bibinfo{author}{M.~H. Anderson}, \bibinfo{author}{J.~R. Ensher},
  \bibinfo{author}{M.~R. Matthews}, \bibinfo{author}{C.~E. Wieman},
  \bibinfo{author}{E.~A. Cornell},
\newblock \bibinfo{title}{{Observation of Bose-Einstein Condensation in a
  Dilute Atomic Vapor}},
\newblock \bibinfo{journal}{Science} \bibinfo{volume}{269}
  (\bibinfo{year}{1995}) \bibinfo{pages}{198--201}. \URLprefix
  \url{https://www.science.org/doi/abs/10.1126/science.269.5221.198}.
  \DOIprefix\doi{10.1126/science.269.5221.198}.
\bibitem[{Davis et~al.(1995)Davis, Mewes, Andrews, van Druten, Durfee, Kurn,
  and Ketterle}]{davis1995bose-einstein}
\bibinfo{author}{K.~B. Davis}, \bibinfo{author}{M.~O. Mewes},
  \bibinfo{author}{M.~R. Andrews}, \bibinfo{author}{N.~J. van Druten},
  \bibinfo{author}{D.~S. Durfee}, \bibinfo{author}{D.~M. Kurn},
  \bibinfo{author}{W.~Ketterle},
\newblock \bibinfo{title}{{Bose-Einstein Condensation in a Gas of Sodium
  Atoms}},
\newblock \bibinfo{journal}{Physical Review Letters} \bibinfo{volume}{75}
  (\bibinfo{year}{1995}) \bibinfo{pages}{3969--3973}. \URLprefix
  \url{https://link.aps.org/doi/10.1103/PhysRevLett.75.3969}.
  \DOIprefix\doi{10.1103/PhysRevLett.75.3969}.
\bibitem[{Cornell and Wieman(2002)}]{cornell2002nobel}
\bibinfo{author}{E.~A. Cornell}, \bibinfo{author}{C.~E. Wieman},
\newblock \bibinfo{title}{Nobel lecture: Bose-einstein condensation in a dilute
  gas, the first 70 years and some recent experiments},
\newblock \bibinfo{journal}{Reviews of Modern Physics} \bibinfo{volume}{74}
  (\bibinfo{year}{2002}) \bibinfo{pages}{875--893}. \URLprefix
  \url{https://link.aps.org/doi/10.1103/RevModPhys.74.875}.
  \DOIprefix\doi{10.1103/RevModPhys.74.875}.
\bibitem[{Ketterle(2002)}]{ketterle2002nobel}
\bibinfo{author}{W.~Ketterle},
\newblock \bibinfo{title}{{Nobel lecture: When atoms behave as waves:
  Bose-Einstein condensation and the atom laser}},
\newblock \bibinfo{journal}{Reviews of Modern Physics} \bibinfo{volume}{74}
  (\bibinfo{year}{2002}) \bibinfo{pages}{1131--1151}. \URLprefix
  \url{https://link.aps.org/doi/10.1103/RevModPhys.74.1131}.
  \DOIprefix\doi{10.1103/RevModPhys.74.1131}.
\bibitem[{Pethick and Smith(2008)}]{pethick2008bose}
\bibinfo{author}{C.~J. Pethick}, \bibinfo{author}{H.~Smith},
  \bibinfo{title}{{Bose--Einstein condensation in dilute gases}},
  \bibinfo{publisher}{Cambridge university press}, \bibinfo{year}{2008}.
\bibitem[{Pitaevskii and Stringari(2016)}]{pitaevskii2016bose}
\bibinfo{author}{L.~Pitaevskii}, \bibinfo{author}{S.~Stringari},
  \bibinfo{title}{{Bose-Einstein condensation and superfluidity}}, volume
  \bibinfo{volume}{164}, \bibinfo{publisher}{Oxford University Press},
  \bibinfo{year}{2016}.
\bibitem[{Ginzburg(2004)}]{ginzburg2004nobel}
\bibinfo{author}{V.~L. Ginzburg},
\newblock \bibinfo{title}{{Nobel Lecture: On superconductivity and
  superfluidity (what I have and have not managed to do) as well as on the
  ``physical minimum'' at the beginning of the XXI century}},
\newblock \bibinfo{journal}{Reviews of Modern Physics} \bibinfo{volume}{76}
  (\bibinfo{year}{2004}) \bibinfo{pages}{981--998}. \URLprefix
  \url{https://link.aps.org/doi/10.1103/RevModPhys.76.981}.
  \DOIprefix\doi{10.1103/RevModPhys.76.981}.
\bibitem[{Leggett(2004)}]{leggett2004nobel}
\bibinfo{author}{A.~J. Leggett},
\newblock \bibinfo{title}{{Nobel Lecture: Superfluid $^{3}\mathrm{He}$: the
  early days as seen by a theorist}},
\newblock \bibinfo{journal}{Reviews of Modern Physics} \bibinfo{volume}{76}
  (\bibinfo{year}{2004}) \bibinfo{pages}{999--1011}. \URLprefix
  \url{https://link.aps.org/doi/10.1103/RevModPhys.76.999}.
  \DOIprefix\doi{10.1103/RevModPhys.76.999}.
\bibitem[{Abrikosov(2004)}]{abrikosov2004nobel}
\bibinfo{author}{A.~A. Abrikosov},
\newblock \bibinfo{title}{{Nobel Lecture: Type-II superconductors and the
  vortex lattice}},
\newblock \bibinfo{journal}{Reviews of Modern Physics} \bibinfo{volume}{76}
  (\bibinfo{year}{2004}) \bibinfo{pages}{975--979}. \URLprefix
  \url{https://link.aps.org/doi/10.1103/RevModPhys.76.975}.
  \DOIprefix\doi{10.1103/RevModPhys.76.975}.
\bibitem[{Leggett(2008)}]{leggett2008quantum}
\bibinfo{author}{A.~J. Leggett},
\newblock \bibinfo{title}{Quantum liquids},
\newblock \bibinfo{journal}{Science} \bibinfo{volume}{319}
  (\bibinfo{year}{2008}) \bibinfo{pages}{1203--1205}.
\bibitem[{Dalibard et~al.(2011)Dalibard, Gerbier, Juzeli\ifmmode~\bar{u}\else
  \={u}\fi{}nas, and \"Ohberg}]{dalibard2011artificial}
\bibinfo{author}{J.~Dalibard}, \bibinfo{author}{F.~Gerbier},
  \bibinfo{author}{G.~Juzeli\ifmmode~\bar{u}\else \={u}\fi{}nas},
  \bibinfo{author}{P.~\"Ohberg},
\newblock \bibinfo{title}{{Colloquium: Artificial gauge potentials for neutral
  atoms}},
\newblock \bibinfo{journal}{Reviews of Modern Physics} \bibinfo{volume}{83}
  (\bibinfo{year}{2011}) \bibinfo{pages}{1523--1543}. \URLprefix
  \url{https://link.aps.org/doi/10.1103/RevModPhys.83.1523}.
  \DOIprefix\doi{10.1103/RevModPhys.83.1523}.
\bibitem[{Goldman et~al.(2014)Goldman, Juzeliūnas, Öhberg, and
  Spielman}]{goldman2014light}
\bibinfo{author}{N.~Goldman}, \bibinfo{author}{G.~Juzeliūnas},
  \bibinfo{author}{P.~Öhberg}, \bibinfo{author}{I.~B. Spielman},
\newblock \bibinfo{title}{Light-induced gauge fields for ultracold atoms},
\newblock \bibinfo{journal}{Reports on Progress in Physics}
  \bibinfo{volume}{77} (\bibinfo{year}{2014}) \bibinfo{pages}{126401}.
  \URLprefix \url{https://dx.doi.org/10.1088/0034-4885/77/12/126401}.
  \DOIprefix\doi{10.1088/0034-4885/77/12/126401}.
\bibitem[{Fetter(2009)}]{fetter2009rotating}
\bibinfo{author}{A.~L. Fetter},
\newblock \bibinfo{title}{{Rotating trapped Bose-Einstein condensates}},
\newblock \bibinfo{journal}{Reviews of Modern Physics} \bibinfo{volume}{81}
  (\bibinfo{year}{2009}) \bibinfo{pages}{647--691}.
\bibitem[{Dalibard(2016)}]{dalibard2016introduction}
\bibinfo{author}{J.~Dalibard},
\newblock \bibinfo{title}{Introduction to the physics of artificial gauge
  fields},
\newblock in: \bibinfo{booktitle}{Quantum Matter at Ultralow Temperatures},
  \bibinfo{publisher}{IOS Press}, \bibinfo{year}{2016}, pp.
  \bibinfo{pages}{1--61}.
\bibitem[{Wright et~al.(2013)Wright, Blakestad, Lobb, Phillips, and
  Campbell}]{wright2013driving}
\bibinfo{author}{K.~C. Wright}, \bibinfo{author}{R.~B. Blakestad},
  \bibinfo{author}{C.~J. Lobb}, \bibinfo{author}{W.~D. Phillips},
  \bibinfo{author}{G.~K. Campbell},
\newblock \bibinfo{title}{{Driving Phase Slips in a Superfluid Atom Circuit
  with a Rotating Weak Link}},
\newblock \bibinfo{journal}{Physical Review Letters} \bibinfo{volume}{110}
  (\bibinfo{year}{2013}) \bibinfo{pages}{025302}. \URLprefix
  \url{https://link.aps.org/doi/10.1103/PhysRevLett.110.025302}.
  \DOIprefix\doi{10.1103/PhysRevLett.110.025302}.
\bibitem[{Cai et~al.(2022)Cai, Allman, Sabharwal, and
  Wright}]{cai2022persistent}
\bibinfo{author}{Y.~Cai}, \bibinfo{author}{D.~G. Allman},
  \bibinfo{author}{P.~Sabharwal}, \bibinfo{author}{K.~C. Wright},
\newblock \bibinfo{title}{{Persistent Currents in Rings of Ultracold Fermionic
  Atoms}},
\newblock \bibinfo{journal}{Physical Review Letters} \bibinfo{volume}{128}
  (\bibinfo{year}{2022}) \bibinfo{pages}{150401}. \URLprefix
  \url{https://link.aps.org/doi/10.1103/PhysRevLett.128.150401}.
  \DOIprefix\doi{10.1103/PhysRevLett.128.150401}.
\bibitem[{Andersen et~al.(2006)Andersen, Ryu, Cladé, Natarajan, Vaziri,
  Helmerson, and Phillips}]{andersen2006quantized}
\bibinfo{author}{M.~F. Andersen}, \bibinfo{author}{C.~Ryu},
  \bibinfo{author}{P.~Cladé}, \bibinfo{author}{V.~Natarajan},
  \bibinfo{author}{A.~Vaziri}, \bibinfo{author}{K.~Helmerson},
  \bibinfo{author}{W.~D. Phillips},
\newblock \bibinfo{title}{Quantized {Rotation} of {Atoms} from {Photons} with
  {Orbital} {Angular} {Momentum}},
\newblock \bibinfo{journal}{Physical Review Letters} \bibinfo{volume}{97}
  (\bibinfo{year}{2006}) \bibinfo{pages}{170406}. \URLprefix
  \url{https://link.aps.org/doi/10.1103/PhysRevLett.97.170406}.
  \DOIprefix\doi{10.1103/PhysRevLett.97.170406}, \bibinfo{note}{publisher:
  American Physical Society}.
\bibitem[{Ramanathan et~al.(2011)Ramanathan, Wright, Muniz, Zelan, Hill, Lobb,
  Helmerson, Phillips, and Campbell}]{ramanathan2011superflow}
\bibinfo{author}{A.~Ramanathan}, \bibinfo{author}{K.~C. Wright},
  \bibinfo{author}{S.~R. Muniz}, \bibinfo{author}{M.~Zelan},
  \bibinfo{author}{W.~T. Hill}, \bibinfo{author}{C.~J. Lobb},
  \bibinfo{author}{K.~Helmerson}, \bibinfo{author}{W.~D. Phillips},
  \bibinfo{author}{G.~K. Campbell},
\newblock \bibinfo{title}{{Superflow in a Toroidal {Bose-Einstein} Condensate:
  An Atom Circuit with a Tunable Weak Link}},
\newblock \bibinfo{journal}{Physical Review Letters} \bibinfo{volume}{106}
  (\bibinfo{year}{2011}) \bibinfo{pages}{130401}. \URLprefix
  \url{https://link.aps.org/doi/10.1103/PhysRevLett.106.130401}.
  \DOIprefix\doi{10.1103/PhysRevLett.106.130401}.
\bibitem[{Moulder et~al.(2012)Moulder, Beattie, Smith, Tammuz, and
  Hadzibabic}]{moulder2012quantized}
\bibinfo{author}{S.~Moulder}, \bibinfo{author}{S.~Beattie},
  \bibinfo{author}{R.~P. Smith}, \bibinfo{author}{N.~Tammuz},
  \bibinfo{author}{Z.~Hadzibabic},
\newblock \bibinfo{title}{Quantized supercurrent decay in an annular
  {Bose-Einstein} condensate},
\newblock \bibinfo{journal}{Physical Review A} \bibinfo{volume}{86}
  (\bibinfo{year}{2012}) \bibinfo{pages}{013629}. \URLprefix
  \url{https://link.aps.org/doi/10.1103/PhysRevA.86.013629}.
  \DOIprefix\doi{10.1103/PhysRevA.86.013629}.
\bibitem[{Ryu et~al.(2007)Ryu, Andersen, Clad\'e, Natarajan, Helmerson, and
  Phillips}]{ryu2007observation}
\bibinfo{author}{C.~Ryu}, \bibinfo{author}{M.~F. Andersen},
  \bibinfo{author}{P.~Clad\'e}, \bibinfo{author}{V.~Natarajan},
  \bibinfo{author}{K.~Helmerson}, \bibinfo{author}{W.~D. Phillips},
\newblock \bibinfo{title}{{Observation of Persistent Flow of a {Bose-Einstein}
  Condensate in a Toroidal Trap}},
\newblock \bibinfo{journal}{Physical Review Letters} \bibinfo{volume}{99}
  (\bibinfo{year}{2007}) \bibinfo{pages}{260401}. \URLprefix
  \url{https://link.aps.org/doi/10.1103/PhysRevLett.99.260401}.
  \DOIprefix\doi{10.1103/PhysRevLett.99.260401}.
\bibitem[{Kumar et~al.(2018)Kumar, Dubessy, Badr, De~Rossi, de~Go\"er~de Herve,
  Longchambon, and Perrin}]{kumar2018producing}
\bibinfo{author}{A.~Kumar}, \bibinfo{author}{R.~Dubessy},
  \bibinfo{author}{T.~Badr}, \bibinfo{author}{C.~De~Rossi},
  \bibinfo{author}{M.~de~Go\"er~de Herve}, \bibinfo{author}{L.~Longchambon},
  \bibinfo{author}{H.~Perrin},
\newblock \bibinfo{title}{Producing superfluid circulation states using phase
  imprinting},
\newblock \bibinfo{journal}{Physical Review A} \bibinfo{volume}{97}
  (\bibinfo{year}{2018}) \bibinfo{pages}{043615}. \URLprefix
  \url{https://link.aps.org/doi/10.1103/PhysRevA.97.043615}.
  \DOIprefix\doi{10.1103/PhysRevA.97.043615}.
\bibitem[{Del~Pace et~al.(2022)Del~Pace, Xhani, Muzi~Falconi, Fedrizzi, Grani,
  Hernandez~Rajkov, Inguscio, Scazza, Kwon, and Roati}]{del_pace2022imprinting}
\bibinfo{author}{G.~Del~Pace}, \bibinfo{author}{K.~Xhani},
  \bibinfo{author}{A.~Muzi~Falconi}, \bibinfo{author}{M.~Fedrizzi},
  \bibinfo{author}{N.~Grani}, \bibinfo{author}{D.~Hernandez~Rajkov},
  \bibinfo{author}{M.~Inguscio}, \bibinfo{author}{F.~Scazza},
  \bibinfo{author}{W.~Kwon}, \bibinfo{author}{G.~Roati},
\newblock \bibinfo{title}{Imprinting {Persistent} {Currents} in {Tunable}
  {Fermionic} {Rings}},
\newblock \bibinfo{journal}{Physical Review X} \bibinfo{volume}{12}
  (\bibinfo{year}{2022}) \bibinfo{pages}{041037}. \URLprefix
  \url{https://link.aps.org/doi/10.1103/PhysRevX.12.041037}.
\bibitem[{Franke-Arnold et~al.(2008)Franke-Arnold, Allen, and
  Padgett}]{franke2008advances}
\bibinfo{author}{S.~Franke-Arnold}, \bibinfo{author}{L.~Allen},
  \bibinfo{author}{M.~Padgett},
\newblock \bibinfo{title}{Advances in optical angular momentum},
\newblock \bibinfo{journal}{Laser \& Photonics Reviews} \bibinfo{volume}{2}
  (\bibinfo{year}{2008}) \bibinfo{pages}{299--313}.
\bibitem[{Gauthier et~al.(2021)Gauthier, Bell, Stilgoe, Baker,
  Rubinsztein-Dunlop, and Neely}]{gauthier2021dynamic}
\bibinfo{author}{G.~Gauthier}, \bibinfo{author}{T.~A. Bell},
  \bibinfo{author}{A.~B. Stilgoe}, \bibinfo{author}{M.~Baker},
  \bibinfo{author}{H.~Rubinsztein-Dunlop}, \bibinfo{author}{T.~W. Neely},
\newblock \bibinfo{title}{Dynamic high-resolution optical trapping of ultracold
  atoms},
\newblock \bibinfo{journal}{Advances In Atomic, Molecular, and Optical Physics}
   (\bibinfo{year}{2021}) \bibinfo{pages}{1--101}.
\bibitem[{Gauthier et~al.(2019)Gauthier, Szigeti, Reeves, Baker, Bell,
  Rubinsztein-Dunlop, Davis, and Neely}]{gauthier2019quantitative}
\bibinfo{author}{G.~Gauthier}, \bibinfo{author}{S.~S. Szigeti},
  \bibinfo{author}{M.~T. Reeves}, \bibinfo{author}{M.~Baker},
  \bibinfo{author}{T.~A. Bell}, \bibinfo{author}{H.~Rubinsztein-Dunlop},
  \bibinfo{author}{M.~J. Davis}, \bibinfo{author}{T.~W. Neely},
\newblock \bibinfo{title}{{Quantitative Acoustic Models for Superfluid
  Circuits}},
\newblock \bibinfo{journal}{Physical Review Letters} \bibinfo{volume}{123}
  (\bibinfo{year}{2019}) \bibinfo{pages}{260402}. \URLprefix
  \url{https://doi.org/10.1103/PhysRevLett.123.260402}.
  \DOIprefix\doi{10.1103/PhysRevLett.123.260402}.
\bibitem[{Rubinsztein-Dunlop et~al.(2016)Rubinsztein-Dunlop, Forbes, Berry,
  Dennis, Andrews, Mansuripur et~al.}]{rubinsztein2016roadmap}
\bibinfo{author}{H.~Rubinsztein-Dunlop}, \bibinfo{author}{A.~Forbes},
  \bibinfo{author}{M.~V. Berry}, \bibinfo{author}{M.~R. Dennis},
  \bibinfo{author}{D.~L. Andrews}, \bibinfo{author}{M.~Mansuripur}, et~al.,
\newblock \bibinfo{title}{Roadmap on structured light},
\newblock \bibinfo{journal}{Journal of Optics} \bibinfo{volume}{19}
  (\bibinfo{year}{2016}) \bibinfo{pages}{013001}. \URLprefix
  \url{https://dx.doi.org/10.1088/2040-8978/19/1/013001}.
  \DOIprefix\doi{10.1088/2040-8978/19/1/013001}.
\bibitem[{Henderson et~al.(2009)Henderson, Ryu, MacCormick, and
  Boshier}]{henderson2009experimental}
\bibinfo{author}{K.~Henderson}, \bibinfo{author}{C.~Ryu},
  \bibinfo{author}{C.~MacCormick}, \bibinfo{author}{M.~G. Boshier},
\newblock \bibinfo{title}{Experimental demonstration of painting arbitrary and
  dynamic potentials for {Bose}–{Einstein} condensates},
\newblock \bibinfo{journal}{New Journal of Physics} \bibinfo{volume}{11}
  (\bibinfo{year}{2009}) \bibinfo{pages}{043030}. \URLprefix
  \url{https://dx.doi.org/10.1088/1367-2630/11/4/043030}.
  \DOIprefix\doi{10.1088/1367-2630/11/4/043030}.
\bibitem[{Bloch(2008)}]{bloch2008quantum}
\bibinfo{author}{I.~Bloch},
\newblock \bibinfo{title}{Quantum coherence and entanglement with ultracold
  atoms in optical lattices},
\newblock \bibinfo{journal}{Nature} \bibinfo{volume}{453}
  (\bibinfo{year}{2008}) \bibinfo{pages}{1016--1022}.
\bibitem[{Gardiner and Zoller(2017)}]{gardiner2017quantum}
\bibinfo{author}{C.~W. Gardiner}, \bibinfo{author}{P.~Zoller},
  \bibinfo{title}{{Quantum World Of Ultra-cold Atoms And Light, The-Book Iii:
  Ultra-cold Atoms}}, volume~\bibinfo{volume}{5}, \bibinfo{publisher}{World
  Scientific}, \bibinfo{year}{2017}.
\bibitem[{Lewenstein et~al.(2012)Lewenstein, Sanpera, and
  Ahufinger}]{lewenstein2012ultracold}
\bibinfo{author}{M.~Lewenstein}, \bibinfo{author}{A.~Sanpera},
  \bibinfo{author}{V.~Ahufinger}, \bibinfo{title}{{Ultracold Atoms in Optical
  Lattices: Simulating quantum many-body systems}}, \bibinfo{publisher}{OUP
  Oxford}, \bibinfo{year}{2012}.
\bibitem[{Bakr et~al.(2009)Bakr, Gillen, Peng, F{\"o}lling, and
  Greiner}]{bakr2009quantum}
\bibinfo{author}{W.~S. Bakr}, \bibinfo{author}{J.~I. Gillen},
  \bibinfo{author}{A.~Peng}, \bibinfo{author}{S.~F{\"o}lling},
  \bibinfo{author}{M.~Greiner},
\newblock \bibinfo{title}{{A quantum gas microscope for detecting single atoms
  in a Hubbard-regime optical lattice}},
\newblock \bibinfo{journal}{Nature} \bibinfo{volume}{462}
  (\bibinfo{year}{2009}) \bibinfo{pages}{74--77}.
\bibitem[{Sherson et~al.(2010)Sherson, Weitenberg, Endres, Cheneau, Bloch, and
  Kuhr}]{sherson2010single}
\bibinfo{author}{J.~F. Sherson}, \bibinfo{author}{C.~Weitenberg},
  \bibinfo{author}{M.~Endres}, \bibinfo{author}{M.~Cheneau},
  \bibinfo{author}{I.~Bloch}, \bibinfo{author}{S.~Kuhr},
\newblock \bibinfo{title}{{Single-atom-resolved fluorescence imaging of an
  atomic Mott insulator}},
\newblock \bibinfo{journal}{Nature} \bibinfo{volume}{467}
  (\bibinfo{year}{2010}) \bibinfo{pages}{68--72}.
\bibitem[{Endres et~al.(2016)Endres, Bernien, Keesling, Levine, Anschuetz,
  Krajenbrink, Senko, Vuletic, Greiner, and Lukin}]{Endres2016atom}
\bibinfo{author}{M.~Endres}, \bibinfo{author}{H.~Bernien},
  \bibinfo{author}{A.~Keesling}, \bibinfo{author}{H.~Levine},
  \bibinfo{author}{E.~R. Anschuetz}, \bibinfo{author}{A.~Krajenbrink},
  \bibinfo{author}{C.~Senko}, \bibinfo{author}{V.~Vuletic},
  \bibinfo{author}{M.~Greiner}, \bibinfo{author}{M.~D. Lukin},
\newblock \bibinfo{title}{Atom-by-atom assembly of defect-free one-dimensional
  cold atom arrays},
\newblock \bibinfo{journal}{Science} \bibinfo{volume}{354}
  (\bibinfo{year}{2016}) \bibinfo{pages}{1024}. \URLprefix
  \url{http://dx.doi.org/10.1126/science.aah3752}.
  \DOIprefix\doi{10.1126/science.aah3752}.
\bibitem[{Bergschneider et~al.(2018)Bergschneider, Klinkhamer, Becher, Klemt,
  Z\"urn, Preiss, and Jochim}]{bergschneider2018spin}
\bibinfo{author}{A.~Bergschneider}, \bibinfo{author}{V.~M. Klinkhamer},
  \bibinfo{author}{J.~H. Becher}, \bibinfo{author}{R.~Klemt},
  \bibinfo{author}{G.~Z\"urn}, \bibinfo{author}{P.~M. Preiss},
  \bibinfo{author}{S.~Jochim},
\newblock \bibinfo{title}{Spin-resolved single-atom imaging of
  $^{6}\mathrm{Li}$ in free space},
\newblock \bibinfo{journal}{Physical Review A} \bibinfo{volume}{97}
  (\bibinfo{year}{2018}) \bibinfo{pages}{063613}. \URLprefix
  \url{https://link.aps.org/doi/10.1103/PhysRevA.97.063613}.
  \DOIprefix\doi{10.1103/PhysRevA.97.063613}.
\bibitem[{Qian et~al.(2021)Qian, Cui, Luo, Zheng, Huang, Ai, He, Li, and
  Guo}]{qian2021super}
\bibinfo{author}{Z.-H. Qian}, \bibinfo{author}{J.-M. Cui},
  \bibinfo{author}{X.-W. Luo}, \bibinfo{author}{Y.-X. Zheng},
  \bibinfo{author}{Y.-F. Huang}, \bibinfo{author}{M.-Z. Ai},
  \bibinfo{author}{R.~He}, \bibinfo{author}{C.-F. Li}, \bibinfo{author}{G.-C.
  Guo},
\newblock \bibinfo{title}{{Super-resolved Imaging of a Single Cold Atom on a
  Nanosecond Timescale}},
\newblock \bibinfo{journal}{Physical Review Letters} \bibinfo{volume}{127}
  (\bibinfo{year}{2021}) \bibinfo{pages}{263603}. \URLprefix
  \url{https://link.aps.org/doi/10.1103/PhysRevLett.127.263603}.
  \DOIprefix\doi{10.1103/PhysRevLett.127.263603}.
\bibitem[{Kwon et~al.(2020)Kwon, Pace, Panza, Inguscio, Zwerger, Zaccanti,
  Scazza, and Roati}]{kwon2020strongly}
\bibinfo{author}{W.~J. Kwon}, \bibinfo{author}{G.~D. Pace},
  \bibinfo{author}{R.~Panza}, \bibinfo{author}{M.~Inguscio},
  \bibinfo{author}{W.~Zwerger}, \bibinfo{author}{M.~Zaccanti},
  \bibinfo{author}{F.~Scazza}, \bibinfo{author}{G.~Roati},
\newblock \bibinfo{title}{{Strongly correlated superfluid order parameters from
  dc Josephson supercurrents}},
\newblock \bibinfo{journal}{Science} \bibinfo{volume}{369}
  (\bibinfo{year}{2020}) \bibinfo{pages}{84--88}. \URLprefix
  \url{https://www.science.org/doi/abs/10.1126/science.aaz2463}.
  \DOIprefix\doi{10.1126/science.aaz2463}.
  \href{http://arxiv.org/abs/https://www.science.org/doi/pdf/10.1126/science.aaz2463}{{\tt
  arXiv:https://www.science.org/doi/pdf/10.1126/science.aaz2463}}.
\bibitem[{Barredo et~al.(2018)Barredo, Lienhard, de~Léséleuc, Lahaye, and
  Browaeys}]{barredo2018synthetic}
\bibinfo{author}{D.~Barredo}, \bibinfo{author}{V.~Lienhard},
  \bibinfo{author}{S.~de~Léséleuc}, \bibinfo{author}{T.~Lahaye},
  \bibinfo{author}{A.~Browaeys},
\newblock \bibinfo{title}{Synthetic three-dimensional atomic structures
  assembled atom by atom},
\newblock \bibinfo{journal}{Nature} \bibinfo{volume}{561}
  (\bibinfo{year}{2018}) \bibinfo{pages}{79–82}. \URLprefix
  \url{http://dx.doi.org/10.1038/s41586-018-0450-2}.
  \DOIprefix\doi{10.1038/s41586-018-0450-2}.
\bibitem[{Gaunt and Hadzibabic(2012)}]{gaunt2012robust}
\bibinfo{author}{A.~L. Gaunt}, \bibinfo{author}{Z.~Hadzibabic},
\newblock \bibinfo{title}{{Robust Digital Holography For Ultracold Atom
  Trapping}},
\newblock \bibinfo{journal}{Scientific Reports} \bibinfo{volume}{2}
  (\bibinfo{year}{2012}) \bibinfo{pages}{721}. \URLprefix
  \url{https://doi.org/10.1038/srep00721}. \DOIprefix\doi{10.1038/srep00721}.
\bibitem[{Scazza et~al.(2014)Scazza, Hofrichter, H\"{o}fer, Groot, Bloch, and
  F\"{o}lling}]{scazza2014observation}
\bibinfo{author}{F.~Scazza}, \bibinfo{author}{C.~Hofrichter},
  \bibinfo{author}{M.~H\"{o}fer}, \bibinfo{author}{P.~C.~D. Groot},
  \bibinfo{author}{I.~Bloch}, \bibinfo{author}{S.~F\"{o}lling},
\newblock \bibinfo{title}{Observation of two-orbital spin-exchange interactions
  with ultracold {SU}(n)-symmetric fermions},
\newblock \bibinfo{journal}{Nature Physics} \bibinfo{volume}{10}
  (\bibinfo{year}{2014}) \bibinfo{pages}{779--784}. \URLprefix
  \url{https://doi.org/10.1038/nphys3061}. \DOIprefix\doi{10.1038/nphys3061}.
\bibitem[{Taie et~al.(2022)Taie, Ibarra-García-Padilla, Nishizawa, Takasu,
  Kuno, Wei, Scalettar, Hazzard, and Takahashi}]{taie2022observation}
\bibinfo{author}{S.~Taie}, \bibinfo{author}{E.~Ibarra-García-Padilla},
  \bibinfo{author}{N.~Nishizawa}, \bibinfo{author}{Y.~Takasu},
  \bibinfo{author}{Y.~Kuno}, \bibinfo{author}{H.-T. Wei},
  \bibinfo{author}{R.~T. Scalettar}, \bibinfo{author}{K.~R.~A. Hazzard},
  \bibinfo{author}{Y.~Takahashi},
\newblock \bibinfo{title}{{Observation of antiferromagnetic correlations in an
  ultracold SU(N) Hubbard model}},
\newblock \bibinfo{journal}{Nature Physics} \bibinfo{volume}{18}
  (\bibinfo{year}{2022}) \bibinfo{pages}{1356–1361}. \URLprefix
  \url{http://dx.doi.org/10.1038/s41567-022-01725-6}.
  \DOIprefix\doi{10.1038/s41567-022-01725-6}.
\bibitem[{Takahashi(2022)}]{takahashi2022quantum}
\bibinfo{author}{Y.~Takahashi},
\newblock \bibinfo{title}{Quantum simulation of quantum many-body systems with
  ultracold two-electron atoms in an optical lattice},
\newblock \bibinfo{journal}{Proceedings of the Japan Academy, Series B}
  \bibinfo{volume}{98} (\bibinfo{year}{2022}) \bibinfo{pages}{141–160}.
  \URLprefix \url{http://dx.doi.org/10.2183/pjab.98.010}.
  \DOIprefix\doi{10.2183/pjab.98.010}.
\bibitem[{Myatt et~al.(1997)Myatt, Burt, Ghrist, Cornell, and
  Wieman}]{myatt1997production}
\bibinfo{author}{C.~J. Myatt}, \bibinfo{author}{E.~A. Burt},
  \bibinfo{author}{R.~W. Ghrist}, \bibinfo{author}{E.~A. Cornell},
  \bibinfo{author}{C.~E. Wieman},
\newblock \bibinfo{title}{{Production of Two Overlapping Bose-Einstein
  Condensates by Sympathetic Cooling}},
\newblock \bibinfo{journal}{Physical Review Letters} \bibinfo{volume}{78}
  (\bibinfo{year}{1997}) \bibinfo{pages}{586--589}. \URLprefix
  \url{https://link.aps.org/doi/10.1103/PhysRevLett.78.586}.
  \DOIprefix\doi{10.1103/PhysRevLett.78.586}.
\bibitem[{Hall et~al.(1998)Hall, Matthews, Wieman, and
  Cornell}]{hall1998measurements}
\bibinfo{author}{D.~S. Hall}, \bibinfo{author}{M.~R. Matthews},
  \bibinfo{author}{C.~E. Wieman}, \bibinfo{author}{E.~A. Cornell},
\newblock \bibinfo{title}{{Measurements of Relative Phase in Two-Component
  Bose-Einstein Condensates}},
\newblock \bibinfo{journal}{Physical Review Letters} \bibinfo{volume}{81}
  (\bibinfo{year}{1998}) \bibinfo{pages}{1543--1546}. \URLprefix
  \url{https://link.aps.org/doi/10.1103/PhysRevLett.81.1543}.
  \DOIprefix\doi{10.1103/PhysRevLett.81.1543}.
\bibitem[{Schreck et~al.(2001)Schreck, Khaykovich, Corwin, Ferrari, Bourdel,
  Cubizolles, and Salomon}]{schreck2001quasipure}
\bibinfo{author}{F.~Schreck}, \bibinfo{author}{L.~Khaykovich},
  \bibinfo{author}{K.~L. Corwin}, \bibinfo{author}{G.~Ferrari},
  \bibinfo{author}{T.~Bourdel}, \bibinfo{author}{J.~Cubizolles},
  \bibinfo{author}{C.~Salomon},
\newblock \bibinfo{title}{{Quasipure Bose-Einstein Condensate Immersed in a
  Fermi Sea}},
\newblock \bibinfo{journal}{Physical Review Letters} \bibinfo{volume}{87}
  (\bibinfo{year}{2001}) \bibinfo{pages}{080403}. \URLprefix
  \url{https://link.aps.org/doi/10.1103/PhysRevLett.87.080403}.
  \DOIprefix\doi{10.1103/PhysRevLett.87.080403}.
\bibitem[{Roati et~al.(2002)Roati, Riboli, Modugno, and
  Inguscio}]{roati2002fermi}
\bibinfo{author}{G.~Roati}, \bibinfo{author}{F.~Riboli},
  \bibinfo{author}{G.~Modugno}, \bibinfo{author}{M.~Inguscio},
\newblock \bibinfo{title}{{Fermi-Bose Quantum Degenerate
  $^{\mathrm{40}}\mathrm{K}\mathrm{\text{\ensuremath{-}}}^{\mathrm{87}}\mathrm{R}\mathrm{b}$
  Mixture with Attractive Interaction}},
\newblock \bibinfo{journal}{Physical Review Letters} \bibinfo{volume}{89}
  (\bibinfo{year}{2002}) \bibinfo{pages}{150403}. \URLprefix
  \url{https://link.aps.org/doi/10.1103/PhysRevLett.89.150403}.
  \DOIprefix\doi{10.1103/PhysRevLett.89.150403}.
\bibitem[{Lahaye et~al.(2009)Lahaye, Menotti, Santos, Lewenstein, and
  Pfau}]{lahaye2009physics}
\bibinfo{author}{T.~Lahaye}, \bibinfo{author}{C.~Menotti},
  \bibinfo{author}{L.~Santos}, \bibinfo{author}{M.~Lewenstein},
  \bibinfo{author}{T.~Pfau},
\newblock \bibinfo{title}{The physics of dipolar bosonic quantum gases},
\newblock \bibinfo{journal}{Reports on Progress in Physics}
  \bibinfo{volume}{72} (\bibinfo{year}{2009}) \bibinfo{pages}{126401}.
  \URLprefix \url{https://dx.doi.org/10.1088/0034-4885/72/12/126401}.
  \DOIprefix\doi{10.1088/0034-4885/72/12/126401}.
\bibitem[{Neely et~al.(2013)Neely, Bradley, Samson, Rooney, Wright, Law,
  Carretero-Gonz\'alez, Kevrekidis, Davis, and
  Anderson}]{neely2013characteristics}
\bibinfo{author}{T.~W. Neely}, \bibinfo{author}{A.~S. Bradley},
  \bibinfo{author}{E.~C. Samson}, \bibinfo{author}{S.~J. Rooney},
  \bibinfo{author}{E.~M. Wright}, \bibinfo{author}{K.~J.~H. Law},
  \bibinfo{author}{R.~Carretero-Gonz\'alez}, \bibinfo{author}{P.~G.
  Kevrekidis}, \bibinfo{author}{M.~J. Davis}, \bibinfo{author}{B.~P. Anderson},
\newblock \bibinfo{title}{C{haracteristics of Two-Dimensional Quantum
  Turbulence in a Compressible Superfluid}},
\newblock \bibinfo{journal}{Physical Review Letters} \bibinfo{volume}{111}
  (\bibinfo{year}{2013}) \bibinfo{pages}{235301}. \URLprefix
  \url{https://link.aps.org/doi/10.1103/PhysRevLett.111.235301}.
  \DOIprefix\doi{10.1103/PhysRevLett.111.235301}.
\bibitem[{Pandey et~al.(2019)Pandey, Mas, Drougakis, Thekkeppatt, Bolpasi,
  Vasilakis, Poulios, and von Klitzing}]{pandey2019hyper}
\bibinfo{author}{S.~Pandey}, \bibinfo{author}{H.~Mas},
  \bibinfo{author}{G.~Drougakis}, \bibinfo{author}{P.~Thekkeppatt},
  \bibinfo{author}{V.~Bolpasi}, \bibinfo{author}{G.~Vasilakis},
  \bibinfo{author}{K.~Poulios}, \bibinfo{author}{W.~von Klitzing},
\newblock \bibinfo{title}{{Hypersonic Bose{\textendash}Einstein condensates in
  accelerator rings}},
\newblock \bibinfo{journal}{Nature} \bibinfo{volume}{570}
  (\bibinfo{year}{2019}) \bibinfo{pages}{205--209}. \URLprefix
  \url{https://doi.org/10.1038/s41586-019-1273-5}.
  \DOIprefix\doi{10.1038/s41586-019-1273-5}.
\bibitem[{Ögren et~al.(2021)Ögren, Drougakis, Vasilakis, von Klitzing, and
  Kavoulakis}]{wolf2021stationary}
\bibinfo{author}{M.~Ögren}, \bibinfo{author}{G.~Drougakis},
  \bibinfo{author}{G.~Vasilakis}, \bibinfo{author}{W.~von Klitzing},
  \bibinfo{author}{G.~M. Kavoulakis},
\newblock \bibinfo{title}{{Stationary states of Bose{\textendash}Einstein
  condensed atoms rotating in an asymmetric ring potential}},
\newblock \bibinfo{journal}{Journal of Physics B: Atomic, Molecular and Optical
  Physics} \bibinfo{volume}{54} (\bibinfo{year}{2021}) \bibinfo{pages}{145303}.
  \URLprefix \url{https://doi.org/10.1088/1361-6455/ac1647}.
  \DOIprefix\doi{10.1088/1361-6455/ac1647}.
\bibitem[{Pecci et~al.(2021)Pecci, Naldesi, Amico, and
  Minguzzi}]{pecci2021probing}
\bibinfo{author}{G.~Pecci}, \bibinfo{author}{P.~Naldesi},
  \bibinfo{author}{L.~Amico}, \bibinfo{author}{A.~Minguzzi},
\newblock \bibinfo{title}{Probing the {BCS-BEC} crossover with persistent
  currents},
\newblock
  \bibinfo{journal}{\href{https://link.aps.org/doi/10.1103/PhysRevResearch.3.L032064}{Physical
  Review Research}} \bibinfo{volume}{\textbf{3}} (\bibinfo{year}{(2021)})
  \bibinfo{pages}{L032064}. \URLprefix
  \url{https://link.aps.org/doi/10.1103/PhysRevResearch.3.L032064}.
  \DOIprefix\doi{10.1103/PhysRevResearch.3.L032064}.
\bibitem[{Pecci(2023)}]{pecci2022coherence}
\bibinfo{author}{G.~Pecci}, \bibinfo{title}{{Coherence and dynamics of
  ultracold atomic Fermi gases}}, Ph.D. thesis, Université Grenoble Alpes,
  \bibinfo{year}{2023}. \URLprefix
  \url{https://theses.hal.science/tel-04003403/}.
\bibitem[{White et~al.(2016)White, Hennessy, and Busch}]{white20161emergence}
\bibinfo{author}{A.~White}, \bibinfo{author}{T.~Hennessy},
  \bibinfo{author}{T.~Busch},
\newblock \bibinfo{title}{{Emergence of classical rotation in superfluid
  Bose-Einstein condensates}},
\newblock \bibinfo{journal}{Physical Review A} \bibinfo{volume}{93}
  (\bibinfo{year}{2016}) \bibinfo{pages}{033601}. \URLprefix
  \url{https://link.aps.org/doi/10.1103/PhysRevA.93.033601}.
  \DOIprefix\doi{10.1103/PhysRevA.93.033601}.
\bibitem[{Chetcuti et~al.(2022)Chetcuti, Haug, Kwek, and
  Amico}]{chetcuti2021persistent}
\bibinfo{author}{W.~J. Chetcuti}, \bibinfo{author}{T.~Haug},
  \bibinfo{author}{L.-C. Kwek}, \bibinfo{author}{L.~Amico},
\newblock \bibinfo{title}{{Persistent Current of {SU($N$)} Fermions}},
\newblock \bibinfo{journal}{SciPost Physics} \bibinfo{volume}{12}
  (\bibinfo{year}{2022}) \bibinfo{pages}{33}. \URLprefix
  \url{https://scipost.org/10.21468/SciPostPhys.12.1.033}.
  \DOIprefix\doi{10.21468/SciPostPhys.12.1.033}.
\bibitem[{Chetcuti et~al.(2023{\natexlab{a}})Chetcuti, Polo, Osterloh,
  Castorina, and Amico}]{chetcuti2023probe}
\bibinfo{author}{W.~J. Chetcuti}, \bibinfo{author}{J.~Polo},
  \bibinfo{author}{A.~Osterloh}, \bibinfo{author}{P.~Castorina},
  \bibinfo{author}{L.~Amico},
\newblock \bibinfo{title}{Probe for bound states of {SU}(3) fermions and colour
  deconfinement},
\newblock \bibinfo{journal}{Communications Physics} \bibinfo{volume}{6}
  (\bibinfo{year}{2023}{\natexlab{a}}). \URLprefix
  \url{https://doi.org/10.1038/s42005-023-01256-3}.
  \DOIprefix\doi{10.1038/s42005-023-01256-3}.
\bibitem[{Chetcuti et~al.(2023{\natexlab{b}})Chetcuti, Osterloh, Amico, and
  Polo}]{chetcuti2022interference}
\bibinfo{author}{W.~J. Chetcuti}, \bibinfo{author}{A.~Osterloh},
  \bibinfo{author}{L.~Amico}, \bibinfo{author}{J.~Polo},
\newblock \bibinfo{title}{{Interference dynamics of matter-waves of SU($N$)
  fermions}},
\newblock \bibinfo{journal}{SciPost Physics} \bibinfo{volume}{15}
  (\bibinfo{year}{2023}{\natexlab{b}}) \bibinfo{pages}{181}. \URLprefix
  \url{https://scipost.org/10.21468/SciPostPhys.15.4.181}.
  \DOIprefix\doi{10.21468/SciPostPhys.15.4.181}.
\bibitem[{Osterloh et~al.(2023)Osterloh, Polo, Chetcuti, and
  Amico}]{osterloh2023exact}
\bibinfo{author}{A.~Osterloh}, \bibinfo{author}{J.~Polo},
  \bibinfo{author}{W.~J. Chetcuti}, \bibinfo{author}{L.~Amico},
\newblock \bibinfo{title}{{Exact one-particle density matrix for SU(N)
  fermionic matter-waves in the strong repulsive limit}},
\newblock \bibinfo{journal}{SciPost Physics} \bibinfo{volume}{15}
  (\bibinfo{year}{2023}) \bibinfo{pages}{006}. \URLprefix
  \url{https://scipost.org/10.21468/SciPostPhys.15.1.006}.
  \DOIprefix\doi{10.21468/SciPostPhys.15.1.006}.
\bibitem[{Consiglio et~al.(2022)Consiglio, Chetcuti, Bravo-Prieto,
  Ramos-Calderer, Minguzzi, Latorre, Amico, and
  Apollaro}]{consiglio2022variational}
\bibinfo{author}{M.~Consiglio}, \bibinfo{author}{W.~J. Chetcuti},
  \bibinfo{author}{C.~Bravo-Prieto}, \bibinfo{author}{S.~Ramos-Calderer},
  \bibinfo{author}{A.~Minguzzi}, \bibinfo{author}{J.~I. Latorre},
  \bibinfo{author}{L.~Amico}, \bibinfo{author}{T.~J.~G. Apollaro},
\newblock \bibinfo{title}{{Variational quantum eigensolver for SU(N)
  fermions}},
\newblock \bibinfo{journal}{Journal of Physics A: Mathematical and Theoretical}
  \bibinfo{volume}{55} (\bibinfo{year}{2022}) \bibinfo{pages}{265301}.
  \URLprefix \url{https://dx.doi.org/10.1088/1751-8121/ac7016}.
  \DOIprefix\doi{10.1088/1751-8121/ac7016}.
\bibitem[{Chetcuti(2023)}]{chetcuti2023persistent}
\bibinfo{author}{W.~J. Chetcuti}, \bibinfo{title}{{Persistent Currents in
  Atomtronic Circuits of SU($N$) Fermions}}, Ph.D. thesis, University of
  Catania, \bibinfo{year}{2023}. \href{http://arxiv.org/abs/2311.03072}{{\tt
  arXiv:2311.03072}}.
\bibitem[{Pecci et~al.(2023)Pecci, Aupetit-Diallo, Albert, Vignolo, and
  Minguzzi}]{pecci2023persistent}
\bibinfo{author}{G.~Pecci}, \bibinfo{author}{G.~Aupetit-Diallo},
  \bibinfo{author}{M.~Albert}, \bibinfo{author}{P.~Vignolo},
  \bibinfo{author}{A.~Minguzzi},
\newblock \bibinfo{title}{{Persistent currents in a strongly interacting
  multicomponent Bose gas on a ring}},
\newblock \bibinfo{journal}{Comptes Rendus. Physique} \bibinfo{volume}{24}
  (\bibinfo{year}{2023}) \bibinfo{pages}{1--13}. \URLprefix
  \url{https://doi.org/10.5802/crphys.157}. \DOIprefix\doi{10.5802/crphys.157}.
\bibitem[{Ferraretto et~al.(2023)Ferraretto, Richaud, Re, Fallani, and
  Capone}]{ferraretto2023enhancement}
\bibinfo{author}{M.~Ferraretto}, \bibinfo{author}{A.~Richaud},
  \bibinfo{author}{L.~D. Re}, \bibinfo{author}{L.~Fallani},
  \bibinfo{author}{M.~Capone},
\newblock \bibinfo{title}{{Enhancement of chiral edge currents in
  ($d$+1)-dimensional atomic Mott-band hybrid insulators}},
\newblock \bibinfo{journal}{SciPost Physics} \bibinfo{volume}{14}
  (\bibinfo{year}{2023}) \bibinfo{pages}{048}. \URLprefix
  \url{https://scipost.org/10.21468/SciPostPhys.14.3.048}.
  \DOIprefix\doi{10.21468/SciPostPhys.14.3.048}.
\bibitem[{Richaud et~al.(2021)Richaud, Ferraretto, and
  Capone}]{richaud2021interaction}
\bibinfo{author}{A.~Richaud}, \bibinfo{author}{M.~Ferraretto},
  \bibinfo{author}{M.~Capone},
\newblock \bibinfo{title}{{Interaction-resistant metals in multicomponent Fermi
  systems}},
\newblock \bibinfo{journal}{Physical Review B} \bibinfo{volume}{103}
  (\bibinfo{year}{2021}) \bibinfo{pages}{205132}. \URLprefix
  \url{https://link.aps.org/doi/10.1103/PhysRevB.103.205132}.
  \DOIprefix\doi{10.1103/PhysRevB.103.205132}.
\bibitem[{Richaud et~al.(2022)Richaud, Ferraretto, and
  Capone}]{richaud2022mimicking}
\bibinfo{author}{A.~Richaud}, \bibinfo{author}{M.~Ferraretto},
  \bibinfo{author}{M.~Capone},
\newblock \bibinfo{title}{{Mimicking Multiorbital Systems with SU(N) Atoms:
  Hund's Physics and Beyond}},
\newblock \bibinfo{journal}{Condensed Matter} \bibinfo{volume}{7}
  (\bibinfo{year}{2022}). \URLprefix
  \url{https://www.mdpi.com/2410-3896/7/1/18}.
  \DOIprefix\doi{10.3390/condmat7010018}.
\bibitem[{Wu and Zaremba(2013)}]{wu2013mean}
\bibinfo{author}{Z.~Wu}, \bibinfo{author}{E.~Zaremba},
\newblock \bibinfo{title}{{Mean-field yrast spectrum of a two-component Bose
  gas in ring geometry: Persistent currents at higher angular momentum}},
\newblock \bibinfo{journal}{Physical Review A} \bibinfo{volume}{88}
  (\bibinfo{year}{2013}) \bibinfo{pages}{063640}. \URLprefix
  \url{https://link.aps.org/doi/10.1103/PhysRevA.88.063640}.
  \DOIprefix\doi{10.1103/PhysRevA.88.063640}.
\bibitem[{Smyrnakis et~al.(2014)Smyrnakis, Magiropoulos, Efremidis, and
  Kavoulakis}]{smyrnakis2014persistent}
\bibinfo{author}{J.~Smyrnakis}, \bibinfo{author}{M.~Magiropoulos},
  \bibinfo{author}{N.~K. Efremidis}, \bibinfo{author}{G.~M. Kavoulakis},
\newblock \bibinfo{title}{{Persistent currents in a two-component
  Bose–Einstein condensate confined in a ring potential}},
\newblock \bibinfo{journal}{Journal of Physics B: Atomic, Molecular and Optical
  Physics} \bibinfo{volume}{47} (\bibinfo{year}{2014}) \bibinfo{pages}{215302}.
  \URLprefix \url{https://dx.doi.org/10.1088/0953-4075/47/21/215302}.
  \DOIprefix\doi{10.1088/0953-4075/47/21/215302}.
\bibitem[{Polo et~al.(2024)Polo, Chetcuti, Minguzzi, Osterloh, and
  Amico}]{polo2024static}
\bibinfo{author}{J.~Polo}, \bibinfo{author}{W.~J. Chetcuti},
  \bibinfo{author}{A.~Minguzzi}, \bibinfo{author}{A.~Osterloh},
  \bibinfo{author}{L.~Amico}, \bibinfo{title}{{Static impurity in a mesoscopic
  system of SU($N$) fermionic matter-waves}}, \bibinfo{year}{2024}. \URLprefix
  \url{https://arxiv.org/abs/2411.14546}.
  \href{http://arxiv.org/abs/2411.14546}{{\tt arXiv:2411.14546}}.
\bibitem[{Matsushita and De~Passos(2018)}]{matsushita2018mixture}
\bibinfo{author}{E.~Matsushita}, \bibinfo{author}{E.~De~Passos},
\newblock \bibinfo{title}{Mixture of two ultra cold bosonic atoms confined in a
  ring: stability and persistent currents},
\newblock \bibinfo{journal}{Journal of Physics Communications}
  \bibinfo{volume}{2} (\bibinfo{year}{2018}) \bibinfo{pages}{035023}.
\bibitem[{Spehner et~al.(2021)Spehner, Morales-Molina, and
  Reyes}]{spehner2021persistent}
\bibinfo{author}{D.~Spehner}, \bibinfo{author}{L.~Morales-Molina},
  \bibinfo{author}{S.~Reyes},
\newblock \bibinfo{title}{{Persistent currents in Bose--Bose mixtures after an
  interspecies interaction quench}},
\newblock \bibinfo{journal}{New Journal of Physics} \bibinfo{volume}{23}
  (\bibinfo{year}{2021}) \bibinfo{pages}{063025}.
\bibitem[{Brauneis et~al.(2023)Brauneis, Ghazaryan, Hammer, and
  Volosniev}]{brauneis2023emergence}
\bibinfo{author}{F.~Brauneis}, \bibinfo{author}{A.~Ghazaryan},
  \bibinfo{author}{H.-W. Hammer}, \bibinfo{author}{A.~G. Volosniev},
\newblock \bibinfo{title}{{Emergence of a Bose polaron in a small ring threaded
  by the Aharonov-Bohm flux}},
\newblock \bibinfo{journal}{Communications Physics} \bibinfo{volume}{6}
  (\bibinfo{year}{2023}). \URLprefix
  \url{http://dx.doi.org/10.1038/s42005-023-01281-2}.
  \DOIprefix\doi{10.1038/s42005-023-01281-2}.
\bibitem[{Suga et~al.(2014)Suga, Egawa, Masaki, and Mori}]{suga2014persistent}
\bibinfo{author}{S.~Suga}, \bibinfo{author}{T.~Egawa},
  \bibinfo{author}{A.~Masaki}, \bibinfo{author}{H.~Mori},
\newblock \bibinfo{title}{{Persistent Current in Bose--Fermi Mixture on a
  Ring}},
\newblock \bibinfo{journal}{Journal of the Physical Society of Japan}
  \bibinfo{volume}{83} (\bibinfo{year}{2014}) \bibinfo{pages}{024007}.
\bibitem[{Schymik et~al.(2020)Schymik, Lienhard, Barredo, Scholl, Williams,
  Browaeys, and Lahaye}]{schymik2020enhanced}
\bibinfo{author}{K.-N. Schymik}, \bibinfo{author}{V.~Lienhard},
  \bibinfo{author}{D.~Barredo}, \bibinfo{author}{P.~Scholl},
  \bibinfo{author}{H.~Williams}, \bibinfo{author}{A.~Browaeys},
  \bibinfo{author}{T.~Lahaye},
\newblock \bibinfo{title}{Enhanced atom-by-atom assembly of arbitrary tweezer
  arrays},
\newblock \bibinfo{journal}{Physical Review A} \bibinfo{volume}{102}
  (\bibinfo{year}{2020}) \bibinfo{pages}{063107}. \URLprefix
  \url{https://link.aps.org/doi/10.1103/PhysRevA.102.063107}.
  \DOIprefix\doi{10.1103/PhysRevA.102.063107}.
\bibitem[{Barredo et~al.(2015)Barredo, Labuhn, Ravets, Lahaye, Browaeys, and
  Adams}]{barredo2015coherent}
\bibinfo{author}{D.~Barredo}, \bibinfo{author}{H.~Labuhn},
  \bibinfo{author}{S.~Ravets}, \bibinfo{author}{T.~Lahaye},
  \bibinfo{author}{A.~Browaeys}, \bibinfo{author}{C.~S. Adams},
\newblock \bibinfo{title}{{Coherent excitation transfer in a spin chain of
  three Rydberg atoms}},
\newblock \bibinfo{journal}{Physical review letters} \bibinfo{volume}{114}
  (\bibinfo{year}{2015}) \bibinfo{pages}{113002}.
  \DOIprefix\doi{10.1103/PhysRevLett.114.113002}.
\bibitem[{Perciavalle et~al.(2023)Perciavalle, Rossini, Haug, Morsch, and
  Amico}]{perciavalle2022controlled}
\bibinfo{author}{F.~Perciavalle}, \bibinfo{author}{D.~Rossini},
  \bibinfo{author}{T.~Haug}, \bibinfo{author}{O.~Morsch},
  \bibinfo{author}{L.~Amico},
\newblock \bibinfo{title}{Controlled flow of excitations in a ring-shaped
  network of rydberg atoms},
\newblock \bibinfo{journal}{Physical Review A} \bibinfo{volume}{108}
  (\bibinfo{year}{2023}) \bibinfo{pages}{023305}.
\bibitem[{Perciavalle et~al.(2024{\natexlab{a}})Perciavalle, Rossini, Polo,
  Morsch, and Amico}]{perciavalle2024quantum}
\bibinfo{author}{F.~Perciavalle}, \bibinfo{author}{D.~Rossini},
  \bibinfo{author}{J.~Polo}, \bibinfo{author}{O.~Morsch},
  \bibinfo{author}{L.~Amico},
\newblock \bibinfo{title}{{Quantum superpositions of current states in
  Rydberg-atom networks}},
\newblock \bibinfo{journal}{Physical Review Research} \bibinfo{volume}{6}
  (\bibinfo{year}{2024}{\natexlab{a}}) \bibinfo{pages}{043025}. \URLprefix
  \url{https://link.aps.org/doi/10.1103/PhysRevResearch.6.043025}.
  \DOIprefix\doi{10.1103/PhysRevResearch.6.043025}.
\bibitem[{Perciavalle et~al.(2024{\natexlab{b}})Perciavalle, Morsch, Rossini,
  and Amico}]{perciavalle2023coherent}
\bibinfo{author}{F.~Perciavalle}, \bibinfo{author}{O.~Morsch},
  \bibinfo{author}{D.~Rossini}, \bibinfo{author}{L.~Amico},
\newblock \bibinfo{title}{Coherent excitation transport through ring-shaped
  networks},
\newblock \bibinfo{journal}{Physical Review A} \bibinfo{volume}{109}
  (\bibinfo{year}{2024}{\natexlab{b}}) \bibinfo{pages}{062619}. \URLprefix
  \url{https://link.aps.org/doi/10.1103/PhysRevA.109.062619}.
  \DOIprefix\doi{10.1103/PhysRevA.109.062619}.
\bibitem[{Lienhard et~al.(2020)Lienhard, Scholl, Weber, Barredo,
  de~L{\'e}s{\'e}leuc, Bai, Lang, Fleischhauer, B{\"u}chler, Lahaye
  et~al.}]{lienhard2020realization}
\bibinfo{author}{V.~Lienhard}, \bibinfo{author}{P.~Scholl},
  \bibinfo{author}{S.~Weber}, \bibinfo{author}{D.~Barredo},
  \bibinfo{author}{S.~de~L{\'e}s{\'e}leuc}, \bibinfo{author}{R.~Bai},
  \bibinfo{author}{N.~Lang}, \bibinfo{author}{M.~Fleischhauer},
  \bibinfo{author}{H.~P. B{\"u}chler}, \bibinfo{author}{T.~Lahaye}, et~al.,
\newblock \bibinfo{title}{{Realization of a density-dependent Peierls phase in
  a synthetic, spin-orbit coupled Rydberg system}},
\newblock \bibinfo{journal}{Physical Review X} \bibinfo{volume}{10}
  (\bibinfo{year}{2020}) \bibinfo{pages}{021031}.
  \DOIprefix\doi{10.1103/PhysRevX.10.021031}.
\bibitem[{Wu et~al.(2022)Wu, Yang, Yang, M{\o}lmer, Pohl, Tey, and
  You}]{wu2022manipulating}
\bibinfo{author}{X.~Wu}, \bibinfo{author}{F.~Yang}, \bibinfo{author}{S.~Yang},
  \bibinfo{author}{K.~M{\o}lmer}, \bibinfo{author}{T.~Pohl},
  \bibinfo{author}{M.~K. Tey}, \bibinfo{author}{L.~You},
\newblock \bibinfo{title}{{Manipulating synthetic gauge fluxes via multicolor
  dressing of Rydberg-atom arrays}},
\newblock \bibinfo{journal}{Physical Review Research} \bibinfo{volume}{4}
  (\bibinfo{year}{2022}) \bibinfo{pages}{L032046}.
  \DOIprefix\doi{10.1103/PhysRevResearch.4.L032046}.
\bibitem[{Li et~al.(2022)Li, You, Shao, and Li}]{li2022coherent}
\bibinfo{author}{X.~X. Li}, \bibinfo{author}{J.~B. You}, \bibinfo{author}{X.~Q.
  Shao}, \bibinfo{author}{W.~Li},
\newblock \bibinfo{title}{Coherent ground-state transport of neutral atoms},
\newblock \bibinfo{journal}{Physical Review A} \bibinfo{volume}{105}
  (\bibinfo{year}{2022}) \bibinfo{pages}{032417}. \URLprefix
  \url{https://link.aps.org/doi/10.1103/PhysRevA.105.032417}.
  \DOIprefix\doi{10.1103/PhysRevA.105.032417}.
\bibitem[{Bornet et~al.(2024)Bornet, Emperauger, Chen, Machado, Chern, Leclerc,
  G\'ely, Chew, Barredo, Lahaye, Yao, and Browaeys}]{bornet2024enhancing}
\bibinfo{author}{G.~Bornet}, \bibinfo{author}{G.~Emperauger},
  \bibinfo{author}{C.~Chen}, \bibinfo{author}{F.~Machado},
  \bibinfo{author}{S.~Chern}, \bibinfo{author}{L.~Leclerc},
  \bibinfo{author}{B.~G\'ely}, \bibinfo{author}{Y.~T. Chew},
  \bibinfo{author}{D.~Barredo}, \bibinfo{author}{T.~Lahaye},
  \bibinfo{author}{N.~Y. Yao}, \bibinfo{author}{A.~Browaeys},
\newblock \bibinfo{title}{{Enhancing a Many-Body Dipolar Rydberg Tweezer Array
  with Arbitrary Local Controls}},
\newblock \bibinfo{journal}{Physical Review Letters} \bibinfo{volume}{132}
  (\bibinfo{year}{2024}) \bibinfo{pages}{263601}. \URLprefix
  \url{https://link.aps.org/doi/10.1103/PhysRevLett.132.263601}.
  \DOIprefix\doi{10.1103/PhysRevLett.132.263601}.
\bibitem[{Han and Yi(2024)}]{han2024tuning}
\bibinfo{author}{Y.~Han}, \bibinfo{author}{W.~Yi},
\newblock \bibinfo{title}{{Tuning Excitation Transport in a Dissipative Rydberg
  Ring}},
\newblock \bibinfo{journal}{Chinese Physics Letters} \bibinfo{volume}{41}
  (\bibinfo{year}{2024}) \bibinfo{pages}{033201}. \URLprefix
  \url{https://dx.doi.org/10.1088/0256-307X/41/3/033201}.
  \DOIprefix\doi{10.1088/0256-307X/41/3/033201}.
\bibitem[{Polo et~al.(2018)Polo, Ahufinger, Hekking, and
  Minguzzi}]{polo2018damping}
\bibinfo{author}{J.~Polo}, \bibinfo{author}{V.~Ahufinger},
  \bibinfo{author}{F.~W.~J. Hekking}, \bibinfo{author}{A.~Minguzzi},
\newblock \bibinfo{title}{{Damping of Josephson Oscillations in Strongly
  Correlated One-Dimensional Atomic Gases}},
\newblock \bibinfo{journal}{Physical Review Letters} \bibinfo{volume}{121}
  (\bibinfo{year}{2018}) \bibinfo{pages}{090404}. \URLprefix
  \url{https://link.aps.org/doi/10.1103/PhysRevLett.121.090404}.
  \DOIprefix\doi{10.1103/PhysRevLett.121.090404}.
\bibitem[{Polo et~al.(2019)Polo, Dubessy, Pedri, Perrin, and
  Minguzzi}]{polo2019oscillations}
\bibinfo{author}{J.~Polo}, \bibinfo{author}{R.~Dubessy},
  \bibinfo{author}{P.~Pedri}, \bibinfo{author}{H.~Perrin},
  \bibinfo{author}{A.~Minguzzi},
\newblock \bibinfo{title}{{Oscillations and Decay of Superfluid Currents in a
  One-Dimensional Bose Gas on a Ring}},
\newblock \bibinfo{journal}{Physical Review Letters} \bibinfo{volume}{123}
  (\bibinfo{year}{2019}) \bibinfo{pages}{195301}. \URLprefix
  \url{https://link.aps.org/doi/10.1103/PhysRevLett.123.195301}.
  \DOIprefix\doi{10.1103/PhysRevLett.123.195301}.
\bibitem[{Kohn et~al.(2020)Kohn, Silvi, Gerster, Keck, Fazio, Santoro, and
  Montangero}]{kohn2020superfluid}
\bibinfo{author}{L.~Kohn}, \bibinfo{author}{P.~Silvi},
  \bibinfo{author}{M.~Gerster}, \bibinfo{author}{M.~Keck},
  \bibinfo{author}{R.~Fazio}, \bibinfo{author}{G.~E. Santoro},
  \bibinfo{author}{S.~Montangero},
\newblock \bibinfo{title}{{Superfluid-to-Mott transition in a Bose-Hubbard
  ring: Persistent currents and defect formation}},
\newblock \bibinfo{journal}{Physical Review A} \bibinfo{volume}{101}
  (\bibinfo{year}{2020}) \bibinfo{pages}{023617}.
\bibitem[{Domanti et~al.(2024)Domanti, Castorina, Zappal\`a, and
  Amico}]{domanti2023coherence}
\bibinfo{author}{E.~C. Domanti}, \bibinfo{author}{P.~Castorina},
  \bibinfo{author}{D.~Zappal\`a}, \bibinfo{author}{L.~Amico},
\newblock \bibinfo{title}{{Aharonov-Bohm effect for confined matter in lattice
  gauge theories}},
\newblock \bibinfo{journal}{Physical Review Research} \bibinfo{volume}{6}
  (\bibinfo{year}{2024}) \bibinfo{pages}{013268}. \URLprefix
  \url{https://link.aps.org/doi/10.1103/PhysRevResearch.6.013268}.
  \DOIprefix\doi{10.1103/PhysRevResearch.6.013268}.
\bibitem[{Tengstrand et~al.(2021)Tengstrand, Boholm, Sachdeva, Bengtsson, and
  Reimann}]{tengstrand2021persistent}
\bibinfo{author}{M.~N. Tengstrand}, \bibinfo{author}{D.~Boholm},
  \bibinfo{author}{R.~Sachdeva}, \bibinfo{author}{J.~Bengtsson},
  \bibinfo{author}{S.~M. Reimann},
\newblock \bibinfo{title}{Persistent currents in toroidal dipolar supersolids},
\newblock \bibinfo{journal}{Physical Review A} \bibinfo{volume}{103}
  (\bibinfo{year}{2021}) \bibinfo{pages}{013313}. \URLprefix
  \url{https://link.aps.org/doi/10.1103/PhysRevA.103.013313}.
  \DOIprefix\doi{10.1103/PhysRevA.103.013313}.
\bibitem[{P\'erez-Obiol et~al.(2022)P\'erez-Obiol, Polo, and
  Amico}]{perezobiol2022coherent}
\bibinfo{author}{A.~P\'erez-Obiol}, \bibinfo{author}{J.~Polo},
  \bibinfo{author}{L.~Amico},
\newblock \bibinfo{title}{Coherent phase slips in coupled matter-wave
  circuits},
\newblock \bibinfo{journal}{Physical Review Research} \bibinfo{volume}{4}
  (\bibinfo{year}{2022}) \bibinfo{pages}{L022038}. \URLprefix
  \url{https://link.aps.org/doi/10.1103/PhysRevResearch.4.L022038}.
  \DOIprefix\doi{10.1103/PhysRevResearch.4.L022038}.
\bibitem[{Haug et~al.(2019{\natexlab{a}})Haug, Heimonen, Dumke, Kwek, and
  Amico}]{haug2019aharonov}
\bibinfo{author}{T.~Haug}, \bibinfo{author}{H.~Heimonen},
  \bibinfo{author}{R.~Dumke}, \bibinfo{author}{L.-C. Kwek},
  \bibinfo{author}{L.~Amico},
\newblock \bibinfo{title}{{Aharonov-Bohm effect in mesoscopic Bose-Einstein
  condensates}},
\newblock \bibinfo{journal}{Physical Review A} \bibinfo{volume}{100}
  (\bibinfo{year}{2019}{\natexlab{a}}) \bibinfo{pages}{041601}.
  \DOIprefix\doi{10.1103/PhysRevA.100.041601}.
\bibitem[{Haug et~al.(2019{\natexlab{b}})Haug, Dumke, Kwek, and
  Amico}]{haug2019andreev}
\bibinfo{author}{T.~Haug}, \bibinfo{author}{R.~Dumke}, \bibinfo{author}{L.-C.
  Kwek}, \bibinfo{author}{L.~Amico},
\newblock \bibinfo{title}{{Andreev-reflection and Aharonov--Bohm dynamics in
  atomtronic circuits}},
\newblock \bibinfo{journal}{Quantum Science and Technology} \bibinfo{volume}{4}
  (\bibinfo{year}{2019}{\natexlab{b}}) \bibinfo{pages}{045001}.
  \DOIprefix\doi{10.1088/2058-9565/ab2e61}.
\bibitem[{Haug et~al.(2018)Haug, Amico, Dumke, and Kwek}]{haug2018mesoscopic}
\bibinfo{author}{T.~Haug}, \bibinfo{author}{L.~Amico},
  \bibinfo{author}{R.~Dumke}, \bibinfo{author}{L.-C. Kwek},
\newblock \bibinfo{title}{Mesoscopic vortex–meissner currents in ring
  ladders},
\newblock \bibinfo{journal}{Quantum Science and Technology} \bibinfo{volume}{3}
  (\bibinfo{year}{2018}) \bibinfo{pages}{035006}. \URLprefix
  \url{https://dx.doi.org/10.1088/2058-9565/aaa8c6}.
  \DOIprefix\doi{10.1088/2058-9565/aaa8c6}.
\bibitem[{Cherny et~al.(2009)Cherny, Caux, and Brand}]{cherny2009decay}
\bibinfo{author}{A.~Y. Cherny}, \bibinfo{author}{J.-S. Caux},
  \bibinfo{author}{J.~Brand},
\newblock \bibinfo{title}{{Decay of superfluid currents in the interacting
  one-dimensional Bose gas}},
\newblock \bibinfo{journal}{Physical Review A} \bibinfo{volume}{80}
  (\bibinfo{year}{2009}) \bibinfo{pages}{043604}. \URLprefix
  \url{https://link.aps.org/doi/10.1103/PhysRevA.80.043604}.
  \DOIprefix\doi{10.1103/PhysRevA.80.043604}.
\bibitem[{Naldesi et~al.(2019)Naldesi, Gomez, Malomed, Olshanii, Minguzzi, and
  Amico}]{naldesi2019rise}
\bibinfo{author}{P.~Naldesi}, \bibinfo{author}{J.~P. Gomez},
  \bibinfo{author}{B.~Malomed}, \bibinfo{author}{M.~Olshanii},
  \bibinfo{author}{A.~Minguzzi}, \bibinfo{author}{L.~Amico},
\newblock \bibinfo{title}{{Rise and Fall of a Bright Soliton in an Optical
  Lattice}},
\newblock
  \bibinfo{journal}{\href{https://link.aps.org/doi/10.1103/PhysRevLett.122.053001}{Physical
  Review Letters}} \bibinfo{volume}{\textbf{122}} (\bibinfo{year}{(2019)})
  \bibinfo{pages}{053001}. \URLprefix
  \url{https://link.aps.org/doi/10.1103/PhysRevLett.122.053001}.
  \DOIprefix\doi{10.1103/PhysRevLett.122.053001}.
\bibitem[{Naldesi et~al.(2022)Naldesi, Polo, Dunjko, Perrin, Olshanii, Amico,
  and Minguzzi}]{naldesi2022enhancing}
\bibinfo{author}{P.~Naldesi}, \bibinfo{author}{J.~Polo},
  \bibinfo{author}{V.~Dunjko}, \bibinfo{author}{H.~Perrin},
  \bibinfo{author}{M.~Olshanii}, \bibinfo{author}{L.~Amico},
  \bibinfo{author}{A.~Minguzzi},
\newblock \bibinfo{title}{{Enhancing sensitivity to rotations with quantum
  solitonic currents}},
\newblock \bibinfo{journal}{SciPost Physics} \bibinfo{volume}{12}
  (\bibinfo{year}{2022}) \bibinfo{pages}{138}. \URLprefix
  \url{https://scipost.org/10.21468/SciPostPhys.12.4.138}.
  \DOIprefix\doi{10.21468/SciPostPhys.12.4.138}.
\bibitem[{Cominotti et~al.(2014)Cominotti, Rossini, Rizzi, Hekking, and
  Minguzzi}]{cominotti2014optimal}
\bibinfo{author}{M.~Cominotti}, \bibinfo{author}{D.~Rossini},
  \bibinfo{author}{M.~Rizzi}, \bibinfo{author}{F.~Hekking},
  \bibinfo{author}{A.~Minguzzi},
\newblock \bibinfo{title}{{Optimal Persistent Currents for Interacting Bosons
  on a Ring with a Gauge Field}},
\newblock \bibinfo{journal}{Physical Review Letters} \bibinfo{volume}{113}
  (\bibinfo{year}{2014}) \bibinfo{pages}{025301}. \URLprefix
  \url{https://link.aps.org/doi/10.1103/PhysRevLett.113.025301}.
  \DOIprefix\doi{10.1103/PhysRevLett.113.025301}.
\bibitem[{Aghamalyan et~al.(2015)Aghamalyan, Cominotti, Rizzi, Rossini,
  Hekking, Minguzzi, Kwek, and Amico}]{aghamalyan2015coherent}
\bibinfo{author}{D.~Aghamalyan}, \bibinfo{author}{M.~Cominotti},
  \bibinfo{author}{M.~Rizzi}, \bibinfo{author}{D.~Rossini},
  \bibinfo{author}{F.~Hekking}, \bibinfo{author}{A.~Minguzzi},
  \bibinfo{author}{L.-C. Kwek}, \bibinfo{author}{L.~Amico},
\newblock \bibinfo{title}{Coherent superposition of current flows in an
  atomtronic quantum interference device},
\newblock \bibinfo{journal}{New Journal of Physics} \bibinfo{volume}{17}
  (\bibinfo{year}{2015}) \bibinfo{pages}{045023}. \URLprefix
  \url{https://doi.org/10.1088/1367-2630/17/4/045023}.
  \DOIprefix\doi{10.1088/1367-2630/17/4/045023}.
\bibitem[{Haug et~al.(2018)Haug, Tan, Theng, Dumke, Kwek, and
  Amico}]{haug2018readout}
\bibinfo{author}{T.~Haug}, \bibinfo{author}{J.~Tan},
  \bibinfo{author}{M.~Theng}, \bibinfo{author}{R.~Dumke},
  \bibinfo{author}{L.-C. Kwek}, \bibinfo{author}{L.~Amico},
\newblock \bibinfo{title}{Readout of the atomtronic quantum interference
  device},
\newblock \bibinfo{journal}{Physical Review A} \bibinfo{volume}{97}
  (\bibinfo{year}{2018}) \bibinfo{pages}{013633}. \URLprefix
  \url{https://link.aps.org/doi/10.1103/PhysRevA.97.013633}.
  \DOIprefix\doi{10.1103/PhysRevA.97.013633}.
\bibitem[{Polo et~al.(2021)Polo, Naldesi, Minguzzi, and
  Amico}]{polo2022quantum}
\bibinfo{author}{J.~Polo}, \bibinfo{author}{P.~Naldesi},
  \bibinfo{author}{A.~Minguzzi}, \bibinfo{author}{L.~Amico},
\newblock \bibinfo{title}{The quantum solitons atomtronic interference device},
\newblock \bibinfo{journal}{Quantum Science and Technology} \bibinfo{volume}{7}
  (\bibinfo{year}{2021}) \bibinfo{pages}{015015}. \URLprefix
  \url{https://dx.doi.org/10.1088/2058-9565/ac39f6}.
  \DOIprefix\doi{10.1088/2058-9565/ac39f6}.
\bibitem[{Fisher and Barber(1972)}]{fisher1972scaling}
\bibinfo{author}{M.~E. Fisher}, \bibinfo{author}{M.~N. Barber},
\newblock \bibinfo{title}{Scaling theory for finite-size effects in the
  critical region},
\newblock \bibinfo{journal}{Physical Review Letters} \bibinfo{volume}{28}
  (\bibinfo{year}{1972}) \bibinfo{pages}{1516}.
\bibitem[{Amico et~al.(2017)Amico, Birkl, Boshier, and Kwek}]{amico2017focus}
\bibinfo{author}{L.~Amico}, \bibinfo{author}{G.~Birkl},
  \bibinfo{author}{M.~Boshier}, \bibinfo{author}{L.-C. Kwek},
\newblock \bibinfo{title}{Focus on atomtronics-enabled quantum technologies},
\newblock \bibinfo{journal}{New Journal of Physics} \bibinfo{volume}{19}
  (\bibinfo{year}{2017}) \bibinfo{pages}{020201}. \URLprefix
  \url{https://dx.doi.org/10.1088/1367-2630/aa5a6d}.
  \DOIprefix\doi{10.1088/1367-2630/aa5a6d}.
\bibitem[{Amico et~al.(2021)Amico, Boshier, Birkl, Minguzzi, Miniatura, Kwek,
  Aghamalyan, Ahufinger, Anderson, Andrei et~al.}]{amico2021roadmap}
\bibinfo{author}{L.~Amico}, \bibinfo{author}{M.~Boshier},
  \bibinfo{author}{G.~Birkl}, \bibinfo{author}{A.~Minguzzi},
  \bibinfo{author}{C.~Miniatura}, \bibinfo{author}{L.~C. Kwek},
  \bibinfo{author}{D.~Aghamalyan}, \bibinfo{author}{V.~Ahufinger},
  \bibinfo{author}{D.~Anderson}, \bibinfo{author}{N.~Andrei}, et~al.,
\newblock \bibinfo{title}{{Roadmap on Atomtronics: State of the art and
  perspective}},
\newblock \bibinfo{journal}{AVS Quantum Science} \bibinfo{volume}{3}
  (\bibinfo{year}{2021}) \bibinfo{pages}{039201}. \URLprefix
  \url{https://doi.org/10.1116/5.0026178}. \DOIprefix\doi{10.1116/5.0026178}.
\bibitem[{Amico et~al.(2022)Amico, Anderson, Boshier, Brantut, Kwek, Minguzzi,
  and von Klitzing}]{amico2022atomtronic}
\bibinfo{author}{L.~Amico}, \bibinfo{author}{D.~Anderson},
  \bibinfo{author}{M.~Boshier}, \bibinfo{author}{J.-P. Brantut},
  \bibinfo{author}{L.-C. Kwek}, \bibinfo{author}{A.~Minguzzi},
  \bibinfo{author}{W.~von Klitzing},
\newblock \bibinfo{title}{{Colloquium: Atomtronic circuits: From many-body
  physics to quantum technologies}},
\newblock \bibinfo{journal}{Reviews of Modern Physics} \bibinfo{volume}{94}
  (\bibinfo{year}{2022}) \bibinfo{pages}{041001}. \URLprefix
  \url{https://link.aps.org/doi/10.1103/RevModPhys.94.041001}.
  \DOIprefix\doi{10.1103/RevModPhys.94.041001}.
\bibitem[{Pepino(2021)}]{pepino2021entropy}
\bibinfo{author}{R.~A. Pepino},
\newblock \bibinfo{title}{{Advances in Atomtronics}},
\newblock \bibinfo{journal}{Entropy} \bibinfo{volume}{23}
  (\bibinfo{year}{2021}). \URLprefix
  \url{https://www.mdpi.com/1099-4300/23/5/534}.
  \DOIprefix\doi{10.3390/e23050534}.
\bibitem[{Polo et~al.(2024)Polo, Chetcuti, Domanti, Kitson, Osterloh,
  Perciavalle, Singh, and Amico}]{polo2024perspective}
\bibinfo{author}{J.~Polo}, \bibinfo{author}{W.~J. Chetcuti},
  \bibinfo{author}{E.~C. Domanti}, \bibinfo{author}{P.~Kitson},
  \bibinfo{author}{A.~Osterloh}, \bibinfo{author}{F.~Perciavalle},
  \bibinfo{author}{V.~P. Singh}, \bibinfo{author}{L.~Amico},
\newblock \bibinfo{title}{Perspective on new implementations of atomtronic
  circuits},
\newblock \bibinfo{journal}{Quantum Science and Technology} \bibinfo{volume}{9}
  (\bibinfo{year}{2024}) \bibinfo{pages}{030501}. \URLprefix
  \url{https://dx.doi.org/10.1088/2058-9565/ad48b2}.
  \DOIprefix\doi{10.1088/2058-9565/ad48b2}.
\bibitem[{Tinkham(2004)}]{tinkham2004introduction}
\bibinfo{author}{M.~Tinkham}, \bibinfo{title}{Introduction to
  superconductivity}, \bibinfo{publisher}{Courier Corporation},
  \bibinfo{year}{2004}.
\bibitem[{Ryu et~al.(2013)Ryu, Blackburn, Blinova, and
  Boshier}]{ryu2013experimental}
\bibinfo{author}{C.~Ryu}, \bibinfo{author}{P.~W. Blackburn},
  \bibinfo{author}{A.~A. Blinova}, \bibinfo{author}{M.~G. Boshier},
\newblock \bibinfo{title}{{Experimental Realization of Josephson Junctions for
  an Atom SQUID}},
\newblock \bibinfo{journal}{Physical Review Letters} \bibinfo{volume}{111}
  (\bibinfo{year}{2013}) \bibinfo{pages}{205301}. \URLprefix
  \url{https://link.aps.org/doi/10.1103/PhysRevLett.111.205301}.
  \DOIprefix\doi{10.1103/PhysRevLett.111.205301}.
\bibitem[{Ryu et~al.(2020)Ryu, Samson, and Boshier}]{ryu2020quantum}
\bibinfo{author}{C.~Ryu}, \bibinfo{author}{E.~C. Samson},
  \bibinfo{author}{M.~G. Boshier},
\newblock \bibinfo{title}{Quantum interference of currents in an atomtronic
  {SQUID}},
\newblock \bibinfo{journal}{Nature Communications} \bibinfo{volume}{11}
  (\bibinfo{year}{2020}). \DOIprefix\doi{10.1038/s41467-020-17185-6}.
\bibitem[{Eckel et~al.(2014)Eckel, Jendrzejewski, Kumar, Lobb, and
  Campbell}]{eckel2014interferometric}
\bibinfo{author}{S.~Eckel}, \bibinfo{author}{F.~Jendrzejewski},
  \bibinfo{author}{A.~Kumar}, \bibinfo{author}{C.~J. Lobb},
  \bibinfo{author}{G.~K. Campbell},
\newblock \bibinfo{title}{{Interferometric Measurement of the Current-Phase
  Relationship of a Superfluid Weak Link}},
\newblock \bibinfo{journal}{Physical Review X} \bibinfo{volume}{4}
  (\bibinfo{year}{2014}) \bibinfo{pages}{031052}. \URLprefix
  \url{https://link.aps.org/doi/10.1103/PhysRevX.4.031052}.
  \DOIprefix\doi{10.1103/PhysRevX.4.031052}.
\bibitem[{Leggett(1987)}]{leggett1987macroscopic}
\bibinfo{author}{A.~J. Leggett},
\newblock \bibinfo{title}{Macroscopic quantum tunnelling and related matters},
\newblock \bibinfo{journal}{Japanese Journal of Applied Physics}
  \bibinfo{volume}{26} (\bibinfo{year}{1987}) \bibinfo{pages}{1986}.
\bibitem[{Hallwood et~al.(2006)Hallwood, Burnett, and
  Dunningham}]{hallwood2006macroscopic}
\bibinfo{author}{D.~W. Hallwood}, \bibinfo{author}{K.~Burnett},
  \bibinfo{author}{J.~Dunningham},
\newblock \bibinfo{title}{Macroscopic superpositions of superfluid flows},
\newblock \bibinfo{journal}{New Journal of Physics} \bibinfo{volume}{8}
  (\bibinfo{year}{2006}) \bibinfo{pages}{180}. \URLprefix
  \url{https://iopscience.iop.org/article/10.1088/1367-2630/8/9/180/pdf}.
  \DOIprefix\doi{10.1088/1367-2630/8/9/180}.
\bibitem[{Hallwood et~al.(2010)Hallwood, Ernst, and Brand}]{hallwood2010robust}
\bibinfo{author}{D.~W. Hallwood}, \bibinfo{author}{T.~Ernst},
  \bibinfo{author}{J.~Brand},
\newblock \bibinfo{title}{Robust mesoscopic superposition of strongly
  correlated ultracold atoms},
\newblock \bibinfo{journal}{Physical Review A} \bibinfo{volume}{82}
  (\bibinfo{year}{2010}) \bibinfo{pages}{063623}. \URLprefix
  \url{https://link.aps.org/doi/10.1103/PhysRevA.82.063623}.
  \DOIprefix\doi{10.1103/PhysRevA.82.063623}.
\bibitem[{Amico et~al.(2014)Amico, Aghamalyan, Auksztol, Crepaz, Dumke, and
  Kwek}]{amico2014superfluid}
\bibinfo{author}{L.~Amico}, \bibinfo{author}{D.~Aghamalyan},
  \bibinfo{author}{F.~Auksztol}, \bibinfo{author}{H.~Crepaz},
  \bibinfo{author}{R.~Dumke}, \bibinfo{author}{L.~C. Kwek},
\newblock \bibinfo{title}{Superfluid qubit systems with ring shaped optical
  lattices},
\newblock \bibinfo{journal}{Scientific reports} \bibinfo{volume}{4}
  (\bibinfo{year}{2014}) \bibinfo{pages}{4298}.
\bibitem[{Solenov and Mozyrsky(2010)}]{solenov2010macroscopic}
\bibinfo{author}{D.~Solenov}, \bibinfo{author}{D.~Mozyrsky},
\newblock \bibinfo{title}{Macroscopic two-state systems in trapped atomic
  condensates},
\newblock \bibinfo{journal}{Physical Review A—Atomic, Molecular, and Optical
  Physics} \bibinfo{volume}{82} (\bibinfo{year}{2010}) \bibinfo{pages}{061601}.
\bibitem[{Nunnenkamp et~al.(2011)Nunnenkamp, Rey, and
  Burnett}]{nunnenkamp2011superposition}
\bibinfo{author}{A.~Nunnenkamp}, \bibinfo{author}{A.~M. Rey},
  \bibinfo{author}{K.~Burnett},
\newblock \bibinfo{title}{Superposition states of ultracold bosons in rotating
  rings with a realistic potential barrier},
\newblock \bibinfo{journal}{Physical Review A} \bibinfo{volume}{84}
  (\bibinfo{year}{2011}) \bibinfo{pages}{053604}. \URLprefix
  \url{https://link.aps.org/doi/10.1103/PhysRevA.84.053604}.
  \DOIprefix\doi{10.1103/PhysRevA.84.053604}.
\bibitem[{Schenke et~al.(2011)Schenke, Minguzzi, and
  Hekking}]{schenke2011nonadiabatic}
\bibinfo{author}{C.~Schenke}, \bibinfo{author}{A.~Minguzzi},
  \bibinfo{author}{F.~Hekking},
\newblock \bibinfo{title}{Nonadiabatic creation of macroscopic superpositions
  with strongly correlated one-dimensional bosons in a ring trap},
\newblock \bibinfo{journal}{Physical Review A—Atomic, Molecular, and Optical
  Physics} \bibinfo{volume}{84} (\bibinfo{year}{2011}) \bibinfo{pages}{053636}.
\bibitem[{Aghamalyan et~al.(2016)Aghamalyan, Nguyen, Auksztol, Gan, Valado,
  Condylis, Kwek, Dumke, and Amico}]{aghamalyan2016atomtronic}
\bibinfo{author}{D.~Aghamalyan}, \bibinfo{author}{N.~T. Nguyen},
  \bibinfo{author}{F.~Auksztol}, \bibinfo{author}{K.~S. Gan},
  \bibinfo{author}{M.~M. Valado}, \bibinfo{author}{P.~C. Condylis},
  \bibinfo{author}{L.-C. Kwek}, \bibinfo{author}{R.~Dumke},
  \bibinfo{author}{L.~Amico},
\newblock \bibinfo{title}{{An atomtronic flux qubit: a ring lattice of
  Bose–Einstein condensates interrupted by three weak links}},
\newblock \bibinfo{journal}{New Journal of Physics} \bibinfo{volume}{18}
  (\bibinfo{year}{2016}) \bibinfo{pages}{075013}. \URLprefix
  \url{https://dx.doi.org/10.1088/1367-2630/18/7/075013}.
  \DOIprefix\doi{10.1088/1367-2630/18/7/075013}.
\bibitem[{Degen et~al.(2017)Degen, Reinhard, and Cappellaro}]{degen2017quantum}
\bibinfo{author}{C.~L. Degen}, \bibinfo{author}{F.~Reinhard},
  \bibinfo{author}{P.~Cappellaro},
\newblock \bibinfo{title}{Quantum sensing},
\newblock \bibinfo{journal}{Reviews of modern physics} \bibinfo{volume}{89}
  (\bibinfo{year}{2017}) \bibinfo{pages}{035002}.
\bibitem[{Barrett et~al.(2014)Barrett, Geiger, Dutta, Meunier, Canuel, Gauguet,
  Bouyer, and Landragin}]{barrett2014sagnac}
\bibinfo{author}{B.~Barrett}, \bibinfo{author}{R.~Geiger},
  \bibinfo{author}{I.~Dutta}, \bibinfo{author}{M.~Meunier},
  \bibinfo{author}{B.~Canuel}, \bibinfo{author}{A.~Gauguet},
  \bibinfo{author}{P.~Bouyer}, \bibinfo{author}{A.~Landragin},
\newblock \bibinfo{title}{The sagnac effect: 20 years of development in
  matter-wave interferometry},
\newblock \bibinfo{journal}{Comptes Rendus Physique} \bibinfo{volume}{15}
  (\bibinfo{year}{2014}) \bibinfo{pages}{875--883}. \URLprefix
  \url{https://www.sciencedirect.com/science/article/pii/S1631070514001467}.
  \DOIprefix\doi{https://doi.org/10.1016/j.crhy.2014.10.009},
  \bibinfo{note}{the Sagnac effect: 100 years later / L'effet Sagnac : 100 ans
  après}.
\bibitem[{Olson et~al.(2007)Olson, Terraciano, Bashkansky, and
  Fatemi}]{olson2007cold-atom}
\bibinfo{author}{S.~E. Olson}, \bibinfo{author}{M.~L. Terraciano},
  \bibinfo{author}{M.~Bashkansky}, \bibinfo{author}{F.~K. Fatemi},
\newblock \bibinfo{title}{Cold-atom confinement in an all-optical dark ring
  trap},
\newblock \bibinfo{journal}{Physical Review A} \bibinfo{volume}{76}
  (\bibinfo{year}{2007}) \bibinfo{pages}{061404}. \URLprefix
  \url{https://link.aps.org/doi/10.1103/PhysRevA.76.061404}.
  \DOIprefix\doi{10.1103/PhysRevA.76.061404}.
\bibitem[{Helm et~al.(2015)Helm, Cornish, and Gardiner}]{helm2015sagnac}
\bibinfo{author}{J.~L. Helm}, \bibinfo{author}{S.~L. Cornish},
  \bibinfo{author}{S.~A. Gardiner},
\newblock \bibinfo{title}{{Sagnac Interferometry Using Bright Matter-Wave
  Solitons}},
\newblock \bibinfo{journal}{Physical Review Letters} \bibinfo{volume}{114}
  (\bibinfo{year}{2015}) \bibinfo{pages}{134101}. \URLprefix
  \url{https://link.aps.org/doi/10.1103/PhysRevLett.114.134101}.
  \DOIprefix\doi{10.1103/PhysRevLett.114.134101}.
\bibitem[{Kim et~al.(2022)Kim, Krzyzanowska, Henderson, Ryu, Timmermans, and
  Boshier}]{kim2022one}
\bibinfo{author}{H.~Kim}, \bibinfo{author}{K.~Krzyzanowska},
  \bibinfo{author}{K.~C. Henderson}, \bibinfo{author}{C.~Ryu},
  \bibinfo{author}{E.~Timmermans}, \bibinfo{author}{M.~Boshier},
  \bibinfo{title}{One second interrogation time in a 200 round-trip waveguide
  atom interferometer}, \bibinfo{year}{2022}. \URLprefix
  \url{https://arxiv.org/abs/2201.11888}.
  \DOIprefix\doi{10.48550/ARXIV.2201.11888}.
\bibitem[{Gustavson et~al.(1997)Gustavson, Bouyer, and
  Kasevich}]{kasevich1997precision}
\bibinfo{author}{T.~L. Gustavson}, \bibinfo{author}{P.~Bouyer},
  \bibinfo{author}{M.~A. Kasevich},
\newblock \bibinfo{title}{{Precision Rotation Measurements with an Atom
  Interferometer Gyroscope}},
\newblock \bibinfo{journal}{Physical Review Letters} \bibinfo{volume}{78}
  (\bibinfo{year}{1997}) \bibinfo{pages}{2046--2049}. \URLprefix
  \url{https://link.aps.org/doi/10.1103/PhysRevLett.78.2046}.
  \DOIprefix\doi{10.1103/PhysRevLett.78.2046}.
\bibitem[{Krzyzanowska et~al.(2023)Krzyzanowska, Ferreras, Ryu, Samson, and
  Boshier}]{krzyzanowska2022matter}
\bibinfo{author}{K.~A. Krzyzanowska}, \bibinfo{author}{J.~Ferreras},
  \bibinfo{author}{C.~Ryu}, \bibinfo{author}{E.~C. Samson},
  \bibinfo{author}{M.~G. Boshier},
\newblock \bibinfo{title}{Matter-wave analog of a fiber-optic gyroscope},
\newblock \bibinfo{journal}{Physical Review A} \bibinfo{volume}{108}
  (\bibinfo{year}{2023}) \bibinfo{pages}{043305}. \URLprefix
  \url{https://link.aps.org/doi/10.1103/PhysRevA.108.043305}.
  \DOIprefix\doi{10.1103/PhysRevA.108.043305}.
\bibitem[{Burke and Sackett(2009)}]{burke2009scalable}
\bibinfo{author}{J.~Burke}, \bibinfo{author}{C.~Sackett},
\newblock \bibinfo{title}{{Scalable Bose-Einstein-condensate Sagnac
  interferometer in a linear trap}},
\newblock \bibinfo{journal}{Physical Review A—Atomic, Molecular, and Optical
  Physics} \bibinfo{volume}{80} (\bibinfo{year}{2009}) \bibinfo{pages}{061603}.
\bibitem[{Moan et~al.(2020)Moan, Horne, Arpornthip, Luo, Fallon, Berl, and
  Sackett}]{moan2020quantum}
\bibinfo{author}{E.~R. Moan}, \bibinfo{author}{R.~A. Horne},
  \bibinfo{author}{T.~Arpornthip}, \bibinfo{author}{Z.~Luo},
  \bibinfo{author}{A.~J. Fallon}, \bibinfo{author}{S.~J. Berl},
  \bibinfo{author}{C.~A. Sackett},
\newblock \bibinfo{title}{{Quantum Rotation Sensing with Dual Sagnac
  Interferometers in an Atom-Optical Waveguide}},
\newblock \bibinfo{journal}{Physical Review Letters} \bibinfo{volume}{124}
  (\bibinfo{year}{2020}).
\bibitem[{Qi et~al.(2017)Qi, Hu, Valenzuela, Zhang, Zhai, Quan, Waltham, and
  Fang}]{qi2017magnetically}
\bibinfo{author}{L.~Qi}, \bibinfo{author}{Z.~Hu},
  \bibinfo{author}{T.~Valenzuela}, \bibinfo{author}{Y.~Zhang},
  \bibinfo{author}{Y.~Zhai}, \bibinfo{author}{W.~Quan},
  \bibinfo{author}{N.~Waltham}, \bibinfo{author}{J.~Fang},
\newblock \bibinfo{title}{{Magnetically guided Cesium interferometer for
  inertial sensing}},
\newblock \bibinfo{journal}{Applied Physics Letters} \bibinfo{volume}{110}
  (\bibinfo{year}{2017}) \bibinfo{pages}{153502}.
\bibitem[{Wu et~al.(2007)Wu, Su, and Prentiss}]{wu2007demonstration}
\bibinfo{author}{S.~Wu}, \bibinfo{author}{E.~Su},
  \bibinfo{author}{M.~Prentiss},
\newblock \bibinfo{title}{{Demonstration of an Area-Enclosing Guided-Atom
  Interferometer for Rotation Sensing}},
\newblock \bibinfo{journal}{Physical Review Letters} \bibinfo{volume}{99}
  (\bibinfo{year}{2007}) \bibinfo{pages}{173201}. \URLprefix
  \url{https://link.aps.org/doi/10.1103/PhysRevLett.99.173201}.
  \DOIprefix\doi{10.1103/PhysRevLett.99.173201}.
\bibitem[{Beydler et~al.(2024)Beydler, Moan, Luo, Chu, and
  Sackett}]{beydler2024guided}
\bibinfo{author}{M.~Beydler}, \bibinfo{author}{E.~Moan},
  \bibinfo{author}{Z.~Luo}, \bibinfo{author}{Z.~Chu},
  \bibinfo{author}{C.~Sackett},
\newblock \bibinfo{title}{{Guided-wave Sagnac atom interferometer with large
  area and multiple orbits}},
\newblock \bibinfo{journal}{AVS Quantum Science} \bibinfo{volume}{6}
  (\bibinfo{year}{2024}).
\bibitem[{Viefers et~al.(2004)Viefers, Koskinen, {Singha Deo}, and
  Manninen}]{viefers2004quantum}
\bibinfo{author}{S.~Viefers}, \bibinfo{author}{P.~Koskinen},
  \bibinfo{author}{P.~{Singha Deo}}, \bibinfo{author}{M.~Manninen},
\newblock \bibinfo{title}{Quantum rings for beginners: energy spectra and
  persistent currents},
\newblock \bibinfo{journal}{Physica E: Low-dimensional Systems and
  Nanostructures} \bibinfo{volume}{21} (\bibinfo{year}{2004})
  \bibinfo{pages}{1--35}. \URLprefix
  \url{https://www.sciencedirect.com/science/article/pii/S1386947703005186}.
  \DOIprefix\doi{https://doi.org/10.1016/j.physe.2003.08.076}.
\bibitem[{Landau(2013)}]{landau2013classical}
\bibinfo{author}{L.~D. Landau}, \bibinfo{title}{The classical theory of
  fields}, volume~\bibinfo{volume}{2}, \bibinfo{publisher}{Elsevier},
  \bibinfo{year}{2013}.
\bibitem[{Lin et~al.(2009)Lin, Compton, Jiménez-García, Porto, and
  Spielman}]{lin2009synthetic}
\bibinfo{author}{Y.-J. Lin}, \bibinfo{author}{R.~L. Compton},
  \bibinfo{author}{K.~Jiménez-García}, \bibinfo{author}{J.~V. Porto},
  \bibinfo{author}{I.~B. Spielman},
\newblock \bibinfo{title}{Synthetic magnetic fields for ultracold neutral
  atoms},
\newblock \bibinfo{journal}{Nature} \bibinfo{volume}{462}
  (\bibinfo{year}{2009}). \URLprefix
  \url{https://www.nature.com/articles/nature08609}.
  \DOIprefix\doi{10.1038/nature08609}.
\bibitem[{Ehrenberg and Siday(1949)}]{ehrenberg1949refractive}
\bibinfo{author}{W.~Ehrenberg}, \bibinfo{author}{R.~E. Siday},
\newblock \bibinfo{title}{{The Refractive Index in Electron Optics and the
  Principles of Dynamics}},
\newblock \bibinfo{journal}{Proceedings of the Physical Society. Section B}
  \bibinfo{volume}{62} (\bibinfo{year}{1949}) \bibinfo{pages}{8}. \URLprefix
  \url{https://dx.doi.org/10.1088/0370-1301/62/1/303}.
  \DOIprefix\doi{10.1088/0370-1301/62/1/303}.
\bibitem[{Chambers(1960)}]{chambers1960shift}
\bibinfo{author}{R.~G. Chambers},
\newblock \bibinfo{title}{{Shift of an Electron Interference Pattern by
  Enclosed Magnetic Flux}},
\newblock \bibinfo{journal}{Physical Review Letters} \bibinfo{volume}{5}
  (\bibinfo{year}{1960}) \bibinfo{pages}{3--5}. \URLprefix
  \url{https://link.aps.org/doi/10.1103/PhysRevLett.5.3}.
  \DOIprefix\doi{10.1103/PhysRevLett.5.3}.
\bibitem[{Tonomura et~al.(1986)Tonomura, Osakabe, Matsuda, Kawasaki, Endo,
  Yano, and Yamada}]{tonomura1986evidence}
\bibinfo{author}{A.~Tonomura}, \bibinfo{author}{N.~Osakabe},
  \bibinfo{author}{T.~Matsuda}, \bibinfo{author}{T.~Kawasaki},
  \bibinfo{author}{J.~Endo}, \bibinfo{author}{S.~Yano},
  \bibinfo{author}{H.~Yamada},
\newblock \bibinfo{title}{{Evidence for Aharonov-Bohm effect with magnetic
  field completely shielded from electron wave}},
\newblock \bibinfo{journal}{Physical Review Letters} \bibinfo{volume}{56}
  (\bibinfo{year}{1986}) \bibinfo{pages}{792--795}. \URLprefix
  \url{https://link.aps.org/doi/10.1103/PhysRevLett.56.792}.
  \DOIprefix\doi{10.1103/PhysRevLett.56.792}.
\bibitem[{Kohn(1964)}]{kohn1964theory}
\bibinfo{author}{W.~Kohn},
\newblock \bibinfo{title}{Theory of the insulating state},
\newblock \bibinfo{journal}{Physical Review} \bibinfo{volume}{133}
  (\bibinfo{year}{1964}) \bibinfo{pages}{A171}.
\bibitem[{Thouless(1977)}]{thouless1977maximum}
\bibinfo{author}{D.~J. Thouless},
\newblock \bibinfo{title}{Maximum metallic resistance in thin wires},
\newblock \bibinfo{journal}{Phys. Rev. Lett.} \bibinfo{volume}{39}
  (\bibinfo{year}{1977}) \bibinfo{pages}{1167--1169}. \URLprefix
  \url{https://link.aps.org/doi/10.1103/PhysRevLett.39.1167}.
  \DOIprefix\doi{10.1103/PhysRevLett.39.1167}.
\bibitem[{Edwards and Thouless(1972)}]{edwards1972numerical}
\bibinfo{author}{J.~T. Edwards}, \bibinfo{author}{D.~J. Thouless},
\newblock \bibinfo{title}{Numerical studies of localization in disordered
  systems},
\newblock \bibinfo{journal}{Journal of Physics C: Solid State Physics}
  \bibinfo{volume}{5} (\bibinfo{year}{1972}) \bibinfo{pages}{807}. \URLprefix
  \url{https://dx.doi.org/10.1088/0022-3719/5/8/007}.
  \DOIprefix\doi{10.1088/0022-3719/5/8/007}.
\bibitem[{Ashcroft and Mermin(1976)}]{ashcroft1976solid}
\bibinfo{author}{N.~W. Ashcroft}, \bibinfo{author}{N.~Mermin},
\newblock \bibinfo{title}{Solid state},
\newblock \bibinfo{journal}{Physics (New York: Holt, Rinehart and Winston)
  Appendix C} \bibinfo{volume}{1} (\bibinfo{year}{1976}).
\bibitem[{Leggett(1991)}]{leggett1991theorem}
\bibinfo{author}{A.~J. Leggett}, \bibinfo{title}{{Dephasing and Non-Dephasing
  Collisions in Nanostructures}}, \bibinfo{publisher}{Springer US},
  \bibinfo{address}{Boston, MA}, \bibinfo{year}{(1991)}, pp.
  \bibinfo{pages}{297--311}.
\bibitem[{Imry(1986)}]{imry1986physics}
\bibinfo{author}{Y.~Imry},
\newblock \bibinfo{title}{Physics of mesoscopic systems},
\newblock in: \bibinfo{booktitle}{Directions in Condensed Matter Physics:
  Memorial Volume in Honor of Shang-keng Ma}, \bibinfo{publisher}{World
  Scientific}, \bibinfo{year}{1986}, pp. \bibinfo{pages}{101--163}.
\bibitem[{Leggett(1991)}]{leggett1991dephasing}
\bibinfo{author}{A.~J. Leggett}, \bibinfo{title}{{Dephasing and Non-Dephasing
  Collisions in Nanostructures}}, \bibinfo{publisher}{Springer US},
  \bibinfo{address}{Boston, MA}, \bibinfo{year}{1991}, pp.
  \bibinfo{pages}{297--311}.
\bibitem[{Peierls(1933)}]{peierls1933zur}
\bibinfo{author}{R.~Peierls},
\newblock \bibinfo{title}{{Zur Theorie des Diamagnetismus von
  Leitungselektronen}},
\newblock
  \bibinfo{journal}{\href{https://doi.org/10.1007/bf01342591}{Zeitschrift f\"ur
  Physik}} \bibinfo{volume}{\textbf{80}} (\bibinfo{year}{(1933)})
  \bibinfo{pages}{763--791}. \URLprefix
  \url{https://doi.org/10.1007/bf01342591}. \DOIprefix\doi{10.1007/bf01342591}.
\bibitem[{Essler et~al.(2005)Essler, Frahm, Göhmann, Klümper, and
  Korepin}]{essler2005one}
\bibinfo{author}{F.~H.~L. Essler}, \bibinfo{author}{H.~Frahm},
  \bibinfo{author}{F.~Göhmann}, \bibinfo{author}{A.~Klümper},
  \bibinfo{author}{V.~E. Korepin}, \bibinfo{title}{{The One-Dimensional Hubbard
  Model}}, \bibinfo{publisher}{Cambridge University Press},
  \bibinfo{year}{(2005)}. \DOIprefix\doi{10.1017/CBO9780511534843}.
\bibitem[{Leggett and Sols(1991)}]{leggett1991concept}
\bibinfo{author}{A.~J. Leggett}, \bibinfo{author}{F.~Sols},
\newblock \bibinfo{title}{On the concept of spontaneously broken gauge symmetry
  in condensed matter physics},
\newblock \bibinfo{journal}{Foundations of physics} \bibinfo{volume}{21}
  (\bibinfo{year}{1991}) \bibinfo{pages}{353--364}.
\bibitem[{Pecci et~al.(2022)Pecci, Naldesi, Minguzzi, and
  Amico}]{pecci2022single}
\bibinfo{author}{G.~Pecci}, \bibinfo{author}{P.~Naldesi},
  \bibinfo{author}{A.~Minguzzi}, \bibinfo{author}{L.~Amico},
\newblock \bibinfo{title}{{Single-particle versus many-body phase coherence in
  an interacting Fermi gas}},
\newblock \bibinfo{journal}{Quantum Science and Technology} \bibinfo{volume}{8}
  (\bibinfo{year}{2022}) \bibinfo{pages}{01LT03}.
\bibitem[{Leggett(1973)}]{leggett1973topics}
\bibinfo{author}{A.~Leggett}, \bibinfo{title}{Topics in the theory of helium},
  \bibinfo{type}{Technical Report}, Univ. of Sussex, Brighton, Eng.,
  \bibinfo{year}{1973}.
\bibitem[{Fisher et~al.(1973)Fisher, Barber, and Jasnow}]{fisher1973helicity}
\bibinfo{author}{M.~E. Fisher}, \bibinfo{author}{M.~N. Barber},
  \bibinfo{author}{D.~Jasnow},
\newblock \bibinfo{title}{Helicity modulus, superfluidity, and scaling in
  isotropic systems},
\newblock \bibinfo{journal}{Physical Review A} \bibinfo{volume}{8}
  (\bibinfo{year}{1973}) \bibinfo{pages}{1111}.
\bibitem[{Roth and Burnett(2003)}]{roth2003superfluidity}
\bibinfo{author}{R.~Roth}, \bibinfo{author}{K.~Burnett},
\newblock \bibinfo{title}{Superfluidity and interference pattern of ultracold
  bosons in optical lattices},
\newblock \bibinfo{journal}{Physical Review A} \bibinfo{volume}{67}
  (\bibinfo{year}{2003}) \bibinfo{pages}{031602}.
\bibitem[{Lieb et~al.(2005)Lieb, Lieb, Seiringer, and
  Yngvason}]{lieb2005superfluidity}
\bibinfo{author}{E.~H. Lieb}, \bibinfo{author}{E.~H. Lieb},
  \bibinfo{author}{R.~Seiringer}, \bibinfo{author}{J.~Yngvason},
\newblock \bibinfo{title}{{Superfluidity in dilute trapped Bose gases}},
\newblock \bibinfo{journal}{{The Stability of Matter: From Atoms to Stars:
  Selecta of Elliott H. Lieb}}  (\bibinfo{year}{2005})
  \bibinfo{pages}{903--908}.
\bibitem[{Het{\'e}nyi(2012)}]{hetenyi2012current}
\bibinfo{author}{B.~Het{\'e}nyi},
\newblock \bibinfo{title}{Current response in extended systems as a geometric
  phase: Application to variational wavefunctions},
\newblock \bibinfo{journal}{Journal of the Physical Society of Japan}
  \bibinfo{volume}{81} (\bibinfo{year}{2012}) \bibinfo{pages}{124711}.
\bibitem[{Het{\'e}nyi(2014)}]{hetenyi2014drude}
\bibinfo{author}{B.~Het{\'e}nyi},
\newblock \bibinfo{title}{Drude weight, meissner weight, rotational inertia of
  bosonic superfluids: How are they distinguished?},
\newblock \bibinfo{journal}{Journal of the Physical Society of Japan}
  \bibinfo{volume}{83} (\bibinfo{year}{2014}) \bibinfo{pages}{034711}.
\bibitem[{Shastry and Sutherland(1990)}]{shastry1990twisted}
\bibinfo{author}{B.~S. Shastry}, \bibinfo{author}{B.~Sutherland},
\newblock \bibinfo{title}{{Twisted boundary conditions and effective mass in
  Heisenberg-Ising and Hubbard rings}},
\newblock \bibinfo{journal}{Physical Review letters} \bibinfo{volume}{65}
  (\bibinfo{year}{1990}) \bibinfo{pages}{243}.
\bibitem[{Castella et~al.(1995)Castella, Zotos, and
  Prelov{\v{s}}ek}]{castella1995integrability}
\bibinfo{author}{H.~Castella}, \bibinfo{author}{X.~Zotos},
  \bibinfo{author}{P.~Prelov{\v{s}}ek},
\newblock \bibinfo{title}{Integrability and ideal conductance at finite
  temperatures},
\newblock \bibinfo{journal}{Physical review letters} \bibinfo{volume}{74}
  (\bibinfo{year}{1995}) \bibinfo{pages}{972}.
\bibitem[{Fye et~al.(1991)Fye, Martins, Scalapino, Wagner, and
  Hanke}]{fye1991drude}
\bibinfo{author}{R.~Fye}, \bibinfo{author}{M.~Martins},
  \bibinfo{author}{D.~Scalapino}, \bibinfo{author}{J.~Wagner},
  \bibinfo{author}{W.~Hanke},
\newblock \bibinfo{title}{{Drude weight, optical conductivity, and flux
  properties of one-dimensional Hubbard rings}},
\newblock \bibinfo{journal}{Physical Review B} \bibinfo{volume}{44}
  (\bibinfo{year}{1991}) \bibinfo{pages}{6909}.
\bibitem[{Amico et~al.(2005)Amico, Osterloh, and Cataliotti}]{amico2005quantum}
\bibinfo{author}{L.~Amico}, \bibinfo{author}{A.~Osterloh},
  \bibinfo{author}{F.~Cataliotti},
\newblock \bibinfo{title}{Quantum many particle systems in ring-shaped optical
  lattices},
\newblock \bibinfo{journal}{Physical review letters} \bibinfo{volume}{95}
  (\bibinfo{year}{2005}) \bibinfo{pages}{063201}.
\bibitem[{Sch{\"u}ttelkopf et~al.(2024)Sch{\"u}ttelkopf, Tajik, Bazhan,
  Cataldini, Ji, Schmiedmayer, and M{\o}ller}]{schuttelkopf2024characterising}
\bibinfo{author}{P.~Sch{\"u}ttelkopf}, \bibinfo{author}{M.~Tajik},
  \bibinfo{author}{N.~Bazhan}, \bibinfo{author}{F.~Cataldini},
  \bibinfo{author}{S.-C. Ji}, \bibinfo{author}{J.~Schmiedmayer},
  \bibinfo{author}{F.~M{\o}ller},
\newblock \bibinfo{title}{Characterising transport in a quantum gas by
  measuring drude weights},
\newblock \bibinfo{journal}{arXiv preprint arXiv:2406.17569}
  (\bibinfo{year}{2024}).
\bibitem[{Watanabe et~al.(2020)Watanabe, Liu, and
  Oshikawa}]{watanabe2020general}
\bibinfo{author}{H.~Watanabe}, \bibinfo{author}{Y.~Liu},
  \bibinfo{author}{M.~Oshikawa},
\newblock \bibinfo{title}{On the general properties of non-linear optical
  conductivities},
\newblock \bibinfo{journal}{Journal of Statistical Physics}
  \bibinfo{volume}{181} (\bibinfo{year}{2020}) \bibinfo{pages}{2050--2070}.
\bibitem[{Watanabe and Oshikawa(2020)}]{watanabe2020generalized}
\bibinfo{author}{H.~Watanabe}, \bibinfo{author}{M.~Oshikawa},
\newblock \bibinfo{title}{Generalized f-sum rules and kohn formulas on
  nonlinear conductivities},
\newblock \bibinfo{journal}{Physical Review B} \bibinfo{volume}{102}
  (\bibinfo{year}{2020}) \bibinfo{pages}{165137}.
\bibitem[{Urichuk et~al.(2022)Urichuk, Kl{\"u}mper, and
  Sirker}]{urichuk2022nonlinear}
\bibinfo{author}{A.~Urichuk}, \bibinfo{author}{A.~Kl{\"u}mper},
  \bibinfo{author}{J.~Sirker},
\newblock \bibinfo{title}{Nonlinear transport by bethe bound states},
\newblock \bibinfo{journal}{Physical Review Letters} \bibinfo{volume}{129}
  (\bibinfo{year}{2022}) \bibinfo{pages}{096602}.
\bibitem[{Salerno et~al.(2023)Salerno, Ozawa, and
  T{\"o}rm{\"a}}]{salerno2023drude}
\bibinfo{author}{G.~Salerno}, \bibinfo{author}{T.~Ozawa},
  \bibinfo{author}{P.~T{\"o}rm{\"a}},
\newblock \bibinfo{title}{{Drude weight and the many-body quantum metric in
  one-dimensional Bose systems}},
\newblock \bibinfo{journal}{arXiv preprint arXiv:2307.10012}
  (\bibinfo{year}{2023}).
\bibitem[{Polo et~al.(2020)Polo, Naldesi, Minguzzi, and Amico}]{polo2020exact}
\bibinfo{author}{J.~Polo}, \bibinfo{author}{P.~Naldesi},
  \bibinfo{author}{A.~Minguzzi}, \bibinfo{author}{L.~Amico},
\newblock \bibinfo{title}{Exact results for persistent currents of two bosons
  in a ring lattice},
\newblock \bibinfo{journal}{Physical Review A} \bibinfo{volume}{101}
  (\bibinfo{year}{2020}) \bibinfo{pages}{043418}. \URLprefix
  \url{https://link.aps.org/doi/10.1103/PhysRevA.101.043418}.
  \DOIprefix\doi{10.1103/PhysRevA.101.043418}.
\bibitem[{Donnelly and Fetter(1966)}]{donnelly1966stability}
\bibinfo{author}{R.~J. Donnelly}, \bibinfo{author}{A.~L. Fetter},
\newblock \bibinfo{title}{{Stability of Superfluid Flow in an Annulus}},
\newblock \bibinfo{journal}{Physical Review Letters} \bibinfo{volume}{17}
  (\bibinfo{year}{1966}) \bibinfo{pages}{747--750}. \URLprefix
  \url{https://link.aps.org/doi/10.1103/PhysRevLett.17.747}.
  \DOIprefix\doi{10.1103/PhysRevLett.17.747}.
\bibitem[{Dalfovo et~al.(1999)Dalfovo, Giorgini, Pitaevskii, and
  Stringari}]{dalfovo1999theory}
\bibinfo{author}{F.~Dalfovo}, \bibinfo{author}{S.~Giorgini},
  \bibinfo{author}{L.~P. Pitaevskii}, \bibinfo{author}{S.~Stringari},
\newblock \bibinfo{title}{{Theory of Bose-Einstein condensation in trapped
  gases}},
\newblock \bibinfo{journal}{Review of Modern Physics} \bibinfo{volume}{71}
  (\bibinfo{year}{1999}) \bibinfo{pages}{463--512}. \URLprefix
  \url{https://link.aps.org/doi/10.1103/RevModPhys.71.463}.
  \DOIprefix\doi{10.1103/RevModPhys.71.463}.
\bibitem[{Bolda and Walls(1998)}]{bolda1998detection}
\bibinfo{author}{E.~L. Bolda}, \bibinfo{author}{D.~F. Walls},
\newblock \bibinfo{title}{{Detection of Vorticity in Bose-Einstein Condensed
  Gases by Matter-Wave Interference}},
\newblock \bibinfo{journal}{Physical Review Letters} \bibinfo{volume}{81}
  (\bibinfo{year}{1998}) \bibinfo{pages}{5477--5480}. \URLprefix
  \url{https://link.aps.org/doi/10.1103/PhysRevLett.81.5477}.
  \DOIprefix\doi{10.1103/PhysRevLett.81.5477}.
\bibitem[{Dobrek et~al.(1999)Dobrek, Gajda, Lewenstein, Sengstock, Birkl, and
  Ertmer}]{dobrek1999optical}
\bibinfo{author}{L.~Dobrek}, \bibinfo{author}{M.~Gajda},
  \bibinfo{author}{M.~Lewenstein}, \bibinfo{author}{K.~Sengstock},
  \bibinfo{author}{G.~Birkl}, \bibinfo{author}{W.~Ertmer},
\newblock \bibinfo{title}{{Optical generation of vortices in trapped
  Bose-Einstein condensates}},
\newblock \bibinfo{journal}{Physical Review A} \bibinfo{volume}{60}
  (\bibinfo{year}{1999}) \bibinfo{pages}{R3381--R3384}. \URLprefix
  \url{https://link.aps.org/doi/10.1103/PhysRevA.60.R3381}.
  \DOIprefix\doi{10.1103/PhysRevA.60.R3381}.
\bibitem[{Andrelczyk et~al.(2001)Andrelczyk, Brewczyk, Gajda, Lewenstein
  et~al.}]{andrelczyk2001optical}
\bibinfo{author}{G.~Andrelczyk}, \bibinfo{author}{M.~Brewczyk},
  \bibinfo{author}{M.~Gajda}, \bibinfo{author}{M.~Lewenstein}, et~al.,
\newblock \bibinfo{title}{{Optical generation of vortices in trapped
  Bose-Einstein condensates}},
\newblock \bibinfo{journal}{Physical Review A} \bibinfo{volume}{64}
  (\bibinfo{year}{2001}) \bibinfo{pages}{043601}.
\bibitem[{Burger et~al.(1999)Burger, Bongs, Dettmer, Ertmer, Sengstock,
  Sanpera, Shlyapnikov, and Lewenstein}]{burger1999dark}
\bibinfo{author}{S.~Burger}, \bibinfo{author}{K.~Bongs},
  \bibinfo{author}{S.~Dettmer}, \bibinfo{author}{W.~Ertmer},
  \bibinfo{author}{K.~Sengstock}, \bibinfo{author}{A.~Sanpera},
  \bibinfo{author}{G.~V. Shlyapnikov}, \bibinfo{author}{M.~Lewenstein},
\newblock \bibinfo{title}{{Dark solitons in Bose-Einstein condensates}},
\newblock \bibinfo{journal}{Physical Review Letters} \bibinfo{volume}{83}
  (\bibinfo{year}{1999}) \bibinfo{pages}{5198}.
\bibitem[{Denschlag et~al.(2000)Denschlag, Simsarian, Feder, Clark, Collins,
  Cubizolles, Deng, Hagley, Helmerson, Reinhardt
  et~al.}]{denschlag2000generating}
\bibinfo{author}{J.~Denschlag}, \bibinfo{author}{J.~E. Simsarian},
  \bibinfo{author}{D.~L. Feder}, \bibinfo{author}{C.~W. Clark},
  \bibinfo{author}{L.~A. Collins}, \bibinfo{author}{J.~Cubizolles},
  \bibinfo{author}{L.~Deng}, \bibinfo{author}{E.~W. Hagley},
  \bibinfo{author}{K.~Helmerson}, \bibinfo{author}{W.~P. Reinhardt}, et~al.,
\newblock \bibinfo{title}{{Generating solitons by phase engineering of a
  Bose-Einstein condensate}},
\newblock \bibinfo{journal}{Science} \bibinfo{volume}{287}
  (\bibinfo{year}{2000}) \bibinfo{pages}{97--101}.
\bibitem[{Gajda et~al.(1999)Gajda, Lewenstein, Sengstock, Birkl, Ertmer
  et~al.}]{gajda1999optical}
\bibinfo{author}{M.~Gajda}, \bibinfo{author}{M.~Lewenstein},
  \bibinfo{author}{K.~Sengstock}, \bibinfo{author}{G.~Birkl},
  \bibinfo{author}{W.~Ertmer}, et~al.,
\newblock \bibinfo{title}{{Optical generation of vortices in trapped
  Bose-Einstein condensates}},
\newblock \bibinfo{journal}{Physical Review A} \bibinfo{volume}{60}
  (\bibinfo{year}{1999}) \bibinfo{pages}{R3381}.
\bibitem[{Eckardt(2017)}]{eckardt2017colloquium}
\bibinfo{author}{A.~Eckardt},
\newblock \bibinfo{title}{Colloquium: Atomic quantum gases in periodically
  driven optical lattices},
\newblock \bibinfo{journal}{Reviews of Modern Physics} \bibinfo{volume}{89}
  (\bibinfo{year}{2017}) \bibinfo{pages}{011004}.
\bibitem[{Weitenberg and Simonet(2021)}]{weitenberg2021tailoring}
\bibinfo{author}{C.~Weitenberg}, \bibinfo{author}{J.~Simonet},
\newblock \bibinfo{title}{{Tailoring quantum gases by Floquet engineering}},
\newblock \bibinfo{journal}{Nature Physics} \bibinfo{volume}{17}
  (\bibinfo{year}{2021}) \bibinfo{pages}{1342--1348}.
\bibitem[{Aidelsburger et~al.(2013)Aidelsburger, Atala, Lohse, Barreiro,
  Paredes, and Bloch}]{aidelsburger2013realization}
\bibinfo{author}{M.~Aidelsburger}, \bibinfo{author}{M.~Atala},
  \bibinfo{author}{M.~Lohse}, \bibinfo{author}{J.~T. Barreiro},
  \bibinfo{author}{B.~Paredes}, \bibinfo{author}{I.~Bloch},
\newblock \bibinfo{title}{{Realization of the Hofstadter Hamiltonian with
  ultracold atoms in optical lattices}},
\newblock \bibinfo{journal}{Physical Review Letters} \bibinfo{volume}{111}
  (\bibinfo{year}{2013}) \bibinfo{pages}{185301}.
\bibitem[{Roushan et~al.(2017)Roushan, Neill, Megrant, Chen, Babbush, Barends,
  Campbell, Chen, Chiaro, Dunsworth et~al.}]{roushan2017chiral}
\bibinfo{author}{P.~Roushan}, \bibinfo{author}{C.~Neill},
  \bibinfo{author}{A.~Megrant}, \bibinfo{author}{Y.~Chen},
  \bibinfo{author}{R.~Babbush}, \bibinfo{author}{R.~Barends},
  \bibinfo{author}{B.~Campbell}, \bibinfo{author}{Z.~Chen},
  \bibinfo{author}{B.~Chiaro}, \bibinfo{author}{A.~Dunsworth}, et~al.,
\newblock \bibinfo{title}{Chiral ground-state currents of interacting photons
  in a synthetic magnetic field},
\newblock \bibinfo{journal}{Nature Physics} \bibinfo{volume}{13}
  (\bibinfo{year}{2017}) \bibinfo{pages}{146--151}.
\bibitem[{Eckardt et~al.(2010)Eckardt, Hauke, Soltan-Panahi, Becker, Sengstock,
  and Lewenstein}]{eckardt2010frustrated}
\bibinfo{author}{A.~Eckardt}, \bibinfo{author}{P.~Hauke},
  \bibinfo{author}{P.~Soltan-Panahi}, \bibinfo{author}{C.~Becker},
  \bibinfo{author}{K.~Sengstock}, \bibinfo{author}{M.~Lewenstein},
\newblock \bibinfo{title}{Frustrated quantum antiferromagnetism with ultracold
  bosons in a triangular lattice},
\newblock \bibinfo{journal}{Europhysics Letters} \bibinfo{volume}{89}
  (\bibinfo{year}{2010}) \bibinfo{pages}{10010}.
\bibitem[{Struck et~al.(2012{\natexlab{a}})Struck, {\"O}lschl{\"a}ger,
  Weinberg, Hauke, Simonet, Eckardt, Lewenstein, Sengstock, and
  Windpassinger}]{struck2012tunable}
\bibinfo{author}{J.~Struck}, \bibinfo{author}{C.~{\"O}lschl{\"a}ger},
  \bibinfo{author}{M.~Weinberg}, \bibinfo{author}{P.~Hauke},
  \bibinfo{author}{J.~Simonet}, \bibinfo{author}{A.~Eckardt},
  \bibinfo{author}{M.~Lewenstein}, \bibinfo{author}{K.~Sengstock},
  \bibinfo{author}{P.~Windpassinger},
\newblock \bibinfo{title}{Tunable gauge potential for neutral and spinless
  particles in driven optical lattices},
\newblock \bibinfo{journal}{Physical review letters} \bibinfo{volume}{108}
  (\bibinfo{year}{2012}{\natexlab{a}}) \bibinfo{pages}{225304}.
\bibitem[{Struck et~al.(2012{\natexlab{b}})Struck, \"Olschl\"ager, Weinberg,
  Hauke, Simonet, Eckardt, Lewenstein, Sengstock, and
  Windpassinger}]{maciej2012tunable}
\bibinfo{author}{J.~Struck}, \bibinfo{author}{C.~\"Olschl\"ager},
  \bibinfo{author}{M.~Weinberg}, \bibinfo{author}{P.~Hauke},
  \bibinfo{author}{J.~Simonet}, \bibinfo{author}{A.~Eckardt},
  \bibinfo{author}{M.~Lewenstein}, \bibinfo{author}{K.~Sengstock},
  \bibinfo{author}{P.~Windpassinger},
\newblock \bibinfo{title}{{Tunable Gauge Potential for Neutral and Spinless
  Particles in Driven Optical Lattices}},
\newblock \bibinfo{journal}{Physical Review Letters}
  \bibinfo{volume}{\textbf{108}} (\bibinfo{year}{2012}{\natexlab{b}})
  \bibinfo{pages}{225304}. \URLprefix
  \url{https://link.aps.org/doi/10.1103/PhysRevLett.108.225304}.
  \DOIprefix\doi{10.1103/PhysRevLett.108.225304}.
\bibitem[{Atala et~al.(2014)Atala, Aidelsburger, Lohse, Barreiro, Paredes, and
  Bloch}]{atala2014observation}
\bibinfo{author}{M.~Atala}, \bibinfo{author}{M.~Aidelsburger},
  \bibinfo{author}{M.~Lohse}, \bibinfo{author}{J.~T. Barreiro},
  \bibinfo{author}{B.~Paredes}, \bibinfo{author}{I.~Bloch},
\newblock \bibinfo{title}{Observation of chiral currents with ultracold atoms
  in bosonic ladders},
\newblock \bibinfo{journal}{Nature Physics} \bibinfo{volume}{\textbf{10}}
  (\bibinfo{year}{2014}) \bibinfo{pages}{588--593}. \URLprefix
  \url{https://doi.org/10.1038/nphys2998}. \DOIprefix\doi{10.1038/nphys2998}.
\bibitem[{Kennedy et~al.(2015)Kennedy, Burton, Chung, and
  Ketterle}]{kennedy2015observation}
\bibinfo{author}{C.~J. Kennedy}, \bibinfo{author}{W.~C. Burton},
  \bibinfo{author}{W.~C. Chung}, \bibinfo{author}{W.~Ketterle},
\newblock \bibinfo{title}{{Observation of Bose--Einstein condensation in a
  strong synthetic magnetic field}},
\newblock \bibinfo{journal}{Nature Physics} \bibinfo{volume}{11}
  (\bibinfo{year}{2015}) \bibinfo{pages}{859--864}.
\bibitem[{Simjanovski et~al.(2023)Simjanovski, Gauthier, Davis,
  Rubinsztein-Dunlop, and Neely}]{simjanovski2023optimizing}
\bibinfo{author}{S.~Simjanovski}, \bibinfo{author}{G.~Gauthier},
  \bibinfo{author}{M.~J. Davis}, \bibinfo{author}{H.~Rubinsztein-Dunlop},
  \bibinfo{author}{T.~W. Neely},
\newblock \bibinfo{title}{{Optimizing persistent currents in a ring-shaped
  Bose-Einstein condensate using machine learning}},
\newblock \bibinfo{journal}{arXiv preprint arXiv:2304.06199}
  (\bibinfo{year}{2023}).
\bibitem[{Mathey and Mathey(2016)}]{mathey2016realizing}
\bibinfo{author}{A.~C. Mathey}, \bibinfo{author}{L.~Mathey},
\newblock \bibinfo{title}{{Realizing and optimizing an atomtronic SQUID}},
\newblock \bibinfo{journal}{New Journal of Physics} \bibinfo{volume}{18}
  (\bibinfo{year}{2016}) \bibinfo{pages}{055016}. \URLprefix
  \url{https://dx.doi.org/10.1088/1367-2630/18/5/055016}.
  \DOIprefix\doi{10.1088/1367-2630/18/5/055016}.
\bibitem[{Goodfellow et~al.(2016)Goodfellow, Bengio, and
  Courville}]{goodfellow2016deep}
\bibinfo{author}{I.~Goodfellow}, \bibinfo{author}{Y.~Bengio},
  \bibinfo{author}{A.~Courville}, \bibinfo{title}{Deep learning},
  \bibinfo{publisher}{MIT press}, \bibinfo{year}{2016}.
\bibitem[{Carleo et~al.(2019)Carleo, Cirac, Cranmer, Daudet, Schuld, Tishby,
  Vogt-Maranto, and Zdeborov{\'a}}]{carleo2019machine}
\bibinfo{author}{G.~Carleo}, \bibinfo{author}{I.~Cirac},
  \bibinfo{author}{K.~Cranmer}, \bibinfo{author}{L.~Daudet},
  \bibinfo{author}{M.~Schuld}, \bibinfo{author}{N.~Tishby},
  \bibinfo{author}{L.~Vogt-Maranto}, \bibinfo{author}{L.~Zdeborov{\'a}},
\newblock \bibinfo{title}{Machine learning and the physical sciences},
\newblock \bibinfo{journal}{Reviews of Modern Physics} \bibinfo{volume}{91}
  (\bibinfo{year}{2019}) \bibinfo{pages}{045002}.
\bibitem[{Krenn et~al.(2020)Krenn, Erhard, and Zeilinger}]{krenn2020computer}
\bibinfo{author}{M.~Krenn}, \bibinfo{author}{M.~Erhard},
  \bibinfo{author}{A.~Zeilinger},
\newblock \bibinfo{title}{Computer-inspired quantum experiments},
\newblock \bibinfo{journal}{Nature Reviews Physics} \bibinfo{volume}{2}
  (\bibinfo{year}{2020}) \bibinfo{pages}{649--661}.
\bibitem[{Bharti et~al.(2020)Bharti, Haug, Vedral, and
  Kwek}]{bharti2020machine}
\bibinfo{author}{K.~Bharti}, \bibinfo{author}{T.~Haug},
  \bibinfo{author}{V.~Vedral}, \bibinfo{author}{L.-C. Kwek},
\newblock \bibinfo{title}{{Machine learning meets quantum foundations: A brief
  survey}},
\newblock \bibinfo{journal}{AVS Quantum Science} \bibinfo{volume}{2}
  (\bibinfo{year}{2020}).
\bibitem[{Chih et~al.(2022)Chih, Anderson, and Holland}]{chih2022train}
\bibinfo{author}{L.-Y. Chih}, \bibinfo{author}{D.~Z. Anderson},
  \bibinfo{author}{M.~Holland},
\newblock \bibinfo{title}{{How to Train Your Gyro: Reinforcement Learning for
  Rotation Sensing with a Shaken Optical Lattice}},
\newblock \bibinfo{journal}{arXiv preprint arXiv:2212.14473}
  (\bibinfo{year}{2022}).
\bibitem[{Metz et~al.(2021)Metz, Polo, Weber, and Busch}]{metz2021deep}
\bibinfo{author}{F.~Metz}, \bibinfo{author}{J.~Polo},
  \bibinfo{author}{N.~Weber}, \bibinfo{author}{T.~Busch},
\newblock \bibinfo{title}{{Deep-learning-based quantum vortex detection in
  atomic Bose--Einstein condensates}},
\newblock \bibinfo{journal}{Machine Learning: Science and Technology}
  \bibinfo{volume}{2} (\bibinfo{year}{2021}) \bibinfo{pages}{035019}.
  \DOIprefix\doi{10.1088/2632-2153/abea6a}.
\bibitem[{Wigley et~al.(2016)Wigley, Everitt, van~den Hengel, Bastian,
  Sooriyabandara, McDonald, Hardman, Quinlivan, Manju, Kuhn
  et~al.}]{wigley2016fast}
\bibinfo{author}{P.~B. Wigley}, \bibinfo{author}{P.~J. Everitt},
  \bibinfo{author}{A.~van~den Hengel}, \bibinfo{author}{J.~W. Bastian},
  \bibinfo{author}{M.~A. Sooriyabandara}, \bibinfo{author}{G.~D. McDonald},
  \bibinfo{author}{K.~S. Hardman}, \bibinfo{author}{C.~D. Quinlivan},
  \bibinfo{author}{P.~Manju}, \bibinfo{author}{C.~C. Kuhn}, et~al.,
\newblock \bibinfo{title}{Fast machine-learning online optimization of
  ultra-cold-atom experiments},
\newblock \bibinfo{journal}{Scientific reports} \bibinfo{volume}{6}
  (\bibinfo{year}{2016}) \bibinfo{pages}{25890}.
\bibitem[{Haug et~al.(2021)Haug, Dumke, Kwek, Miniatura, and
  Amico}]{haug2021machine}
\bibinfo{author}{T.~Haug}, \bibinfo{author}{R.~Dumke}, \bibinfo{author}{L.-C.
  Kwek}, \bibinfo{author}{C.~Miniatura}, \bibinfo{author}{L.~Amico},
\newblock \bibinfo{title}{Machine-learning engineering of quantum currents},
\newblock \bibinfo{journal}{Physical Review Research} \bibinfo{volume}{3}
  (\bibinfo{year}{2021}) \bibinfo{pages}{013034}. \URLprefix
  \url{https://link.aps.org/doi/10.1103/PhysRevResearch.3.013034}.
  \DOIprefix\doi{10.1103/PhysRevResearch.3.013034}.
\bibitem[{Gauthier et~al.(2016)Gauthier, Lenton, Parry, Baker, Davis,
  Rubinsztein-Dunlop, and Neely}]{gauthier2016direct}
\bibinfo{author}{G.~Gauthier}, \bibinfo{author}{I.~Lenton},
  \bibinfo{author}{N.~M. Parry}, \bibinfo{author}{M.~Baker},
  \bibinfo{author}{M.~Davis}, \bibinfo{author}{H.~Rubinsztein-Dunlop},
  \bibinfo{author}{T.~Neely},
\newblock \bibinfo{title}{Direct imaging of a digital-micromirror device for
  configurable microscopic optical potentials},
\newblock \bibinfo{journal}{Optica} \bibinfo{volume}{3} (\bibinfo{year}{2016})
  \bibinfo{pages}{1136--1143}. \DOIprefix\doi{10.1364/OPTICA.3.001136}.
\bibitem[{Schulman et~al.(2017)Schulman, Wolski, Dhariwal, Radford, and
  Klimov}]{schulman2017proximal}
\bibinfo{author}{J.~Schulman}, \bibinfo{author}{F.~Wolski},
  \bibinfo{author}{P.~Dhariwal}, \bibinfo{author}{A.~Radford},
  \bibinfo{author}{O.~Klimov},
\newblock \bibinfo{title}{Proximal policy optimization algorithms},
\newblock \bibinfo{journal}{arXiv preprint arXiv:1707.06347}
  (\bibinfo{year}{2017}). \DOIprefix\doi{10.48550/arXiv.1707.06347}.
\bibitem[{Lin et~al.(2011)Lin, Compton, Jimenez-Garcia, Phillips, Porto, and
  Spielman}]{lin2011synthetic}
\bibinfo{author}{Y.-J. Lin}, \bibinfo{author}{R.~L. Compton},
  \bibinfo{author}{K.~Jimenez-Garcia}, \bibinfo{author}{W.~D. Phillips},
  \bibinfo{author}{J.~V. Porto}, \bibinfo{author}{I.~B. Spielman},
\newblock \bibinfo{title}{A synthetic electric force acting on neutral atoms},
\newblock \bibinfo{journal}{Nature Physics} \bibinfo{volume}{7}
  (\bibinfo{year}{2011}) \bibinfo{pages}{531--534}.
\bibitem[{Kozuma et~al.(1999)Kozuma, Deng, Hagley, Wen, Lutwak, Helmerson,
  Rolston, and Phillips}]{kozuma1999coherent}
\bibinfo{author}{M.~Kozuma}, \bibinfo{author}{L.~Deng}, \bibinfo{author}{E.~W.
  Hagley}, \bibinfo{author}{J.~Wen}, \bibinfo{author}{R.~Lutwak},
  \bibinfo{author}{K.~Helmerson}, \bibinfo{author}{S.~L. Rolston},
  \bibinfo{author}{W.~D. Phillips},
\newblock \bibinfo{title}{Coherent splitting of bose-einstein condensed atoms
  with optically induced bragg diffraction},
\newblock \bibinfo{journal}{Physical Review Letters} \bibinfo{volume}{82}
  (\bibinfo{year}{1999}) \bibinfo{pages}{871--875}. \URLprefix
  \url{https://link.aps.org/doi/10.1103/PhysRevLett.82.871}.
  \DOIprefix\doi{10.1103/PhysRevLett.82.871}.
\bibitem[{Stamper-Kurn et~al.(1999)Stamper-Kurn, Chikkatur, G\"orlitz, Inouye,
  Gupta, Pritchard, and Ketterle}]{stamperkurn1999phonons}
\bibinfo{author}{D.~M. Stamper-Kurn}, \bibinfo{author}{A.~P. Chikkatur},
  \bibinfo{author}{A.~G\"orlitz}, \bibinfo{author}{S.~Inouye},
  \bibinfo{author}{S.~Gupta}, \bibinfo{author}{D.~E. Pritchard},
  \bibinfo{author}{W.~Ketterle},
\newblock \bibinfo{title}{{Excitation of Phonons in a Bose-Einstein Condensate
  by Light Scattering}},
\newblock \bibinfo{journal}{Physical Review Letters} \bibinfo{volume}{83}
  (\bibinfo{year}{1999}) \bibinfo{pages}{2876--2879}. \URLprefix
  \url{https://link.aps.org/doi/10.1103/PhysRevLett.83.2876}.
  \DOIprefix\doi{10.1103/PhysRevLett.83.2876}.
\bibitem[{Wright et~al.(2009)Wright, Leslie, Hansen, and
  Bigelow}]{WrightSculptingPRL2009}
\bibinfo{author}{K.~C. Wright}, \bibinfo{author}{L.~S. Leslie},
  \bibinfo{author}{A.~Hansen}, \bibinfo{author}{N.~P. Bigelow},
\newblock \bibinfo{title}{{Sculpting the Vortex State of a Spinor BEC}},
\newblock \bibinfo{journal}{Physical Review Letters} \bibinfo{volume}{102}
  (\bibinfo{year}{2009}) \bibinfo{pages}{030405}. \URLprefix
  \url{https://link.aps.org/doi/10.1103/PhysRevLett.102.030405}.
  \DOIprefix\doi{10.1103/PhysRevLett.102.030405}.
\bibitem[{Dobrek et~al.(1999)Dobrek, Gajda, Lewenstein, Sengstock, Birkl, and
  Ertmer}]{DobrekOpticalPRA1999}
\bibinfo{author}{L.~Dobrek}, \bibinfo{author}{M.~Gajda},
  \bibinfo{author}{M.~Lewenstein}, \bibinfo{author}{K.~Sengstock},
  \bibinfo{author}{G.~Birkl}, \bibinfo{author}{W.~Ertmer},
\newblock \bibinfo{title}{{Optical generation of vortices in trapped
  Bose-Einstein condensates}},
\newblock \bibinfo{journal}{Physical Review A} \bibinfo{volume}{60}
  (\bibinfo{year}{1999}) \bibinfo{pages}{R3381--R3384}. \URLprefix
  \url{https://link.aps.org/doi/10.1103/PhysRevA.60.R3381}.
  \DOIprefix\doi{10.1103/PhysRevA.60.R3381}.
\bibitem[{Zheng and Javanainen(2003)}]{zheng2003classical}
\bibinfo{author}{Y.~Zheng}, \bibinfo{author}{J.~Javanainen},
\newblock \bibinfo{title}{Classical and quantum models for phase imprinting},
\newblock \bibinfo{journal}{Physical Review A} \bibinfo{volume}{67}
  (\bibinfo{year}{2003}) \bibinfo{pages}{035602}. \URLprefix
  \url{https://link.aps.org/doi/10.1103/PhysRevA.67.035602}.
  \DOIprefix\doi{10.1103/PhysRevA.67.035602}.
\bibitem[{Swingle and Kennedy(2005)}]{swingle2005generation}
\bibinfo{author}{B.~G. Swingle}, \bibinfo{author}{T.~A.~B. Kennedy},
\newblock \bibinfo{title}{Generation of topological flows by phase imprinting},
\newblock \bibinfo{journal}{Journal of Physics B: Atomic, Molecular and Optical
  Physics} \bibinfo{volume}{38} (\bibinfo{year}{2005}) \bibinfo{pages}{3503}.
  \URLprefix \url{https://dx.doi.org/10.1088/0953-4075/38/19/004}.
\bibitem[{Isoshima et~al.(2000)Isoshima, Nakahara, Ohmi, and
  Machida}]{isoshima2000creation}
\bibinfo{author}{T.~Isoshima}, \bibinfo{author}{M.~Nakahara},
  \bibinfo{author}{T.~Ohmi}, \bibinfo{author}{K.~Machida},
\newblock \bibinfo{title}{{Creation of a persistent current and vortex in a
  Bose-Einstein condensate of alkali-metal atoms}},
\newblock \bibinfo{journal}{Physical Review A} \bibinfo{volume}{61}
  (\bibinfo{year}{2000}) \bibinfo{pages}{063610}. \URLprefix
  \url{https://link.aps.org/doi/10.1103/PhysRevA.61.063610}.
  \DOIprefix\doi{10.1103/PhysRevA.61.063610}.
\bibitem[{Leanhardt et~al.(2002)Leanhardt, G\"orlitz, Chikkatur, Kielpinski,
  Shin, Pritchard, and Ketterle}]{leanhardt2002imprinting}
\bibinfo{author}{A.~E. Leanhardt}, \bibinfo{author}{A.~G\"orlitz},
  \bibinfo{author}{A.~P. Chikkatur}, \bibinfo{author}{D.~Kielpinski},
  \bibinfo{author}{Y.~Shin}, \bibinfo{author}{D.~E. Pritchard},
  \bibinfo{author}{W.~Ketterle},
\newblock \bibinfo{title}{{Imprinting Vortices in a Bose-Einstein Condensate
  using Topological Phases}},
\newblock \bibinfo{journal}{Physical Review Letters} \bibinfo{volume}{89}
  (\bibinfo{year}{2002}) \bibinfo{pages}{190403}. \URLprefix
  \url{https://link.aps.org/doi/10.1103/PhysRevLett.89.190403}.
  \DOIprefix\doi{10.1103/PhysRevLett.89.190403}.
\bibitem[{Inouye et~al.(1998)Inouye, Andrews, Stenger, Miesner, Stamper-Kurn,
  and Ketterle}]{inouye1998observation}
\bibinfo{author}{S.~Inouye}, \bibinfo{author}{M.~Andrews},
  \bibinfo{author}{J.~Stenger}, \bibinfo{author}{H.-J. Miesner},
  \bibinfo{author}{D.~M. Stamper-Kurn}, \bibinfo{author}{W.~Ketterle},
\newblock \bibinfo{title}{{Observation of Feshbach resonances in a
  Bose--Einstein condensate}},
\newblock \bibinfo{journal}{Nature} \bibinfo{volume}{392}
  (\bibinfo{year}{1998}) \bibinfo{pages}{151--154}.
\bibitem[{Roberts et~al.(1998)Roberts, Claussen, Burke, Greene, Cornell, and
  Wieman}]{roberts1998resonant}
\bibinfo{author}{J.~L. Roberts}, \bibinfo{author}{N.~R. Claussen},
  \bibinfo{author}{J.~P. Burke}, \bibinfo{author}{C.~H. Greene},
  \bibinfo{author}{E.~A. Cornell}, \bibinfo{author}{C.~E. Wieman},
\newblock \bibinfo{title}{{Resonant Magnetic Field Control of Elastic
  Scattering in Cold $^{85}Rb$}},
\newblock \bibinfo{journal}{Physical Review Letters} \bibinfo{volume}{81}
  (\bibinfo{year}{1998}) \bibinfo{pages}{5109--5112}. \URLprefix
  \url{https://link.aps.org/doi/10.1103/PhysRevLett.81.5109}.
  \DOIprefix\doi{10.1103/PhysRevLett.81.5109}.
\bibitem[{Theis et~al.(2004)Theis, Thalhammer, Winkler, Hellwig, Ruff, Grimm,
  and Denschlag}]{theis2004tuning}
\bibinfo{author}{M.~Theis}, \bibinfo{author}{G.~Thalhammer},
  \bibinfo{author}{K.~Winkler}, \bibinfo{author}{M.~Hellwig},
  \bibinfo{author}{G.~Ruff}, \bibinfo{author}{R.~Grimm}, \bibinfo{author}{J.~H.
  Denschlag},
\newblock \bibinfo{title}{{Tuning the Scattering Length with an Optically
  Induced Feshbach Resonance}},
\newblock \bibinfo{journal}{Physical Review Letters} \bibinfo{volume}{93}
  (\bibinfo{year}{2004}) \bibinfo{pages}{123001}. \URLprefix
  \url{https://link.aps.org/doi/10.1103/PhysRevLett.93.123001}.
  \DOIprefix\doi{10.1103/PhysRevLett.93.123001}.
\bibitem[{Fedichev et~al.(1996)Fedichev, Kagan, Shlyapnikov, and
  Walraven}]{fedichev1996influence}
\bibinfo{author}{P.~O. Fedichev}, \bibinfo{author}{Y.~Kagan},
  \bibinfo{author}{G.~V. Shlyapnikov}, \bibinfo{author}{J.~T.~M. Walraven},
\newblock \bibinfo{title}{{Influence of Nearly Resonant Light on the Scattering
  Length in Low-Temperature Atomic Gases}},
\newblock \bibinfo{journal}{Physical Review Letters} \bibinfo{volume}{77}
  (\bibinfo{year}{1996}) \bibinfo{pages}{2913--2916}. \URLprefix
  \url{https://link.aps.org/doi/10.1103/PhysRevLett.77.2913}.
  \DOIprefix\doi{10.1103/PhysRevLett.77.2913}.
\bibitem[{Gross(1957)}]{gross1957unified}
\bibinfo{author}{E.~P. Gross},
\newblock \bibinfo{title}{{Unified Theory of Interacting Bosons}},
\newblock \bibinfo{journal}{Physical Review} \bibinfo{volume}{106}
  (\bibinfo{year}{1957}) \bibinfo{pages}{161--162}. \URLprefix
  \url{https://link.aps.org/doi/10.1103/PhysRev.106.161}.
  \DOIprefix\doi{10.1103/PhysRev.106.161}.
\bibitem[{Ginzburg and Pitaevskii(1958)}]{ginzburg1958sov}
\bibinfo{author}{V.~Ginzburg}, \bibinfo{author}{L.~Pitaevskii},
\newblock \bibinfo{title}{Sov. phys. jetp, 7: 858, 1958},
\newblock \bibinfo{journal}{Zh. Eksp. Teor. Fiz} \bibinfo{volume}{34}
  (\bibinfo{year}{1958}) \bibinfo{pages}{1240}.
\bibitem[{Petrov et~al.(2003)}]{petrov2003bose}
\bibinfo{author}{D.~S. Petrov}, et~al., \bibinfo{title}{{Bose-Einstein
  condensation in low-dimensional trapped gases}}, Ph.D. thesis, Universiteit
  van Amsterdam, \bibinfo{year}{2003}.
\bibitem[{Petrov et~al.(2000)Petrov, Shlyapnikov, and
  Walraven}]{petrov2000regimes}
\bibinfo{author}{D.~S. Petrov}, \bibinfo{author}{G.~V. Shlyapnikov},
  \bibinfo{author}{J.~T.~M. Walraven},
\newblock \bibinfo{title}{{Regimes of Quantum Degeneracy in Trapped 1D Gases}},
\newblock \bibinfo{journal}{Physical Review Letters} \bibinfo{volume}{85}
  (\bibinfo{year}{2000}) \bibinfo{pages}{3745--3749}. \URLprefix
  \url{https://link.aps.org/doi/10.1103/PhysRevLett.85.3745}.
  \DOIprefix\doi{10.1103/PhysRevLett.85.3745}.
\bibitem[{Lieb and Liniger(1963)}]{lieb1963exact}
\bibinfo{author}{E.~H. Lieb}, \bibinfo{author}{W.~Liniger},
\newblock \bibinfo{title}{{Exact Analysis of an Interacting Bose Gas. I. The
  General Solution and the Ground State}},
\newblock \bibinfo{journal}{Physical Review} \bibinfo{volume}{130}
  (\bibinfo{year}{1963}) \bibinfo{pages}{1605--1616}. \URLprefix
  \url{https://link.aps.org/doi/10.1103/PhysRev.130.1605}.
  \DOIprefix\doi{10.1103/PhysRev.130.1605}.
\bibitem[{Piroli and Calabrese(2016)}]{piroli2016local}
\bibinfo{author}{L.~Piroli}, \bibinfo{author}{P.~Calabrese},
\newblock \bibinfo{title}{{Local correlations in the attractive one-dimensional
  Bose gas: From Bethe ansatz to the Gross-Pitaevskii equation}},
\newblock \bibinfo{journal}{Physical Review A} \bibinfo{volume}{94}
  (\bibinfo{year}{2016}) \bibinfo{pages}{053620}. \URLprefix
  \url{https://link.aps.org/doi/10.1103/PhysRevA.94.053620}.
  \DOIprefix\doi{10.1103/PhysRevA.94.053620}.
\bibitem[{Lang et~al.(2017)Lang, Hekking, and Minguzzi}]{guillaume2017ground}
\bibinfo{author}{G.~Lang}, \bibinfo{author}{F.~Hekking},
  \bibinfo{author}{A.~Minguzzi},
\newblock \bibinfo{title}{{Ground-state energy and excitation spectrum of the
  Lieb-Liniger model : accurate analytical results and conjectures about the
  exact solution}},
\newblock \bibinfo{journal}{SciPost Physics} \bibinfo{volume}{3}
  (\bibinfo{year}{2017}) \bibinfo{pages}{003}. \URLprefix
  \url{https://scipost.org/10.21468/SciPostPhys.3.1.003}.
  \DOIprefix\doi{10.21468/SciPostPhys.3.1.003}.
\bibitem[{Piroli et~al.(2016)Piroli, Calabrese, and
  Essler}]{piroli2016multiparticle}
\bibinfo{author}{L.~Piroli}, \bibinfo{author}{P.~Calabrese},
  \bibinfo{author}{F.~H.~L. Essler},
\newblock \bibinfo{title}{{Multiparticle Bound-State Formation following a
  Quantum Quench to the One-Dimensional Bose Gas with Attractive
  Interactions}},
\newblock \bibinfo{journal}{Physical Review Letters} \bibinfo{volume}{116}
  (\bibinfo{year}{2016}) \bibinfo{pages}{070408}. \URLprefix
  \url{https://link.aps.org/doi/10.1103/PhysRevLett.116.070408}.
  \DOIprefix\doi{10.1103/PhysRevLett.116.070408}.
\bibitem[{Van~Dyke et~al.(2021)Van~Dyke, Barron, Mayhall, Barnes, and
  Economou}]{dyke2021preparing}
\bibinfo{author}{J.~S. Van~Dyke}, \bibinfo{author}{G.~S. Barron},
  \bibinfo{author}{N.~J. Mayhall}, \bibinfo{author}{E.~Barnes},
  \bibinfo{author}{S.~E. Economou},
\newblock \bibinfo{title}{{Preparing Bethe Ansatz Eigenstates on a Quantum
  Computer}},
\newblock \bibinfo{journal}{PRX Quantum} \bibinfo{volume}{2}
  (\bibinfo{year}{2021}) \bibinfo{pages}{040329}. \URLprefix
  \url{https://link.aps.org/doi/10.1103/PRXQuantum.2.040329}.
  \DOIprefix\doi{10.1103/PRXQuantum.2.040329}.
\bibitem[{Sopena et~al.(2022)Sopena, Gordon, Garc{\'{i}}a-Mart{\'{i}}n, Sierra,
  and L{\'{o}}pez}]{sopena2022algebraic}
\bibinfo{author}{A.~Sopena}, \bibinfo{author}{M.~H. Gordon},
  \bibinfo{author}{D.~Garc{\'{i}}a-Mart{\'{i}}n}, \bibinfo{author}{G.~Sierra},
  \bibinfo{author}{E.~L{\'{o}}pez},
\newblock \bibinfo{title}{Algebraic {B}ethe {C}ircuits},
\newblock \bibinfo{journal}{{Quantum}} \bibinfo{volume}{6}
  (\bibinfo{year}{2022}) \bibinfo{pages}{796}. \URLprefix
  \url{https://doi.org/10.22331/q-2022-09-08-796}.
  \DOIprefix\doi{10.22331/q-2022-09-08-796}.
\bibitem[{Ruiz et~al.(2024)Ruiz, Sopena, Gordon, Sierra, and
  L{\'{o}}pez}]{ruiz2024bethe}
\bibinfo{author}{R.~Ruiz}, \bibinfo{author}{A.~Sopena}, \bibinfo{author}{M.~H.
  Gordon}, \bibinfo{author}{G.~Sierra}, \bibinfo{author}{E.~L{\'{o}}pez},
\newblock \bibinfo{title}{The {B}ethe {A}nsatz as a {Q}uantum {C}ircuit},
\newblock \bibinfo{journal}{{Quantum}} \bibinfo{volume}{8}
  (\bibinfo{year}{2024}) \bibinfo{pages}{1356}. \URLprefix
  \url{https://doi.org/10.22331/q-2024-05-23-1356}.
  \DOIprefix\doi{10.22331/q-2024-05-23-1356}.
\bibitem[{Kanamoto et~al.(2003)Kanamoto, Saito, and Ueda}]{kanamoto2003quantum}
\bibinfo{author}{R.~Kanamoto}, \bibinfo{author}{H.~Saito},
  \bibinfo{author}{M.~Ueda},
\newblock \bibinfo{title}{{Quantum phase transition in one-dimensional
  Bose-Einstein condensates with attractive interactions}},
\newblock \bibinfo{journal}{Physical Review A} \bibinfo{volume}{67}
  (\bibinfo{year}{2003}) \bibinfo{pages}{013608}. \URLprefix
  \url{https://link.aps.org/doi/10.1103/PhysRevA.67.013608}.
  \DOIprefix\doi{10.1103/PhysRevA.67.013608}.
\bibitem[{Wilczek(2012)}]{wilczek2012quantum}
\bibinfo{author}{F.~Wilczek},
\newblock \bibinfo{title}{Quantum time crystals},
\newblock \bibinfo{journal}{Phys. Rev. Lett.} \bibinfo{volume}{109}
  (\bibinfo{year}{2012}) \bibinfo{pages}{160401}. \URLprefix
  \url{https://link.aps.org/doi/10.1103/PhysRevLett.109.160401}.
  \DOIprefix\doi{10.1103/PhysRevLett.109.160401}.
\bibitem[{Bruno(2013)}]{patrick2016comment}
\bibinfo{author}{P.~Bruno},
\newblock \bibinfo{title}{Comment on ``quantum time crystals''},
\newblock \bibinfo{journal}{Phys. Rev. Lett.} \bibinfo{volume}{110}
  (\bibinfo{year}{2013}) \bibinfo{pages}{118901}. \URLprefix
  \url{https://link.aps.org/doi/10.1103/PhysRevLett.110.118901}.
  \DOIprefix\doi{10.1103/PhysRevLett.110.118901}.
\bibitem[{Calogero and Degasperis(1975)}]{calogero1975comparison}
\bibinfo{author}{F.~Calogero}, \bibinfo{author}{A.~Degasperis},
\newblock \bibinfo{title}{Comparison between the exact and hartree solutions of
  a one-dimensional many-body problem},
\newblock \bibinfo{journal}{Physical Review A} \bibinfo{volume}{11}
  (\bibinfo{year}{1975}) \bibinfo{pages}{265--269}. \URLprefix
  \url{https://link.aps.org/doi/10.1103/PhysRevA.11.265}.
  \DOIprefix\doi{10.1103/PhysRevA.11.265}.
\bibitem[{Walters et~al.(2013)Walters, Cotugno, Johnson, Clark, and
  Jaksch}]{walters2013ab}
\bibinfo{author}{R.~Walters}, \bibinfo{author}{G.~Cotugno},
  \bibinfo{author}{T.~H. Johnson}, \bibinfo{author}{S.~R. Clark},
  \bibinfo{author}{D.~Jaksch},
\newblock \bibinfo{title}{{Ab initio derivation of Hubbard models for cold
  atoms in optical lattices}},
\newblock \bibinfo{journal}{Physical Review A} \bibinfo{volume}{87}
  (\bibinfo{year}{2013}) \bibinfo{pages}{043613}. \URLprefix
  \url{https://link.aps.org/doi/10.1103/PhysRevA.87.043613}.
  \DOIprefix\doi{10.1103/PhysRevA.87.043613}.
\bibitem[{White(1992)}]{white1992density}
\bibinfo{author}{S.~R. White},
\newblock \bibinfo{title}{Density matrix formulation for quantum
  renormalization groups},
\newblock \bibinfo{journal}{Physical Review Letters} \bibinfo{volume}{69}
  (\bibinfo{year}{1992}) \bibinfo{pages}{2863--2866}. \URLprefix
  \url{https://link.aps.org/doi/10.1103/PhysRevLett.69.2863}.
  \DOIprefix\doi{10.1103/PhysRevLett.69.2863}.
\bibitem[{Vidal(2003)}]{vidal2003efficient}
\bibinfo{author}{G.~Vidal},
\newblock \bibinfo{title}{{Efficient Classical Simulation of Slightly Entangled
  Quantum Computations}},
\newblock \bibinfo{journal}{Physical Review Letters} \bibinfo{volume}{91}
  (\bibinfo{year}{2003}) \bibinfo{pages}{147902}. \URLprefix
  \url{https://link.aps.org/doi/10.1103/PhysRevLett.91.147902}.
  \DOIprefix\doi{10.1103/PhysRevLett.91.147902}.
\bibitem[{Haug et~al.(2019)Haug, Dumke, Kwek, and Amico}]{haug2019topological}
\bibinfo{author}{T.~Haug}, \bibinfo{author}{R.~Dumke}, \bibinfo{author}{L.-C.
  Kwek}, \bibinfo{author}{L.~Amico},
\newblock \bibinfo{title}{{Topological pumping in Aharonov--Bohm rings}},
\newblock \bibinfo{journal}{Communications Physics} \bibinfo{volume}{2}
  (\bibinfo{year}{2019}) \bibinfo{pages}{127}.
  \DOIprefix\doi{10.1038/s42005-019-0229-2}.
\bibitem[{Amico and Korepin(2004)}]{amico2004universality}
\bibinfo{author}{L.~Amico}, \bibinfo{author}{V.~Korepin},
\newblock \bibinfo{title}{Universality of the one-dimensional bose gas with
  delta interaction},
\newblock \bibinfo{journal}{Annals of Physics} \bibinfo{volume}{314}
  (\bibinfo{year}{2004}) \bibinfo{pages}{496--507}. \URLprefix
  \url{https://www.sciencedirect.com/science/article/pii/S0003491604001447}.
  \DOIprefix\doi{https://doi.org/10.1016/j.aop.2004.08.001}.
\bibitem[{Fazio and {van der Zant}(2001)}]{fazio2001quantum}
\bibinfo{author}{R.~Fazio}, \bibinfo{author}{H.~{van der Zant}},
\newblock \bibinfo{title}{Quantum phase transitions and vortex dynamics in
  superconducting networks},
\newblock \bibinfo{journal}{Physics Reports} \bibinfo{volume}{355}
  (\bibinfo{year}{2001}) \bibinfo{pages}{235--334}. \URLprefix
  \url{https://www.sciencedirect.com/science/article/pii/S0370157301000229}.
  \DOIprefix\doi{https://doi.org/10.1016/S0370-1573(01)00022-9}.
\bibitem[{Garc\'{i}a-Ripoll et~al.(2004)Garc\'{i}a-Ripoll, Cirac, Zoller,
  Kollath, Schollw\"{o}ck, and von Delft}]{garcia2004variational}
\bibinfo{author}{J.~J. Garc\'{i}a-Ripoll}, \bibinfo{author}{J.~I. Cirac},
  \bibinfo{author}{P.~Zoller}, \bibinfo{author}{C.~Kollath},
  \bibinfo{author}{U.~Schollw\"{o}ck}, \bibinfo{author}{J.~von Delft},
\newblock \bibinfo{title}{{Variational ansatz for the superfluid Mott-insulator
  transition in optical lattices}},
\newblock \bibinfo{journal}{Opt. Express} \bibinfo{volume}{12}
  (\bibinfo{year}{2004}) \bibinfo{pages}{42--54}. \URLprefix
  \url{https://opg.optica.org/oe/abstract.cfm?URI=oe-12-1-42}.
  \DOIprefix\doi{10.1364/OPEX.12.000042}.
\bibitem[{Dubin et~al.(1995)Dubin, Hennings, and Smith}]{dubin1995mathematical}
\bibinfo{author}{D.~Dubin}, \bibinfo{author}{M.~Hennings},
  \bibinfo{author}{T.~Smith},
\newblock \bibinfo{title}{Mathematical aspects of quantum phase},
\newblock \bibinfo{journal}{International Journal of Modern Physics B}
  \bibinfo{volume}{9} (\bibinfo{year}{1995}) \bibinfo{pages}{2597--2687}.
\bibitem[{Fisher and Grinstein(1988)}]{fisher1988quantum}
\bibinfo{author}{M.~P.~A. Fisher}, \bibinfo{author}{G.~Grinstein},
\newblock \bibinfo{title}{{Quantum Critical Phenomena in Charged
  Superconductors}},
\newblock \bibinfo{journal}{Physical Review Letters} \bibinfo{volume}{60}
  (\bibinfo{year}{1988}) \bibinfo{pages}{208--211}. \URLprefix
  \url{https://link.aps.org/doi/10.1103/PhysRevLett.60.208}.
  \DOIprefix\doi{10.1103/PhysRevLett.60.208}.
\bibitem[{Sturm et~al.(2017)Sturm, Schlosser, Walser, and
  Birkl}]{sturm2017quantum}
\bibinfo{author}{M.~R. Sturm}, \bibinfo{author}{M.~Schlosser},
  \bibinfo{author}{R.~Walser}, \bibinfo{author}{G.~Birkl},
\newblock \bibinfo{title}{{Quantum simulators by design: Many-body physics in
  reconfigurable arrays of tunnel-coupled traps}},
\newblock \bibinfo{journal}{Physical Review A} \bibinfo{volume}{95}
  (\bibinfo{year}{2017}) \bibinfo{pages}{063625}. \URLprefix
  \url{https://link.aps.org/doi/10.1103/PhysRevA.95.063625}.
  \DOIprefix\doi{10.1103/PhysRevA.95.063625}.
\bibitem[{Pezz{\`e} et~al.(2024)Pezz{\`e}, Xhani, Daix, Grani, Donelli, Scazza,
  Hernandez-Rajkov, Kwon, Del~Pace, and Roati}]{pezze2023stabilizing}
\bibinfo{author}{L.~Pezz{\`e}}, \bibinfo{author}{K.~Xhani},
  \bibinfo{author}{C.~Daix}, \bibinfo{author}{N.~Grani},
  \bibinfo{author}{B.~Donelli}, \bibinfo{author}{F.~Scazza},
  \bibinfo{author}{D.~Hernandez-Rajkov}, \bibinfo{author}{W.~J. Kwon},
  \bibinfo{author}{G.~Del~Pace}, \bibinfo{author}{G.~Roati},
\newblock \bibinfo{title}{{Stabilizing persistent currents in an atomtronic
  Josephson junction necklace}},
\newblock \bibinfo{journal}{Nature Communications} \bibinfo{volume}{15}
  (\bibinfo{year}{2024}) \bibinfo{pages}{4831}.
\bibitem[{Lieb(1963)}]{lieb1963exactII}
\bibinfo{author}{E.~H. Lieb},
\newblock \bibinfo{title}{{Exact Analysis of an Interacting Bose Gas. II. The
  Excitation Spectrum}},
\newblock \bibinfo{journal}{Physical Review} \bibinfo{volume}{130}
  (\bibinfo{year}{1963}) \bibinfo{pages}{1616--1624}. \URLprefix
  \url{https://link.aps.org/doi/10.1103/PhysRev.130.1616}.
  \DOIprefix\doi{10.1103/PhysRev.130.1616}.
\bibitem[{Minguzzi and Vignolo(2022)}]{minguzzi2022strongly}
\bibinfo{author}{A.~Minguzzi}, \bibinfo{author}{P.~Vignolo},
\newblock \bibinfo{title}{{Strongly interacting trapped one-dimensional quantum
  gases: Exact solution}},
\newblock \bibinfo{journal}{AVS Quantum Science} \bibinfo{volume}{4}
  (\bibinfo{year}{2022}) \bibinfo{pages}{027102}. \URLprefix
  \url{https://doi.org/10.1116/5.0077423}. \DOIprefix\doi{10.1116/5.0077423}.
\bibitem[{Haldane(1981{\natexlab{a}})}]{haldane1981luttinger}
\bibinfo{author}{F.~D.~M. Haldane},
\newblock \bibinfo{title}{{'Luttinger liquid theory' of one-dimensional quantum
  fluids. I. Properties of the Luttinger model and their extension to the
  general 1D interacting spinless Fermi gas}},
\newblock \bibinfo{journal}{Journal of Physics C: Solid State Physics}
  \bibinfo{volume}{14} (\bibinfo{year}{1981}{\natexlab{a}})
  \bibinfo{pages}{2585}. \URLprefix
  \url{https://dx.doi.org/10.1088/0022-3719/14/19/010}.
  \DOIprefix\doi{10.1088/0022-3719/14/19/010}.
\bibitem[{Haldane(1981{\natexlab{b}})}]{haldane1981effective}
\bibinfo{author}{F.~D.~M. Haldane},
\newblock \bibinfo{title}{{Effective Harmonic-Fluid Approach to Low-Energy
  Properties of One-Dimensional Quantum Fluids}},
\newblock \bibinfo{journal}{Physical Review Letters} \bibinfo{volume}{47}
  (\bibinfo{year}{1981}{\natexlab{b}}) \bibinfo{pages}{1840--1843}. \URLprefix
  \url{https://link.aps.org/doi/10.1103/PhysRevLett.47.1840}.
  \DOIprefix\doi{10.1103/PhysRevLett.47.1840}.
\bibitem[{Cazalilla(2004)}]{cazalilla2004bosonizing}
\bibinfo{author}{M.~A. Cazalilla},
\newblock \bibinfo{title}{Bosonizing one-dimensional cold atomic gases},
\newblock \bibinfo{journal}{Journal of Physics B: Atomic, Molecular and Optical
  Physics} \bibinfo{volume}{37} (\bibinfo{year}{2004}) \bibinfo{pages}{S1}.
  \URLprefix \url{https://dx.doi.org/10.1088/0953-4075/37/7/051}.
  \DOIprefix\doi{10.1088/0953-4075/37/7/051}.
\bibitem[{Tomonaga(1950)}]{tomonaga1950remarks}
\bibinfo{author}{S.-i. Tomonaga},
\newblock \bibinfo{title}{{Remarks on Bloch's Method of Sound Waves applied to
  Many-Fermion Problems}},
\newblock \bibinfo{journal}{Progress of Theoretical Physics}
  \bibinfo{volume}{5} (\bibinfo{year}{1950}) \bibinfo{pages}{544--569}.
  \URLprefix \url{https://doi.org/10.1143/ptp/5.4.544}.
  \DOIprefix\doi{10.1143/ptp/5.4.544}.
  \href{http://arxiv.org/abs/https://academic.oup.com/ptp/article-pdf/5/4/544/5430161/5-4-544.pdf}{{\tt
  arXiv:https://academic.oup.com/ptp/article-pdf/5/4/544/5430161/5-4-544.pdf}}.
\bibitem[{{Mattis} and {Lieb}(1965)}]{mattis1965exact}
\bibinfo{author}{D.~C. {Mattis}}, \bibinfo{author}{E.~H. {Lieb}},
\newblock \bibinfo{title}{{Exact Solution of a Many-Fermion System and Its
  Associated Boson Field}},
\newblock \bibinfo{journal}{Journal of Mathematical Physics}
  \bibinfo{volume}{6} (\bibinfo{year}{1965}) \bibinfo{pages}{304--312}.
  \DOIprefix\doi{10.1063/1.1704281}.
\bibitem[{Luther and Peschel(1974)}]{luther1974single}
\bibinfo{author}{A.~Luther}, \bibinfo{author}{I.~Peschel},
\newblock \bibinfo{title}{{Single-particle states, Kohn anomaly, and pairing
  fluctuations in one dimension}},
\newblock \bibinfo{journal}{Physical Review B} \bibinfo{volume}{9}
  (\bibinfo{year}{1974}) \bibinfo{pages}{2911--2919}. \URLprefix
  \url{https://link.aps.org/doi/10.1103/PhysRevB.9.2911}.
  \DOIprefix\doi{10.1103/PhysRevB.9.2911}.
\bibitem[{Efetov and Larkin(1976)}]{efetov1976correlation}
\bibinfo{author}{K.~Efetov}, \bibinfo{author}{A.~Larkin},
\newblock \bibinfo{title}{Correlation functions in one-dimensional systems with
  a strong interaction},
\newblock \bibinfo{journal}{Sov. Phys. JETP} \bibinfo{volume}{42}
  (\bibinfo{year}{1976}) \bibinfo{pages}{11}.
\bibitem[{Abrikosov and Silverman(1964)}]{abrikosov1964methods}
\bibinfo{author}{A.~A. Abrikosov}, \bibinfo{author}{R.~A. Silverman},
\newblock \bibinfo{title}{{Methods of Quantum Field Theory in Statistical
  Physics}},
\newblock \bibinfo{journal}{Mathematical Foundations of Quantum Field Theory}
  (\bibinfo{year}{1964}). \URLprefix
  \url{https://api.semanticscholar.org/CorpusID:115314524}.
\bibitem[{Bruun et~al.(1999)Bruun, Castin, Dum, and Burnett}]{bruun1999BCS}
\bibinfo{author}{G.~Bruun}, \bibinfo{author}{Y.~Castin},
  \bibinfo{author}{R.~Dum}, \bibinfo{author}{K.~Burnett},
\newblock \bibinfo{title}{{BCS theory for trapped ultracold fermions}},
\newblock \bibinfo{journal}{The European Physical Journal D - Atomic,
  Molecular, Optical and Plasma Physics} \bibinfo{volume}{7}
  (\bibinfo{year}{1999}) \bibinfo{pages}{433--439}. \URLprefix
  \url{https://doi.org/10.1007/s100530050587}.
  \DOIprefix\doi{10.1007/s100530050587}.
\bibitem[{Grasso and Urban(2003)}]{grasso2003hartree}
\bibinfo{author}{M.~Grasso}, \bibinfo{author}{M.~Urban},
\newblock \bibinfo{title}{{Hartree-Fock-Bogoliubov theory versus local-density
  approximation for superfluid trapped fermionic atoms}},
\newblock \bibinfo{journal}{Physical Review A} \bibinfo{volume}{68}
  (\bibinfo{year}{2003}) \bibinfo{pages}{033610}. \URLprefix
  \url{https://link.aps.org/doi/10.1103/PhysRevA.68.033610}.
  \DOIprefix\doi{10.1103/PhysRevA.68.033610}.
\bibitem[{Simonucci et~al.(2013)Simonucci, Pieri, and
  Strinati}]{simonucci2013temperature}
\bibinfo{author}{S.~Simonucci}, \bibinfo{author}{P.~Pieri},
  \bibinfo{author}{G.~C. Strinati},
\newblock \bibinfo{title}{{Temperature dependence of a vortex in a superfluid
  Fermi gas}},
\newblock \bibinfo{journal}{Physical Review B} \bibinfo{volume}{87}
  (\bibinfo{year}{2013}) \bibinfo{pages}{214507}. \URLprefix
  \url{https://link.aps.org/doi/10.1103/PhysRevB.87.214507}.
  \DOIprefix\doi{10.1103/PhysRevB.87.214507}.
\bibitem[{Gorkov(1959)}]{gorkov1959microscopic}
\bibinfo{author}{L.~P. Gorkov},
\newblock \bibinfo{title}{{Microscopic derivation of the Ginzburg-Landau
  equations in the theory of superconductivity}},
\newblock \bibinfo{journal}{Sov. Phys.—JETP} \bibinfo{volume}{9}
  (\bibinfo{year}{1959}) \bibinfo{pages}{1364–1367}.
\bibitem[{Pieri and Strinati(2003)}]{pieri2003derivation}
\bibinfo{author}{P.~Pieri}, \bibinfo{author}{G.~C. Strinati},
\newblock \bibinfo{title}{{Derivation of the Gross-Pitaevskii Equation for
  Condensed Bosons from the Bogoliubov--de Gennes Equations for Superfluid
  Fermions}},
\newblock \bibinfo{journal}{Physical Review Letters} \bibinfo{volume}{91}
  (\bibinfo{year}{2003}) \bibinfo{pages}{030401}. \URLprefix
  \url{https://link.aps.org/doi/10.1103/PhysRevLett.91.030401}.
  \DOIprefix\doi{10.1103/PhysRevLett.91.030401}.
\bibitem[{Spuntarelli et~al.(2010)Spuntarelli, Pieri, and
  Strinati}]{spuntarelli2010solution}
\bibinfo{author}{A.~Spuntarelli}, \bibinfo{author}{P.~Pieri},
  \bibinfo{author}{G.~Strinati},
\newblock \bibinfo{title}{{Solution of the Bogoliubov–de Gennes equations at
  zero temperature throughout the BCS–BEC crossover: Josephson and related
  effects}},
\newblock \bibinfo{journal}{Physics Reports} \bibinfo{volume}{488}
  (\bibinfo{year}{2010}) \bibinfo{pages}{111--167}. \URLprefix
  \url{https://www.sciencedirect.com/science/article/pii/S0370157309002890}.
  \DOIprefix\doi{https://doi.org/10.1016/j.physrep.2009.12.005}.
\bibitem[{Spuntarelli et~al.(2007)Spuntarelli, Pieri, and
  Strinati}]{spuntarelli2007josephson}
\bibinfo{author}{A.~Spuntarelli}, \bibinfo{author}{P.~Pieri},
  \bibinfo{author}{G.~C. Strinati},
\newblock \bibinfo{title}{{Josephson Effect throughout the BCS-BEC Crossover}},
\newblock \bibinfo{journal}{Physical Review Letters} \bibinfo{volume}{99}
  (\bibinfo{year}{2007}) \bibinfo{pages}{040401}. \URLprefix
  \url{https://link.aps.org/doi/10.1103/PhysRevLett.99.040401}.
  \DOIprefix\doi{10.1103/PhysRevLett.99.040401}.
\bibitem[{Piselli et~al.(2020)Piselli, Simonucci, and
  Strinati}]{piselli2020josephson}
\bibinfo{author}{V.~Piselli}, \bibinfo{author}{S.~Simonucci},
  \bibinfo{author}{G.~C. Strinati},
\newblock \bibinfo{title}{{Josephson effect at finite temperature along the
  BCS-BEC crossover}},
\newblock \bibinfo{journal}{Physical Review B} \bibinfo{volume}{102}
  (\bibinfo{year}{2020}) \bibinfo{pages}{144517}. \URLprefix
  \url{https://link.aps.org/doi/10.1103/PhysRevB.102.144517}.
  \DOIprefix\doi{10.1103/PhysRevB.102.144517}.
\bibitem[{Leggett(1980)}]{leggett1980diatomic}
\bibinfo{author}{A.~Leggett},
\newblock \bibinfo{title}{{Diatomic molecules and Cooper pairs}},
\newblock in: \bibinfo{booktitle}{, Modern Trends in the Theory of Condensed
  Matter}, \bibinfo{publisher}{Springer-Verlag}, \bibinfo{year}{1980}.
\bibitem[{Nozi{\`e}res and Schmitt-Rink(1985)}]{nozieres1985bose}
\bibinfo{author}{P.~Nozi{\`e}res}, \bibinfo{author}{S.~Schmitt-Rink},
\newblock \bibinfo{title}{Bose condensation in an attractive fermion gas: From
  weak to strong coupling superconductivity},
\newblock \bibinfo{journal}{Journal of Low Temperature Physics}
  \bibinfo{volume}{59} (\bibinfo{year}{1985}) \bibinfo{pages}{195--211}.
  \URLprefix \url{https://doi.org/10.1007/BF00683774}.
  \DOIprefix\doi{10.1007/BF00683774}.
\bibitem[{S\'a~de Melo et~al.(1993)S\'a~de Melo, Randeria, and
  Engelbrecht}]{sademelo1993bose}
\bibinfo{author}{C.~A.~R. S\'a~de Melo}, \bibinfo{author}{M.~Randeria},
  \bibinfo{author}{J.~R. Engelbrecht},
\newblock \bibinfo{title}{{Crossover from BCS to Bose superconductivity:
  Transition temperature and time-dependent Ginzburg-Landau theory}},
\newblock \bibinfo{journal}{Physical Review Letters} \bibinfo{volume}{71}
  (\bibinfo{year}{1993}) \bibinfo{pages}{3202--3205}. \URLprefix
  \url{https://link.aps.org/doi/10.1103/PhysRevLett.71.3202}.
  \DOIprefix\doi{10.1103/PhysRevLett.71.3202}.
\bibitem[{Piselli et~al.(2023)Piselli, Pisani, and
  Strinati}]{piselli2023josephson}
\bibinfo{author}{V.~Piselli}, \bibinfo{author}{L.~Pisani},
  \bibinfo{author}{G.~C. Strinati},
\newblock \bibinfo{title}{{Josephson current flowing through a nontrivial
  geometry: Role of pairing fluctuations across the BCS-BEC crossove}r},
\newblock \bibinfo{journal}{Physical Review B} \bibinfo{volume}{108}
  (\bibinfo{year}{2023}) \bibinfo{pages}{214504}. \URLprefix
  \url{https://link.aps.org/doi/10.1103/PhysRevB.108.214504}.
  \DOIprefix\doi{10.1103/PhysRevB.108.214504}.
\bibitem[{Pisani et~al.(2023)Pisani, Piselli, and
  Strinati}]{pisani2023inclusion}
\bibinfo{author}{L.~Pisani}, \bibinfo{author}{V.~Piselli},
  \bibinfo{author}{G.~C. Strinati},
\newblock \bibinfo{title}{Inclusion of pairing fluctuations in the differential
  equation for the gap parameter for superfluid fermions in the presence of
  nontrivial spatial constraints},
\newblock \bibinfo{journal}{Physical Review B} \bibinfo{volume}{108}
  (\bibinfo{year}{2023}) \bibinfo{pages}{214503}. \URLprefix
  \url{https://link.aps.org/doi/10.1103/PhysRevB.108.214503}.
  \DOIprefix\doi{10.1103/PhysRevB.108.214503}.
\bibitem[{Pisani et~al.(2024)Pisani, Piselli, and
  Strinati}]{pisani2024critical}
\bibinfo{author}{L.~Pisani}, \bibinfo{author}{V.~Piselli},
  \bibinfo{author}{G.~C. Strinati},
\newblock \bibinfo{title}{{Critical current throughout the BCS-BEC crossover
  with the inclusion of pairing fluctuations}},
\newblock \bibinfo{journal}{Physical Review A} \bibinfo{volume}{109}
  (\bibinfo{year}{2024}) \bibinfo{pages}{033306}. \URLprefix
  \url{https://link.aps.org/doi/10.1103/PhysRevA.109.033306}.
  \DOIprefix\doi{10.1103/PhysRevA.109.033306}.
\bibitem[{Boulet et~al.(2022)Boulet, Wlaz\l{}owski, and
  Magierski}]{boulet2022local}
\bibinfo{author}{A.~Boulet}, \bibinfo{author}{G.~Wlaz\l{}owski},
  \bibinfo{author}{P.~Magierski},
\newblock \bibinfo{title}{{Local energy density functional for superfluid Fermi
  gases from effective field theory}},
\newblock \bibinfo{journal}{Physical Review A} \bibinfo{volume}{106}
  (\bibinfo{year}{2022}) \bibinfo{pages}{013306}. \URLprefix
  \url{https://link.aps.org/doi/10.1103/PhysRevA.106.013306}.
  \DOIprefix\doi{10.1103/PhysRevA.106.013306}.
\bibitem[{Astrakharchik et~al.(2004)Astrakharchik, Boronat, Casulleras, and
  Giorgini}]{astrakharchik2004equation}
\bibinfo{author}{G.~E. Astrakharchik}, \bibinfo{author}{J.~Boronat},
  \bibinfo{author}{J.~Casulleras}, \bibinfo{author}{S.~Giorgini},
\newblock \bibinfo{title}{{Equation of State of a Fermi Gas in the BEC-BCS
  Crossover: A Quantum Monte Carlo Study}},
\newblock \bibinfo{journal}{Physical Review Letters} \bibinfo{volume}{93}
  (\bibinfo{year}{2004}) \bibinfo{pages}{200404}. \URLprefix
  \url{https://link.aps.org/doi/10.1103/PhysRevLett.93.200404}.
  \DOIprefix\doi{10.1103/PhysRevLett.93.200404}.
\bibitem[{Gandolfi(2014)}]{gandolfi2014quantum}
\bibinfo{author}{S.~Gandolfi},
\newblock \bibinfo{title}{{Quantum Monte Carlo study of strongly interacting
  Fermi gases}},
\newblock \bibinfo{journal}{Journal of Physics: Conference Series}
  \bibinfo{volume}{529} (\bibinfo{year}{2014}) \bibinfo{pages}{012011}.
  \URLprefix \url{https://dx.doi.org/10.1088/1742-6596/529/1/012011}.
  \DOIprefix\doi{10.1088/1742-6596/529/1/012011}.
\bibitem[{Strinati et~al.(2018)Strinati, Pieri, Röpke, Schuck, and
  Urban}]{strinati2018BCS-BEC}
\bibinfo{author}{G.~C. Strinati}, \bibinfo{author}{P.~Pieri},
  \bibinfo{author}{G.~Röpke}, \bibinfo{author}{P.~Schuck},
  \bibinfo{author}{M.~Urban},
\newblock \bibinfo{title}{{The BCS–BEC crossover: From ultra-cold Fermi gases
  to nuclear systems}},
\newblock \bibinfo{journal}{Physics Reports} \bibinfo{volume}{738}
  (\bibinfo{year}{2018}) \bibinfo{pages}{1--76}. \URLprefix
  \url{https://www.sciencedirect.com/science/article/pii/S0370157318300267}.
  \DOIprefix\doi{https://doi.org/10.1016/j.physrep.2018.02.004},
  \bibinfo{note}{the BCS–BEC crossover: From ultra-cold Fermi gases to
  nuclear systems}.
\bibitem[{Montorsi(1991)}]{rasetti1991hubbard}
\bibinfo{author}{A.~Montorsi}, \bibinfo{title}{{The Hubbard Model}},
  \bibinfo{publisher}{WORLD SCIENTIFIC}, \bibinfo{year}{1991}. \URLprefix
  \url{https://www.worldscientific.com/doi/abs/10.1142/1346}.
  \DOIprefix\doi{10.1142/1346}.
  \href{http://arxiv.org/abs/https://www.worldscientific.com/doi/pdf/10.1142/1346}{{\tt
  arXiv:https://www.worldscientific.com/doi/pdf/10.1142/1346}}.
\bibitem[{Rasetti(1992)}]{montorsi1992hubbard}
\bibinfo{author}{M.~Rasetti}, \bibinfo{title}{{The Hubbard Model}},
  \bibinfo{publisher}{WORLD SCIENTIFIC}, \bibinfo{year}{1992}. \URLprefix
  \url{https://www.worldscientific.com/doi/abs/10.1142/1346}.
  \DOIprefix\doi{10.1142/1346}.
  \href{http://arxiv.org/abs/https://www.worldscientific.com/doi/pdf/10.1142/1346}{{\tt
  arXiv:https://www.worldscientific.com/doi/pdf/10.1142/1346}}.
\bibitem[{Tarruell and Sanchez-Palencia(2018)}]{tarruell2918quantum}
\bibinfo{author}{L.~Tarruell}, \bibinfo{author}{L.~Sanchez-Palencia},
\newblock \bibinfo{title}{{Quantum simulation of the Hubbard model with
  ultracold fermions in optical lattices}},
\newblock \bibinfo{journal}{Comptes Rendus Physique} \bibinfo{volume}{19}
  (\bibinfo{year}{2018}) \bibinfo{pages}{365--393}. \URLprefix
  \url{https://www.sciencedirect.com/science/article/pii/S1631070518300926}.
  \DOIprefix\doi{https://doi.org/10.1016/j.crhy.2018.10.013},
  \bibinfo{note}{quantum simulation / Simulation quantique}.
\bibitem[{Esslinger(2010)}]{esslinger2010fermi}
\bibinfo{author}{T.~Esslinger},
\newblock \bibinfo{title}{{Fermi-Hubbard Physics with Atoms in an Optical
  Lattice}},
\newblock \bibinfo{journal}{Annual Review of Condensed Matter Physics}
  \bibinfo{volume}{1} (\bibinfo{year}{2010}) \bibinfo{pages}{129--152}.
  \URLprefix \url{https://doi.org/10.1146/annurev-conmatphys-070909-104059}.
  \DOIprefix\doi{10.1146/annurev-conmatphys-070909-104059}.
  \href{http://arxiv.org/abs/https://doi.org/10.1146/annurev-conmatphys-070909-104059}{{\tt
  arXiv:https://doi.org/10.1146/annurev-conmatphys-070909-104059}}.
\bibitem[{Lewenstein et~al.(2007)Lewenstein, Sanpera, Ahufinger, Damski,
  Sen(De), and Sen}]{lewenstein2007ultracold}
\bibinfo{author}{M.~Lewenstein}, \bibinfo{author}{A.~Sanpera},
  \bibinfo{author}{V.~Ahufinger}, \bibinfo{author}{B.~Damski},
  \bibinfo{author}{A.~Sen(De)}, \bibinfo{author}{U.~Sen},
\newblock \bibinfo{title}{Ultracold atomic gases in optical lattices: mimicking
  condensed matter physics and beyond},
\newblock \bibinfo{journal}{Advances in Physics} \bibinfo{volume}{56}
  (\bibinfo{year}{2007}) \bibinfo{pages}{243--379}. \URLprefix
  \url{https://doi.org/10.1080/00018730701223200eprint = {
  https://doi.org/10.1080/00018730701223200}}.
  \DOIprefix\doi{10.1080/00018730701223200}.
\bibitem[{Jaksch and Zoller(2005)}]{jaksch2005cold}
\bibinfo{author}{D.~Jaksch}, \bibinfo{author}{P.~Zoller},
\newblock \bibinfo{title}{{The cold atom Hubbard toolbox}},
\newblock \bibinfo{journal}{Annals of Physics} \bibinfo{volume}{315}
  (\bibinfo{year}{2005}) \bibinfo{pages}{52--79}. \URLprefix
  \url{https://www.sciencedirect.com/science/article/pii/S0003491604001782}.
  \DOIprefix\doi{https://doi.org/10.1016/j.aop.2004.09.010},
  \bibinfo{note}{special Issue}.
\bibitem[{Gorshkov et~al.(2010)Gorshkov, Hermele, Gurarie, Xu, Julienne, Ye,
  Zoller, Demler, Lukin, and Rey}]{gorshkov2010two}
\bibinfo{author}{A.~V. Gorshkov}, \bibinfo{author}{M.~Hermele},
  \bibinfo{author}{V.~Gurarie}, \bibinfo{author}{C.~Xu}, \bibinfo{author}{P.~S.
  Julienne}, \bibinfo{author}{J.~Ye}, \bibinfo{author}{P.~Zoller},
  \bibinfo{author}{E.~Demler}, \bibinfo{author}{M.~D. Lukin},
  \bibinfo{author}{A.~M. Rey},
\newblock \bibinfo{title}{{Two-orbital SU(N) magnetism with ultracold
  alkaline-earth atoms}},
\newblock \bibinfo{journal}{Nature Physics} \bibinfo{volume}{6}
  (\bibinfo{year}{2010}) \bibinfo{pages}{289--295}. \URLprefix
  \url{https://doi.org/10.1038/nphys1535}. \DOIprefix\doi{10.1038/nphys1535}.
\bibitem[{Cazalilla and Rey(2014)}]{cazalilla2014ultracold}
\bibinfo{author}{M.~A. Cazalilla}, \bibinfo{author}{A.~M. Rey},
\newblock \bibinfo{title}{{Ultracold Fermi gases with emergent SU(N)
  symmetry}},
\newblock \bibinfo{journal}{Reports on Progress in Physics}
  \bibinfo{volume}{77} (\bibinfo{year}{2014}) \bibinfo{pages}{124401}.
  \URLprefix \url{https://dx.doi.org/10.1088/0034-4885/77/12/124401}.
  \DOIprefix\doi{10.1088/0034-4885/77/12/124401}.
\bibitem[{Capponi et~al.(2016)Capponi, Lecheminant, and
  Totsuka}]{capponi2016phases}
\bibinfo{author}{S.~Capponi}, \bibinfo{author}{P.~Lecheminant},
  \bibinfo{author}{K.~Totsuka},
\newblock \bibinfo{title}{{Phases of one-dimensional SU(N) cold atomic Fermi
  gases—From molecular Luttinger liquids to topological phases}},
\newblock \bibinfo{journal}{Annals of Physics} \bibinfo{volume}{367}
  (\bibinfo{year}{2016}) \bibinfo{pages}{50--95}. \URLprefix
  \url{https://www.sciencedirect.com/science/article/pii/S0003491616000130}.
  \DOIprefix\doi{https://doi.org/10.1016/j.aop.2016.01.011}.
\bibitem[{Frahm and Schadschneider(1995)}]{frahm1995on}
\bibinfo{author}{H.~Frahm}, \bibinfo{author}{A.~Schadschneider},
  \bibinfo{title}{{On the Bethe Ansatz Soluble Degenerate Hubbard Model}},
  \bibinfo{publisher}{Springer US}, \bibinfo{address}{Boston, MA},
  \bibinfo{year}{1995}, pp. \bibinfo{pages}{21--28}. \URLprefix
  \url{https://doi.org/10.1007/978-1-4899-1042-4_2}.
  \DOIprefix\doi{10.1007/978-1-4899-1042-4_2}.
\bibitem[{Taie et~al.(2010)Taie, Takasu, Sugawa, Yamazaki, Tsujimoto, Murakami,
  and Takahashi}]{taie2010realization}
\bibinfo{author}{S.~Taie}, \bibinfo{author}{Y.~Takasu},
  \bibinfo{author}{S.~Sugawa}, \bibinfo{author}{R.~Yamazaki},
  \bibinfo{author}{T.~Tsujimoto}, \bibinfo{author}{R.~Murakami},
  \bibinfo{author}{Y.~Takahashi},
\newblock \bibinfo{title}{{Realization of a
  $\mathrm{SU}(2)\ifmmode\times\else\texttimes\fi{}\mathrm{SU}(6)$ System of
  Fermions in a Cold Atomic Gas}},
\newblock \bibinfo{journal}{Physical Review Letters} \bibinfo{volume}{105}
  (\bibinfo{year}{2010}) \bibinfo{pages}{190401}. \URLprefix
  \url{https://link.aps.org/doi/10.1103/PhysRevLett.105.190401}.
  \DOIprefix\doi{10.1103/PhysRevLett.105.190401}.
\bibitem[{Pagano et~al.(2014)Pagano, Mancini, Cappellini, Lombardi,
  Sch\"{a}fer, Hu, Liu, Catani, Sias, Inguscio, and Fallani}]{pagano2014one}
\bibinfo{author}{G.~Pagano}, \bibinfo{author}{M.~Mancini},
  \bibinfo{author}{G.~Cappellini}, \bibinfo{author}{P.~Lombardi},
  \bibinfo{author}{F.~Sch\"{a}fer}, \bibinfo{author}{H.~Hu},
  \bibinfo{author}{X.-J. Liu}, \bibinfo{author}{J.~Catani},
  \bibinfo{author}{C.~Sias}, \bibinfo{author}{M.~Inguscio},
  \bibinfo{author}{L.~Fallani},
\newblock \bibinfo{title}{A one-dimensional liquid of fermions with tunable
  spin},
\newblock \bibinfo{journal}{Nature Physics} \bibinfo{volume}{10}
  (\bibinfo{year}{2014}) \bibinfo{pages}{198–201}. \URLprefix
  \url{http://dx.doi.org/10.1038/nphys2878}. \DOIprefix\doi{10.1038/nphys2878}.
\bibitem[{Taie et~al.(2012)Taie, Yamazaki, Sugawa, and
  Takahashi}]{taie2012mott}
\bibinfo{author}{S.~Taie}, \bibinfo{author}{R.~Yamazaki},
  \bibinfo{author}{S.~Sugawa}, \bibinfo{author}{Y.~Takahashi},
\newblock \bibinfo{title}{{An SU(6) Mott insulator of an atomic Fermi gas
  realized by large-spin Pomeranchuk cooling}},
\newblock \bibinfo{journal}{Nature Physics} \bibinfo{volume}{8}
  (\bibinfo{year}{2012}) \bibinfo{pages}{825–830}. \URLprefix
  \url{http://dx.doi.org/10.1038/nphys2430}. \DOIprefix\doi{10.1038/nphys2430}.
\bibitem[{Sonderhouse et~al.(2020)Sonderhouse, Sanner, Hutson, Goban,
  Bilitewski, Yan, Milner, Rey, and Ye}]{sonderhouse2020thermodynamics}
\bibinfo{author}{L.~Sonderhouse}, \bibinfo{author}{C.~Sanner},
  \bibinfo{author}{R.~B. Hutson}, \bibinfo{author}{A.~Goban},
  \bibinfo{author}{T.~Bilitewski}, \bibinfo{author}{L.~Yan},
  \bibinfo{author}{W.~R. Milner}, \bibinfo{author}{A.~M. Rey},
  \bibinfo{author}{J.~Ye},
\newblock \bibinfo{title}{{Thermodynamics of a deeply degenerate
  SU(N)-symmetric Fermi gas}},
\newblock \bibinfo{journal}{Nature Physics} \bibinfo{volume}{16}
  (\bibinfo{year}{2020}) \bibinfo{pages}{1216–1221}. \URLprefix
  \url{http://dx.doi.org/10.1038/s41567-020-0986-6}.
  \DOIprefix\doi{10.1038/s41567-020-0986-6}.
\bibitem[{Hofrichter et~al.(2016)Hofrichter, Riegger, Scazza, H\"ofer,
  Fernandes, Bloch, and F\"olling}]{hofrichter2016direct}
\bibinfo{author}{C.~Hofrichter}, \bibinfo{author}{L.~Riegger},
  \bibinfo{author}{F.~Scazza}, \bibinfo{author}{M.~H\"ofer},
  \bibinfo{author}{D.~R. Fernandes}, \bibinfo{author}{I.~Bloch},
  \bibinfo{author}{S.~F\"olling},
\newblock \bibinfo{title}{{Direct Probing of the Mott Crossover in the
  $\mathrm{SU}(N)$ Fermi-Hubbard Model}},
\newblock \bibinfo{journal}{Physical Review X} \bibinfo{volume}{6}
  (\bibinfo{year}{2016}) \bibinfo{pages}{021030}. \URLprefix
  \url{https://link.aps.org/doi/10.1103/PhysRevX.6.021030}.
  \DOIprefix\doi{10.1103/PhysRevX.6.021030}.
\bibitem[{Yip and Ho(1999)}]{yip19999zero}
\bibinfo{author}{S.-K. Yip}, \bibinfo{author}{T.-L. Ho},
\newblock \bibinfo{title}{{Zero sound modes of dilute Fermi gases with
  arbitrary spin}},
\newblock \bibinfo{journal}{Physical Review A} \bibinfo{volume}{59}
  (\bibinfo{year}{1999}) \bibinfo{pages}{4653--4656}. \URLprefix
  \url{https://link.aps.org/doi/10.1103/PhysRevA.59.4653}.
  \DOIprefix\doi{10.1103/PhysRevA.59.4653}.
\bibitem[{Stamper-Kurn and Ueda(2013)}]{stamper2013spinor}
\bibinfo{author}{D.~M. Stamper-Kurn}, \bibinfo{author}{M.~Ueda},
\newblock \bibinfo{title}{{Spinor Bose gases: Symmetries, magnetism, and
  quantum dynamics}},
\newblock \bibinfo{journal}{Reviews of Modern Physics} \bibinfo{volume}{85}
  (\bibinfo{year}{2013}) \bibinfo{pages}{1191--1244}. \URLprefix
  \url{https://link.aps.org/doi/10.1103/RevModPhys.85.1191}.
  \DOIprefix\doi{10.1103/RevModPhys.85.1191}.
\bibitem[{Sutherland(1968)}]{sutherland1968further}
\bibinfo{author}{B.~Sutherland},
\newblock \bibinfo{title}{{Further Results for the Many-Body Problem in One
  Dimension}},
\newblock \bibinfo{journal}{Physical Review Letters} \bibinfo{volume}{20}
  (\bibinfo{year}{1968}) \bibinfo{pages}{98--100}. \URLprefix
  \url{https://link.aps.org/doi/10.1103/PhysRevLett.20.98}.
  \DOIprefix\doi{10.1103/PhysRevLett.20.98}.
\bibitem[{Naldesi et~al.(2023)Naldesi, Polo, Drummond, Dunjko, Amico, Minguzzi,
  and Olshanii}]{naldesi2023massive}
\bibinfo{author}{P.~Naldesi}, \bibinfo{author}{J.~Polo}, \bibinfo{author}{P.~D.
  Drummond}, \bibinfo{author}{V.~Dunjko}, \bibinfo{author}{L.~Amico},
  \bibinfo{author}{A.~Minguzzi}, \bibinfo{author}{M.~Olshanii},
\newblock \bibinfo{title}{{Massive particle interferometry with lattice
  solitons}},
\newblock \bibinfo{journal}{SciPost Physics} \bibinfo{volume}{15}
  (\bibinfo{year}{2023}) \bibinfo{pages}{187}. \URLprefix
  \url{https://scipost.org/10.21468/SciPostPhys.15.5.187}.
  \DOIprefix\doi{10.21468/SciPostPhys.15.5.187}.
\bibitem[{Sutherland(2004)}]{sutherland2004beautiful}
\bibinfo{author}{B.~Sutherland}, \bibinfo{title}{{Beautiful Models}},
  \bibinfo{publisher}{WORLD SCIENTIFIC}, \bibinfo{year}{2004}. \URLprefix
  \url{https://www.worldscientific.com/doi/abs/10.1142/5552}.
  \DOIprefix\doi{10.1142/5552}.
  \href{http://arxiv.org/abs/https://www.worldscientific.com/doi/pdf/10.1142/5552}{{\tt
  arXiv:https://www.worldscientific.com/doi/pdf/10.1142/5552}}.
\bibitem[{Korepin et~al.(1993)Korepin, Bogoliubov, and
  Izergin}]{korepin1993quantum}
\bibinfo{author}{V.~E. Korepin}, \bibinfo{author}{N.~M. Bogoliubov},
  \bibinfo{author}{A.~G. Izergin}, \bibinfo{title}{{Quantum Inverse Scattering
  Method and Correlation Functions}}, Cambridge Monographs on Mathematical
  Physics, \bibinfo{publisher}{Cambridge University Press},
  \bibinfo{year}{1993}. \DOIprefix\doi{10.1017/CBO9780511628832}.
\bibitem[{Faddeev(1996)}]{faddeev1996algebraic}
\bibinfo{author}{L.~D. Faddeev}, \bibinfo{title}{{How Algebraic Bethe Ansatz
  works for integrable model}}, \bibinfo{year}{1996}.
  \href{http://arxiv.org/abs/hep-th/9605187}{{\tt arXiv:hep-th/9605187}}.
\bibitem[{Deguchi et~al.(2000)Deguchi, Essler, Göhmann, Klümper, Korepin, and
  Kusakabe}]{deguchi2000thermodynamics}
\bibinfo{author}{T.~Deguchi}, \bibinfo{author}{F.~Essler},
  \bibinfo{author}{F.~Göhmann}, \bibinfo{author}{A.~Klümper},
  \bibinfo{author}{V.~Korepin}, \bibinfo{author}{K.~Kusakabe},
\newblock \bibinfo{title}{{Thermodynamics and excitations of the
  one-dimensional Hubbard model}},
\newblock \bibinfo{journal}{Physics Reports} \bibinfo{volume}{331}
  (\bibinfo{year}{2000}) \bibinfo{pages}{197--281}. \URLprefix
  \url{https://www.sciencedirect.com/science/article/pii/S0370157300000107}.
  \DOIprefix\doi{https://doi.org/10.1016/S0370-1573(00)00010-7}.
\bibitem[{Lieb and Wu(1968)}]{lieb1968absence}
\bibinfo{author}{E.~H. Lieb}, \bibinfo{author}{F.~Y. Wu},
\newblock \bibinfo{title}{{Absence of Mott Transition in an Exact Solution of
  the Short-Range, One-Band Model in One Dimension}},
\newblock \bibinfo{journal}{Physical Review Letters} \bibinfo{volume}{20}
  (\bibinfo{year}{1968}) \bibinfo{pages}{1445--1448}. \URLprefix
  \url{https://link.aps.org/doi/10.1103/PhysRevLett.20.1445}.
  \DOIprefix\doi{10.1103/PhysRevLett.20.1445}.
\bibitem[{Gaudin(1967)}]{gaudin1967un}
\bibinfo{author}{M.~Gaudin},
\newblock \bibinfo{title}{Un systeme a une dimension de fermions en
  interaction},
\newblock \bibinfo{journal}{Physics Letters A} \bibinfo{volume}{24}
  (\bibinfo{year}{1967}) \bibinfo{pages}{55--56}. \URLprefix
  \url{https://www.sciencedirect.com/science/article/pii/0375960167901934}.
  \DOIprefix\doi{https://doi.org/10.1016/0375-9601(67)90193-4}.
\bibitem[{Yang(1967)}]{yang1967some}
\bibinfo{author}{C.~N. Yang},
\newblock \bibinfo{title}{{Some Exact Results for the Many-Body Problem in one
  Dimension with Repulsive Delta-Function Interaction}},
\newblock \bibinfo{journal}{Physical Review Letters} \bibinfo{volume}{19}
  (\bibinfo{year}{1967}) \bibinfo{pages}{1312--1315}. \URLprefix
  \url{https://link.aps.org/doi/10.1103/PhysRevLett.19.1312}.
  \DOIprefix\doi{10.1103/PhysRevLett.19.1312}.
\bibitem[{Haldane(1980)}]{haldane1980solidification}
\bibinfo{author}{F.~Haldane},
\newblock \bibinfo{title}{{“Solidification” in a soluble model of bosons on
  a one-dimensional lattice: The “Boson-Hubbard chain”}, journal = {Physics
  Letters A}} \bibinfo{volume}{80} (\bibinfo{year}{1980})
  \bibinfo{pages}{281--283}. \URLprefix
  \url{https://www.sciencedirect.com/science/article/pii/0375960180900225}.
  \DOIprefix\doi{https://doi.org/10.1016/0375-9601(80)90022-5}.
\bibitem[{Choy(1980)}]{choy1980some}
\bibinfo{author}{T.~Choy},
\newblock \bibinfo{title}{{Some exact results for a degenerate Hubbard model in
  one dimension}},
\newblock \bibinfo{journal}{Physics Letters A} \bibinfo{volume}{80}
  (\bibinfo{year}{1980}) \bibinfo{pages}{49--52}. \URLprefix
  \url{https://www.sciencedirect.com/science/article/pii/037596018090451X}.
  \DOIprefix\doi{https://doi.org/10.1016/0375-9601(80)90451-X}.
\bibitem[{Choy and Haldane(1982)}]{choy1982failure}
\bibinfo{author}{T.~Choy}, \bibinfo{author}{F.~Haldane},
\newblock \bibinfo{title}{{Failure of Bethe-Ansatz solutions of generalisations
  of the Hubbard chain to arbitrary permutation symmetry}},
\newblock \bibinfo{journal}{Physics Letters A} \bibinfo{volume}{90}
  (\bibinfo{year}{1982}) \bibinfo{pages}{83--84}. \URLprefix
  \url{https://www.sciencedirect.com/science/article/pii/0375960182900573}.
  \DOIprefix\doi{https://doi.org/10.1016/0375-9601(82)90057-3}.
\bibitem[{Schlottmann(1991)}]{schlottmann1991spin}
\bibinfo{author}{P.~Schlottmann},
\newblock \bibinfo{title}{{Spin and charge excitations of the degenerate
  Hubbard model in one dimension}},
\newblock \bibinfo{journal}{Physical Review B} \bibinfo{volume}{43}
  (\bibinfo{year}{1991}) \bibinfo{pages}{3101--3116}. \URLprefix
  \url{https://link.aps.org/doi/10.1103/PhysRevB.43.3101}.
  \DOIprefix\doi{10.1103/PhysRevB.43.3101}.
\bibitem[{Gu and Yang(1989)}]{gu1989a}
\bibinfo{author}{C.~H. Gu}, \bibinfo{author}{C.~N. Yang},
\newblock \bibinfo{title}{{A one-{dimensional $N$} Fermion problem with
  {factorized $S$} matrix}},
\newblock \bibinfo{journal}{Communications in Mathematical Physics}
  \bibinfo{volume}{122} (\bibinfo{year}{1989}) \bibinfo{pages}{105--116}.
  \URLprefix \url{https://doi.org/10.1007/bf01221409}.
  \DOIprefix\doi{10.1007/bf01221409}.
\bibitem[{Lai(1974)}]{lai1974lattice}
\bibinfo{author}{C.~K. Lai},
\newblock \bibinfo{title}{{Lattice gas with nearest‐neighbor interaction in
  one dimension with arbitrary statistics}},
\newblock \bibinfo{journal}{Journal of Mathematical Physics}
  \bibinfo{volume}{15} (\bibinfo{year}{1974}) \bibinfo{pages}{1675--1676}.
  \URLprefix \url{https://doi.org/10.1063/1.1666522}.
  \DOIprefix\doi{10.1063/1.1666522}.
  \href{http://arxiv.org/abs/https://pubs.aip.org/aip/jmp/article-pdf/15/10/1675/8146940/1675\_1\_online.pdf}{{\tt
  arXiv:https://pubs.aip.org/aip/jmp/article-pdf/15/10/1675/8146940/1675\_1\_online.pdf}}.
\bibitem[{Sutherland(1975)}]{sutherland1975model}
\bibinfo{author}{B.~Sutherland},
\newblock \bibinfo{title}{Model for a multicomponent quantum system},
\newblock \bibinfo{journal}{Physical Review B} \bibinfo{volume}{12}
  (\bibinfo{year}{1975}) \bibinfo{pages}{3795--3805}. \URLprefix
  \url{https://link.aps.org/doi/10.1103/PhysRevB.12.3795}.
  \DOIprefix\doi{10.1103/PhysRevB.12.3795}.
\bibitem[{Takahashi(1999)}]{takahashi1999thermodynamics}
\bibinfo{author}{M.~Takahashi}, \bibinfo{title}{{Thermodynamics of
  One-Dimensional Solvable Models}}, \bibinfo{publisher}{Cambridge University
  Press}, \bibinfo{year}{1999}. \DOIprefix\doi{10.1017/CBO9780511524332}.
\bibitem[{Amico et~al.(1998)Amico, Osterloh, and Eckern}]{amico1998one}
\bibinfo{author}{L.~Amico}, \bibinfo{author}{A.~Osterloh},
  \bibinfo{author}{U.~Eckern},
\newblock \bibinfo{title}{{One-dimensional $\mathrm{XXZ}$ model for particles
  obeying fractional statistics}},
\newblock \bibinfo{journal}{Physical Review B} \bibinfo{volume}{58}
  (\bibinfo{year}{1998}) \bibinfo{pages}{R1703--R1706}. \URLprefix
  \url{https://link.aps.org/doi/10.1103/PhysRevB.58.R1703}.
  \DOIprefix\doi{10.1103/PhysRevB.58.R1703}.
\bibitem[{Takahashi(1970)}]{takahashi1970many}
\bibinfo{author}{M.~Takahashi},
\newblock \bibinfo{title}{{Many-Body Problem of Attractive Fermions with
  Arbitrary Spin in One Dimension}},
\newblock \bibinfo{journal}{Progress of Theoretical Physics}
  \bibinfo{volume}{44} (\bibinfo{year}{1970}) \bibinfo{pages}{899--904}.
  \URLprefix \url{https://doi.org/10.1143/PTP.44.899}.
  \DOIprefix\doi{10.1143/PTP.44.899}.
  \href{http://arxiv.org/abs/https://academic.oup.com/ptp/article-pdf/44/4/899/5386649/44-4-899.pdf}{{\tt
  arXiv:https://academic.oup.com/ptp/article-pdf/44/4/899/5386649/44-4-899.pdf}}.
\bibitem[{Takahashi(1971)}]{takashi1971one}
\bibinfo{author}{M.~Takahashi},
\newblock \bibinfo{title}{{One-Dimensional Electron Gas with Delta-Function
  Interaction at Finite Temperature}},
\newblock \bibinfo{journal}{Progress of Theoretical Physics}
  \bibinfo{volume}{46} (\bibinfo{year}{1971}) \bibinfo{pages}{1388--1406}.
  \URLprefix \url{https://doi.org/10.1143/PTP.46.1388}.
  \DOIprefix\doi{10.1143/PTP.46.1388}.
  \href{http://arxiv.org/abs/https://academic.oup.com/ptp/article-pdf/46/5/1388/5269149/46-5-1388.pdf}{{\tt
  arXiv:https://academic.oup.com/ptp/article-pdf/46/5/1388/5269149/46-5-1388.pdf}}.
\bibitem[{Oelkers and Links(2007)}]{oelkers2007ground}
\bibinfo{author}{N.~Oelkers}, \bibinfo{author}{J.~Links},
\newblock \bibinfo{title}{{Ground-state properties of the attractive
  one-dimensional Bose-Hubbard model}},
\newblock \bibinfo{journal}{Physical Review B} \bibinfo{volume}{75}
  (\bibinfo{year}{2007}) \bibinfo{pages}{115119}. \URLprefix
  \url{https://link.aps.org/doi/10.1103/PhysRevB.75.115119}.
  \DOIprefix\doi{10.1103/PhysRevB.75.115119}.
\bibitem[{SCHULZ(1991)}]{schulz1991correlated}
\bibinfo{author}{H.~SCHULZ},
\newblock \bibinfo{title}{{Correlated Fermions In One Dimension}},
\newblock \bibinfo{journal}{International Journal of Modern Physics B}
  \bibinfo{volume}{05} (\bibinfo{year}{1991}) \bibinfo{pages}{57--74}.
  \URLprefix \url{https://doi.org/10.1142/S0217979291000055}.
  \DOIprefix\doi{10.1142/S0217979291000055}.
  \href{http://arxiv.org/abs/https://doi.org/10.1142/S0217979291000055}{{\tt
  arXiv:https://doi.org/10.1142/S0217979291000055}}.
\bibitem[{Loss(1992)}]{loss1992parity}
\bibinfo{author}{D.~Loss},
\newblock \bibinfo{title}{{Parity effects in a Luttinger liquid: Diamagnetic
  and paramagnetic ground states}},
\newblock \bibinfo{journal}{Physical Review Letters} \bibinfo{volume}{69}
  (\bibinfo{year}{1992}) \bibinfo{pages}{343--346}. \URLprefix
  \url{https://link.aps.org/doi/10.1103/PhysRevLett.69.343}.
  \DOIprefix\doi{10.1103/PhysRevLett.69.343}.
\bibitem[{Schmeltzer(1993)}]{schmeltzer1993persistent}
\bibinfo{author}{D.~Schmeltzer},
\newblock \bibinfo{title}{{Persistent current for a Luttinger liquid}},
\newblock \bibinfo{journal}{Physical Review B} \bibinfo{volume}{47}
  (\bibinfo{year}{1993}) \bibinfo{pages}{7591--7593}. \URLprefix
  \url{https://link.aps.org/doi/10.1103/PhysRevB.47.7591}.
  \DOIprefix\doi{10.1103/PhysRevB.47.7591}.
\bibitem[{Waintal et~al.(2008)Waintal, Fleury, Kazymyrenko, Houzet,
  Schmitteckert, and Weinmann}]{waintal2008persistent}
\bibinfo{author}{X.~Waintal}, \bibinfo{author}{G.~Fleury},
  \bibinfo{author}{K.~Kazymyrenko}, \bibinfo{author}{M.~Houzet},
  \bibinfo{author}{P.~Schmitteckert}, \bibinfo{author}{D.~Weinmann},
\newblock \bibinfo{title}{{Persistent Currents in One Dimension: The
  Counterpart of Leggett's Theorem}},
\newblock \bibinfo{journal}{Physical Review Letters} \bibinfo{volume}{101}
  (\bibinfo{year}{2008}) \bibinfo{pages}{106804}. \URLprefix
  \url{https://link.aps.org/doi/10.1103/PhysRevLett.101.106804}.
  \DOIprefix\doi{10.1103/PhysRevLett.101.106804}.
\bibitem[{Kusmartsev et~al.(1994)Kusmartsev, Weisz, Kishore, and
  Takahashi}]{kusmartsev1994strong}
\bibinfo{author}{F.~V. Kusmartsev}, \bibinfo{author}{J.~F. Weisz},
  \bibinfo{author}{R.~Kishore}, \bibinfo{author}{M.~Takahashi},
\newblock \bibinfo{title}{{Strong correlations versus U-center pairing and
  fractional Aharonov-Bohm effect}},
\newblock \bibinfo{journal}{Physical Review B} \bibinfo{volume}{49}
  (\bibinfo{year}{1994}) \bibinfo{pages}{16234--16247}. \URLprefix
  \url{https://link.aps.org/doi/10.1103/PhysRevB.49.16234}.
  \DOIprefix\doi{10.1103/PhysRevB.49.16234}.
\bibitem[{Pradhan et~al.(2023)Pradhan, Kumar, Kanamoto, Dey, Bhattacharya, and
  Mishra}]{pradhan2023cavity}
\bibinfo{author}{N.~Pradhan}, \bibinfo{author}{P.~Kumar},
  \bibinfo{author}{R.~Kanamoto}, \bibinfo{author}{T.~N. Dey},
  \bibinfo{author}{M.~Bhattacharya}, \bibinfo{author}{P.~K. Mishra},
\newblock \bibinfo{title}{Cavity optomechanical detection of persistent
  currents and solitons in a bosonic ring condensate},
\newblock \bibinfo{journal}{arXiv:2306.06720}  (\bibinfo{year}{2023}).
\bibitem[{Pradhan et~al.(2024{\natexlab{a}})Pradhan, Kumar, Kanamoto, Dey,
  Bhattacharya, and Mishra}]{pradhan2024cavity}
\bibinfo{author}{N.~Pradhan}, \bibinfo{author}{P.~Kumar},
  \bibinfo{author}{R.~Kanamoto}, \bibinfo{author}{T.~N. Dey},
  \bibinfo{author}{M.~Bhattacharya}, \bibinfo{author}{P.~K. Mishra},
\newblock \bibinfo{title}{Cavity optomechanical detection of persistent
  currents and solitons in a bosonic ring condensate},
\newblock \bibinfo{journal}{Physical Review Res.} \bibinfo{volume}{6}
  (\bibinfo{year}{2024}{\natexlab{a}}) \bibinfo{pages}{013104}. \URLprefix
  \url{https://link.aps.org/doi/10.1103/PhysRevResearch.6.013104}.
  \DOIprefix\doi{10.1103/PhysRevResearch.6.013104}.
\bibitem[{Pradhan et~al.(2024{\natexlab{b}})Pradhan, Kanamoto, Bhattacharya,
  and Mishra}]{pradhan2024detection}
\bibinfo{author}{N.~Pradhan}, \bibinfo{author}{R.~Kanamoto},
  \bibinfo{author}{M.~Bhattacharya}, \bibinfo{author}{P.~K. Mishra},
  \bibinfo{title}{{Detection of Andreev-Bashkin superfluid drag using Cavity
  Optomechanics}}, \bibinfo{year}{2024}{\natexlab{b}}. \URLprefix
  \url{https://arxiv.org/abs/2410.21015}.
  \href{http://arxiv.org/abs/2410.21015}{{\tt arXiv:2410.21015}}.
\bibitem[{Safaei et~al.(2019)Safaei, Kwek, Dumke, and
  Amico}]{safaei2019monitoring}
\bibinfo{author}{S.~Safaei}, \bibinfo{author}{L.-C. Kwek},
  \bibinfo{author}{R.~Dumke}, \bibinfo{author}{L.~Amico},
\newblock \bibinfo{title}{Monitoring currents in cold-atom circuis},
\newblock \bibinfo{journal}{Physical Review A} \bibinfo{volume}{100}
  (\bibinfo{year}{2019}) \bibinfo{pages}{013621}.
\bibitem[{Kumar et~al.(2016)Kumar, Anderson, Phillips, Eckel, Campbell, and
  Stringari}]{kumar2016minimally}
\bibinfo{author}{A.~Kumar}, \bibinfo{author}{N.~Anderson},
  \bibinfo{author}{W.~Phillips}, \bibinfo{author}{S.~Eckel},
  \bibinfo{author}{G.~Campbell}, \bibinfo{author}{S.~Stringari},
\newblock \bibinfo{title}{{Minimally destructive, Doppler measurement of a
  quantized flow in a ring-shaped Bose--Einstein condensate}},
\newblock \bibinfo{journal}{New Journal of Physics} \bibinfo{volume}{18}
  (\bibinfo{year}{2016}) \bibinfo{pages}{025001}.
\bibitem[{Marti et~al.(2015)Marti, Olf, and Stamper-Kurn}]{marti2015collective}
\bibinfo{author}{G.~E. Marti}, \bibinfo{author}{R.~Olf}, \bibinfo{author}{D.~M.
  Stamper-Kurn},
\newblock \bibinfo{title}{{Collective excitation interferometry with a toroidal
  Bose-Einstein condensate}},
\newblock \bibinfo{journal}{Physical Review A} \bibinfo{volume}{91}
  (\bibinfo{year}{2015}) \bibinfo{pages}{013602}. \URLprefix
  \url{https://link.aps.org/doi/10.1103/PhysRevA.91.013602}.
  \DOIprefix\doi{10.1103/PhysRevA.91.013602}.
\bibitem[{Davis et~al.(1995)Davis, Mewes, Andrews, van Druten, Durfee, Kurn,
  and Ketterle}]{davis1995bose}
\bibinfo{author}{K.~B. Davis}, \bibinfo{author}{M.~O. Mewes},
  \bibinfo{author}{M.~R. Andrews}, \bibinfo{author}{N.~J. van Druten},
  \bibinfo{author}{D.~S. Durfee}, \bibinfo{author}{D.~M. Kurn},
  \bibinfo{author}{W.~Ketterle},
\newblock \bibinfo{title}{Bose-einstein condensation in a gas of sodium atoms},
\newblock \bibinfo{journal}{Physical Review Letters} \bibinfo{volume}{75}
  (\bibinfo{year}{1995}) \bibinfo{pages}{3969--3973}. \URLprefix
  \url{https://link.aps.org/doi/10.1103/PhysRevLett.75.3969}.
  \DOIprefix\doi{10.1103/PhysRevLett.75.3969}.
\bibitem[{Lett et~al.(1988)Lett, Watts, Westbrook, Phillips, Gould, and
  Metcalf}]{Lettobservation1988}
\bibinfo{author}{P.~D. Lett}, \bibinfo{author}{R.~N. Watts},
  \bibinfo{author}{C.~I. Westbrook}, \bibinfo{author}{W.~D. Phillips},
  \bibinfo{author}{P.~L. Gould}, \bibinfo{author}{H.~J. Metcalf},
\newblock \bibinfo{title}{{Observation of Atoms Laser Cooled below the Doppler
  Limit}},
\newblock \bibinfo{journal}{Physical Review Letters} \bibinfo{volume}{61}
  (\bibinfo{year}{1988}) \bibinfo{pages}{169--172}. \URLprefix
  \url{https://link.aps.org/doi/10.1103/PhysRevLett.61.169}.
  \DOIprefix\doi{10.1103/PhysRevLett.61.169}.
\bibitem[{Westbrook et~al.(1990)Westbrook, Watts, Tanner, Rolston, Phillips,
  Lett, and Gould}]{westbrooklocalization1990}
\bibinfo{author}{C.~I. Westbrook}, \bibinfo{author}{R.~N. Watts},
  \bibinfo{author}{C.~E. Tanner}, \bibinfo{author}{S.~L. Rolston},
  \bibinfo{author}{W.~D. Phillips}, \bibinfo{author}{P.~D. Lett},
  \bibinfo{author}{P.~L. Gould},
\newblock \bibinfo{title}{Localization of atoms in a three-dimensional standing
  wave},
\newblock \bibinfo{journal}{Physical Review Letters} \bibinfo{volume}{65}
  (\bibinfo{year}{1990}) \bibinfo{pages}{33--36}. \URLprefix
  \url{https://link.aps.org/doi/10.1103/PhysRevLett.65.33}.
  \DOIprefix\doi{10.1103/PhysRevLett.65.33}.
\bibitem[{Kohns et~al.(1993)Kohns, Buch, Süptitz, Csambal, and
  Ertmer}]{Kohnsonline1993}
\bibinfo{author}{P.~Kohns}, \bibinfo{author}{P.~Buch},
  \bibinfo{author}{W.~Süptitz}, \bibinfo{author}{C.~Csambal},
  \bibinfo{author}{W.~Ertmer},
\newblock \bibinfo{title}{{On-Line Measurement of Sub-Doppler Temperatures in a
  Rb Magneto-optical Trap-by-Trap Centre Oscillations}},
\newblock \bibinfo{journal}{Europhysics Letters} \bibinfo{volume}{22}
  (\bibinfo{year}{1993}) \bibinfo{pages}{517}. \URLprefix
  \url{https://dx.doi.org/10.1209/0295-5075/22/7/007}.
  \DOIprefix\doi{10.1209/0295-5075/22/7/007}.
\bibitem[{Courtois et~al.(1994)Courtois, Grynberg, Lounis, and
  Verkerk}]{courtoisrecoil1994}
\bibinfo{author}{J.-Y. Courtois}, \bibinfo{author}{G.~Grynberg},
  \bibinfo{author}{B.~Lounis}, \bibinfo{author}{P.~Verkerk},
\newblock \bibinfo{title}{{Recoil-induced resonances in cesium: An atomic
  analog to the free-electron laser}},
\newblock \bibinfo{journal}{Physical Review Letters} \bibinfo{volume}{72}
  (\bibinfo{year}{1994}) \bibinfo{pages}{3017--3020}. \URLprefix
  \url{https://link.aps.org/doi/10.1103/PhysRevLett.72.3017}.
  \DOIprefix\doi{10.1103/PhysRevLett.72.3017}.
\bibitem[{Read and Cooper(2003{\natexlab{a}})}]{read2003free}
\bibinfo{author}{N.~Read}, \bibinfo{author}{N.~R. Cooper},
\newblock \bibinfo{title}{{Free expansion of lowest-Landau-level states of
  trapped atoms: A wave-function microscope}},
\newblock \bibinfo{journal}{Physical Review A} \bibinfo{volume}{68}
  (\bibinfo{year}{2003}{\natexlab{a}}) \bibinfo{pages}{035601}. \URLprefix
  \url{https://link.aps.org/doi/10.1103/PhysRevA.68.035601}.
  \DOIprefix\doi{10.1103/PhysRevA.68.035601}.
\bibitem[{Read and Cooper(2003{\natexlab{b}})}]{readfree2003}
\bibinfo{author}{N.~Read}, \bibinfo{author}{N.~R. Cooper},
\newblock \bibinfo{title}{{Free expansion of lowest-Landau-level states of
  trapped atoms: A wave-function microscope}},
\newblock \bibinfo{journal}{Physical Review A} \bibinfo{volume}{68}
  (\bibinfo{year}{2003}{\natexlab{b}}) \bibinfo{pages}{035601}. \URLprefix
  \url{https://link.aps.org/doi/10.1103/PhysRevA.68.035601}.
  \DOIprefix\doi{10.1103/PhysRevA.68.035601}.
\bibitem[{Berman et~al.(2004)Berman, Borgonovi, Izrailev, and
  Smerzi}]{berman2004irregular}
\bibinfo{author}{G.~P. Berman}, \bibinfo{author}{F.~Borgonovi},
  \bibinfo{author}{F.~M. Izrailev}, \bibinfo{author}{A.~Smerzi},
\newblock \bibinfo{title}{{Irregular Dynamics in a One-Dimensional Bose
  System}},
\newblock \bibinfo{journal}{Physical Review Letters} \bibinfo{volume}{92}
  (\bibinfo{year}{2004}) \bibinfo{pages}{030404}. \URLprefix
  \url{https://link.aps.org/doi/10.1103/PhysRevLett.92.030404}.
  \DOIprefix\doi{10.1103/PhysRevLett.92.030404}.
\bibitem[{Allman(2023)}]{allman2023equilibrium}
\bibinfo{author}{D.~G. Allman},
\newblock \bibinfo{title}{Equilibrium and quench-dynamical studies of ultracold
  fermions in ring-shaped optical traps}  (\bibinfo{year}{2023}). \URLprefix
  \url{https://digitalcommons.dartmouth.edu/dissertations/158/}.
\bibitem[{Ryu et~al.(2014)Ryu, Henderson, and Boshier}]{ryu2014creation}
\bibinfo{author}{C.~Ryu}, \bibinfo{author}{K.~C. Henderson},
  \bibinfo{author}{M.~G. Boshier},
\newblock \bibinfo{title}{Creation of matter wave {Bessel} beams and
  observation of quantized circulation in a {Bose}–{Einstein} condensate},
\newblock \bibinfo{journal}{New Journal of Physics} \bibinfo{volume}{16}
  (\bibinfo{year}{2014}) \bibinfo{pages}{013046}. \URLprefix
  \url{https://dx.doi.org/10.1088/1367-2630/16/1/013046},
  \bibinfo{note}{publisher: IOP Publishing}.
\bibitem[{Murray et~al.(2013)Murray, Krygier, Edwards, Wright, Campbell, and
  Clark}]{murray2013probing}
\bibinfo{author}{N.~Murray}, \bibinfo{author}{M.~Krygier},
  \bibinfo{author}{M.~Edwards}, \bibinfo{author}{K.~C. Wright},
  \bibinfo{author}{G.~K. Campbell}, \bibinfo{author}{C.~W. Clark},
\newblock \bibinfo{title}{Probing the circulation of ring-shaped bose-einstein
  condensates},
\newblock \bibinfo{journal}{Physical Review A} \bibinfo{volume}{88}
  (\bibinfo{year}{2013}) \bibinfo{pages}{053615}. \URLprefix
  \url{https://link.aps.org/doi/10.1103/PhysRevA.88.053615}.
  \DOIprefix\doi{10.1103/PhysRevA.88.053615}.
\bibitem[{Corman et~al.(2014)Corman, Chomaz, Bienaim\'e, Desbuquois,
  Weitenberg, Nascimb\`ene, Dalibard, and Beugnon}]{corman2014quench}
\bibinfo{author}{L.~Corman}, \bibinfo{author}{L.~Chomaz},
  \bibinfo{author}{T.~Bienaim\'e}, \bibinfo{author}{R.~Desbuquois},
  \bibinfo{author}{C.~Weitenberg}, \bibinfo{author}{S.~Nascimb\`ene},
  \bibinfo{author}{J.~Dalibard}, \bibinfo{author}{J.~Beugnon},
\newblock \bibinfo{title}{{Quench-Induced Supercurrents in an Annular Bose
  Gas}},
\newblock \bibinfo{journal}{Physical Review Letters} \bibinfo{volume}{113}
  (\bibinfo{year}{2014}) \bibinfo{pages}{135302}. \URLprefix
  \url{https://link.aps.org/doi/10.1103/PhysRevLett.113.135302}.
  \DOIprefix\doi{10.1103/PhysRevLett.113.135302}.
\bibitem[{Mathew et~al.(2015)Mathew, Kumar, Eckel, Jendrzejewski, Campbell,
  Edwards, and Tiesinga}]{mathew2015selfheterodyne}
\bibinfo{author}{R.~Mathew}, \bibinfo{author}{A.~Kumar},
  \bibinfo{author}{S.~Eckel}, \bibinfo{author}{F.~Jendrzejewski},
  \bibinfo{author}{G.~K. Campbell}, \bibinfo{author}{M.~Edwards},
  \bibinfo{author}{E.~Tiesinga},
\newblock \bibinfo{title}{Self-heterodyne detection of the in situ phase of an
  atomic superconducting quantum interference device},
\newblock \bibinfo{journal}{Physical Review A} \bibinfo{volume}{92}
  (\bibinfo{year}{2015}) \bibinfo{pages}{033602}. \URLprefix
  \url{https://link.aps.org/doi/10.1103/PhysRevA.92.033602}.
  \DOIprefix\doi{10.1103/PhysRevA.92.033602}.
\bibitem[{Roscilde et~al.(2016)Roscilde, Faulkner, Bramwell, and
  Holdsworth}]{roscilde2016quantum}
\bibinfo{author}{T.~Roscilde}, \bibinfo{author}{M.~F. Faulkner},
  \bibinfo{author}{S.~T. Bramwell}, \bibinfo{author}{P.~C. Holdsworth},
\newblock \bibinfo{title}{From quantum to thermal topological-sector
  fluctuations of strongly interacting bosons in a ring lattice},
\newblock \bibinfo{journal}{New Journal of Physics} \bibinfo{volume}{18}
  (\bibinfo{year}{2016}) \bibinfo{pages}{075003}.
\bibitem[{Stamper-Kurn et~al.(1998)Stamper-Kurn, Andrews, Chikkatur, Inouye,
  Miesner, Stenger, and Ketterle}]{stamper1998optical}
\bibinfo{author}{D.~Stamper-Kurn}, \bibinfo{author}{M.~Andrews},
  \bibinfo{author}{A.~Chikkatur}, \bibinfo{author}{S.~Inouye},
  \bibinfo{author}{H.-J. Miesner}, \bibinfo{author}{J.~Stenger},
  \bibinfo{author}{W.~Ketterle},
\newblock \bibinfo{title}{{Optical confinement of a Bose-Einstein condensate}},
\newblock \bibinfo{journal}{Physical Review Letters} \bibinfo{volume}{80}
  (\bibinfo{year}{1998}) \bibinfo{pages}{2027}.
\bibitem[{Ernst et~al.(1998)Ernst, Schuster, Schreck, Marte, Kuhn, and
  Rempe}]{ernst1998free}
\bibinfo{author}{U.~Ernst}, \bibinfo{author}{J.~Schuster},
  \bibinfo{author}{F.~Schreck}, \bibinfo{author}{A.~Marte},
  \bibinfo{author}{A.~Kuhn}, \bibinfo{author}{G.~Rempe},
\newblock \bibinfo{title}{{Free expansion of a Bose--Einstein condensate from
  an Ioffe--Pritchard magnetic trap}},
\newblock \bibinfo{journal}{Applied Physics B} \bibinfo{volume}{67}
  (\bibinfo{year}{1998}) \bibinfo{pages}{719--722}.
\bibitem[{Kagan et~al.(1996)Kagan, Surkov, and
  Shlyapnikov}]{kagan1996evolution}
\bibinfo{author}{Y.~Kagan}, \bibinfo{author}{E.~Surkov},
  \bibinfo{author}{G.~Shlyapnikov},
\newblock \bibinfo{title}{{Evolution of a Bose-condensed gas under variations
  of the confining potential}},
\newblock \bibinfo{journal}{Physical Review A} \bibinfo{volume}{54}
  (\bibinfo{year}{1996}) \bibinfo{pages}{R1753}.
\bibitem[{Castin and Dum(1996)}]{castin1996bose}
\bibinfo{author}{Y.~Castin}, \bibinfo{author}{R.~Dum},
\newblock \bibinfo{title}{{Bose-Einstein condensates in time dependent traps}},
\newblock \bibinfo{journal}{Physical Review Letters} \bibinfo{volume}{77}
  (\bibinfo{year}{1996}) \bibinfo{pages}{5315}.
\bibitem[{Dalfovo et~al.(1997)Dalfovo, Minniti, and
  Pitaevskii}]{dalfovo1997frequency}
\bibinfo{author}{F.~Dalfovo}, \bibinfo{author}{C.~Minniti},
  \bibinfo{author}{L.~Pitaevskii},
\newblock \bibinfo{title}{{Frequency shift and mode coupling in the nonlinear
  dynamics of a Bose-condensed gas}},
\newblock \bibinfo{journal}{Physical Review A} \bibinfo{volume}{56}
  (\bibinfo{year}{1997}) \bibinfo{pages}{4855}.
\bibitem[{Ketterle and Zwierlein(2008)}]{ketterle2008making}
\bibinfo{author}{W.~Ketterle}, \bibinfo{author}{M.~W. Zwierlein},
\newblock \bibinfo{title}{{Making, probing and understanding ultracold Fermi
  gases}},
\newblock \bibinfo{journal}{La Rivista del Nuovo Cimento} \bibinfo{volume}{31}
  (\bibinfo{year}{2008}) \bibinfo{pages}{247--422}.
\bibitem[{Menotti et~al.(2002)Menotti, Pedri, and
  Stringari}]{menotti2002expansion}
\bibinfo{author}{C.~Menotti}, \bibinfo{author}{P.~Pedri},
  \bibinfo{author}{S.~Stringari},
\newblock \bibinfo{title}{{Expansion of an interacting Fermi gas}},
\newblock \bibinfo{journal}{Physical Review letters} \bibinfo{volume}{89}
  (\bibinfo{year}{2002}) \bibinfo{pages}{250402}.
\bibitem[{Allman et~al.(2024)Allman, Sabharwal, and
  Wright}]{allman2023quench-induced}
\bibinfo{author}{D.~G. Allman}, \bibinfo{author}{P.~Sabharwal},
  \bibinfo{author}{K.~C. Wright},
\newblock \bibinfo{title}{Quench-induced spontaneous currents in rings of
  ultracold fermionic atoms},
\newblock \bibinfo{journal}{Physical Review A} \bibinfo{volume}{109}
  (\bibinfo{year}{2024}) \bibinfo{pages}{053320}.
\bibitem[{Xhani et~al.(2023)Xhani, Del~Pace, Scazza, and
  Roati}]{xhani2023decay}
\bibinfo{author}{K.~Xhani}, \bibinfo{author}{G.~Del~Pace},
  \bibinfo{author}{F.~Scazza}, \bibinfo{author}{G.~Roati},
\newblock \bibinfo{title}{Decay of {Persistent} {Currents} in {Annular}
  {Atomic} {Superfluids}},
\newblock \bibinfo{journal}{Atoms} \bibinfo{volume}{11} (\bibinfo{year}{2023})
  \bibinfo{pages}{109}. \URLprefix
  \url{https://www.mdpi.com/2218-2004/11/8/109}.
  \DOIprefix\doi{10.3390/atoms11080109}.
\bibitem[{Piazza et~al.(2009)Piazza, Collins, and Smerzi}]{piazza2009vortex}
\bibinfo{author}{F.~Piazza}, \bibinfo{author}{L.~A. Collins},
  \bibinfo{author}{A.~Smerzi},
\newblock \bibinfo{title}{{Vortex-induced phase-slip dissipation in a toroidal
  Bose-Einstein condensate flowing through a barrier}},
\newblock \bibinfo{journal}{Physical Review A} \bibinfo{volume}{80}
  (\bibinfo{year}{2009}) \bibinfo{pages}{021601}. \URLprefix
  \url{https://link.aps.org/doi/10.1103/PhysRevA.80.021601}.
  \DOIprefix\doi{10.1103/PhysRevA.80.021601}.
\bibitem[{Gupta et~al.(2005)Gupta, Murch, Moore, Purdy, and
  Stamper-Kurn}]{gupta2005circularwaveguide}
\bibinfo{author}{S.~Gupta}, \bibinfo{author}{K.~W. Murch},
  \bibinfo{author}{K.~L. Moore}, \bibinfo{author}{T.~P. Purdy},
  \bibinfo{author}{D.~M. Stamper-Kurn},
\newblock \bibinfo{title}{{Bose-Einstein Condensation in a Circular
  Waveguide}},
\newblock \bibinfo{journal}{Physical Review Letters} \bibinfo{volume}{95}
  (\bibinfo{year}{2005}) \bibinfo{pages}{143201}. \URLprefix
  \url{https://link.aps.org/doi/10.1103/PhysRevLett.95.143201}.
  \DOIprefix\doi{10.1103/PhysRevLett.95.143201}.
\bibitem[{Arnold et~al.(2006)Arnold, Garvie, and Riis}]{arnold2006ring}
\bibinfo{author}{A.~S. Arnold}, \bibinfo{author}{C.~S. Garvie},
  \bibinfo{author}{E.~Riis},
\newblock \bibinfo{title}{{Large magnetic storage ring for Bose-Einstein
  condensates}},
\newblock \bibinfo{journal}{Physical Review A} \bibinfo{volume}{73}
  (\bibinfo{year}{2006}) \bibinfo{pages}{041606}. \URLprefix
  \url{https://link.aps.org/doi/10.1103/PhysRevA.73.041606}.
  \DOIprefix\doi{10.1103/PhysRevA.73.041606}.
\bibitem[{Morizot et~al.(2006)Morizot, Colombe, Lorent, Perrin, and
  Garraway}]{morizot2006ring}
\bibinfo{author}{O.~Morizot}, \bibinfo{author}{Y.~Colombe},
  \bibinfo{author}{V.~Lorent}, \bibinfo{author}{H.~Perrin},
  \bibinfo{author}{B.~M. Garraway},
\newblock \bibinfo{title}{Ring trap for ultracold atoms},
\newblock \bibinfo{journal}{Physical Review A} \bibinfo{volume}{74}
  (\bibinfo{year}{2006}) \bibinfo{pages}{023617}. \URLprefix
  \url{https://link.aps.org/doi/10.1103/PhysRevA.74.023617}.
  \DOIprefix\doi{10.1103/PhysRevA.74.023617}.
\bibitem[{Lesanovsky and von Klitzing(2007)}]{lesanovsky2007time}
\bibinfo{author}{I.~Lesanovsky}, \bibinfo{author}{W.~von Klitzing},
\newblock \bibinfo{title}{{Time-Averaged Adiabatic Potentials: Versatile
  Matter-Wave Guides and Atom Traps}},
\newblock \bibinfo{journal}{Physical Review Letters} \bibinfo{volume}{99}
  (\bibinfo{year}{2007}) \bibinfo{pages}{083001}. \URLprefix
  \url{https://link.aps.org/doi/10.1103/PhysRevLett.99.083001}.
  \DOIprefix\doi{10.1103/PhysRevLett.99.083001}.
\bibitem[{Heathcote et~al.(2008)Heathcote, Nugent, Sheard, and
  Foot}]{heathcote2008ring}
\bibinfo{author}{W.~H. Heathcote}, \bibinfo{author}{E.~Nugent},
  \bibinfo{author}{B.~T. Sheard}, \bibinfo{author}{C.~J. Foot},
\newblock \bibinfo{title}{{A ring trap for ultracold atoms in an RF-dressed
  state}},
\newblock \bibinfo{journal}{New Journal of Physics} \bibinfo{volume}{10}
  (\bibinfo{year}{2008}) \bibinfo{pages}{043012}. \URLprefix
  \url{https://dx.doi.org/10.1088/1367-2630/10/4/043012}.
  \DOIprefix\doi{10.1088/1367-2630/10/4/043012}.
\bibitem[{Schnelle et~al.(2008)Schnelle, Ooijen, Davis, Heckenberg, and
  Rubinsztein-Dunlop}]{schnelle2008versatile}
\bibinfo{author}{S.~K. Schnelle}, \bibinfo{author}{E.~D.~v. Ooijen},
  \bibinfo{author}{M.~J. Davis}, \bibinfo{author}{N.~R. Heckenberg},
  \bibinfo{author}{H.~Rubinsztein-Dunlop},
\newblock \bibinfo{title}{Versatile two-dimensional potentials for ultra-cold
  atoms},
\newblock \bibinfo{journal}{Optics Express} \bibinfo{volume}{16}
  (\bibinfo{year}{2008}) \bibinfo{pages}{1405--1412}. \URLprefix
  \url{https://opg.optica.org/oe/abstract.cfm?uri=oe-16-3-1405}.
  \DOIprefix\doi{10.1364/OE.16.001405}, \bibinfo{note}{publisher: Optica
  Publishing Group}.
\bibitem[{Bruce et~al.(2011)Bruce, Mayoh, Smirne, Torralbo-Campo, and
  Cassettari}]{bruce2011smooth}
\bibinfo{author}{G.~D. Bruce}, \bibinfo{author}{J.~Mayoh},
  \bibinfo{author}{G.~Smirne}, \bibinfo{author}{L.~Torralbo-Campo},
  \bibinfo{author}{D.~Cassettari},
\newblock \bibinfo{title}{A smooth, holographically generated ring trap for the
  investigation of superfluidity in ultracold atoms},
\newblock \bibinfo{journal}{Physica Scripta} \bibinfo{volume}{2011}
  (\bibinfo{year}{2011}) \bibinfo{pages}{014008}. \URLprefix
  \url{https://dx.doi.org/10.1088/0031-8949/2011/T143/014008}.
  \DOIprefix\doi{10.1088/0031-8949/2011/T143/014008}.
\bibitem[{Sherlock et~al.(2011)Sherlock, Gildemeister, Owen, Nugent, and
  Foot}]{sherlock2011time}
\bibinfo{author}{B.~E. Sherlock}, \bibinfo{author}{M.~Gildemeister},
  \bibinfo{author}{E.~Owen}, \bibinfo{author}{E.~Nugent},
  \bibinfo{author}{C.~J. Foot},
\newblock \bibinfo{title}{Time-averaged adiabatic ring potential for ultracold
  atoms},
\newblock \bibinfo{journal}{Physical Review A} \bibinfo{volume}{83}
  (\bibinfo{year}{2011}) \bibinfo{pages}{043408}. \URLprefix
  \url{https://link.aps.org/doi/10.1103/PhysRevA.83.043408}.
  \DOIprefix\doi{10.1103/PhysRevA.83.043408}.
\bibitem[{Navez et~al.(2016)Navez, Pandey, Mas, Poulios, Fernholz, and von
  Klitzing}]{navez2016matter}
\bibinfo{author}{P.~Navez}, \bibinfo{author}{S.~Pandey},
  \bibinfo{author}{H.~Mas}, \bibinfo{author}{K.~Poulios},
  \bibinfo{author}{T.~Fernholz}, \bibinfo{author}{W.~von Klitzing},
\newblock \bibinfo{title}{Matter-wave interferometers using taap rings},
\newblock \bibinfo{journal}{New Journal of Physics} \bibinfo{volume}{18}
  (\bibinfo{year}{2016}) \bibinfo{pages}{075014}. \URLprefix
  \url{https://dx.doi.org/10.1088/1367-2630/18/7/075014}.
  \DOIprefix\doi{10.1088/1367-2630/18/7/075014}.
\bibitem[{Bell et~al.(2016)Bell, Glidden, Humbert, Bromley, Haine, Davis,
  Neely, Baker, and Rubinsztein-Dunlop}]{bell2016bose-einstein}
\bibinfo{author}{T.~A. Bell}, \bibinfo{author}{J.~A.~P. Glidden},
  \bibinfo{author}{L.~Humbert}, \bibinfo{author}{M.~W.~J. Bromley},
  \bibinfo{author}{S.~A. Haine}, \bibinfo{author}{M.~J. Davis},
  \bibinfo{author}{T.~W. Neely}, \bibinfo{author}{M.~A. Baker},
  \bibinfo{author}{H.~Rubinsztein-Dunlop},
\newblock \bibinfo{title}{Bose–{Einstein} condensation in large time-averaged
  optical ring potentials},
\newblock \bibinfo{journal}{New Journal of Physics} \bibinfo{volume}{18}
  (\bibinfo{year}{2016}) \bibinfo{pages}{035003}. \URLprefix
  \url{https://dx.doi.org/10.1088/1367-2630/18/3/035003}.
  \DOIprefix\doi{10.1088/1367-2630/18/3/035003}.
\bibitem[{Weiler et~al.(2008)Weiler, Neely, Scherer, Bradley, Davis, and
  Anderson}]{weiler2008spontaneous}
\bibinfo{author}{C.~N. Weiler}, \bibinfo{author}{T.~W. Neely},
  \bibinfo{author}{D.~R. Scherer}, \bibinfo{author}{A.~S. Bradley},
  \bibinfo{author}{M.~J. Davis}, \bibinfo{author}{B.~P. Anderson},
\newblock \bibinfo{title}{Spontaneous vortices in the formation of
  {Bose}–{Einstein} condensates},
\newblock \bibinfo{journal}{Nature} \bibinfo{volume}{455}
  (\bibinfo{year}{2008}) \bibinfo{pages}{948--951}. \URLprefix
  \url{https://www.nature.com/articles/nature07334}.
  \DOIprefix\doi{10.1038/nature07334}.
\bibitem[{Wright et~al.(2000)Wright, Arlt, and Dholakia}]{wright2000toroidal}
\bibinfo{author}{E.~M. Wright}, \bibinfo{author}{J.~Arlt},
  \bibinfo{author}{K.~Dholakia},
\newblock \bibinfo{title}{{Toroidal optical dipole traps for atomic
  Bose-Einstein condensates using Laguerre-Gaussian beams}},
\newblock \bibinfo{journal}{Physical Review A} \bibinfo{volume}{63}
  (\bibinfo{year}{2000}) \bibinfo{pages}{013608}. \URLprefix
  \url{https://link.aps.org/doi/10.1103/PhysRevA.63.013608}.
\bibitem[{Brand and Reinhardt(2001)}]{brand2001generating}
\bibinfo{author}{J.~Brand}, \bibinfo{author}{W.~P. Reinhardt},
\newblock \bibinfo{title}{Generating ring currents, solitons and svortices by
  stirring a {Bose}-{Einstein} condensate in a toroidal trap},
\newblock \bibinfo{journal}{Journal of Physics B: Atomic, Molecular and Optical
  Physics} \bibinfo{volume}{34} (\bibinfo{year}{2001}) \bibinfo{pages}{L113}.
  \URLprefix \url{https://dx.doi.org/10.1088/0953-4075/34/4/105}.
  \DOIprefix\doi{10.1088/0953-4075/34/4/105}.
\bibitem[{Franke-Arnold et~al.(2007)Franke-Arnold, Leach, Padgett, Lembessis,
  Ellinas, Wright, Girkin, Öhberg, and Arnold}]{franke-arnold2007optical}
\bibinfo{author}{S.~Franke-Arnold}, \bibinfo{author}{J.~Leach},
  \bibinfo{author}{M.~J. Padgett}, \bibinfo{author}{V.~E. Lembessis},
  \bibinfo{author}{D.~Ellinas}, \bibinfo{author}{A.~J. Wright},
  \bibinfo{author}{J.~M. Girkin}, \bibinfo{author}{P.~Öhberg},
  \bibinfo{author}{A.~S. Arnold},
\newblock \bibinfo{title}{Optical ferris wheel for ultracold atoms},
\newblock \bibinfo{journal}{Optics Express} \bibinfo{volume}{15}
  (\bibinfo{year}{2007}) \bibinfo{pages}{8619--8625}. \URLprefix
  \url{https://opg.optica.org/oe/abstract.cfm?uri=oe-15-14-8619}.
  \DOIprefix\doi{10.1364/OE.15.008619}.
\bibitem[{Gauthier et~al.(2021)Gauthier, Bell, Stilgoe, Baker,
  Rubinsztein-Dunlop, and Neely}]{gauthier2021chapter}
\bibinfo{author}{G.~Gauthier}, \bibinfo{author}{T.~A. Bell},
  \bibinfo{author}{A.~B. Stilgoe}, \bibinfo{author}{M.~Baker},
  \bibinfo{author}{H.~Rubinsztein-Dunlop}, \bibinfo{author}{T.~W. Neely},
\newblock \bibinfo{title}{{Chapter One - Dynamic high-resolution optical
  trapping of ultracold atoms}},
\newblock volume~\bibinfo{volume}{70} of \textit{\bibinfo{series}{Advances In
  Atomic, Molecular, and Optical Physics}}, \bibinfo{publisher}{Academic
  Press}, \bibinfo{year}{2021}, pp. \bibinfo{pages}{1--101}. \URLprefix
  \url{https://www.sciencedirect.com/science/article/pii/S1049250X2100001X}.
  \DOIprefix\doi{https://doi.org/10.1016/bs.aamop.2021.04.001}.
\bibitem[{Turpin et~al.(2015)Turpin, Polo, Loiko, K\"{u}ber, Schmaltz,
  Kalkandjiev, Ahufinger, Birkl, and Mompart}]{turpin2015blue}
\bibinfo{author}{A.~Turpin}, \bibinfo{author}{J.~Polo}, \bibinfo{author}{Y.~V.
  Loiko}, \bibinfo{author}{J.~K\"{u}ber}, \bibinfo{author}{F.~Schmaltz},
  \bibinfo{author}{T.~K. Kalkandjiev}, \bibinfo{author}{V.~Ahufinger},
  \bibinfo{author}{G.~Birkl}, \bibinfo{author}{J.~Mompart},
\newblock \bibinfo{title}{{Blue-detuned optical ring trap for Bose-Einstein
  condensates based on conical refraction}},
\newblock \bibinfo{journal}{Optics Express} \bibinfo{volume}{23}
  (\bibinfo{year}{2015}) \bibinfo{pages}{1638}. \URLprefix
  \url{http://dx.doi.org/10.1364/OE.23.001638}.
  \DOIprefix\doi{10.1364/oe.23.001638}.
\bibitem[{Dubessy et~al.(2012)Dubessy, Liennard, Pedri, and
  Perrin}]{dubessy2012critical}
\bibinfo{author}{R.~Dubessy}, \bibinfo{author}{T.~Liennard},
  \bibinfo{author}{P.~Pedri}, \bibinfo{author}{H.~Perrin},
\newblock \bibinfo{title}{{Critical rotation of an annular superfluid
  Bose-Einstein condensate}},
\newblock \bibinfo{journal}{Physical Review A} \bibinfo{volume}{86}
  (\bibinfo{year}{2012}) \bibinfo{pages}{011602}.
\bibitem[{Beattie et~al.(2013)Beattie, Moulder, Fletcher, and
  Hadzibabic}]{beattie2013spinor}
\bibinfo{author}{S.~Beattie}, \bibinfo{author}{S.~Moulder},
  \bibinfo{author}{R.~J. Fletcher}, \bibinfo{author}{Z.~Hadzibabic},
\newblock \bibinfo{title}{Persistent {Currents} in {Spinor} {Condensates}},
\newblock \bibinfo{journal}{Physical Review Letters} \bibinfo{volume}{110}
  (\bibinfo{year}{2013}) \bibinfo{pages}{025301}. \URLprefix
  \url{https://link.aps.org/doi/10.1103/PhysRevLett.110.025301}.
  \DOIprefix\doi{10.1103/PhysRevLett.110.025301}, \bibinfo{note}{publisher:
  American Physical Society}.
\bibitem[{Smyrnakis et~al.(2009)Smyrnakis, Bargi, Kavoulakis, Magiropoulos,
  K\"arkk\"ainen, and Reimann}]{smyrnakis2009mixtures}
\bibinfo{author}{J.~Smyrnakis}, \bibinfo{author}{S.~Bargi},
  \bibinfo{author}{G.~M. Kavoulakis}, \bibinfo{author}{M.~Magiropoulos},
  \bibinfo{author}{K.~K\"arkk\"ainen}, \bibinfo{author}{S.~M. Reimann},
\newblock \bibinfo{title}{{Mixtures of Bose Gases Confined in a Ring
  Potential}},
\newblock \bibinfo{journal}{Physical Review Letters} \bibinfo{volume}{103}
  (\bibinfo{year}{2009}) \bibinfo{pages}{100404}. \URLprefix
  \url{https://link.aps.org/doi/10.1103/PhysRevLett.103.100404}.
  \DOIprefix\doi{10.1103/PhysRevLett.103.100404}.
\bibitem[{Bargi et~al.(2010)Bargi, Malet, Kavoulakis, and
  Reimann}]{bargi2010persistent}
\bibinfo{author}{S.~Bargi}, \bibinfo{author}{F.~Malet}, \bibinfo{author}{G.~M.
  Kavoulakis}, \bibinfo{author}{S.~M. Reimann},
\newblock \bibinfo{title}{{Persistent currents in Bose gases confined in
  annular traps}},
\newblock \bibinfo{journal}{Physical Review A} \bibinfo{volume}{82}
  (\bibinfo{year}{2010}) \bibinfo{pages}{043631}. \URLprefix
  \url{https://link.aps.org/doi/10.1103/PhysRevA.82.043631}.
  \DOIprefix\doi{10.1103/PhysRevA.82.043631}.
\bibitem[{Anoshkin et~al.(2013)Anoshkin, Wu, and
  Zaremba}]{anoshkin2013persistent}
\bibinfo{author}{K.~Anoshkin}, \bibinfo{author}{Z.~Wu},
  \bibinfo{author}{E.~Zaremba},
\newblock \bibinfo{title}{Persistent currents in a bosonic mixture in the ring
  geometry},
\newblock \bibinfo{journal}{Physical Review A} \bibinfo{volume}{88}
  (\bibinfo{year}{2013}) \bibinfo{pages}{013609}. \URLprefix
  \url{https://link.aps.org/doi/10.1103/PhysRevA.88.013609}.
  \DOIprefix\doi{10.1103/PhysRevA.88.013609}.
\bibitem[{Eckel et~al.(2014)Eckel, Lee, Jendrzejewski, Murray, Clark, Lobb,
  Phillips, Edwards, and Campbell}]{eckel2014hysteresis}
\bibinfo{author}{S.~Eckel}, \bibinfo{author}{J.~G. Lee},
  \bibinfo{author}{F.~Jendrzejewski}, \bibinfo{author}{N.~Murray},
  \bibinfo{author}{C.~W. Clark}, \bibinfo{author}{C.~J. Lobb},
  \bibinfo{author}{W.~D. Phillips}, \bibinfo{author}{M.~Edwards},
  \bibinfo{author}{G.~K. Campbell},
\newblock \bibinfo{title}{Hysteresis in a quantized, superfluid atomtronic
  circuit},
\newblock \bibinfo{journal}{Nature} \bibinfo{volume}{506}
  (\bibinfo{year}{2014}) \bibinfo{pages}{200--203}. \URLprefix
  \url{http://arxiv.org/abs/1402.2958}. \DOIprefix\doi{10.1038/nature12958}.
\bibitem[{Wright et~al.(2013)Wright, Blakestad, Lobb, Phillips, and
  Campbell}]{wright2013threshold}
\bibinfo{author}{K.~C. Wright}, \bibinfo{author}{R.~B. Blakestad},
  \bibinfo{author}{C.~J. Lobb}, \bibinfo{author}{W.~D. Phillips},
  \bibinfo{author}{G.~K. Campbell},
\newblock \bibinfo{title}{Threshold for creating excitations in a stirred
  superfluid ring},
\newblock \bibinfo{journal}{Physical Review A} \bibinfo{volume}{88}
  (\bibinfo{year}{2013}) \bibinfo{pages}{063633}. \URLprefix
  \url{https://link.aps.org/doi/10.1103/PhysRevA.88.063633}.
  \DOIprefix\doi{10.1103/PhysRevA.88.063633}.
\bibitem[{Neely et~al.(2013)Neely, Bradley, Samson, Rooney, Wright, Law,
  Carretero-Gonz\'alez, Kevrekidis, Davis, and Anderson}]{neely2013turbulence}
\bibinfo{author}{T.~W. Neely}, \bibinfo{author}{A.~S. Bradley},
  \bibinfo{author}{E.~C. Samson}, \bibinfo{author}{S.~J. Rooney},
  \bibinfo{author}{E.~M. Wright}, \bibinfo{author}{K.~J.~H. Law},
  \bibinfo{author}{R.~Carretero-Gonz\'alez}, \bibinfo{author}{P.~G.
  Kevrekidis}, \bibinfo{author}{M.~J. Davis}, \bibinfo{author}{B.~P. Anderson},
\newblock \bibinfo{title}{{Characteristics of Two-Dimensional Quantum
  Turbulence in a Compressible Superfluid}},
\newblock \bibinfo{journal}{Physical Review Letters} \bibinfo{volume}{111}
  (\bibinfo{year}{2013}) \bibinfo{pages}{235301}. \URLprefix
  \url{https://link.aps.org/doi/10.1103/PhysRevLett.111.235301}.
  \DOIprefix\doi{10.1103/PhysRevLett.111.235301}.
\bibitem[{Jendrzejewski et~al.(2014)Jendrzejewski, Eckel, Murray, Lanier,
  Edwards, Lobb, and Campbell}]{jendrzejewski2014resistive}
\bibinfo{author}{F.~Jendrzejewski}, \bibinfo{author}{S.~Eckel},
  \bibinfo{author}{N.~Murray}, \bibinfo{author}{C.~Lanier},
  \bibinfo{author}{M.~Edwards}, \bibinfo{author}{C.~J. Lobb},
  \bibinfo{author}{G.~K. Campbell},
\newblock \bibinfo{title}{Resistive flow in a weakly interacting
  {Bose}-{Einstein} condensate},
\newblock \bibinfo{journal}{Physical Review Letters} \bibinfo{volume}{113}
  (\bibinfo{year}{2014}) \bibinfo{pages}{045305}. \URLprefix
  \url{http://arxiv.org/abs/1402.3335}.
  \DOIprefix\doi{10.1103/PhysRevLett.113.045305}.
\bibitem[{Aidelsburger et~al.(2017)Aidelsburger, Ville, Saint-Jalm,
  Nascimb\`ene, Dalibard, and Beugnon}]{aidelsburger2017relaxation}
\bibinfo{author}{M.~Aidelsburger}, \bibinfo{author}{J.~L. Ville},
  \bibinfo{author}{R.~Saint-Jalm}, \bibinfo{author}{S.~Nascimb\`ene},
  \bibinfo{author}{J.~Dalibard}, \bibinfo{author}{J.~Beugnon},
\newblock \bibinfo{title}{{Relaxation Dynamics in the Merging of $N$
  Independent Condensates}},
\newblock \bibinfo{journal}{Physical Review Letters} \bibinfo{volume}{119}
  (\bibinfo{year}{2017}) \bibinfo{pages}{190403}. \URLprefix
  \url{https://link.aps.org/doi/10.1103/PhysRevLett.119.190403}.
  \DOIprefix\doi{10.1103/PhysRevLett.119.190403}.
\bibitem[{Eckel et~al.(2018)Eckel, Kumar, Jacobson, Spielman, and
  Campbell}]{eckel2018rapidly}
\bibinfo{author}{S.~Eckel}, \bibinfo{author}{A.~Kumar},
  \bibinfo{author}{T.~Jacobson}, \bibinfo{author}{I.~B. Spielman},
  \bibinfo{author}{G.~K. Campbell},
\newblock \bibinfo{title}{{A Rapidly Expanding Bose-Einstein Condensate: An
  Expanding Universe in the Lab}},
\newblock \bibinfo{journal}{Physical Review X} \bibinfo{volume}{8}
  (\bibinfo{year}{2018}) \bibinfo{pages}{021021}. \URLprefix
  \url{https://link.aps.org/doi/10.1103/PhysRevX.8.021021}.
  \DOIprefix\doi{10.1103/PhysRevX.8.021021}.
\bibitem[{Kumar et~al.(2017)Kumar, Eckel, Jendrzejewski, and
  Campbell}]{kumar2017temperature}
\bibinfo{author}{A.~Kumar}, \bibinfo{author}{S.~Eckel},
  \bibinfo{author}{F.~Jendrzejewski}, \bibinfo{author}{G.~K. Campbell},
\newblock \bibinfo{title}{Temperature-induced decay of persistent currents in a
  superfluid ultracold gas},
\newblock \bibinfo{journal}{Physical Review A} \bibinfo{volume}{95}
  (\bibinfo{year}{2017}) \bibinfo{pages}{021602}. \URLprefix
  \url{https://link.aps.org/doi/10.1103/PhysRevA.95.021602}.
  \DOIprefix\doi{10.1103/PhysRevA.95.021602}.
\bibitem[{Mathey et~al.(2014)Mathey, Clark, and Mathey}]{mathey2014decay}
\bibinfo{author}{A.~C. Mathey}, \bibinfo{author}{C.~W. Clark},
  \bibinfo{author}{L.~Mathey},
\newblock \bibinfo{title}{Decay of a superfluid current of ultracold atoms in a
  toroidal trap},
\newblock \bibinfo{journal}{Physical Review A} \bibinfo{volume}{90}
  (\bibinfo{year}{2014}) \bibinfo{pages}{023604}. \URLprefix
  \url{https://link.aps.org/doi/10.1103/PhysRevA.90.023604}.
  \DOIprefix\doi{10.1103/PhysRevA.90.023604}.
\bibitem[{Snizhko et~al.(2016)Snizhko, Isaieva, Kuriatnikov, Bidasyuk,
  Vilchinskii, and Yakimenko}]{snizhko2016stochastic}
\bibinfo{author}{K.~Snizhko}, \bibinfo{author}{K.~Isaieva},
  \bibinfo{author}{Y.~Kuriatnikov}, \bibinfo{author}{Y.~Bidasyuk},
  \bibinfo{author}{S.~Vilchinskii}, \bibinfo{author}{A.~Yakimenko},
\newblock \bibinfo{title}{Stochastic phase slips in toroidal bose-einstein
  condensates},
\newblock \bibinfo{journal}{Physical Review A} \bibinfo{volume}{94}
  (\bibinfo{year}{2016}) \bibinfo{pages}{063642}.
\bibitem[{Kunimi and Danshita(2019)}]{kunimi2019decay}
\bibinfo{author}{M.~Kunimi}, \bibinfo{author}{I.~Danshita},
\newblock \bibinfo{title}{{Decay mechanisms of superflow of Bose-Einstein
  condensates in ring traps}},
\newblock \bibinfo{journal}{Physical Review A} \bibinfo{volume}{99}
  (\bibinfo{year}{2019}) \bibinfo{pages}{043613}. \URLprefix
  \url{https://link.aps.org/doi/10.1103/PhysRevA.99.043613}.
  \DOIprefix\doi{10.1103/PhysRevA.99.043613}.
\bibitem[{Mehdi et~al.(2021)Mehdi, Bradley, Hope, and
  Szigeti}]{mehdi2021superflow}
\bibinfo{author}{Z.~Mehdi}, \bibinfo{author}{A.~Bradley},
  \bibinfo{author}{J.~Hope}, \bibinfo{author}{S.~Szigeti},
\newblock \bibinfo{title}{{Superflow decay in a toroidal Bose gas: The effect
  of quantum and thermal fluctuations}},
\newblock \bibinfo{journal}{SciPost Physics} \bibinfo{volume}{11}
  (\bibinfo{year}{2021}) \bibinfo{pages}{080}.
\bibitem[{Law et~al.(2014)Law, Neely, Kevrekidis, Anderson, Bradley, and
  Carretero-González}]{law2014dynamic}
\bibinfo{author}{K.~J.~H. Law}, \bibinfo{author}{T.~W. Neely},
  \bibinfo{author}{P.~G. Kevrekidis}, \bibinfo{author}{B.~P. Anderson},
  \bibinfo{author}{A.~S. Bradley}, \bibinfo{author}{R.~Carretero-González},
\newblock \bibinfo{title}{Dynamic and energetic stabilization of persistent
  currents in {Bose}-{Einstein} condensates},
\newblock \bibinfo{journal}{Physical Review A} \bibinfo{volume}{89}
  (\bibinfo{year}{2014}) \bibinfo{pages}{053606}. \URLprefix
  \url{https://link.aps.org/doi/10.1103/PhysRevA.89.053606}.
  \DOIprefix\doi{10.1103/PhysRevA.89.053606}.
\bibitem[{Wilson et~al.(2022)Wilson, Samson, Newman, and
  Anderson}]{wilson2022generation}
\bibinfo{author}{K.~E. Wilson}, \bibinfo{author}{E.~C. Samson},
  \bibinfo{author}{Z.~L. Newman}, \bibinfo{author}{B.~P. Anderson},
\newblock \bibinfo{title}{Generation of high-winding-number superfluid
  circulation in {Bose}-{Einstein} condensates},
\newblock \bibinfo{journal}{Physical Review A} \bibinfo{volume}{106}
  (\bibinfo{year}{2022}) \bibinfo{pages}{033319}. \URLprefix
  \url{https://link.aps.org/doi/10.1103/PhysRevA.106.033319}.
  \DOIprefix\doi{10.1103/PhysRevA.106.033319}.
\bibitem[{{de Go\"er de Herve} et~al.(2021){de Go\"er de Herve}, Guo, {De
  Rossi}, Kumar, Badr, Dubessy, Longchambon, and Perrin}]{herve2021versatile}
\bibinfo{author}{M.~{de Go\"er de Herve}}, \bibinfo{author}{Y.~Guo},
  \bibinfo{author}{C.~{De Rossi}}, \bibinfo{author}{A.~Kumar},
  \bibinfo{author}{T.~Badr}, \bibinfo{author}{R.~Dubessy},
  \bibinfo{author}{L.~Longchambon}, \bibinfo{author}{H.~Perrin},
\newblock \bibinfo{title}{A versatile ring trap for quantum gases},
\newblock \bibinfo{journal}{Journal of Physics B: Atomic, Molecular and Optical
  Physics} \bibinfo{volume}{54} (\bibinfo{year}{2021}) \bibinfo{pages}{125302}.
  \URLprefix \url{https://doi.org/10.1088/1361-6455/ac0579}.
  \DOIprefix\doi{10.1088/1361-6455/ac0579}.
\bibitem[{Dubessy and Perrin(2024)}]{dubessy2024perspective}
\bibinfo{author}{R.~Dubessy}, \bibinfo{author}{H.~Perrin},
  \bibinfo{title}{{Perspective: Quantum gases in bubble traps}},
  \bibinfo{year}{2024}. \URLprefix \url{https://arxiv.org/abs/2410.10268}.
  \href{http://arxiv.org/abs/2410.10268}{{\tt arXiv:2410.10268}}.
\bibitem[{Chen et~al.(2011)Chen, Torrontegui, Stefanatos, Li, and
  Muga}]{chen2014optimal}
\bibinfo{author}{X.~Chen}, \bibinfo{author}{E.~Torrontegui},
  \bibinfo{author}{D.~Stefanatos}, \bibinfo{author}{J.-S. Li},
  \bibinfo{author}{J.~G. Muga},
\newblock \bibinfo{title}{Optimal trajectories for efficient atomic transport
  without final excitation},
\newblock \bibinfo{journal}{Physical Review A} \bibinfo{volume}{84}
  (\bibinfo{year}{2011}) \bibinfo{pages}{043415}. \URLprefix
  \url{https://link.aps.org/doi/10.1103/PhysRevA.84.043415}.
  \DOIprefix\doi{10.1103/PhysRevA.84.043415}.
\bibitem[{Guo et~al.(2020)Guo, Dubessy, {de Go\"er de Herve}, Kumar, Badr,
  Perrin, Longchambon, and Perrin}]{guo2020supersonic}
\bibinfo{author}{Y.~Guo}, \bibinfo{author}{R.~Dubessy}, \bibinfo{author}{M.~{de
  Go\"er de Herve}}, \bibinfo{author}{A.~Kumar}, \bibinfo{author}{T.~Badr},
  \bibinfo{author}{A.~Perrin}, \bibinfo{author}{L.~Longchambon},
  \bibinfo{author}{H.~Perrin},
\newblock \bibinfo{title}{{Supersonic Rotation of a Superfluid: A Long-Lived
  Dynamical Ring}},
\newblock \bibinfo{journal}{Physical Review Letters} \bibinfo{volume}{124}
  (\bibinfo{year}{2020}) \bibinfo{pages}{025301}. \URLprefix
  \url{https://link.aps.org/doi/10.1103/PhysRevLett.124.025301}.
  \DOIprefix\doi{10.1103/PhysRevLett.124.025301}.
\bibitem[{Girardeau(1960)}]{girardeau1960relationship}
\bibinfo{author}{M.~Girardeau},
\newblock \bibinfo{title}{{Relationship between Systems of Impenetrable Bosons
  and Fermions in One Dimension}},
\newblock \bibinfo{journal}{Journal of Mathematical Physics}
  \bibinfo{volume}{1} (\bibinfo{year}{1960}) \bibinfo{pages}{516--523}.
  \URLprefix \url{https://doi.org/10.1063/1.1703687}.
  \DOIprefix\doi{10.1063/1.1703687}.
  \href{http://arxiv.org/abs/https://pubs.aip.org/aip/jmp/article-pdf/1/6/516/19055341/516\_1\_online.pdf}{{\tt
  arXiv:https://pubs.aip.org/aip/jmp/article-pdf/1/6/516/19055341/516\_1\_online.pdf}}.
\bibitem[{Kanamoto et~al.(2010)Kanamoto, Carr, and
  Ueda}]{kanamoto2010metastable}
\bibinfo{author}{R.~Kanamoto}, \bibinfo{author}{L.~D. Carr},
  \bibinfo{author}{M.~Ueda},
\newblock \bibinfo{title}{{Metastable quantum phase transitions in a periodic
  one-dimensional Bose gas. II. Many-body theory}},
\newblock \bibinfo{journal}{Physical Review A} \bibinfo{volume}{81}
  (\bibinfo{year}{2010}) \bibinfo{pages}{023625}. \URLprefix
  \url{https://link.aps.org/doi/10.1103/PhysRevA.81.023625}.
  \DOIprefix\doi{10.1103/PhysRevA.81.023625}.
\bibitem[{Oelkers et~al.(2006)Oelkers, Batchelor, Bortz, and
  Guan}]{oelkers2006bethe}
\bibinfo{author}{N.~Oelkers}, \bibinfo{author}{M.~T. Batchelor},
  \bibinfo{author}{M.~Bortz}, \bibinfo{author}{X.-W. Guan},
\newblock \bibinfo{title}{{Bethe ansatz study of one-dimensional Bose and Fermi
  gases with periodic and hard wall boundary conditions}},
\newblock \bibinfo{journal}{Journal of Physics A: Mathematical and General}
  \bibinfo{volume}{39} (\bibinfo{year}{2006}) \bibinfo{pages}{1073}. \URLprefix
  \url{https://dx.doi.org/10.1088/0305-4470/39/5/005}.
  \DOIprefix\doi{10.1088/0305-4470/39/5/005}.
\bibitem[{Victorin et~al.(2018)Victorin, Hekking, and
  Minguzzi}]{victorin2018bosonic}
\bibinfo{author}{N.~Victorin}, \bibinfo{author}{F.~Hekking},
  \bibinfo{author}{A.~Minguzzi},
\newblock \bibinfo{title}{Bosonic double ring lattice under artificial gauge
  fields},
\newblock \bibinfo{journal}{Physical Review A} \bibinfo{volume}{98}
  (\bibinfo{year}{2018}) \bibinfo{pages}{053626}. \URLprefix
  \url{https://link.aps.org/doi/10.1103/PhysRevA.98.053626}.
  \DOIprefix\doi{10.1103/PhysRevA.98.053626}.
\bibitem[{Montgomery et~al.(2010)Montgomery, Scott, Lesanovsky, and
  Fromhold}]{montgomery2010spontaneous}
\bibinfo{author}{T.~W.~A. Montgomery}, \bibinfo{author}{R.~G. Scott},
  \bibinfo{author}{I.~Lesanovsky}, \bibinfo{author}{T.~M. Fromhold},
\newblock \bibinfo{title}{Spontaneous creation of nonzero-angular-momentum
  modes in tunnel-coupled two-dimensional degenerate bose gases},
\newblock \bibinfo{journal}{Phys. Rev. A} \bibinfo{volume}{81}
  (\bibinfo{year}{2010}) \bibinfo{pages}{063611}. \URLprefix
  \url{https://link.aps.org/doi/10.1103/PhysRevA.81.063611}.
  \DOIprefix\doi{10.1103/PhysRevA.81.063611}.
\bibitem[{P\'erez-Obiol et~al.(2020)P\'erez-Obiol, Polo, and
  Cheon}]{perez2020current}
\bibinfo{author}{A.~P\'erez-Obiol}, \bibinfo{author}{J.~Polo},
  \bibinfo{author}{T.~Cheon},
\newblock \bibinfo{title}{Current production in ring condensates with a weak
  link},
\newblock \bibinfo{journal}{Physical Review A} \bibinfo{volume}{102}
  (\bibinfo{year}{2020}) \bibinfo{pages}{063302}. \URLprefix
  \url{https://link.aps.org/doi/10.1103/PhysRevA.102.063302}.
  \DOIprefix\doi{10.1103/PhysRevA.102.063302}.
\bibitem[{Polo et~al.(2016{\natexlab{a}})Polo, Benseny, Busch, Ahufinger, and
  Mompart}]{polo2016transport}
\bibinfo{author}{J.~Polo}, \bibinfo{author}{A.~Benseny},
  \bibinfo{author}{T.~Busch}, \bibinfo{author}{V.~Ahufinger},
  \bibinfo{author}{J.~Mompart},
\newblock \bibinfo{title}{Transport of ultracold atoms between concentric traps
  via spatial adiabatic passage},
\newblock \bibinfo{journal}{New Journal of Physics} \bibinfo{volume}{18}
  (\bibinfo{year}{2016}{\natexlab{a}}) \bibinfo{pages}{015010}. \URLprefix
  \url{https://dx.doi.org/10.1088/1367-2630/18/1/015010}.
  \DOIprefix\doi{10.1088/1367-2630/18/1/015010}.
\bibitem[{Polo et~al.(2016{\natexlab{b}})Polo, Mompart, and
  Ahufinger}]{polo2016geometrically}
\bibinfo{author}{J.~Polo}, \bibinfo{author}{J.~Mompart},
  \bibinfo{author}{V.~Ahufinger},
\newblock \bibinfo{title}{Geometrically induced complex tunnelings for
  ultracold atoms carrying orbital angular momentum},
\newblock \bibinfo{journal}{Physical Review A} \bibinfo{volume}{93}
  (\bibinfo{year}{2016}{\natexlab{b}}) \bibinfo{pages}{033613}. \URLprefix
  \url{https://link.aps.org/doi/10.1103/PhysRevA.93.033613}.
  \DOIprefix\doi{10.1103/PhysRevA.93.033613}.
\bibitem[{Menchon-Enrich et~al.(2014)Menchon-Enrich, McEndoo, Mompart,
  Ahufinger, and Busch}]{menchonenrich2014tunneling}
\bibinfo{author}{R.~Menchon-Enrich}, \bibinfo{author}{S.~McEndoo},
  \bibinfo{author}{J.~Mompart}, \bibinfo{author}{V.~Ahufinger},
  \bibinfo{author}{T.~Busch},
\newblock \bibinfo{title}{Tunneling-induced angular momentum for single cold
  atoms},
\newblock \bibinfo{journal}{Physical Review A} \bibinfo{volume}{89}
  (\bibinfo{year}{2014}) \bibinfo{pages}{013626}. \URLprefix
  \url{https://link.aps.org/doi/10.1103/PhysRevA.89.013626}.
  \DOIprefix\doi{10.1103/PhysRevA.89.013626}.
\bibitem[{Bland et~al.(2020)Bland, Marolleau, Comaron, Malomed, and
  Proukakis}]{bland2020persistent}
\bibinfo{author}{T.~Bland}, \bibinfo{author}{Q.~Marolleau},
  \bibinfo{author}{P.~Comaron}, \bibinfo{author}{B.~A. Malomed},
  \bibinfo{author}{N.~P. Proukakis},
\newblock \bibinfo{title}{Persistent current formation in double-ring
  geometries},
\newblock \bibinfo{journal}{Journal of Physics B: Atomic, Molecular and Optical
  Physics} \bibinfo{volume}{53} (\bibinfo{year}{2020}) \bibinfo{pages}{115301}.
  \URLprefix \url{https://dx.doi.org/10.1088/1361-6455/ab81e9}.
  \DOIprefix\doi{10.1088/1361-6455/ab81e9}.
\bibitem[{Bland et~al.(2022)Bland, Yatsuta, Edwards, Nikolaieva, Oliinyk,
  Yakimenko, and Proukakis}]{bland2022persistent}
\bibinfo{author}{T.~Bland}, \bibinfo{author}{I.~V. Yatsuta},
  \bibinfo{author}{M.~Edwards}, \bibinfo{author}{Y.~O. Nikolaieva},
  \bibinfo{author}{A.~O. Oliinyk}, \bibinfo{author}{A.~I. Yakimenko},
  \bibinfo{author}{N.~P. Proukakis},
\newblock \bibinfo{title}{Persistent current oscillations in a double-ring
  quantum gas},
\newblock \bibinfo{journal}{Physical Review Research} \bibinfo{volume}{4}
  (\bibinfo{year}{2022}) \bibinfo{pages}{043171}. \URLprefix
  \url{https://link.aps.org/doi/10.1103/PhysRevResearch.4.043171}.
  \DOIprefix\doi{10.1103/PhysRevResearch.4.043171}.
\bibitem[{Lesanovsky and von Klitzing(2007)}]{lesanovsky2007spontaneous}
\bibinfo{author}{I.~Lesanovsky}, \bibinfo{author}{W.~von Klitzing},
\newblock \bibinfo{title}{{Spontaneous Emergence of Angular Momentum Josephson
  Oscillations in Coupled Annular Bose-Einstein Condensates}},
\newblock \bibinfo{journal}{Physical Review Letters} \bibinfo{volume}{98}
  (\bibinfo{year}{2007}) \bibinfo{pages}{050401}. \URLprefix
  \url{https://link.aps.org/doi/10.1103/PhysRevLett.98.050401}.
  \DOIprefix\doi{10.1103/PhysRevLett.98.050401}.
\bibitem[{Brand et~al.(2010)Brand, Haigh, and Z{\"u}licke}]{brand2010sign}
\bibinfo{author}{J.~Brand}, \bibinfo{author}{T.~J. Haigh},
  \bibinfo{author}{U.~Z{\"u}licke},
\newblock \bibinfo{title}{Sign of coupling in barrier-separated bose-einstein
  condensates and stability of double-ring systems},
\newblock \bibinfo{journal}{Physical Review A—Atomic, Molecular, and Optical
  Physics} \bibinfo{volume}{81} (\bibinfo{year}{2010}) \bibinfo{pages}{025602}.
\bibitem[{Oliinyk et~al.(2019{\natexlab{a}})Oliinyk, Yatsuta, Malomed, and
  Yakimenko}]{oliinyk2019symmetry}
\bibinfo{author}{A.~Oliinyk}, \bibinfo{author}{I.~Yatsuta},
  \bibinfo{author}{B.~Malomed}, \bibinfo{author}{A.~Yakimenko},
\newblock \bibinfo{title}{{Symmetry Breaking in Interacting Ring-Shaped
  Superflows of Bose–Einstein Condensates}},
\newblock \bibinfo{journal}{Symmetry} \bibinfo{volume}{11}
  (\bibinfo{year}{2019}{\natexlab{a}}). \URLprefix
  \url{https://www.mdpi.com/2073-8994/11/10/1312}.
  \DOIprefix\doi{10.3390/sym11101312}.
\bibitem[{Oliinyk et~al.(2019{\natexlab{b}})Oliinyk, Yakimenko, and
  Malomed}]{oliinyk2019tunneling}
\bibinfo{author}{A.~Oliinyk}, \bibinfo{author}{A.~Yakimenko},
  \bibinfo{author}{B.~Malomed},
\newblock \bibinfo{title}{{Tunneling of persistent currents in coupled
  ring-shaped Bose–Einstein condensates}},
\newblock \bibinfo{journal}{Journal of Physics B: Atomic, Molecular and Optical
  Physics} \bibinfo{volume}{52} (\bibinfo{year}{2019}{\natexlab{b}})
  \bibinfo{pages}{225301}. \URLprefix
  \url{https://dx.doi.org/10.1088/1361-6455/ab46f9}.
  \DOIprefix\doi{10.1088/1361-6455/ab46f9}.
\bibitem[{Nicolau et~al.(2020)Nicolau, Mompart, Juli\'a-D\'{\i}az, and
  Ahufinger}]{nicolau2020orbital}
\bibinfo{author}{E.~Nicolau}, \bibinfo{author}{J.~Mompart},
  \bibinfo{author}{B.~Juli\'a-D\'{\i}az}, \bibinfo{author}{V.~Ahufinger},
\newblock \bibinfo{title}{{Orbital angular momentum dynamics of Bose-Einstein
  condensates trapped in two stacked rings}},
\newblock \bibinfo{journal}{Physical Review A} \bibinfo{volume}{102}
  (\bibinfo{year}{2020}) \bibinfo{pages}{023331}. \URLprefix
  \url{https://link.aps.org/doi/10.1103/PhysRevA.102.023331}.
  \DOIprefix\doi{10.1103/PhysRevA.102.023331}.
\bibitem[{Bazhan et~al.(2022)Bazhan, Svetlichnyi, Pfeiffer, Derr, Birkl, and
  Yakimenko}]{bazhan2022generation}
\bibinfo{author}{N.~Bazhan}, \bibinfo{author}{A.~Svetlichnyi},
  \bibinfo{author}{D.~Pfeiffer}, \bibinfo{author}{D.~Derr},
  \bibinfo{author}{G.~Birkl}, \bibinfo{author}{A.~Yakimenko},
\newblock \bibinfo{title}{Generation of josephson vortices in stacked toroidal
  bose-einstein condensates},
\newblock \bibinfo{journal}{Physical Review A} \bibinfo{volume}{106}
  (\bibinfo{year}{2022}) \bibinfo{pages}{043305}.
\bibitem[{Escriv\`a et~al.(2019)Escriv\`a, Mateo, Guilleumas, and
  Juli\'a-D\'{\i}az}]{escriva2019tunneling}
\bibinfo{author}{A.~Escriv\`a}, \bibinfo{author}{A.~M.~n. Mateo},
  \bibinfo{author}{M.~Guilleumas}, \bibinfo{author}{B.~Juli\'a-D\'{\i}az},
\newblock \bibinfo{title}{{Tunneling vortex dynamics in linearly coupled
  Bose-Hubbard rings}},
\newblock \bibinfo{journal}{Physical Review A} \bibinfo{volume}{100}
  (\bibinfo{year}{2019}) \bibinfo{pages}{063621}. \URLprefix
  \url{https://link.aps.org/doi/10.1103/PhysRevA.100.063621}.
  \DOIprefix\doi{10.1103/PhysRevA.100.063621}.
\bibitem[{Hejazi et~al.(2022)Hejazi, Polo, and Tsubota}]{hejazi2022formation}
\bibinfo{author}{S.~S.~S. Hejazi}, \bibinfo{author}{J.~Polo},
  \bibinfo{author}{M.~Tsubota},
\newblock \bibinfo{title}{{Formation of local and global currents in a toroidal
  Bose-Einstein condensate via an inhomogeneous artificial gauge field}},
\newblock \bibinfo{journal}{Physical Review A} \bibinfo{volume}{105}
  (\bibinfo{year}{2022}) \bibinfo{pages}{053307}. \URLprefix
  \url{https://link.aps.org/doi/10.1103/PhysRevA.105.053307}.
  \DOIprefix\doi{10.1103/PhysRevA.105.053307}.
\bibitem[{Piazza et~al.(2011)Piazza, Collins, and
  Smerzi}]{piazza2011instability}
\bibinfo{author}{F.~Piazza}, \bibinfo{author}{L.~A. Collins},
  \bibinfo{author}{A.~Smerzi},
\newblock \bibinfo{title}{Instability and vortex ring dynamics in a
  three-dimensional superfluid flow through a constriction},
\newblock \bibinfo{journal}{New Journal of Physics} \bibinfo{volume}{13}
  (\bibinfo{year}{2011}) \bibinfo{pages}{043008}. \URLprefix
  \url{https://dx.doi.org/10.1088/1367-2630/13/4/043008}.
  \DOIprefix\doi{10.1088/1367-2630/13/4/043008}.
\bibitem[{Yakimenko et~al.(2015{\natexlab{a}})Yakimenko, Bidasyuk, Weyrauch,
  Kuriatnikov, and Vilchinskii}]{yakimenko2015vortices}
\bibinfo{author}{A.~Yakimenko}, \bibinfo{author}{Y.~Bidasyuk},
  \bibinfo{author}{M.~Weyrauch}, \bibinfo{author}{Y.~Kuriatnikov},
  \bibinfo{author}{S.~Vilchinskii},
\newblock \bibinfo{title}{{Vortices in a toroidal Bose-Einstein condensate with
  a rotating weak link}},
\newblock \bibinfo{journal}{Physical Review A} \bibinfo{volume}{91}
  (\bibinfo{year}{2015}{\natexlab{a}}) \bibinfo{pages}{033607}.
\bibitem[{Yakimenko et~al.(2015{\natexlab{b}})Yakimenko, Isaieva, Vilchinskii,
  and Ostrovskaya}]{yakimenko2015vortex}
\bibinfo{author}{A.~Yakimenko}, \bibinfo{author}{K.~Isaieva},
  \bibinfo{author}{S.~Vilchinskii}, \bibinfo{author}{E.~Ostrovskaya},
\newblock \bibinfo{title}{{Vortex excitation in a stirred toroidal
  Bose-Einstein condensate}},
\newblock \bibinfo{journal}{Physical Review A} \bibinfo{volume}{91}
  (\bibinfo{year}{2015}{\natexlab{b}}) \bibinfo{pages}{023607}.
\bibitem[{Yakimenko et~al.(2015{\natexlab{c}})Yakimenko, Vilchinskii,
  Kuriatnikov, Isaieva, Bidasyuk, and Weyrauch}]{yakimenko2014generation}
\bibinfo{author}{A.~Yakimenko}, \bibinfo{author}{S.~Vilchinskii},
  \bibinfo{author}{Y.~Kuriatnikov}, \bibinfo{author}{K.~Isaieva},
  \bibinfo{author}{Y.~Bidasyuk}, \bibinfo{author}{M.~Weyrauch},
\newblock \bibinfo{title}{Generation and decay of persistent currents in a
  toroidal bose-einstein condensate},
\newblock \bibinfo{journal}{Romanian Reports on Physics} \bibinfo{volume}{67}
  (\bibinfo{year}{2015}{\natexlab{c}}) \bibinfo{pages}{249--272}.
\bibitem[{Weimer et~al.(2015)Weimer, Morgener, Singh, Siegl, Hueck, Luick,
  Mathey, and Moritz}]{weimer2015critical}
\bibinfo{author}{W.~Weimer}, \bibinfo{author}{K.~Morgener},
  \bibinfo{author}{V.~P. Singh}, \bibinfo{author}{J.~Siegl},
  \bibinfo{author}{K.~Hueck}, \bibinfo{author}{N.~Luick},
  \bibinfo{author}{L.~Mathey}, \bibinfo{author}{H.~Moritz},
\newblock \bibinfo{title}{{Critical velocity in the BEC-BCS crossover}},
\newblock \bibinfo{journal}{Physical Review Letters} \bibinfo{volume}{114}
  (\bibinfo{year}{2015}) \bibinfo{pages}{095301}.
\bibitem[{Singh et~al.(2016)Singh, Weimer, Morgener, Siegl, Hueck, Luick,
  Moritz, and Mathey}]{singh2016probing}
\bibinfo{author}{V.~P. Singh}, \bibinfo{author}{W.~Weimer},
  \bibinfo{author}{K.~Morgener}, \bibinfo{author}{J.~Siegl},
  \bibinfo{author}{K.~Hueck}, \bibinfo{author}{N.~Luick},
  \bibinfo{author}{H.~Moritz}, \bibinfo{author}{L.~Mathey},
\newblock \bibinfo{title}{{Probing superfluidity of Bose-Einstein condensates
  via laser stirring}},
\newblock \bibinfo{journal}{Physical Review A} \bibinfo{volume}{93}
  (\bibinfo{year}{2016}) \bibinfo{pages}{023634}.
\bibitem[{Eller et~al.(2020)Eller, Oladehin, Fogarty, Heller, Clark, and
  Edwards}]{eller2020producing}
\bibinfo{author}{B.~Eller}, \bibinfo{author}{O.~Oladehin},
  \bibinfo{author}{D.~Fogarty}, \bibinfo{author}{C.~Heller},
  \bibinfo{author}{C.~W. Clark}, \bibinfo{author}{M.~Edwards},
\newblock \bibinfo{title}{Producing flow in racetrack atom circuits by
  stirring},
\newblock \bibinfo{journal}{Physical Review A} \bibinfo{volume}{102}
  (\bibinfo{year}{2020}) \bibinfo{pages}{063324}.
\bibitem[{Nikolaieva et~al.(2023)Nikolaieva, Salasnich, and
  Yakimenko}]{nikolaieva2023engineering}
\bibinfo{author}{Y.~Nikolaieva}, \bibinfo{author}{L.~Salasnich},
  \bibinfo{author}{A.~Yakimenko},
\newblock \bibinfo{title}{{Engineering phase and density of Bose--Einstein
  condensates in curved waveguides with toroidal topology}},
\newblock \bibinfo{journal}{New Journal of Physics} \bibinfo{volume}{25}
  (\bibinfo{year}{2023}) \bibinfo{pages}{103003}.
\bibitem[{Tononi et~al.(2024)Tononi, Salasnich, and
  Yakimenko}]{tononi2024quantum}
\bibinfo{author}{A.~Tononi}, \bibinfo{author}{L.~Salasnich},
  \bibinfo{author}{A.~Yakimenko},
\newblock \bibinfo{title}{Quantum vortices in curved geometries},
\newblock \bibinfo{journal}{AVS Quantum Science} \bibinfo{volume}{6}
  (\bibinfo{year}{2024}). \DOIprefix\doi{10.1116/5.0211426}.
\bibitem[{P\'erez-Obiol and Cheon(2020)}]{perez2020bose}
\bibinfo{author}{A.~P\'erez-Obiol}, \bibinfo{author}{T.~Cheon},
\newblock \bibinfo{title}{{Bose-Einstein condensate confined in a
  one-dimensional ring stirred with a rotating delta link}},
\newblock \bibinfo{journal}{Physical Review E} \bibinfo{volume}{101}
  (\bibinfo{year}{2020}) \bibinfo{pages}{022212}. \URLprefix
  \url{https://link.aps.org/doi/10.1103/PhysRevE.101.022212}.
  \DOIprefix\doi{10.1103/PhysRevE.101.022212}.
\bibitem[{Reeves et~al.(2022)Reeves, Goddard-Lee, Gauthier, Stockdale, Salman,
  Edmonds, Yu, Bradley, Baker, Rubinsztein-Dunlop, Davis, and
  Neely}]{reeves2022turbulent}
\bibinfo{author}{M.~T. Reeves}, \bibinfo{author}{K.~Goddard-Lee},
  \bibinfo{author}{G.~Gauthier}, \bibinfo{author}{O.~R. Stockdale},
  \bibinfo{author}{H.~Salman}, \bibinfo{author}{T.~Edmonds},
  \bibinfo{author}{X.~Yu}, \bibinfo{author}{A.~S. Bradley},
  \bibinfo{author}{M.~Baker}, \bibinfo{author}{H.~Rubinsztein-Dunlop},
  \bibinfo{author}{M.~J. Davis}, \bibinfo{author}{T.~W. Neely},
\newblock \bibinfo{title}{{Turbulent Relaxation to Equilibrium in a
  Two-Dimensional Quantum Vortex Gas}},
\newblock \bibinfo{journal}{Physical Review X} \bibinfo{volume}{12}
  (\bibinfo{year}{2022}) \bibinfo{pages}{011031}. \URLprefix
  \url{https://link.aps.org/doi/10.1103/PhysRevX.12.011031}.
  \DOIprefix\doi{10.1103/PhysRevX.12.011031}.
\bibitem[{Kwon et~al.(2021)Kwon, Del~Pace, Xhani, Galantucci, Muzi~Falconi,
  Inguscio, Scazza, and Roati}]{kwon2021sound}
\bibinfo{author}{W.~J. Kwon}, \bibinfo{author}{G.~Del~Pace},
  \bibinfo{author}{K.~Xhani}, \bibinfo{author}{L.~Galantucci},
  \bibinfo{author}{A.~Muzi~Falconi}, \bibinfo{author}{M.~Inguscio},
  \bibinfo{author}{F.~Scazza}, \bibinfo{author}{G.~Roati},
\newblock \bibinfo{title}{Sound emission and annihilations in a programmable
  quantum vortex collider},
\newblock \bibinfo{journal}{Nature} \bibinfo{volume}{600}
  (\bibinfo{year}{2021}) \bibinfo{pages}{64--69}.
\bibitem[{Chen et~al.(2005)Chen, Stajic, Tan, and Levin}]{chen2005bcs}
\bibinfo{author}{Q.~Chen}, \bibinfo{author}{J.~Stajic},
  \bibinfo{author}{S.~Tan}, \bibinfo{author}{K.~Levin},
\newblock \bibinfo{title}{{BCS--BEC crossover: From high temperature
  superconductors to ultracold superfluids}},
\newblock \bibinfo{journal}{Physics Reports} \bibinfo{volume}{412}
  (\bibinfo{year}{2005}) \bibinfo{pages}{1--88}.
\bibitem[{Zwerger(2011)}]{zwerger2011bcs}
\bibinfo{author}{W.~Zwerger}, \bibinfo{title}{{The BCS-BEC crossover and the
  unitary Fermi gas}}, volume \bibinfo{volume}{836},
  \bibinfo{publisher}{Springer Science \& Business Media},
  \bibinfo{year}{2011}.
\bibitem[{Giorgini et~al.(2008)Giorgini, Pitaevskii, and
  Stringari}]{giorgini2008theory}
\bibinfo{author}{S.~Giorgini}, \bibinfo{author}{L.~P. Pitaevskii},
  \bibinfo{author}{S.~Stringari},
\newblock \bibinfo{title}{{Theory of ultracold atomic Fermi gases}},
\newblock \bibinfo{journal}{Reviews of Modern Physics} \bibinfo{volume}{80}
  (\bibinfo{year}{2008}) \bibinfo{pages}{1215--1274}.
\bibitem[{Allman et~al.(2023)Allman, Sabharwal, and Wright}]{allman2023heating}
\bibinfo{author}{D.~G. Allman}, \bibinfo{author}{P.~Sabharwal},
  \bibinfo{author}{K.~C. Wright},
\newblock \bibinfo{title}{Mitigating heating of degenerate fermions in a
  ring-dimple atomic trap},
\newblock \bibinfo{journal}{Physical Review A} \bibinfo{volume}{107}
  (\bibinfo{year}{2023}) \bibinfo{pages}{043322}. \URLprefix
  \url{https://link.aps.org/doi/10.1103/PhysRevA.107.043322}.
  \DOIprefix\doi{10.1103/PhysRevA.107.043322}.
\bibitem[{Yu and Fowler(1992)}]{yu1992persistent}
\bibinfo{author}{N.~Yu}, \bibinfo{author}{M.~Fowler},
\newblock \bibinfo{title}{{Persistent current of a Hubbard ring threaded with a
  magnetic flux}},
\newblock \bibinfo{journal}{Physical Review B} \bibinfo{volume}{45}
  (\bibinfo{year}{1992}) \bibinfo{pages}{11795--11804}. \URLprefix
  \url{https://link.aps.org/doi/10.1103/PhysRevB.45.11795}.
  \DOIprefix\doi{10.1103/PhysRevB.45.11795}.
\bibitem[{P\^a\ifmmode~\mbox{\c{t}}\else \c{t}\fi{}u and
  Averin(2022)}]{patu2022temperature}
\bibinfo{author}{O.~I. P\^a\ifmmode~\mbox{\c{t}}\else \c{t}\fi{}u},
  \bibinfo{author}{D.~V. Averin},
\newblock \bibinfo{title}{{Temperature-Dependent Periodicity of the Persistent
  Current in Strongly Interacting Systems}},
\newblock \bibinfo{journal}{Physical Review Letters} \bibinfo{volume}{128}
  (\bibinfo{year}{2022}) \bibinfo{pages}{096801}. \URLprefix
  \url{https://link.aps.org/doi/10.1103/PhysRevLett.128.096801}.
  \DOIprefix\doi{10.1103/PhysRevLett.128.096801}.
\bibitem[{Yin et~al.(2011)Yin, Guan, Batchelor, and Chen}]{yin2011effective}
\bibinfo{author}{X.~Yin}, \bibinfo{author}{X.-W. Guan}, \bibinfo{author}{M.~T.
  Batchelor}, \bibinfo{author}{S.~Chen},
\newblock \bibinfo{title}{{Effective super Tonks-Girardeau gases as ground
  states of strongly attractive multicomponent fermions}},
\newblock \bibinfo{journal}{Physical Review A} \bibinfo{volume}{83}
  (\bibinfo{year}{2011}) \bibinfo{pages}{013602}. \URLprefix
  \url{https://link.aps.org/doi/10.1103/PhysRevA.83.013602}.
  \DOIprefix\doi{10.1103/PhysRevA.83.013602}.
\bibitem[{Mooij et~al.(1999)Mooij, Orlando, Levitov, Tian, Van~der Wal, and
  Lloyd}]{mooij1999josephson}
\bibinfo{author}{J.~Mooij}, \bibinfo{author}{T.~Orlando},
  \bibinfo{author}{L.~Levitov}, \bibinfo{author}{L.~Tian},
  \bibinfo{author}{C.~H. Van~der Wal}, \bibinfo{author}{S.~Lloyd},
\newblock \bibinfo{title}{Josephson persistent-current qubit},
\newblock \bibinfo{journal}{Science} \bibinfo{volume}{285}
  (\bibinfo{year}{1999}) \bibinfo{pages}{1036--1039}.
\bibitem[{Chiorescu et~al.(2003)Chiorescu, Nakamura, Harmans, and
  Mooij}]{chiorescu2003coherent}
\bibinfo{author}{I.~Chiorescu}, \bibinfo{author}{Y.~Nakamura},
  \bibinfo{author}{C.~M. Harmans}, \bibinfo{author}{J.~Mooij},
\newblock \bibinfo{title}{Coherent quantum dynamics of a superconducting flux
  qubit},
\newblock \bibinfo{journal}{Science} \bibinfo{volume}{299}
  (\bibinfo{year}{2003}) \bibinfo{pages}{1869--1871}.
\bibitem[{M(2020)}]{travagnin2020cold}
\bibinfo{author}{T.~M},
\newblock \bibinfo{title}{Cold atom interferometry sensors: physics and
  technologies}  (\bibinfo{year}{2020}). \DOIprefix\doi{10.2760/315209
  (online)}.
\bibitem[{Geiger et~al.(2020)Geiger, Landragin, Merlet, and Pereira~dos
  Santos}]{geiger2020high}
\bibinfo{author}{R.~Geiger}, \bibinfo{author}{A.~Landragin},
  \bibinfo{author}{S.~Merlet}, \bibinfo{author}{F.~Pereira~dos Santos},
\newblock \bibinfo{title}{{High-accuracy inertial measurements with cold-atom
  sensors}},
\newblock \bibinfo{journal}{{AVS Quantum Science}} \bibinfo{volume}{2}
  (\bibinfo{year}{2020}) \bibinfo{pages}{024702}. \URLprefix
  \url{https://hal.science/hal-02988722}. \DOIprefix\doi{10.1116/5.0009093}.
\bibitem[{Menoret et~al.(2018)Menoret, Vermeulen, Le~Moigne, Bonvalot, Bouyer,
  Landragin, and Desruelle}]{menoret2018gravity}
\bibinfo{author}{V.~Menoret}, \bibinfo{author}{P.~Vermeulen},
  \bibinfo{author}{N.~Le~Moigne}, \bibinfo{author}{S.~Bonvalot},
  \bibinfo{author}{P.~Bouyer}, \bibinfo{author}{A.~Landragin},
  \bibinfo{author}{B.~Desruelle},
\newblock \bibinfo{title}{Gravity measurements below 10$^{-9}$ g with a
  transportable absolute quantum gravimeter},
\newblock \bibinfo{journal}{Scientific Reports} \bibinfo{volume}{8}
  (\bibinfo{year}{2018}). \URLprefix
  \url{http://dx.doi.org/10.1038/s41598-018-30608-1}.
  \DOIprefix\doi{10.1038/s41598-018-30608-1}.
\bibitem[{Wales et~al.(2020)Wales, Rakonjac, Billam, Helm, Gardiner, and
  Cornish}]{wales2020splitting}
\bibinfo{author}{O.~J. Wales}, \bibinfo{author}{A.~Rakonjac},
  \bibinfo{author}{T.~P. Billam}, \bibinfo{author}{J.~L. Helm},
  \bibinfo{author}{S.~A. Gardiner}, \bibinfo{author}{S.~L. Cornish},
\newblock \bibinfo{title}{Splitting and recombination of bright-solitary-matter
  waves},
\newblock \bibinfo{journal}{Communications Physics} \bibinfo{volume}{3}
  (\bibinfo{year}{2020}) \bibinfo{pages}{51}.
\bibitem[{Gauguet et~al.(2009)Gauguet, Canuel, L\'ev\`eque, Chaibi, and
  Landragin}]{gauguet2009characterization}
\bibinfo{author}{A.~Gauguet}, \bibinfo{author}{B.~Canuel},
  \bibinfo{author}{T.~L\'ev\`eque}, \bibinfo{author}{W.~Chaibi},
  \bibinfo{author}{A.~Landragin},
\newblock \bibinfo{title}{Characterization and limits of a cold-atom sagnac
  interferometer},
\newblock \bibinfo{journal}{Physical Review A} \bibinfo{volume}{80}
  (\bibinfo{year}{2009}) \bibinfo{pages}{063604}. \URLprefix
  \url{https://link.aps.org/doi/10.1103/PhysRevA.80.063604}.
  \DOIprefix\doi{10.1103/PhysRevA.80.063604}.
\bibitem[{McDonald et~al.(2014)McDonald, Kuhn, Hardman, Bennetts, Everitt,
  Altin, Debs, Close, and Robins}]{mcdonald2014bright}
\bibinfo{author}{G.~D. McDonald}, \bibinfo{author}{C.~C.~N. Kuhn},
  \bibinfo{author}{K.~S. Hardman}, \bibinfo{author}{S.~Bennetts},
  \bibinfo{author}{P.~J. Everitt}, \bibinfo{author}{P.~A. Altin},
  \bibinfo{author}{J.~E. Debs}, \bibinfo{author}{J.~D. Close},
  \bibinfo{author}{N.~P. Robins},
\newblock \bibinfo{title}{Bright solitonic matter-wave interferometer},
\newblock \bibinfo{journal}{Physical Review Letters} \bibinfo{volume}{113}
  (\bibinfo{year}{2014}) \bibinfo{pages}{013002}. \URLprefix
  \url{https://link.aps.org/doi/10.1103/PhysRevLett.113.013002}.
  \DOIprefix\doi{10.1103/PhysRevLett.113.013002}.
\bibitem[{Polo and Ahufinger(2013)}]{polo2013soliton}
\bibinfo{author}{J.~Polo}, \bibinfo{author}{V.~Ahufinger},
\newblock \bibinfo{title}{Soliton-based matter-wave interferometer},
\newblock \bibinfo{journal}{Physical Review A} \bibinfo{volume}{88}
  (\bibinfo{year}{2013}) \bibinfo{pages}{053628}. \URLprefix
  \url{https://link.aps.org/doi/10.1103/PhysRevA.88.053628}.
  \DOIprefix\doi{10.1103/PhysRevA.88.053628}.
\bibitem[{Helm et~al.(2014)Helm, Rooney, Weiss, and
  Gardiner}]{helm2014splitting}
\bibinfo{author}{J.~L. Helm}, \bibinfo{author}{S.~J. Rooney},
  \bibinfo{author}{C.~Weiss}, \bibinfo{author}{S.~A. Gardiner},
\newblock \bibinfo{title}{{Splitting bright matter-wave solitons on narrow
  potential barriers: Quantum to classical transition and applications to
  interferometry}},
\newblock \bibinfo{journal}{Physical Review A} \bibinfo{volume}{89}
  (\bibinfo{year}{2014}) \bibinfo{pages}{033610}. \URLprefix
  \url{https://link.aps.org/doi/10.1103/PhysRevA.89.033610}.
  \DOIprefix\doi{10.1103/PhysRevA.89.033610}.
\bibitem[{Helm et~al.(2012)Helm, Billam, and Gardiner}]{helm2012bright}
\bibinfo{author}{J.~L. Helm}, \bibinfo{author}{T.~P. Billam},
  \bibinfo{author}{S.~A. Gardiner},
\newblock \bibinfo{title}{Bright matter-wave soliton collisions at narrow
  barriers},
\newblock \bibinfo{journal}{Physical Review A} \bibinfo{volume}{85}
  (\bibinfo{year}{2012}) \bibinfo{pages}{053621}. \URLprefix
  \url{https://link.aps.org/doi/10.1103/PhysRevA.85.053621}.
  \DOIprefix\doi{10.1103/PhysRevA.85.053621}.
\bibitem[{Streltsov et~al.(2009)Streltsov, Alon, and
  Cederbaum}]{streltsov2009efficient}
\bibinfo{author}{A.~I. Streltsov}, \bibinfo{author}{O.~E. Alon},
  \bibinfo{author}{L.~S. Cederbaum},
\newblock \bibinfo{title}{{Efficient generation and properties of mesoscopic
  quantum superposition states in an attractive Bose–Einstein condensate
  threaded by a potential barrier}},
\newblock \bibinfo{journal}{Journal of Physics B: Atomic, Molecular and Optical
  Physics} \bibinfo{volume}{42} (\bibinfo{year}{2009}) \bibinfo{pages}{091004}.
  \URLprefix \url{https://dx.doi.org/10.1088/0953-4075/42/9/091004}.
  \DOIprefix\doi{10.1088/0953-4075/42/9/091004}.
\bibitem[{Durfee et~al.(2006)Durfee, Shaham, and Kasevich}]{durfee20006long}
\bibinfo{author}{D.~S. Durfee}, \bibinfo{author}{Y.~K. Shaham},
  \bibinfo{author}{M.~A. Kasevich},
\newblock \bibinfo{title}{{Long-Term Stability of an Area-Reversible
  Atom-Interferometer Sagnac Gyroscope}},
\newblock \bibinfo{journal}{Physical Review Letters} \bibinfo{volume}{97}
  (\bibinfo{year}{2006}) \bibinfo{pages}{240801}. \URLprefix
  \url{https://link.aps.org/doi/10.1103/PhysRevLett.97.240801}.
  \DOIprefix\doi{10.1103/PhysRevLett.97.240801}.
\bibitem[{Krzyzanowska et~al.(2023)Krzyzanowska, Ferreras, Ryu, Samson, and
  Boshier}]{krzyzanowska2023matter}
\bibinfo{author}{K.~A. Krzyzanowska}, \bibinfo{author}{J.~Ferreras},
  \bibinfo{author}{C.~Ryu}, \bibinfo{author}{E.~C. Samson},
  \bibinfo{author}{M.~G. Boshier},
\newblock \bibinfo{title}{Matter-wave analog of a fiber-optic gyroscope},
\newblock \bibinfo{journal}{Physical Review A} \bibinfo{volume}{108}
  (\bibinfo{year}{2023}) \bibinfo{pages}{043305}. \URLprefix
  \url{https://link.aps.org/doi/10.1103/PhysRevA.108.043305}.
  \DOIprefix\doi{10.1103/PhysRevA.108.043305}.
\bibitem[{Stevenson et~al.(2015)Stevenson, Hush, Bishop, Lesanovsky, and
  Fernholz}]{stevenson2015sagnac}
\bibinfo{author}{R.~Stevenson}, \bibinfo{author}{M.~R. Hush},
  \bibinfo{author}{T.~Bishop}, \bibinfo{author}{I.~Lesanovsky},
  \bibinfo{author}{T.~Fernholz},
\newblock \bibinfo{title}{{Sagnac Interferometry with a Single Atomic Clock}},
\newblock \bibinfo{journal}{Physical Review Letters} \bibinfo{volume}{115}
  (\bibinfo{year}{2015}) \bibinfo{pages}{163001}. \URLprefix
  \url{https://link.aps.org/doi/10.1103/PhysRevLett.115.163001}.
  \DOIprefix\doi{10.1103/PhysRevLett.115.163001}.
\bibitem[{Beydler et~al.(2024)Beydler, Moan, Luo, Chu, and
  Sackett}]{beydier2024guided}
\bibinfo{author}{M.~M. Beydler}, \bibinfo{author}{E.~R. Moan},
  \bibinfo{author}{Z.~Luo}, \bibinfo{author}{Z.~Chu}, \bibinfo{author}{C.~A.
  Sackett},
\newblock \bibinfo{title}{{Guided-wave Sagnac atom interferometer with large
  area and multiple orbits}},
\newblock \bibinfo{journal}{AVS Quantum Science} \bibinfo{volume}{6}
  (\bibinfo{year}{2024}) \bibinfo{pages}{014401}. \URLprefix
  \url{https://doi.org/10.1116/5.0173769}. \DOIprefix\doi{10.1116/5.0173769}.
\bibitem[{Xhani et~al.(2023)Xhani, Del~Pace, Scazza, and
  Roati}]{klejdja2023decay}
\bibinfo{author}{K.~Xhani}, \bibinfo{author}{G.~Del~Pace},
  \bibinfo{author}{F.~Scazza}, \bibinfo{author}{G.~Roati},
\newblock \bibinfo{title}{{Decay of Persistent Currents in Annular Atomic
  Superfluids}},
\newblock \bibinfo{journal}{Atoms} \bibinfo{volume}{11} (\bibinfo{year}{2023}).
  \URLprefix \url{https://www.mdpi.com/2218-2004/11/8/109}.
  \DOIprefix\doi{10.3390/atoms11080109}.
\bibitem[{Kiehn et~al.(2022)Kiehn, Singh, and Mathey}]{kiehn2022implementation}
\bibinfo{author}{H.~Kiehn}, \bibinfo{author}{V.~P. Singh},
  \bibinfo{author}{L.~Mathey},
\newblock \bibinfo{title}{{Implementation of an atomtronic SQUID in a strongly
  confined toroidal condensate}},
\newblock \bibinfo{journal}{Physical Review Research} \bibinfo{volume}{4}
  (\bibinfo{year}{2022}) \bibinfo{pages}{033024}. \URLprefix
  \url{https://link.aps.org/doi/10.1103/PhysRevResearch.4.033024}.
  \DOIprefix\doi{10.1103/PhysRevResearch.4.033024}.
\bibitem[{Ragole and Taylor(2016)}]{ragole2016interacting}
\bibinfo{author}{S.~Ragole}, \bibinfo{author}{J.~M. Taylor},
\newblock \bibinfo{title}{{Interacting atomic interferometry for rotation
  sensing approaching the Heisenberg Limit}},
\newblock \bibinfo{journal}{Physical Review Letters} \bibinfo{volume}{117}
  (\bibinfo{year}{2016}) \bibinfo{pages}{203002}.
\bibitem[{Pelegrí et~al.(2018)Pelegrí, Mompart, and
  Ahufinger}]{pelegri2018quantum}
\bibinfo{author}{G.~Pelegrí}, \bibinfo{author}{J.~Mompart},
  \bibinfo{author}{V.~Ahufinger},
\newblock \bibinfo{title}{{Quantum sensing using imbalanced counter-rotating
  Bose–Einstein condensate modes}},
\newblock \bibinfo{journal}{New Journal of Physics} \bibinfo{volume}{20}
  (\bibinfo{year}{2018}) \bibinfo{pages}{103001}. \URLprefix
  \url{https://dx.doi.org/10.1088/1367-2630/aae107}.
  \DOIprefix\doi{10.1088/1367-2630/aae107}.
\bibitem[{{Adeniji} et~al.(2024){Adeniji}, {Henry}, {Thomas}, {Colson Sapp},
  {Goyal}, {Clark}, and {Edwards}}]{adeniji2024double}
\bibinfo{author}{O.~{Adeniji}}, \bibinfo{author}{C.~{Henry}},
  \bibinfo{author}{S.~{Thomas}}, \bibinfo{author}{R.~{Colson Sapp}},
  \bibinfo{author}{A.~{Goyal}}, \bibinfo{author}{C.~W. {Clark}},
  \bibinfo{author}{M.~{Edwards}},
\newblock \bibinfo{title}{{Double-target BEC atomtronic rotation sensor}},
\newblock \bibinfo{journal}{arXiv e-prints}  (\bibinfo{year}{2024})
  \bibinfo{pages}{arXiv:2411.06585}. \DOIprefix\doi{10.48550/arXiv.2411.06585}.
  \href{http://arxiv.org/abs/2411.06585}{{\tt arXiv:2411.06585}}.
\bibitem[{{Chaika} et~al.(2024){Chaika}, {Oliinyk}, {Yatsuta}, {Proukakis},
  {Edwards}, {Yakimenko}, and {Bland}}]{chaika2024acceleration}
\bibinfo{author}{A.~{Chaika}}, \bibinfo{author}{A.~O. {Oliinyk}},
  \bibinfo{author}{I.~V. {Yatsuta}}, \bibinfo{author}{N.~P. {Proukakis}},
  \bibinfo{author}{M.~{Edwards}}, \bibinfo{author}{A.~I. {Yakimenko}},
  \bibinfo{author}{T.~{Bland}},
\newblock \bibinfo{title}{{Acceleration-induced transport of quantum vortices
  in joined atomtronic circuits}},
\newblock \bibinfo{journal}{arXiv e-prints}  (\bibinfo{year}{2024})
  \bibinfo{pages}{arXiv:2410.23818}. \DOIprefix\doi{10.48550/arXiv.2410.23818}.
  \href{http://arxiv.org/abs/2410.23818}{{\tt arXiv:2410.23818}}.
\bibitem[{Lau et~al.(2023)Lau, Gan, Dumke, Amico, Kwek, and
  Haug}]{lau2023atomtronic}
\bibinfo{author}{J.~W.~Z. Lau}, \bibinfo{author}{K.~S. Gan},
  \bibinfo{author}{R.~Dumke}, \bibinfo{author}{L.~Amico},
  \bibinfo{author}{L.-C. Kwek}, \bibinfo{author}{T.~Haug},
\newblock \bibinfo{title}{{Atomtronic multiterminal Aharonov-Bohm
  interferometer}},
\newblock \bibinfo{journal}{Physical Review A} \bibinfo{volume}{107}
  (\bibinfo{year}{2023}) \bibinfo{pages}{L051303}.
  \DOIprefix\doi{10.1103/PhysRevA.107.L051303}.
\bibitem[{Makotyn et~al.(2014)Makotyn, Klauss, Goldberger, Cornell, and
  Jin}]{makotyn2014universal}
\bibinfo{author}{P.~Makotyn}, \bibinfo{author}{C.~E. Klauss},
  \bibinfo{author}{D.~L. Goldberger}, \bibinfo{author}{E.~A. Cornell},
  \bibinfo{author}{D.~S. Jin},
\newblock \bibinfo{title}{{Universal dynamics of a degenerate unitary Bose
  gas}},
\newblock \bibinfo{journal}{Nature Physics} \bibinfo{volume}{10}
  (\bibinfo{year}{2014}) \bibinfo{pages}{116--119}. \URLprefix
  \url{https://doi.org/10.1038/nphys2850}. \DOIprefix\doi{10.1038/nphys2850}.
\bibitem[{Hu et~al.(2007)Hu, Drummond, and Liu}]{hu2007universal}
\bibinfo{author}{H.~Hu}, \bibinfo{author}{P.~D. Drummond},
  \bibinfo{author}{X.-J. Liu},
\newblock \bibinfo{title}{{Universal thermodynamics of strongly interacting
  Fermi gases}},
\newblock \bibinfo{journal}{Nature Physics} \bibinfo{volume}{3}
  (\bibinfo{year}{2007}) \bibinfo{pages}{469--472}. \URLprefix
  \url{https://doi.org/10.1038/nphys598}. \DOIprefix\doi{10.1038/nphys598}.
\bibitem[{Nascimb{\`{e}}ne et~al.(2010)Nascimb{\`{e}}ne, Navon, Jiang, Chevy,
  and Salomon}]{nascimbne2010exploring}
\bibinfo{author}{S.~Nascimb{\`{e}}ne}, \bibinfo{author}{N.~Navon},
  \bibinfo{author}{K.~J. Jiang}, \bibinfo{author}{F.~Chevy},
  \bibinfo{author}{C.~Salomon},
\newblock \bibinfo{title}{{Exploring the thermodynamics of a universal Fermi
  gas}},
\newblock \bibinfo{journal}{Nature} \bibinfo{volume}{463}
  (\bibinfo{year}{2010}) \bibinfo{pages}{1057--1060}. \URLprefix
  \url{https://doi.org/10.1038/nature08814}.
  \DOIprefix\doi{10.1038/nature08814}.
\bibitem[{Ho(2004)}]{ho2004universal}
\bibinfo{author}{T.-L. Ho},
\newblock \bibinfo{title}{{Universal Thermodynamics of Degenerate Quantum Gases
  in the Unitarity Limit}},
\newblock \bibinfo{journal}{Physical Review Letters} \bibinfo{volume}{92}
  (\bibinfo{year}{2004}) \bibinfo{pages}{090402}. \URLprefix
  \url{https://link.aps.org/doi/10.1103/PhysRevLett.92.090402}.
  \DOIprefix\doi{10.1103/PhysRevLett.92.090402}.
\bibitem[{Barcel\'o et~al.(2003)Barcel\'o, Liberati, and
  Visser}]{barcelo2003probing}
\bibinfo{author}{C.~Barcel\'o}, \bibinfo{author}{S.~Liberati},
  \bibinfo{author}{M.~Visser},
\newblock \bibinfo{title}{{Probing semiclassical analog gravity in
  Bose-Einstein condensates with widely tunable interactions}},
\newblock \bibinfo{journal}{Physical Review A} \bibinfo{volume}{68}
  (\bibinfo{year}{2003}) \bibinfo{pages}{053613}. \URLprefix
  \url{https://link.aps.org/doi/10.1103/PhysRevA.68.053613}.
  \DOIprefix\doi{10.1103/PhysRevA.68.053613}.
\bibitem[{Giovanazzi(2005)}]{giovanazzi2005hawking}
\bibinfo{author}{S.~Giovanazzi},
\newblock \bibinfo{title}{{Hawking Radiation in Sonic Black Holes}},
\newblock \bibinfo{journal}{Physical Review Letters} \bibinfo{volume}{94}
  (\bibinfo{year}{2005}) \bibinfo{pages}{061302}. \URLprefix
  \url{https://link.aps.org/doi/10.1103/PhysRevLett.94.061302}.
  \DOIprefix\doi{10.1103/PhysRevLett.94.061302}.
\bibitem[{Nguyen et~al.(2015)Nguyen, Gerace, Carusotto, Sanvitto, Galopin,
  Lema\^{\i}tre, Sagnes, Bloch, and Amo}]{nguyen2015acoustic}
\bibinfo{author}{H.~S. Nguyen}, \bibinfo{author}{D.~Gerace},
  \bibinfo{author}{I.~Carusotto}, \bibinfo{author}{D.~Sanvitto},
  \bibinfo{author}{E.~Galopin}, \bibinfo{author}{A.~Lema\^{\i}tre},
  \bibinfo{author}{I.~Sagnes}, \bibinfo{author}{J.~Bloch},
  \bibinfo{author}{A.~Amo},
\newblock \bibinfo{title}{{Acoustic Black Hole in a Stationary Hydrodynamic
  Flow of Microcavity Polaritons}},
\newblock \bibinfo{journal}{Physical Review Letters} \bibinfo{volume}{114}
  (\bibinfo{year}{2015}) \bibinfo{pages}{036402}. \URLprefix
  \url{https://link.aps.org/doi/10.1103/PhysRevLett.114.036402}.
  \DOIprefix\doi{10.1103/PhysRevLett.114.036402}.
\bibitem[{Steinhauer(2016)}]{steinhauer2016observation}
\bibinfo{author}{J.~Steinhauer},
\newblock \bibinfo{title}{{Observation of quantum Hawking radiation and its
  entanglement in an analogue black hole}},
\newblock \bibinfo{journal}{Nature Physics} \bibinfo{volume}{12}
  (\bibinfo{year}{2016}) \bibinfo{pages}{959--965}. \URLprefix
  \url{https://doi.org/10.1038/nphys3863}. \DOIprefix\doi{10.1038/nphys3863}.
\bibitem[{Belgiorno et~al.(2010)Belgiorno, Cacciatori, Clerici, Gorini,
  Ortenzi, Rizzi, Rubino, Sala, and Faccio}]{belgiorno2010hawking}
\bibinfo{author}{F.~Belgiorno}, \bibinfo{author}{S.~L. Cacciatori},
  \bibinfo{author}{M.~Clerici}, \bibinfo{author}{V.~Gorini},
  \bibinfo{author}{G.~Ortenzi}, \bibinfo{author}{L.~Rizzi},
  \bibinfo{author}{E.~Rubino}, \bibinfo{author}{V.~G. Sala},
  \bibinfo{author}{D.~Faccio},
\newblock \bibinfo{title}{{Hawking Radiation from Ultrashort Laser Pulse
  Filaments}},
\newblock \bibinfo{journal}{Physical Review Letters} \bibinfo{volume}{105}
  (\bibinfo{year}{2010}) \bibinfo{pages}{203901}. \URLprefix
  \url{https://link.aps.org/doi/10.1103/PhysRevLett.105.203901}.
  \DOIprefix\doi{10.1103/PhysRevLett.105.203901}.
\bibitem[{Banik et~al.(2022)Banik, Galan, Sosa-Martinez, Anderson, Eckel,
  Spielman, and Campbell}]{banik2022accurate}
\bibinfo{author}{S.~Banik}, \bibinfo{author}{M.~G. Galan},
  \bibinfo{author}{H.~Sosa-Martinez}, \bibinfo{author}{M.~J. Anderson},
  \bibinfo{author}{S.~Eckel}, \bibinfo{author}{I.~B. Spielman},
  \bibinfo{author}{G.~K. Campbell},
\newblock \bibinfo{title}{{Accurate Determination of Hubble Attenuation and
  Amplification in Expanding and Contracting Cold-Atom Universes}},
\newblock \bibinfo{journal}{Physical Review Letters} \bibinfo{volume}{128}
  (\bibinfo{year}{2022}) \bibinfo{pages}{090401}. \URLprefix
  \url{https://link.aps.org/doi/10.1103/PhysRevLett.128.090401}.
  \DOIprefix\doi{10.1103/PhysRevLett.128.090401}.
\bibitem[{Cao et~al.(2011)Cao, Elliott, Joseph, Wu, Petricka, Sch\"{a}fer, and
  Thomas}]{cao2011universal}
\bibinfo{author}{C.~Cao}, \bibinfo{author}{E.~Elliott},
  \bibinfo{author}{J.~Joseph}, \bibinfo{author}{H.~Wu},
  \bibinfo{author}{J.~Petricka}, \bibinfo{author}{T.~Sch\"{a}fer},
  \bibinfo{author}{J.~E. Thomas},
\newblock \bibinfo{title}{{Universal Quantum Viscosity in a Unitary Fermi
  Gas}},
\newblock \bibinfo{journal}{Science} \bibinfo{volume}{331}
  (\bibinfo{year}{2011}) \bibinfo{pages}{58--61}. \URLprefix
  \url{https://doi.org/10.1126/science.1195219}.
  \DOIprefix\doi{10.1126/science.1195219}.
\bibitem[{Cazalilla et~al.(2009)Cazalilla, Ho, and
  Ueda}]{cazalilla2009ultracold}
\bibinfo{author}{M.~A. Cazalilla}, \bibinfo{author}{A.~F. Ho},
  \bibinfo{author}{M.~Ueda},
\newblock \bibinfo{title}{{Ultracold gases of ytterbium: ferromagnetism and
  Mott states in an SU(6) Fermi system}},
\newblock \bibinfo{journal}{New Journal of Physics} \bibinfo{volume}{11}
  (\bibinfo{year}{2009}) \bibinfo{pages}{103033}. \URLprefix
  \url{https://dx.doi.org/10.1088/1367-2630/11/10/103033}.
  \DOIprefix\doi{10.1088/1367-2630/11/10/103033}.
\bibitem[{Guan et~al.(2013)Guan, Batchelor, and Lee}]{guan2013fermi}
\bibinfo{author}{X.-W. Guan}, \bibinfo{author}{M.~T. Batchelor},
  \bibinfo{author}{C.~Lee},
\newblock \bibinfo{title}{{Fermi gases in one dimension: From Bethe ansatz to
  experiments}},
\newblock \bibinfo{journal}{Reviews of Modern Physics} \bibinfo{volume}{85}
  (\bibinfo{year}{2013}) \bibinfo{pages}{1633--1691}. \URLprefix
  \url{https://link.aps.org/doi/10.1103/RevModPhys.85.1633}.
  \DOIprefix\doi{10.1103/RevModPhys.85.1633}.
\bibitem[{Rapp et~al.(2007)Rapp, Zar\'and, Honerkamp, and
  Hofstetter}]{rapp2007color}
\bibinfo{author}{A.~Rapp}, \bibinfo{author}{G.~Zar\'and},
  \bibinfo{author}{C.~Honerkamp}, \bibinfo{author}{W.~Hofstetter},
\newblock \bibinfo{title}{{Color Superfluidity and ``Baryon'' Formation in
  Ultracold Fermions}},
\newblock \bibinfo{journal}{Physical Review Letters} \bibinfo{volume}{98}
  (\bibinfo{year}{2007}) \bibinfo{pages}{160405}. \URLprefix
  \url{https://link.aps.org/doi/10.1103/PhysRevLett.98.160405}.
  \DOIprefix\doi{10.1103/PhysRevLett.98.160405}.
\bibitem[{Ozawa and Baym(2010)}]{ozawa2010population}
\bibinfo{author}{T.~Ozawa}, \bibinfo{author}{G.~Baym},
\newblock \bibinfo{title}{{Population imbalance and pairing in the BCS-BEC
  crossover of three-component ultracold fermions}},
\newblock \bibinfo{journal}{Physical Review A} \bibinfo{volume}{82}
  (\bibinfo{year}{2010}) \bibinfo{pages}{063615}. \URLprefix
  \url{https://link.aps.org/doi/10.1103/PhysRevA.82.063615}.
  \DOIprefix\doi{10.1103/PhysRevA.82.063615}.
\bibitem[{He et~al.(2006)He, Jin, and Zhuang}]{He2006superfluidity}
\bibinfo{author}{L.~He}, \bibinfo{author}{M.~Jin}, \bibinfo{author}{P.~Zhuang},
\newblock \bibinfo{title}{Superfluidity in a three-flavor fermi gas with
  $\mathrm{SU}(3)$ symmetry},
\newblock \bibinfo{journal}{Physical Review A} \bibinfo{volume}{74}
  (\bibinfo{year}{2006}) \bibinfo{pages}{033604}. \URLprefix
  \url{https://link.aps.org/doi/10.1103/PhysRevA.74.033604}.
  \DOIprefix\doi{10.1103/PhysRevA.74.033604}.
\bibitem[{Wiese(2013)}]{wiese2013ultracold}
\bibinfo{author}{U.-J. Wiese},
\newblock \bibinfo{title}{Ultracold quantum gases and lattice systems: quantum
  simulation of lattice gauge theories},
\newblock \bibinfo{journal}{Annalen der Physik} \bibinfo{volume}{525}
  (\bibinfo{year}{2013}) \bibinfo{pages}{777--796}. \URLprefix
  \url{https://doi.org/10.1002/andp.201300104}.
  \DOIprefix\doi{10.1002/andp.201300104}.
\bibitem[{Zohar et~al.(2015)Zohar, Cirac, and Reznik}]{zohar2016quantum}
\bibinfo{author}{E.~Zohar}, \bibinfo{author}{J.~I. Cirac},
  \bibinfo{author}{B.~Reznik},
\newblock \bibinfo{title}{Quantum simulations of lattice gauge theories using
  ultracold atoms in optical lattices},
\newblock \bibinfo{journal}{Reports on Progress in Physics}
  \bibinfo{volume}{79} (\bibinfo{year}{2015}) \bibinfo{pages}{014401}.
  \URLprefix \url{https://dx.doi.org/10.1088/0034-4885/79/1/014401}.
  \DOIprefix\doi{10.1088/0034-4885/79/1/014401}.
\bibitem[{Schweizer et~al.(2019)Schweizer, Grusdt, Berngruber, Barbiero,
  Demler, Goldman, Bloch, and Aidelsburger}]{schweizer2019floquet}
\bibinfo{author}{C.~Schweizer}, \bibinfo{author}{F.~Grusdt},
  \bibinfo{author}{M.~Berngruber}, \bibinfo{author}{L.~Barbiero},
  \bibinfo{author}{E.~Demler}, \bibinfo{author}{N.~Goldman},
  \bibinfo{author}{I.~Bloch}, \bibinfo{author}{M.~Aidelsburger},
\newblock \bibinfo{title}{Floquet approach to $\{BbbZ\}$2 lattice gauge
  theories with ultracold atoms in optical lattices},
\newblock \bibinfo{journal}{Nature Physics} \bibinfo{volume}{15}
  (\bibinfo{year}{2019}) \bibinfo{pages}{1168--1173}. \URLprefix
  \url{https://doi.org/10.1038/s41567-019-0649-7}.
  \DOIprefix\doi{10.1038/s41567-019-0649-7}.
\bibitem[{Aidelsburger et~al.(2021)Aidelsburger, Barbiero, Bermudez, Chanda,
  Dauphin, Gonz{\'{a}}lez-Cuadra, Grzybowski, Hands, Jendrzejewski,
  J\"{u}nemann, Juzeli{\={u}}nas, Kasper, Piga, Ran, Rizzi, Sierra,
  Tagliacozzo, Tirrito, Zache, Zakrzewski, Zohar, and
  Lewenstein}]{aidelsburger2021cold}
\bibinfo{author}{M.~Aidelsburger}, \bibinfo{author}{L.~Barbiero},
  \bibinfo{author}{A.~Bermudez}, \bibinfo{author}{T.~Chanda},
  \bibinfo{author}{A.~Dauphin}, \bibinfo{author}{D.~Gonz{\'{a}}lez-Cuadra},
  \bibinfo{author}{P.~R. Grzybowski}, \bibinfo{author}{S.~Hands},
  \bibinfo{author}{F.~Jendrzejewski}, \bibinfo{author}{J.~J\"{u}nemann},
  \bibinfo{author}{G.~Juzeli{\={u}}nas}, \bibinfo{author}{V.~Kasper},
  \bibinfo{author}{A.~Piga}, \bibinfo{author}{S.-J. Ran},
  \bibinfo{author}{M.~Rizzi}, \bibinfo{author}{G.~Sierra},
  \bibinfo{author}{L.~Tagliacozzo}, \bibinfo{author}{E.~Tirrito},
  \bibinfo{author}{T.~V. Zache}, \bibinfo{author}{J.~Zakrzewski},
  \bibinfo{author}{E.~Zohar}, \bibinfo{author}{M.~Lewenstein},
\newblock \bibinfo{title}{Cold atoms meet lattice gauge theory},
\newblock \bibinfo{journal}{Philosophical Transactions of the Royal Society A:
  Mathematical, Physical and Engineering Sciences} \bibinfo{volume}{380}
  (\bibinfo{year}{2021}). \URLprefix
  \url{https://doi.org/10.1098/rsta.2021.0064}.
  \DOIprefix\doi{10.1098/rsta.2021.0064}.
\bibitem[{Altman et~al.(2021)Altman, Brown, Carleo, Carr, Demler, Chin,
  DeMarco, Economou, Eriksson, Fu, Greiner, Hazzard, Hulet, Koll\'ar, Lev,
  Lukin, Ma, Mi, Misra, Monroe, Murch, Nazario, Ni, Potter, Roushan, Saffman,
  Schleier-Smith, Siddiqi, Simmonds, Singh, Spielman, Temme, Weiss, Vu\ifmmode
  \check{c}\else \v{c}\fi{}kovi\ifmmode~\acute{c}\else \'{c}\fi{},
  Vuleti\ifmmode~\acute{c}\else \'{c}\fi{}, Ye, and
  Zwierlein}]{altman2021quantum}
\bibinfo{author}{E.~Altman}, \bibinfo{author}{K.~R. Brown},
  \bibinfo{author}{G.~Carleo}, \bibinfo{author}{L.~D. Carr},
  \bibinfo{author}{E.~Demler}, \bibinfo{author}{C.~Chin},
  \bibinfo{author}{B.~DeMarco}, \bibinfo{author}{S.~E. Economou},
  \bibinfo{author}{M.~A. Eriksson}, \bibinfo{author}{K.-M.~C. Fu},
  \bibinfo{author}{M.~Greiner}, \bibinfo{author}{K.~R. Hazzard},
  \bibinfo{author}{R.~G. Hulet}, \bibinfo{author}{A.~J. Koll\'ar},
  \bibinfo{author}{B.~L. Lev}, \bibinfo{author}{M.~D. Lukin},
  \bibinfo{author}{R.~Ma}, \bibinfo{author}{X.~Mi}, \bibinfo{author}{S.~Misra},
  \bibinfo{author}{C.~Monroe}, \bibinfo{author}{K.~Murch},
  \bibinfo{author}{Z.~Nazario}, \bibinfo{author}{K.-K. Ni},
  \bibinfo{author}{A.~C. Potter}, \bibinfo{author}{P.~Roushan},
  \bibinfo{author}{M.~Saffman}, \bibinfo{author}{M.~Schleier-Smith},
  \bibinfo{author}{I.~Siddiqi}, \bibinfo{author}{R.~Simmonds},
  \bibinfo{author}{M.~Singh}, \bibinfo{author}{I.~Spielman},
  \bibinfo{author}{K.~Temme}, \bibinfo{author}{D.~S. Weiss},
  \bibinfo{author}{J.~Vu\ifmmode \check{c}\else
  \v{c}\fi{}kovi\ifmmode~\acute{c}\else \'{c}\fi{}},
  \bibinfo{author}{V.~Vuleti\ifmmode~\acute{c}\else \'{c}\fi{}},
  \bibinfo{author}{J.~Ye}, \bibinfo{author}{M.~Zwierlein},
\newblock \bibinfo{title}{{Quantum Simulators: Architectures and
  Opportunities}},
\newblock \bibinfo{journal}{PRX Quantum} \bibinfo{volume}{2}
  (\bibinfo{year}{2021}) \bibinfo{pages}{017003}. \URLprefix
  \url{https://link.aps.org/doi/10.1103/PRXQuantum.2.017003}.
  \DOIprefix\doi{10.1103/PRXQuantum.2.017003}.
\bibitem[{Chien et~al.(2015)Chien, Peotta, and Di~Ventra}]{chien2015quantum}
\bibinfo{author}{C.-C. Chien}, \bibinfo{author}{S.~Peotta},
  \bibinfo{author}{M.~Di~Ventra},
\newblock \bibinfo{title}{Quantum transport in ultracold atoms},
\newblock \bibinfo{journal}{Nature Physics} \bibinfo{volume}{11}
  (\bibinfo{year}{2015}) \bibinfo{pages}{998--1004}.
  \DOIprefix\doi{10.1038/nphys3531}.
\bibitem[{Fagaly(2006)}]{fagaly2006superconducting}
\bibinfo{author}{R.~Fagaly},
\newblock \bibinfo{title}{Superconducting quantum interference device
  instruments and applications},
\newblock \bibinfo{journal}{Review of scientific instruments}
  \bibinfo{volume}{77} (\bibinfo{year}{2006}).
\bibitem[{Krinner et~al.(2017)Krinner, Esslinger, and Brantut}]{krinner2017two}
\bibinfo{author}{S.~Krinner}, \bibinfo{author}{T.~Esslinger},
  \bibinfo{author}{J.-P. Brantut},
\newblock \bibinfo{title}{Two-terminal transport measurements with cold atoms},
\newblock \bibinfo{journal}{Journal of Physics: Condensed Matter}
  \bibinfo{volume}{29} (\bibinfo{year}{2017}) \bibinfo{pages}{343003}.
  \DOIprefix\doi{10.1088/1361-648X/aa74a1}.
\bibitem[{Loiko et~al.(2014)Loiko, Ahufinger, Menchon-Enrich, Birkl, and
  Mompart}]{loiko2014coherent}
\bibinfo{author}{Y.~Loiko}, \bibinfo{author}{V.~Ahufinger},
  \bibinfo{author}{R.~Menchon-Enrich}, \bibinfo{author}{G.~Birkl},
  \bibinfo{author}{J.~Mompart},
\newblock \bibinfo{title}{Coherent injecting, extracting, and velocity
  filtering of neutral atoms in a ring trap via spatial adiabatic passage},
\newblock \bibinfo{journal}{The European Physical Journal D}
  \bibinfo{volume}{68} (\bibinfo{year}{2014}) \bibinfo{pages}{1--5}.
  \DOIprefix\doi{10.1140/epjd/e2014-40696-3}.
\bibitem[{Haug(2021)}]{haug2021quantum}
\bibinfo{author}{T.~F. Haug}, \bibinfo{title}{Quantum transport with cold
  atoms}, Ph.D. thesis, National University of Singapore (Singapore),
  \bibinfo{year}{2021}. \URLprefix
  \url{https://scholarbank.nus.edu.sg/handle/10635/190520}.
\bibitem[{Thouless(1983)}]{thouless1983quantization}
\bibinfo{author}{D.~Thouless},
\newblock \bibinfo{title}{Quantization of particle transport},
\newblock \bibinfo{journal}{Physical Review B} \bibinfo{volume}{27}
  (\bibinfo{year}{1983}) \bibinfo{pages}{6083}.
  \DOIprefix\doi{10.1103/PhysRevB.27.6083}.
\bibitem[{Gou et~al.(2020)Gou, Chen, Xie, Xiao, Deng, Gadway, Yi, and
  Yan}]{gou2020tunable}
\bibinfo{author}{W.~Gou}, \bibinfo{author}{T.~Chen}, \bibinfo{author}{D.~Xie},
  \bibinfo{author}{T.~Xiao}, \bibinfo{author}{T.-S. Deng},
  \bibinfo{author}{B.~Gadway}, \bibinfo{author}{W.~Yi},
  \bibinfo{author}{B.~Yan},
\newblock \bibinfo{title}{{Tunable nonreciprocal quantum transport through a
  dissipative Aharonov-Bohm ring in ultracold atoms}},
\newblock \bibinfo{journal}{Physical review letters} \bibinfo{volume}{124}
  (\bibinfo{year}{2020}) \bibinfo{pages}{070402}.
  \DOIprefix\doi{10.1103/PhysRevLett.124.070402}.
\bibitem[{L{\"o}w et~al.(2012)L{\"o}w, Weimer, Nipper, Balewski, Butscher,
  B{\"u}chler, and Pfau}]{low2012experimental}
\bibinfo{author}{R.~L{\"o}w}, \bibinfo{author}{H.~Weimer},
  \bibinfo{author}{J.~Nipper}, \bibinfo{author}{J.~B. Balewski},
  \bibinfo{author}{B.~Butscher}, \bibinfo{author}{H.~P. B{\"u}chler},
  \bibinfo{author}{T.~Pfau},
\newblock \bibinfo{title}{{An experimental and theoretical guide to strongly
  interacting Rydberg gases}},
\newblock \bibinfo{journal}{Journal of Physics B: Atomic, Molecular and Optical
  Physics} \bibinfo{volume}{45} (\bibinfo{year}{2012}) \bibinfo{pages}{113001}.
  \DOIprefix\doi{10.1088/0953-4075/45/11/113001}.
\bibitem[{Browaeys and Lahaye(2020)}]{browaeys2020many}
\bibinfo{author}{A.~Browaeys}, \bibinfo{author}{T.~Lahaye},
\newblock \bibinfo{title}{{Many-body physics with individually controlled
  Rydberg atoms}},
\newblock \bibinfo{journal}{Nature Physics} \bibinfo{volume}{16}
  (\bibinfo{year}{2020}) \bibinfo{pages}{132--142}.
  \DOIprefix\doi{10.1038/s41567-019-0733-z}.
\bibitem[{Khaneja et~al.(2005)Khaneja, Reiss, Kehlet, Schulte-Herbr{\"u}ggen,
  and Glaser}]{khaneja2005optimal}
\bibinfo{author}{N.~Khaneja}, \bibinfo{author}{T.~Reiss},
  \bibinfo{author}{C.~Kehlet}, \bibinfo{author}{T.~Schulte-Herbr{\"u}ggen},
  \bibinfo{author}{S.~J. Glaser},
\newblock \bibinfo{title}{{Optimal control of coupled spin dynamics: design of
  NMR pulse sequences by gradient ascent algorithms}},
\newblock \bibinfo{journal}{Journal of magnetic resonance}
  \bibinfo{volume}{172} (\bibinfo{year}{2005}) \bibinfo{pages}{296--305}.
\bibitem[{Kitson et~al.(2024)Kitson, Haug, La~Magna, Morsch, and
  Amico}]{kitson2023rydberg}
\bibinfo{author}{P.~Kitson}, \bibinfo{author}{T.~Haug},
  \bibinfo{author}{A.~La~Magna}, \bibinfo{author}{O.~Morsch},
  \bibinfo{author}{L.~Amico},
\newblock \bibinfo{title}{Rydberg atomtronic devices},
\newblock \bibinfo{journal}{Physical Review A} \bibinfo{volume}{110}
  (\bibinfo{year}{2024}) \bibinfo{pages}{043304}. \URLprefix
  \url{https://link.aps.org/doi/10.1103/PhysRevA.110.043304}.
  \DOIprefix\doi{10.1103/PhysRevA.110.043304}.
\bibitem[{Morsch and Lesanovsky(2018)}]{morsch2018dissipative}
\bibinfo{author}{O.~Morsch}, \bibinfo{author}{I.~Lesanovsky},
\newblock \bibinfo{title}{{Dissipative many-body physics of cold Rydberg
  atoms}},
\newblock \bibinfo{journal}{La Rivista del Nuovo Cimento} \bibinfo{volume}{41}
  (\bibinfo{year}{2018}) \bibinfo{pages}{383--414}.
\bibitem[{Valencia-Tortora et~al.(2023)Valencia-Tortora, Pancotti,
  Fleischhauer, Bernien, and Marino}]{valencia2023rydberg}
\bibinfo{author}{R.~J. Valencia-Tortora}, \bibinfo{author}{N.~Pancotti},
  \bibinfo{author}{M.~Fleischhauer}, \bibinfo{author}{H.~Bernien},
  \bibinfo{author}{J.~Marino},
\newblock \bibinfo{title}{A rydberg platform for non-ergodic chiral quantum
  dynamics},
\newblock \bibinfo{journal}{arXiv:2309.12392}  (\bibinfo{year}{2023}).
\bibitem[{Léonard et~al.(2017)Léonard, Morales, Zupancic, Esslinger, and
  Donner}]{Leonard2017}
\bibinfo{author}{J.~Léonard}, \bibinfo{author}{A.~Morales},
  \bibinfo{author}{P.~Zupancic}, \bibinfo{author}{T.~Esslinger},
  \bibinfo{author}{T.~Donner},
\newblock \bibinfo{title}{Supersolid formation in a quantum gas breaking a
  continuous translational symmetry},
\newblock \bibinfo{journal}{Nature} \bibinfo{volume}{543}
  (\bibinfo{year}{2017}) \bibinfo{pages}{87--90}.
  \DOIprefix\doi{10.1038/nature21067}.
\bibitem[{Li et~al.(2017)Li, Lee, Huang, Burchesky, Shteynas, Top, Jamison, and
  Ketterle}]{Li2017}
\bibinfo{author}{J.-R. Li}, \bibinfo{author}{J.~Lee},
  \bibinfo{author}{W.~Huang}, \bibinfo{author}{S.~Burchesky},
  \bibinfo{author}{B.~Shteynas}, \bibinfo{author}{F.~{\c{C}}. Top},
  \bibinfo{author}{A.~O. Jamison}, \bibinfo{author}{W.~Ketterle},
\newblock \bibinfo{title}{A stripe phase with supersolid properties in
  spin-orbit-coupled {Bose}-{Einstein} condensates},
\newblock \bibinfo{journal}{Nature} \bibinfo{volume}{543}
  (\bibinfo{year}{2017}) \bibinfo{pages}{91--94}.
  \DOIprefix\doi{10.1038/nature21431}.
\bibitem[{Chomaz et~al.(2022)Chomaz, Ferrier-Barbut, Ferlaino, Laburthe-Tolra,
  Lev, and Pfau}]{chomaz2022dipolar}
\bibinfo{author}{L.~Chomaz}, \bibinfo{author}{I.~Ferrier-Barbut},
  \bibinfo{author}{F.~Ferlaino}, \bibinfo{author}{B.~Laburthe-Tolra},
  \bibinfo{author}{B.~L. Lev}, \bibinfo{author}{T.~Pfau},
\newblock \bibinfo{title}{Dipolar physics: a review of experiments with
  magnetic quantum gases},
\newblock \bibinfo{journal}{Reports on Progress in Physics}
  \bibinfo{volume}{86} (\bibinfo{year}{2022}) \bibinfo{pages}{026401}.
\bibitem[{Li et~al.(2013)Li, Martone, Pitaevskii, and Stringari}]{Li2013}
\bibinfo{author}{Y.~Li}, \bibinfo{author}{G.~I. Martone},
  \bibinfo{author}{L.~P. Pitaevskii}, \bibinfo{author}{S.~Stringari},
\newblock \bibinfo{title}{{Superstripes and the Excitation Spectrum of a
  Spin-Orbit-Coupled {Bose}-{Einstein} Condensate}},
\newblock \bibinfo{journal}{Physical Review Letters} \bibinfo{volume}{110}
  (\bibinfo{year}{2013}) \bibinfo{pages}{235302}.
  \DOIprefix\doi{10.1103/PhysRevLett.110.235302}.
\bibitem[{Tanzi et~al.(2019)Tanzi, Roccuzzo, Lucioni, Famà, Fioretti,
  Gabbanini, Modugno, Recati, and Stringari}]{Tanzi2019a}
\bibinfo{author}{L.~Tanzi}, \bibinfo{author}{S.~M. Roccuzzo},
  \bibinfo{author}{E.~Lucioni}, \bibinfo{author}{F.~Famà},
  \bibinfo{author}{A.~Fioretti}, \bibinfo{author}{C.~Gabbanini},
  \bibinfo{author}{G.~Modugno}, \bibinfo{author}{A.~Recati},
  \bibinfo{author}{S.~Stringari},
\newblock \bibinfo{title}{Supersolid symmetry breaking from compressional
  oscillations in a dipolar quantum gas},
\newblock \bibinfo{journal}{Nature} \bibinfo{volume}{574}
  (\bibinfo{year}{2019}) \bibinfo{pages}{382--385}.
  \DOIprefix\doi{10.1038/s41586-019-1568-6}.
\bibitem[{Geier et~al.(2023)Geier, Martone, Hauke, Ketterle, and
  Stringari}]{Geier2023}
\bibinfo{author}{K.~T. Geier}, \bibinfo{author}{G.~I. Martone},
  \bibinfo{author}{P.~Hauke}, \bibinfo{author}{W.~Ketterle},
  \bibinfo{author}{S.~Stringari},
\newblock \bibinfo{title}{{Dynamics of Stripe Patterns in Supersolid
  Spin-Orbit-Coupled {Bose} Gases}},
\newblock \bibinfo{journal}{Physical Review Letters} \bibinfo{volume}{130}
  (\bibinfo{year}{2023}) \bibinfo{pages}{156001}.
  \DOIprefix\doi{10.1103/PhysRevLett.130.156001}.
\bibitem[{Windt et~al.(2024)Windt, Bello, Demler, and
  Cirac}]{windt2024fermionic}
\bibinfo{author}{B.~Windt}, \bibinfo{author}{M.~Bello},
  \bibinfo{author}{E.~Demler}, \bibinfo{author}{J.~I. Cirac},
\newblock \bibinfo{title}{Fermionic matter-wave quantum optics with cold-atom
  impurity models},
\newblock \bibinfo{journal}{Physical Review A} \bibinfo{volume}{109}
  (\bibinfo{year}{2024}) \bibinfo{pages}{023306}.
\bibitem[{Hewitt et~al.(2024)Hewitt, Bertheas, Jain, Nishida, and
  Barontini}]{hewitt2024controlling}
\bibinfo{author}{T.~Hewitt}, \bibinfo{author}{T.~Bertheas},
  \bibinfo{author}{M.~Jain}, \bibinfo{author}{Y.~Nishida},
  \bibinfo{author}{G.~Barontini},
\newblock \bibinfo{title}{Controlling the interactions in a cold atom quantum
  impurity system},
\newblock \bibinfo{journal}{Quantum Science and Technology} \bibinfo{volume}{9}
  (\bibinfo{year}{2024}) \bibinfo{pages}{035039}.
\bibitem[{Knap et~al.(2012)Knap, Shashi, Nishida, Imambekov, Abanin, and
  Demler}]{knap2012time}
\bibinfo{author}{M.~Knap}, \bibinfo{author}{A.~Shashi},
  \bibinfo{author}{Y.~Nishida}, \bibinfo{author}{A.~Imambekov},
  \bibinfo{author}{D.~A. Abanin}, \bibinfo{author}{E.~Demler},
\newblock \bibinfo{title}{{Time-dependent impurity in ultracold fermions:
  Orthogonality catastrophe and beyond}},
\newblock \bibinfo{journal}{Physical Review X} \bibinfo{volume}{2}
  (\bibinfo{year}{2012}) \bibinfo{pages}{041020}.
\bibitem[{Spethmann et~al.(2012)Spethmann, Kindermann, John, Weber, Meschede,
  and Widera}]{spethmann2012dynamics}
\bibinfo{author}{N.~Spethmann}, \bibinfo{author}{F.~Kindermann},
  \bibinfo{author}{S.~John}, \bibinfo{author}{C.~Weber},
  \bibinfo{author}{D.~Meschede}, \bibinfo{author}{A.~Widera},
\newblock \bibinfo{title}{{Dynamics of single neutral impurity atoms immersed
  in an ultracold gas}},
\newblock \bibinfo{journal}{Physical Review Letters} \bibinfo{volume}{109}
  (\bibinfo{year}{2012}) \bibinfo{pages}{235301}.
\bibitem[{Bauer et~al.(2023)Bauer, Davoudi, Balantekin, Bhattacharya, Carena,
  de~Jong, Draper, El-Khadra, Gemelke, Hanada, Kharzeev, Lamm, Li, Liu, Lukin,
  Meurice, Monroe, Nachman, Pagano, Preskill, Rinaldi, Roggero, Santiago,
  Savage, Siddiqi, Siopsis, Van~Zanten, Wiebe, Yamauchi, Yeter-Aydeniz, and
  Zorzetti}]{bauer2023quantum}
\bibinfo{author}{C.~W. Bauer}, \bibinfo{author}{Z.~Davoudi},
  \bibinfo{author}{A.~B. Balantekin}, \bibinfo{author}{T.~Bhattacharya},
  \bibinfo{author}{M.~Carena}, \bibinfo{author}{W.~A. de~Jong},
  \bibinfo{author}{P.~Draper}, \bibinfo{author}{A.~El-Khadra},
  \bibinfo{author}{N.~Gemelke}, \bibinfo{author}{M.~Hanada},
  \bibinfo{author}{D.~Kharzeev}, \bibinfo{author}{H.~Lamm},
  \bibinfo{author}{Y.-Y. Li}, \bibinfo{author}{J.~Liu},
  \bibinfo{author}{M.~Lukin}, \bibinfo{author}{Y.~Meurice},
  \bibinfo{author}{C.~Monroe}, \bibinfo{author}{B.~Nachman},
  \bibinfo{author}{G.~Pagano}, \bibinfo{author}{J.~Preskill},
  \bibinfo{author}{E.~Rinaldi}, \bibinfo{author}{A.~Roggero},
  \bibinfo{author}{D.~I. Santiago}, \bibinfo{author}{M.~J. Savage},
  \bibinfo{author}{I.~Siddiqi}, \bibinfo{author}{G.~Siopsis},
  \bibinfo{author}{D.~Van~Zanten}, \bibinfo{author}{N.~Wiebe},
  \bibinfo{author}{Y.~Yamauchi}, \bibinfo{author}{K.~Yeter-Aydeniz},
  \bibinfo{author}{S.~Zorzetti},
\newblock \bibinfo{title}{{Quantum Simulation for High-Energy Physics}},
\newblock \bibinfo{journal}{PRX Quantum} \bibinfo{volume}{4}
  (\bibinfo{year}{2023}). \URLprefix
  \url{http://dx.doi.org/10.1103/PRXQuantum.4.027001}.
  \DOIprefix\doi{10.1103/prxquantum.4.027001}.
\bibitem[{Halimeh et~al.(2023)Halimeh, Aidelsburger, Grusdt, Hauke, and
  Yang}]{halimeh2023cold}
\bibinfo{author}{J.~C. Halimeh}, \bibinfo{author}{M.~Aidelsburger},
  \bibinfo{author}{F.~Grusdt}, \bibinfo{author}{P.~Hauke},
  \bibinfo{author}{B.~Yang}, \bibinfo{title}{Cold-atom quantum simulators of
  gauge theories}, \bibinfo{year}{2023}. \URLprefix
  \url{https://arxiv.org/abs/2310.12201}.
  \DOIprefix\doi{10.48550/ARXIV.2310.12201}.
\bibitem[{Leggett(1980)}]{leggett1980macroscopic}
\bibinfo{author}{A.~J. Leggett},
\newblock \bibinfo{title}{{Macroscopic Quantum Systems and the Quantum Theory
  of Measurement}},
\newblock \bibinfo{journal}{Progress of Theoretical Physics Supplement}
  \bibinfo{volume}{69} (\bibinfo{year}{1980}) \bibinfo{pages}{80--100}.
\bibitem[{Perciavalle et~al.(2024)Perciavalle, Morsch, Rossini, and
  Amico}]{perciavalle2024coherent}
\bibinfo{author}{F.~Perciavalle}, \bibinfo{author}{O.~Morsch},
  \bibinfo{author}{D.~Rossini}, \bibinfo{author}{L.~Amico},
\newblock \bibinfo{title}{Coherent excitation transport through ring-shaped
  networks},
\newblock \bibinfo{journal}{Physical Review A} \bibinfo{volume}{109}
  (\bibinfo{year}{2024}) \bibinfo{pages}{062619}.
\bibitem[{Bluvstein et~al.(2024)Bluvstein, Evered, Geim, Li, Zhou, Manovitz,
  Ebadi, Cain, Kalinowski, Hangleiter et~al.}]{bluvstein2024logical}
\bibinfo{author}{D.~Bluvstein}, \bibinfo{author}{S.~J. Evered},
  \bibinfo{author}{A.~A. Geim}, \bibinfo{author}{S.~H. Li},
  \bibinfo{author}{H.~Zhou}, \bibinfo{author}{T.~Manovitz},
  \bibinfo{author}{S.~Ebadi}, \bibinfo{author}{M.~Cain},
  \bibinfo{author}{M.~Kalinowski}, \bibinfo{author}{D.~Hangleiter}, et~al.,
\newblock \bibinfo{title}{Logical quantum processor based on reconfigurable
  atom arrays},
\newblock \bibinfo{journal}{Nature} \bibinfo{volume}{626}
  (\bibinfo{year}{2024}) \bibinfo{pages}{58--65}.
\bibitem[{Evered et~al.(2023)Evered, Bluvstein, Kalinowski, Ebadi, Manovitz,
  Zhou, Li, Geim, Wang, Maskara et~al.}]{evered2023high}
\bibinfo{author}{S.~J. Evered}, \bibinfo{author}{D.~Bluvstein},
  \bibinfo{author}{M.~Kalinowski}, \bibinfo{author}{S.~Ebadi},
  \bibinfo{author}{T.~Manovitz}, \bibinfo{author}{H.~Zhou},
  \bibinfo{author}{S.~H. Li}, \bibinfo{author}{A.~A. Geim},
  \bibinfo{author}{T.~T. Wang}, \bibinfo{author}{N.~Maskara}, et~al.,
\newblock \bibinfo{title}{High-fidelity parallel entangling gates on a
  neutral-atom quantum computer},
\newblock \bibinfo{journal}{Nature} \bibinfo{volume}{622}
  (\bibinfo{year}{2023}) \bibinfo{pages}{268--272}.
\bibitem[{Xu et~al.(2024)Xu, Bonilla~Ataides, Pattison, Raveendran, Bluvstein,
  Wurtz, Vasi{\'c}, Lukin, Jiang, and Zhou}]{xu2024constant}
\bibinfo{author}{Q.~Xu}, \bibinfo{author}{J.~P. Bonilla~Ataides},
  \bibinfo{author}{C.~A. Pattison}, \bibinfo{author}{N.~Raveendran},
  \bibinfo{author}{D.~Bluvstein}, \bibinfo{author}{J.~Wurtz},
  \bibinfo{author}{B.~Vasi{\'c}}, \bibinfo{author}{M.~D. Lukin},
  \bibinfo{author}{L.~Jiang}, \bibinfo{author}{H.~Zhou},
\newblock \bibinfo{title}{Constant-overhead fault-tolerant quantum computation
  with reconfigurable atom arrays},
\newblock \bibinfo{journal}{Nature Physics}  (\bibinfo{year}{2024})
  \bibinfo{pages}{1--7}.
\bibitem[{Akatsuka et~al.(2017)Akatsuka, Takahashi, and
  Katori}]{akatsuka2017optically}
\bibinfo{author}{T.~Akatsuka}, \bibinfo{author}{T.~Takahashi},
  \bibinfo{author}{H.~Katori},
\newblock \bibinfo{title}{Optically guided atom interferometer tuned to magic
  wavelength},
\newblock \bibinfo{journal}{Applied Physics Express} \bibinfo{volume}{10}
  (\bibinfo{year}{2017}) \bibinfo{pages}{112501}.
\bibitem[{McDonald et~al.(2013{\natexlab{a}})McDonald, Kuhn, Bennetts, Debs,
  Hardman, Johnsson, Close, and Robins}]{mcdonald201380hk}
\bibinfo{author}{G.~D. McDonald}, \bibinfo{author}{C.~C.~N. Kuhn},
  \bibinfo{author}{S.~Bennetts}, \bibinfo{author}{J.~E. Debs},
  \bibinfo{author}{K.~S. Hardman}, \bibinfo{author}{M.~Johnsson},
  \bibinfo{author}{J.~D. Close}, \bibinfo{author}{N.~P. Robins},
\newblock \bibinfo{title}{80hk momentum separation with bloch oscillations in
  an optically guided atom interferometer},
\newblock \bibinfo{journal}{Physical Review A} \bibinfo{volume}{88}
  (\bibinfo{year}{2013}{\natexlab{a}}) \bibinfo{pages}{053620}.
\bibitem[{McDonald et~al.(2013{\natexlab{b}})McDonald, Keal, Altin, Debs,
  Bennetts, Kuhn, Hardman, Johnsson, Close, and Robins}]{mcdonald2013optically}
\bibinfo{author}{G.~D. McDonald}, \bibinfo{author}{H.~Keal},
  \bibinfo{author}{P.~A. Altin}, \bibinfo{author}{J.~E. Debs},
  \bibinfo{author}{S.~Bennetts}, \bibinfo{author}{C.~C.~N. Kuhn},
  \bibinfo{author}{K.~S. Hardman}, \bibinfo{author}{M.~T. Johnsson},
  \bibinfo{author}{J.~D. Close}, \bibinfo{author}{N.~P. Robins},
\newblock \bibinfo{title}{{Optically guided linear Mach-Zehnder atom
  interferometer}},
\newblock \bibinfo{journal}{Physical Review A} \bibinfo{volume}{87}
  (\bibinfo{year}{2013}{\natexlab{b}}) \bibinfo{pages}{013632}.
\bibitem[{Reilly et~al.(2023)Reilly, Wilson, J{\"a}ger, Wilson, and
  Holland}]{reilly2023optimal}
\bibinfo{author}{J.~T. Reilly}, \bibinfo{author}{J.~D. Wilson},
  \bibinfo{author}{S.~B. J{\"a}ger}, \bibinfo{author}{C.~Wilson},
  \bibinfo{author}{M.~J. Holland},
\newblock \bibinfo{title}{Optimal generators for quantum sensing},
\newblock \bibinfo{journal}{Physical Review Letters} \bibinfo{volume}{131}
  (\bibinfo{year}{2023}) \bibinfo{pages}{150802}.
\bibitem[{Alonso et~al.(2022)Alonso, Alpigiani, Altschul, Ara{\'u}jo, Arduini,
  Arlt, Badurina, Bala{\v{z}}, Bandarupally, Barish et~al.}]{alonso2022cold}
\bibinfo{author}{I.~Alonso}, \bibinfo{author}{C.~Alpigiani},
  \bibinfo{author}{B.~Altschul}, \bibinfo{author}{H.~Ara{\'u}jo},
  \bibinfo{author}{G.~Arduini}, \bibinfo{author}{J.~Arlt},
  \bibinfo{author}{L.~Badurina}, \bibinfo{author}{A.~Bala{\v{z}}},
  \bibinfo{author}{S.~Bandarupally}, \bibinfo{author}{B.~C. Barish}, et~al.,
\newblock \bibinfo{title}{Cold atoms in space: community workshop summary and
  proposed road-map},
\newblock \bibinfo{journal}{EPJ Quantum Technology} \bibinfo{volume}{9}
  (\bibinfo{year}{2022}) \bibinfo{pages}{1--55}.
\bibitem[{Thompson et~al.(2023)Thompson, Aveline, Chiow, Elliott, Kellogg,
  Kohel, Sbroscia, Schneider, Williams et~al.}]{thompson2023exploring}
\bibinfo{author}{R.~Thompson}, \bibinfo{author}{D.~Aveline},
  \bibinfo{author}{S.-W. Chiow}, \bibinfo{author}{E.~Elliott},
  \bibinfo{author}{J.~Kellogg}, \bibinfo{author}{J.~Kohel},
  \bibinfo{author}{M.~Sbroscia}, \bibinfo{author}{C.~Schneider},
  \bibinfo{author}{J.~Williams}, et~al.,
\newblock \bibinfo{title}{Exploring the limits of ultracold atoms in space},
\newblock \bibinfo{journal}{Quantum Science and Technology} \bibinfo{volume}{8}
  (\bibinfo{year}{2023}) \bibinfo{pages}{024004}.
\bibitem[{Gaaloul et~al.(2022)Gaaloul, Meister, Corgier, Pichery, Boegel, Herr,
  Ahlers, Charron, Williams, Thompson et~al.}]{gaaloul2022space}
\bibinfo{author}{N.~Gaaloul}, \bibinfo{author}{M.~Meister},
  \bibinfo{author}{R.~Corgier}, \bibinfo{author}{A.~Pichery},
  \bibinfo{author}{P.~Boegel}, \bibinfo{author}{W.~Herr},
  \bibinfo{author}{H.~Ahlers}, \bibinfo{author}{E.~Charron},
  \bibinfo{author}{J.~R. Williams}, \bibinfo{author}{R.~J. Thompson}, et~al.,
\newblock \bibinfo{title}{A space-based quantum gas laboratory at picokelvin
  energy scales},
\newblock \bibinfo{journal}{Nature communications} \bibinfo{volume}{13}
  (\bibinfo{year}{2022}) \bibinfo{pages}{7889}.
\bibitem[{Barrett et~al.(2016)Barrett, Bertoldi, and
  Bouyer}]{barrett2016inertial}
\bibinfo{author}{B.~Barrett}, \bibinfo{author}{A.~Bertoldi},
  \bibinfo{author}{P.~Bouyer},
\newblock \bibinfo{title}{Inertial quantum sensors using light and matter},
\newblock \bibinfo{journal}{Physica Scripta} \bibinfo{volume}{91}
  (\bibinfo{year}{2016}) \bibinfo{pages}{053006}.
\bibitem[{Abend et~al.(2023)Abend, Allard, Arnold, Ban, Barry, Battelier,
  Bawamia, Beaufils, Bernon, Bertoldi et~al.}]{abend2023technology}
\bibinfo{author}{S.~Abend}, \bibinfo{author}{B.~Allard}, \bibinfo{author}{A.~S.
  Arnold}, \bibinfo{author}{T.~Ban}, \bibinfo{author}{L.~Barry},
  \bibinfo{author}{B.~Battelier}, \bibinfo{author}{A.~Bawamia},
  \bibinfo{author}{Q.~Beaufils}, \bibinfo{author}{S.~Bernon},
  \bibinfo{author}{A.~Bertoldi}, et~al.,
\newblock \bibinfo{title}{Technology roadmap for cold-atoms based quantum
  inertial sensor in space},
\newblock \bibinfo{journal}{AVS Quantum Science} \bibinfo{volume}{5}
  (\bibinfo{year}{2023}).
\bibitem[{Hensel et~al.(2021)Hensel, Loriani, Schubert, Fitzek, Abend, Ahlers,
  Siem{\ss}, Hammerer, Rasel, and Gaaloul}]{hensel2021inertial}
\bibinfo{author}{T.~Hensel}, \bibinfo{author}{S.~Loriani},
  \bibinfo{author}{C.~Schubert}, \bibinfo{author}{F.~Fitzek},
  \bibinfo{author}{S.~Abend}, \bibinfo{author}{H.~Ahlers},
  \bibinfo{author}{J.-N. Siem{\ss}}, \bibinfo{author}{K.~Hammerer},
  \bibinfo{author}{E.~M. Rasel}, \bibinfo{author}{N.~Gaaloul},
\newblock \bibinfo{title}{Inertial sensing with quantum gases: a comparative
  performance study of condensed versus thermal sources for atom
  interferometry},
\newblock \bibinfo{journal}{The European Physical Journal D}
  \bibinfo{volume}{75} (\bibinfo{year}{2021}) \bibinfo{pages}{1--13}.
\bibitem[{Richardson et~al.(2020)Richardson, Hines, Schaffer, Anderson, and
  Guzman}]{richardson2020quantum}
\bibinfo{author}{L.~Richardson}, \bibinfo{author}{A.~Hines},
  \bibinfo{author}{A.~Schaffer}, \bibinfo{author}{B.~P. Anderson},
  \bibinfo{author}{F.~Guzman},
\newblock \bibinfo{title}{Quantum hybrid optomechanical inertial sensing},
\newblock \bibinfo{journal}{Applied Optics} \bibinfo{volume}{59}
  (\bibinfo{year}{2020}) \bibinfo{pages}{G160--G166}.
\bibitem[{Templier et~al.(2022)Templier, Cheiney, d’Armagnac~de Castanet,
  Gouraud, Porte, Napolitano, Bouyer, Battelier, and
  Barrett}]{templier2022tracking}
\bibinfo{author}{S.~Templier}, \bibinfo{author}{P.~Cheiney},
  \bibinfo{author}{Q.~d’Armagnac~de Castanet}, \bibinfo{author}{B.~Gouraud},
  \bibinfo{author}{H.~Porte}, \bibinfo{author}{F.~Napolitano},
  \bibinfo{author}{P.~Bouyer}, \bibinfo{author}{B.~Battelier},
  \bibinfo{author}{B.~Barrett},
\newblock \bibinfo{title}{Tracking the vector acceleration with a hybrid
  quantum accelerometer triad},
\newblock \bibinfo{journal}{Science Advances} \bibinfo{volume}{8}
  (\bibinfo{year}{2022}) \bibinfo{pages}{eadd3854}.
\bibitem[{Vahlbruch et~al.(2008)Vahlbruch, Mehmet, Chelkowski, Hage, Franzen,
  Lastzka, Gossler, Danzmann, and Schnabel}]{vahlbruch2008observation}
\bibinfo{author}{H.~Vahlbruch}, \bibinfo{author}{M.~Mehmet},
  \bibinfo{author}{S.~Chelkowski}, \bibinfo{author}{B.~Hage},
  \bibinfo{author}{A.~Franzen}, \bibinfo{author}{N.~Lastzka},
  \bibinfo{author}{S.~Gossler}, \bibinfo{author}{K.~Danzmann},
  \bibinfo{author}{R.~Schnabel},
\newblock \bibinfo{title}{{Observation of squeezed light with 10-dB
  quantum-noise reduction}},
\newblock \bibinfo{journal}{Physical Review Letters} \bibinfo{volume}{100}
  (\bibinfo{year}{2008}) \bibinfo{pages}{033602}.
\bibitem[{Treps et~al.(2002)Treps, Andersen, Buchler, Lam, Maitre, Bachor, and
  Fabre}]{treps2002surpassing}
\bibinfo{author}{N.~Treps}, \bibinfo{author}{U.~Andersen},
  \bibinfo{author}{B.~Buchler}, \bibinfo{author}{P.~K. Lam},
  \bibinfo{author}{A.~Maitre}, \bibinfo{author}{H.-A. Bachor},
  \bibinfo{author}{C.~Fabre},
\newblock \bibinfo{title}{Surpassing the standard quantum limit for optical
  imaging using nonclassical multimode light},
\newblock \bibinfo{journal}{Physical Review Letters} \bibinfo{volume}{88}
  (\bibinfo{year}{2002}) \bibinfo{pages}{203601}.
\bibitem[{Kaubruegger et~al.(2019)Kaubruegger, Silvi, Kokail, van Bijnen, Rey,
  Ye, Kaufman, and Zoller}]{kaubruegger2019variational}
\bibinfo{author}{R.~Kaubruegger}, \bibinfo{author}{P.~Silvi},
  \bibinfo{author}{C.~Kokail}, \bibinfo{author}{R.~van Bijnen},
  \bibinfo{author}{A.~M. Rey}, \bibinfo{author}{J.~Ye}, \bibinfo{author}{A.~M.
  Kaufman}, \bibinfo{author}{P.~Zoller},
\newblock \bibinfo{title}{Variational spin-squeezing algorithms on programmable
  quantum sensors},
\newblock \bibinfo{journal}{Physical Review Letters} \bibinfo{volume}{123}
  (\bibinfo{year}{2019}) \bibinfo{pages}{260505}.
\bibitem[{Likharev and Semenov(1991)}]{likharev1991rsfq}
\bibinfo{author}{K.~K. Likharev}, \bibinfo{author}{V.~K. Semenov},
\newblock \bibinfo{title}{{RSFQ logic/memory family: A new Josephson-junction
  technology for sub-terahertz-clock-frequency digital systems}},
\newblock \bibinfo{journal}{IEEE Transactions on Applied Superconductivity}
  \bibinfo{volume}{1} (\bibinfo{year}{1991}) \bibinfo{pages}{3--28}.
\bibitem[{Kumar et~al.(2021)Kumar, Biswas, Feliz, Kanamoto, Chang, Jha, and
  Bhattacharya}]{kumar2021cavity}
\bibinfo{author}{P.~Kumar}, \bibinfo{author}{T.~Biswas},
  \bibinfo{author}{K.~Feliz}, \bibinfo{author}{R.~Kanamoto},
  \bibinfo{author}{M.-S. Chang}, \bibinfo{author}{A.~K. Jha},
  \bibinfo{author}{M.~Bhattacharya},
\newblock \bibinfo{title}{Cavity optomechanical sensing and manipulation of an
  atomic persistent current},
\newblock \bibinfo{journal}{Physical Review Letters} \bibinfo{volume}{127}
  (\bibinfo{year}{2021}) \bibinfo{pages}{113601}.
\bibitem[{Pradhan et~al.(2023)Pradhan, Kumar, Kanamoto, Dey, Bhattacharya, and
  Mishra}]{pradhan2023ring}
\bibinfo{author}{N.~Pradhan}, \bibinfo{author}{P.~Kumar},
  \bibinfo{author}{R.~Kanamoto}, \bibinfo{author}{T.~N. Dey},
  \bibinfo{author}{M.~Bhattacharya}, \bibinfo{author}{P.~K. Mishra},
  \bibinfo{title}{{Ring Bose-Einstein condensate in a cavity: Chirality
  Detection and Rotation Sensing}}, \bibinfo{year}{2023}.
  \href{http://arxiv.org/abs/2311.15226}{{\tt arXiv:2311.15226}}.
\bibitem[{Das et~al.(2024)Das, Kumar, Bhattacharya, and Dey}]{das2024hybrid}
\bibinfo{author}{S.~Das}, \bibinfo{author}{P.~Kumar},
  \bibinfo{author}{M.~Bhattacharya}, \bibinfo{author}{T.~N. Dey},
  \bibinfo{title}{{Hybrid Rotational Cavity Optomechanics Using Atomic
  Superfluid in a Ring}}, \bibinfo{year}{2024}. \URLprefix
  \url{https://arxiv.org/abs/2407.01990}.
  \href{http://arxiv.org/abs/2407.01990}{{\tt arXiv:2407.01990}}.
\bibitem[{Jing et~al.(2019)Jing, Wang, Yu, Sun, Jiang, Yang, Jiang, Luo, Zhang,
  Jiang, Bao, and Pan}]{jing2019entanglement}
\bibinfo{author}{B.~Jing}, \bibinfo{author}{X.-J. Wang},
  \bibinfo{author}{Y.~Yu}, \bibinfo{author}{P.-F. Sun},
  \bibinfo{author}{Y.~Jiang}, \bibinfo{author}{S.-J. Yang},
  \bibinfo{author}{W.-H. Jiang}, \bibinfo{author}{X.-Y. Luo},
  \bibinfo{author}{J.~Zhang}, \bibinfo{author}{X.~Jiang},
  \bibinfo{author}{X.-H. Bao}, \bibinfo{author}{J.-W. Pan},
\newblock \bibinfo{title}{Entanglement of three quantum memories via
  interference of three single photons},
\newblock \bibinfo{journal}{Nature Photonics} \bibinfo{volume}{13}
  (\bibinfo{year}{2019}) \bibinfo{pages}{210–213}. \URLprefix
  \url{http://dx.doi.org/10.1038/s41566-018-0342-x}.
  \DOIprefix\doi{10.1038/s41566-018-0342-x}.
\bibitem[{Yu et~al.(2020)Yu, Ma, Luo, Jing, Sun, Fang, Yang, Liu, Zheng, Xie,
  Zhang, You, Wang, Chen, Zhang, Bao, and Pan}]{yong2020entanglement}
\bibinfo{author}{Y.~Yu}, \bibinfo{author}{F.~Ma}, \bibinfo{author}{X.-Y. Luo},
  \bibinfo{author}{B.~Jing}, \bibinfo{author}{P.-F. Sun},
  \bibinfo{author}{R.-Z. Fang}, \bibinfo{author}{C.-W. Yang},
  \bibinfo{author}{H.~Liu}, \bibinfo{author}{M.-Y. Zheng},
  \bibinfo{author}{X.-P. Xie}, \bibinfo{author}{W.-J. Zhang},
  \bibinfo{author}{L.-X. You}, \bibinfo{author}{Z.~Wang},
  \bibinfo{author}{T.-Y. Chen}, \bibinfo{author}{Q.~Zhang},
  \bibinfo{author}{X.-H. Bao}, \bibinfo{author}{J.-W. Pan},
\newblock \bibinfo{title}{Entanglement of two quantum memories via fibres over
  dozens of kilometres},
\newblock \bibinfo{journal}{Nature} \bibinfo{volume}{578}
  (\bibinfo{year}{2020}) \bibinfo{pages}{240–245}. \URLprefix
  \url{http://dx.doi.org/10.1038/s41586-020-1976-7}.
  \DOIprefix\doi{10.1038/s41586-020-1976-7}.
\bibitem[{Bao et~al.(2012)Bao, Reingruber, Dietrich, Rui, D\"{u}ck, Strassel,
  Li, Liu, Zhao, and Pan}]{bao2012efficient}
\bibinfo{author}{X.-H. Bao}, \bibinfo{author}{A.~Reingruber},
  \bibinfo{author}{P.~Dietrich}, \bibinfo{author}{J.~Rui},
  \bibinfo{author}{A.~D\"{u}ck}, \bibinfo{author}{T.~Strassel},
  \bibinfo{author}{L.~Li}, \bibinfo{author}{N.-L. Liu},
  \bibinfo{author}{B.~Zhao}, \bibinfo{author}{J.-W. Pan},
\newblock \bibinfo{title}{Efficient and long-lived quantum memory with cold
  atoms inside a ring cavity},
\newblock \bibinfo{journal}{Nature Physics} \bibinfo{volume}{8}
  (\bibinfo{year}{2012}) \bibinfo{pages}{517--521}. \URLprefix
  \url{https://doi.org/10.1038/nphys2324}. \DOIprefix\doi{10.1038/nphys2324}.
\bibitem[{Cho et~al.(2016)Cho, Campbell, Everett, Bernu, Higginbottom, Cao,
  Geng, Robins, Lam, and Buchler}]{cho2016coherent}
\bibinfo{author}{Y.-W. Cho}, \bibinfo{author}{G.~T. Campbell},
  \bibinfo{author}{J.~L. Everett}, \bibinfo{author}{J.~Bernu},
  \bibinfo{author}{D.~B. Higginbottom}, \bibinfo{author}{M.~T. Cao},
  \bibinfo{author}{J.~Geng}, \bibinfo{author}{N.~P. Robins},
  \bibinfo{author}{P.~K. Lam}, \bibinfo{author}{B.~C. Buchler},
\newblock \bibinfo{title}{Highly efficient optical quantum memory with long
  coherence time in cold atoms},
\newblock \bibinfo{journal}{Optica} \bibinfo{volume}{3} (\bibinfo{year}{2016})
  \bibinfo{pages}{100--107}. \URLprefix
  \url{https://opg.optica.org/optica/abstract.cfm?URI=optica-3-1-100}.
  \DOIprefix\doi{10.1364/OPTICA.3.000100}.
\bibitem[{Buob et~al.(2024)Buob, H{\"o}schele, Makhalov, Rubio-Abadal, and
  Tarruell}]{buob2024strontium}
\bibinfo{author}{S.~Buob}, \bibinfo{author}{J.~H{\"o}schele},
  \bibinfo{author}{V.~Makhalov}, \bibinfo{author}{A.~Rubio-Abadal},
  \bibinfo{author}{L.~Tarruell},
\newblock \bibinfo{title}{A strontium quantum-gas microscope},
\newblock \bibinfo{journal}{PRX Quantum} \bibinfo{volume}{5}
  (\bibinfo{year}{2024}) \bibinfo{pages}{020316}.
\bibitem[{Holten et~al.(2022)Holten, Bayha, Subramanian, Brandstetter, Heintze,
  Lunt, Preiss, and Jochim}]{holten2022observation}
\bibinfo{author}{M.~Holten}, \bibinfo{author}{L.~Bayha},
  \bibinfo{author}{K.~Subramanian}, \bibinfo{author}{S.~Brandstetter},
  \bibinfo{author}{C.~Heintze}, \bibinfo{author}{P.~Lunt},
  \bibinfo{author}{P.~M. Preiss}, \bibinfo{author}{S.~Jochim},
\newblock \bibinfo{title}{{Observation of Cooper pairs in a mesoscopic
  two-dimensional Fermi gas}},
\newblock \bibinfo{journal}{Nature} \bibinfo{volume}{606}
  (\bibinfo{year}{2022}) \bibinfo{pages}{287--291}.
\bibitem[{Fedorov et~al.(2021)Fedorov, Remizov, Shapiro, Pogosov, Egorova,
  Tsitsilin, Andronik, Dobronosova, Rodionov, Astafiev, and Ustinov}]{Ustinov}
\bibinfo{author}{G.~P. Fedorov}, \bibinfo{author}{S.~V. Remizov},
  \bibinfo{author}{D.~S. Shapiro}, \bibinfo{author}{W.~V. Pogosov},
  \bibinfo{author}{E.~Egorova}, \bibinfo{author}{I.~Tsitsilin},
  \bibinfo{author}{M.~Andronik}, \bibinfo{author}{A.~A. Dobronosova},
  \bibinfo{author}{I.~A. Rodionov}, \bibinfo{author}{O.~V. Astafiev},
  \bibinfo{author}{A.~V. Ustinov},
\newblock \bibinfo{title}{Photon transport in a bose-hubbard chain of
  superconducting artificial atoms},
\newblock \bibinfo{journal}{Physical Review Letters} \bibinfo{volume}{126}
  (\bibinfo{year}{2021}) \bibinfo{pages}{180503}.
  \DOIprefix\doi{10.1103/PhysRevLett.126.180503}.
\bibitem[{Blain et~al.(2023)Blain, Marchegiani, Polo, Catelani, and
  Amico}]{solitons}
\bibinfo{author}{B.~Blain}, \bibinfo{author}{G.~Marchegiani},
  \bibinfo{author}{J.~Polo}, \bibinfo{author}{G.~Catelani},
  \bibinfo{author}{L.~Amico},
\newblock \bibinfo{title}{Soliton versus single-photon quantum dynamics in
  arrays of superconducting qubits},
\newblock \bibinfo{journal}{Physical Review Research} \bibinfo{volume}{5}
  (\bibinfo{year}{2023}) \bibinfo{pages}{033130}.
  \DOIprefix\doi{10.1103/PhysRevResearch.5.033130}.
\bibitem[{Mansikkam\"aki et~al.(2022)Mansikkam\"aki, Laine, Piltonen, and
  Silveri}]{stacks}
\bibinfo{author}{O.~Mansikkam\"aki}, \bibinfo{author}{S.~Laine},
  \bibinfo{author}{A.~Piltonen}, \bibinfo{author}{M.~Silveri},
\newblock \bibinfo{title}{{Beyond Hard-Core Bosons in Transmon Arrays}},
\newblock \bibinfo{journal}{PRX Quantum} \bibinfo{volume}{3}
  (\bibinfo{year}{2022}) \bibinfo{pages}{040314}.
  \DOIprefix\doi{10.1103/PRXQuantum.3.040314}.
\bibitem[{Carusotto and Ciuti(2013)}]{carusotto2013quantum}
\bibinfo{author}{I.~Carusotto}, \bibinfo{author}{C.~Ciuti},
\newblock \bibinfo{title}{Quantum fluids of light},
\newblock \bibinfo{journal}{Review of Modern Physics} \bibinfo{volume}{85}
  (\bibinfo{year}{2013}) \bibinfo{pages}{299--366}. \URLprefix
  \url{https://link.aps.org/doi/10.1103/RevModPhys.85.299}.
  \DOIprefix\doi{10.1103/RevModPhys.85.299}.
\bibitem[{Bloch et~al.(2022)Bloch, Carusotto, and
  Wouters}]{bloch2022nonequilibrium}
\bibinfo{author}{J.~Bloch}, \bibinfo{author}{I.~Carusotto},
  \bibinfo{author}{M.~Wouters},
\newblock \bibinfo{title}{{Non-equilibrium Bose--Einstein condensation in
  photonic systems}},
\newblock \bibinfo{journal}{Nature Reviews Physics} \bibinfo{volume}{4}
  (\bibinfo{year}{2022}) \bibinfo{pages}{470--488}. \URLprefix
  \url{https://doi.org/10.1038/s42254-022-00464-0}.
  \DOIprefix\doi{10.1038/s42254-022-00464-0}.
\bibitem[{Sanvitto et~al.(2010)Sanvitto, Marchetti, Szyma{\'n}ska, Tosi,
  Baudisch, Laussy, Krizhanovskii, Skolnick, Marrucci, Lemaitre
  et~al.}]{sanvitto2010persistent}
\bibinfo{author}{D.~Sanvitto}, \bibinfo{author}{F.~M. Marchetti},
  \bibinfo{author}{M.~Szyma{\'n}ska}, \bibinfo{author}{G.~Tosi},
  \bibinfo{author}{M.~Baudisch}, \bibinfo{author}{F.~P. Laussy},
  \bibinfo{author}{D.~Krizhanovskii}, \bibinfo{author}{M.~Skolnick},
  \bibinfo{author}{L.~Marrucci}, \bibinfo{author}{A.~Lemaitre}, et~al.,
\newblock \bibinfo{title}{Persistent currents and quantized vortices in a
  polariton superfluid},
\newblock \bibinfo{journal}{Nature Physics} \bibinfo{volume}{6}
  (\bibinfo{year}{2010}) \bibinfo{pages}{527--533}.
\bibitem[{Wouters and Carusotto(2010)}]{wouters2010superfluidity}
\bibinfo{author}{M.~Wouters}, \bibinfo{author}{I.~Carusotto},
\newblock \bibinfo{title}{{Superfluidity and Critical Velocities in
  Nonequilibrium Bose-Einstein Condensates}},
\newblock \bibinfo{journal}{Physical Review Letters} \bibinfo{volume}{105}
  (\bibinfo{year}{2010}) \bibinfo{pages}{020602}. \URLprefix
  \url{https://link.aps.org/doi/10.1103/PhysRevLett.105.020602}.
  \DOIprefix\doi{10.1103/PhysRevLett.105.020602}.
\bibitem[{Li et~al.(2015)Li, Fraser, Yakimenko, and
  Ostrovskaya}]{li2015stability}
\bibinfo{author}{G.~Li}, \bibinfo{author}{M.~D. Fraser},
  \bibinfo{author}{A.~Yakimenko}, \bibinfo{author}{E.~A. Ostrovskaya},
\newblock \bibinfo{title}{Stability of persistent currents in open dissipative
  quantum fluids},
\newblock \bibinfo{journal}{Physical Review B} \bibinfo{volume}{91}
  (\bibinfo{year}{2015}) \bibinfo{pages}{184518}. \URLprefix
  \url{https://link.aps.org/doi/10.1103/PhysRevB.91.184518}.
  \DOIprefix\doi{10.1103/PhysRevB.91.184518}.
\bibitem[{Gallem\'{\i} et~al.(2018)Gallem\'{\i}, Guilleumas, Richard, and
  Minguzzi}]{gallemi2018interaction}
\bibinfo{author}{A.~Gallem\'{\i}}, \bibinfo{author}{M.~Guilleumas},
  \bibinfo{author}{M.~Richard}, \bibinfo{author}{A.~Minguzzi},
\newblock \bibinfo{title}{Interaction-enhanced flow of a polariton superfluid
  current in a ring},
\newblock \bibinfo{journal}{Physical Review B} \bibinfo{volume}{98}
  (\bibinfo{year}{2018}) \bibinfo{pages}{104502}. \URLprefix
  \url{https://link.aps.org/doi/10.1103/PhysRevB.98.104502}.
  \DOIprefix\doi{10.1103/PhysRevB.98.104502}.
\bibitem[{Lukoshkin et~al.(2018)Lukoshkin, Kalevich, Afanasiev, Kavokin,
  Hatzopoulos, Savvidis, Sedov, and Kavokin}]{lukoshkin2018persistent}
\bibinfo{author}{V.~A. Lukoshkin}, \bibinfo{author}{V.~K. Kalevich},
  \bibinfo{author}{M.~M. Afanasiev}, \bibinfo{author}{K.~V. Kavokin},
  \bibinfo{author}{Z.~Hatzopoulos}, \bibinfo{author}{P.~G. Savvidis},
  \bibinfo{author}{E.~S. Sedov}, \bibinfo{author}{A.~V. Kavokin},
\newblock \bibinfo{title}{Persistent circular currents of exciton-polaritons in
  cylindrical pillar microcavities},
\newblock \bibinfo{journal}{Physical Review B} \bibinfo{volume}{97}
  (\bibinfo{year}{2018}) \bibinfo{pages}{195149}. \URLprefix
  \url{https://link.aps.org/doi/10.1103/PhysRevB.97.195149}.
  \DOIprefix\doi{10.1103/PhysRevB.97.195149}.
\bibitem[{Barrat et~al.(2024)Barrat, Cherbunin, Sedov, Aladinskaia, Liubomirov,
  Litvyak, Petrov, Zhou, Hatzopoulos, Kavokin, and
  Savvidis}]{barrat2024stochastic}
\bibinfo{author}{J.~Barrat}, \bibinfo{author}{R.~Cherbunin},
  \bibinfo{author}{E.~Sedov}, \bibinfo{author}{E.~Aladinskaia},
  \bibinfo{author}{A.~Liubomirov}, \bibinfo{author}{V.~Litvyak},
  \bibinfo{author}{M.~Petrov}, \bibinfo{author}{X.~Zhou},
  \bibinfo{author}{Z.~Hatzopoulos}, \bibinfo{author}{A.~Kavokin},
  \bibinfo{author}{P.~G. Savvidis},
\newblock \bibinfo{title}{Stochastic circular persistent currents of exciton
  polaritons},
\newblock \bibinfo{journal}{Scientific Reports} \bibinfo{volume}{14}
  (\bibinfo{year}{2024}) \bibinfo{pages}{12953}. \URLprefix
  \url{https://doi.org/10.1038/s41598-024-63725-1}.
  \DOIprefix\doi{10.1038/s41598-024-63725-1}.
\bibitem[{Chirolli et~al.(2024)Chirolli, Polo, Catelani, and
  Amico}]{chirolli2024synthetic}
\bibinfo{author}{L.~Chirolli}, \bibinfo{author}{J.~Polo},
  \bibinfo{author}{G.~Catelani}, \bibinfo{author}{L.~Amico},
\newblock \bibinfo{title}{Synthetic fractional flux quanta in a ring of
  superconducting qubits},
\newblock \bibinfo{journal}{arXiv preprint arXiv:2409.06511}
  (\bibinfo{year}{2024}).

\end{thebibliography}

\end{document}